\documentclass{aa}  
\newcommand{\HII}{H\,{\scriptsize II}}
\newcommand{\hii}{H\,{\scriptsize II}}
\newcommand{\HI}{H\,{\scriptsize I}}

\newcommand{\av}{A$_{\rm V}$}  
\usepackage[version=3]{mhchem}
\usepackage{graphicx}
\usepackage{subcaption}
\usepackage{txfonts}  
\begin{document}  

\title{Understanding star formation in molecular clouds}  
\subtitle{IV. Column density PDFs from quiescent to massive molecular clouds}   
  
\author{N. Schneider    \inst{1}  % C   
\and V. Ossenkopf-Okada  \inst{1} 
\and S. Clarke          \inst{1,2}
\and R.S. Klessen       \inst{3}  % Institut f¨ur Theoretische Astrophysik, Zentrum f¨ur Astronomie, Universit¨at Heidelberg, Albert-Ueberle-Str. 2, D-69120 Heidelberg, Germany
\and S. Kabanovic       \inst{1}  
\and T. Veltchev        \inst{4}  % Faculty of Physics, University of Sofia, 5 James Bourchier Blvd., 1164 Sofia, Bulgaria
\and S. Bontemps        \inst{5}  % C Laboratoire d’Astrophysique de Bordeaux, Univ. Bordeaux, CNRS, B18N, allée G. Saint-Hilaire, 33615 Pessac, France
\and S. Dib             \inst{5,6}  % (C) 
\and T. Csengeri        \inst{5}  % (C) 
\and C. Federrath       \inst{7}  % Research School of Astronomy and Astrophysics, Australian National University, Canberra, ACT 2611, Australia
\and J. Di Francesco     \inst{8,9}% C Department of Physics and Astronomy, University of Victoria, Victoria, BC, V8P 5C2, Canada
                                    % NRC Herzberg Astronomy and Astrophysics, 5071 West Saanich Road, Victoria, BC, V9E 2E7, Canada
\and F. Motte           \inst{10} % C Université Grenoble Alpes, CNRS, Institut de Planétologie et d’Astrophysique de Grenoble, 38000 Grenoble, France
%  alphabetically    
\and Ph. Andr\'e        \inst{11}  % C Laboratoire AIM, CEA/DSM-CNRS-Université Paris Diderot, IRFU/Service d’Astrophysique, CEA Saclay, 91191 Gif-sur-Yvette, France
\and D. Arzoumanian     \inst{12}  % C Instituto de Astrofísica e Ciências do Espaço, Universidade do Porto, CAUP, Rua das Estrelas, PT4150-762 Porto, Portugal
\and J.R. Beattie       \inst{7}
\and L. Bonne           \inst{5,13}  % (C) 
\and P. Didelon         \inst{11} 
\and D. Elia            \inst{14} % C INAF - IAPS, via Fosso del Cavaliere, 100, I-00133 Roma, Italy
\and V. K\"onyves       \inst{15} % C University of Central Lancashire, Preston, Lancashire PR1 2HE, UK
\and A. Kritsuk         \inst{16} %   Physics Dep. and CASS, University of California, San Diego, La Jolla, CA 92093-0424, USA
\and B. Ladjelate       \inst{17} % C Instituto de Radioastronomía Milimétrica (IRAM), Avda. Divina Pastora 7, Local 20, 18012 Granada, Spain
\and Ph. Myers          \inst{18} %   Center for Astrophysics | Harvard and Smithsonian (CfA), Cambridge, MA 02138, USA 
\and S. Pezzuto         \inst{14} % C 
\and J.F. Robitaille    \inst{10}     
\and A. Roy             \inst{5}  % C 
\and D. Seifried        \inst{1}
\and R. Simon           \inst{1} 
\and J. Soler           \inst{6} % Max Planck Institute for Astronomy, K¨onigstuhl 17, 69117, Heidelberg, Germany
\and D. Ward-Thompson   \inst{15} % C 
  }     
\institute{I. Physikalisches Institut, Universit\"at zu K\"oln, Z\"ulpicher Str. 77, 50937 K\"oln, Germany  %1
\email{nschneid@ph1.uni-koeln.de}
\and Academia Sinica, Institute of Astronomy and Astrophysics, Taipei, Taiwan %2
\and Institut f\"ur Theoretische Astrophysik, Zentrum f\"ur Astronomie, Universit\"at Heidelberg, Albert-Ueberle-Str. 2, 69120 Heidelberg, Germany %3
\and Faculty of Physics, University of Sofia, 5 James Bourchier Blvd., 1164 Sofia, Bulgaria  %4
\and Laboratoire d’Astrophysique de Bordeaux, Univ. Bordeaux, CNRS, B18N, all\'ee G. Saint-Hilaire, 33615 Pessac, France  %5
\and Max Planck Institute for Astronomy, K\"onigstuhl 17, 69117, Heidelberg, Germany %6
\and Research School of Astronomy and Astrophysics, Australian National University, Canberra, ACT 2611, Australia % 7
\and Department of Physics and Astronomy, University of Victoria, Victoria, BC, V8P 5C2, Canada  %8
\and NRC Herzberg Astronomy and Astrophysics, 5071 West Saanich Road, Victoria, BC, V9E 2E7, Canada  %9
\and Universit\'e Grenoble Alpes, CNRS, Institut de Plan\'etologie et d’Astrophysique de Grenoble, 38000 Grenoble, France %10
\and Laboratoire AIM, CEA/DSM-CNRS-Universit\'e Paris Diderot, IRFU/SAP, CEA Saclay, 91191 Gif-sur-Yvette, France  %11
\and Division of Science, National Astronomical Observatory of Japan, 2-21-1 Osawa, Mitaka, Tokyo 181-8588, Japan %12
\and SOFIA Science Center, NASA Ames Research Center, Moffett Field, CA 94 045, USA %13
\and INAF - IAPS, via Fosso del Cavaliere, 100, I-00133 Roma, Italy %14
\and University of Central Lancashire, Preston, Lancashire PR1 2HE, UK %15
\and Physics Dep. and CASS, University of California, San Diego, La Jolla, CA 92093-0424, USA %16
\and IRAM, Avda. Divina Pastora 7, Local 20, 18012 Granada, Spain %17
\and Center for Astrophysics, Harvard and Smithsonian, Cambridge, MA 02138, USA %18
}  
  
%\offprints{}  
 
\date{draft of \today}
 
\titlerunning{Understanding star formation - IV. PDFs}  
\authorrunning{N. Schneider}  
  
%\date{Received September 15, 1996; accepted March 16, 1997}  
  
 \abstract {Probability distribution functions of the total hydrogen
   column density (N-PDFs) are a valuable tool for distinguishing between 
   the various processes (turbulence, gravity, radiative feedback, magnetic
   fields) governing the morphological and dynamical structure of the
   interstellar medium.  We present N-PDFs of 29 Galactic
   regions obtained from {\sl Herschel} imaging at high angular
   resolution (18$''$), covering diffuse and quiescent clouds, and those
   showing low-, intermediate-, and high-mass star formation (SF), and
   characterize the cloud structure using the $\Delta$-variance tool.
   The N-PDFs show a large variety of morphologies. They are all 
   double-log-normal at low column densities, and display one or two 
   power law tails (PLTs) at higher column densities. For diffuse, 
   quiescent, and low-mass SF clouds, we propose that the two 
   log-normals arise from the atomic and molecular phase, respectively. 
   For massive clouds, we suggest that the first log-normal is built up 
   by turbulently mixed H$_2$ and the second one by compressed (via stellar 
   feedback) molecular gas. Nearly all clouds have two PLTs with slopes 
   consistent with self-gravity, where the second one can be flatter or steeper 
   than the first one. A flatter PLT could be caused by stellar
   feedback or other physical processes that slow down collapse and
   reduce the flow of mass toward higher densities. 
   The steeper slope could arise if the magnetic field is oriented perpendicular to the
   LOS column density distribution. The first deviation point (DP),
   where the N-PDF turns from log-normal into a PLT, shows a clustering
   around values of a visual extinction of \av\,(DP1)$\sim$2-5. The
   second DP, which defines the break between the two PLTs, varies
   strongly.  In contrast, the width of the N-PDFs is the most stable
   parameter, with values of $\sigma$ between $\sim$0.5 and 0.6.  Using
   the $\Delta$-variance tool, we observe that the \av\ value, where
   the slope changes between the first and second PLT, increases with
   the characteristic size scale in the $\Delta$-variance spectrum.
   We conclude that at low column densities, atomic and molecular gas
   is turbulently mixed, while at high column densities, the gas is
   fully molecular and dominated by self-gravity. The best
     fitting model N-PDFs of molecular clouds is thus one with
     log-normal low column density distributions, followed by one or
     two PLTs. }

  
\maketitle  
  
%________________________________________________________________  

\section{Introduction} \label{intro}  
  
% PDFs and numerical simulations 
Important tools for characterizing molecular clouds are probability
distribution functions of density ($\rho$-PDF) and column density
(N-PDF) because they can be directly linked to theories of the
star formation process \citep[e.g.,][]{padoan1997,padoan2002,vaz2001,hennebelle2008,hennebelle2009,fed2012,burkhart2018}.
Numerical simulations that include or exclude particular physical
processes (such as solenoidally or compressively driven turbulence,
radiative feedback, gravity, and magnetic fields) show that the shape
of the N-PDF strongly depends on the dominant process and the
  evolutionary state of the cloud.  For example, the N-PDF is purely
log-normal if the cloud structure is governed only by isothermal
supersonic turbulence and develops a power law tail (PLT) under
self-gravity
\citep[e.g.,][]{klessen2000,vaz2001,dib2005,kritsuk2011,collins2012,girichidis2014,ward2014,burkhart2015a, veltchev2016,mocz2017,auddy2018,veltchev2019,koertgen2019,krumholz2020,jaupart2020,donkov2021}.
The slope of the PLT changes during the evolution of the cloud and can
depend on the 2D projection
\citep{ball2011,cho2011,fed2013,burkhart2018}.  In addition,
  \citet{schneider2015c} report the detection of two PLTs in massive
  star-forming regions where the second PLT in the high-column-density
  regime, characterizing small spatial scales (sub-parsec to a few parsec), is
  flatter than the first one. They argue that this is caused by a
  physical process that slows down collapse and reduces the flow of
  mass toward higher densities. Possible processes are rotation of
  collapsing cores, which introduces an angular momentum barrier \citep{khullar2021},
  increasing optical depth and weaker cooling, magnetic fields,
  geometrical effects, and protostellar feedback. Though such a
  flatter PLT was first found in a simulation presented by
  \citet{kritsuk2011}, it is only recently that there are theoretical
  explanations for this phenomenon \citep{jaupart2020,donkov2021}.
  \citet{jaupart2020} develop an analytical theory of the density PDF
  and attribute the second PLT to free-fall collapse of a dense region
  in a cloud. \citet{donkov2021} propose that the thermodynamic state
  of the gas changes from isothermal on large scales to polytropic
  with an exponent larger than 1 on the sub-parsec proto-stellar core
  scale. In the hydrodynamics models of \citet{khullar2021}, the second PLT 
  appears only at much higher densities and small (sub-parsec) scales, and 
  corresponds to rotationally supported material, for example a 
  disc. 

% b-parameter, width of PDF
Numerical simulations of supersonic, isothermal turbulence have
demonstrated that the variance of logarithmic density fluctuations,
expressed by the width of the density PDF, $\sigma_{\rho}$, in a
compressible, turbulent medium correlates with the RMS sonic Mach
number, $\mathcal{M}$, and the type of forcing of the turbulence. The
forcing can be parameterized by the so-called forcing parameter $b$,
which encodes the relative amount of stirring versus compression in
the turbulence, with $\sigma_{\rho}^2$=$\ln$(1\,+\,$b^2
\mathcal{M}^2$) \citep{fed2008}. This variance - Mach relation also
holds for column densities seen in isothermal simulations
\citep{burkhart2012} and hydrodynamic models without self-gravity
\citep{beattie2019}.  \citet{molina2012} extended this expression by
including the ratio between thermal and magnetic energies, expressed
as $\beta_{mag}$, and obtained $\sigma_{\rho}^2$=$\ln$(1\,+\,($b^2
\mathcal{M}^2\,\beta_{mag}/(1+\beta_{mag})))$.

% Interpretation of the PDF shape, PLT
In recent years, observations using extinction maps
\citep[e.g.,][]{lombardi2008,kai2009,froebrich2010,spilker2021} or {\sl Herschel}
column density maps
\citep[e.g.,][]{schneider2012,schneider2013,schneider2015a,russeil2013,catarina2014,tremblin2014,stutz2015,benedettini2015}
started to test the theoretical predictions. The interpretation of the
observed N-PDF shapes, however, varies strongly.  For example, while
\citet{butler2014} propose that N-PDFs of extinction maps of infrared
dark clouds (IRDCs) are best fitted by log-normal distributions,  
\citet{schneider2015b} find a pure power law distribution
for the same clouds. Moreover, \citet{brunt2015}, studying low-mass
clouds, advocate that the PLT is a part of a log-normal N-PDF arising
from the cold, molecular part of the cloud.  Gravity as the dominant
process behind forming PLTs in star-forming regions is suggested by
the observational studies of \citet{froebrich2010} and
\citet{schneider2013,schneider2015a}. In contrast, \citet{kai2011}
propose that pressure due to different phases in the interstellar
medium gives rise to the PLT. \citet{tremblin2014}, on the other hand,
argue that the N-PDF of clouds closely associated with \HII\ regions
can show a more complex shape with several bumps and PLTs due to
radiative feedback effects that cause compression of local gas into
shells and pillars. More recently, Planck polarization observations at
353 GHz have been used to identify that the relative orientations
between the column density structure and the magnetic field
orientation are also related to the PLTs \citep{soler2019}.

% Low column density part
It is not only the nature of the high column density part of the N-PDF
that is strongly debated, but also that of the low column density
range. While the observational studies mentioned above mostly find a
log-normal distribution for star-forming and non-star-forming clouds
for low column densities, \citet{alves2017} claim that there is no
observational evidence for log-normal N-PDFs of molecular clouds but
that they are well described by power laws.  Various authors
\citep{schneider2015a,ossenkopf2016,chen2018,koertgen2019}, however,
discuss the impact of observational limitations such as noise,
line-of-sight (LOS) effects, and incompleteness on the N-PDF but
show that there are efficient methods to correct for noise and
contamination. They conclude that a log-normal and PLT part of the
N-PDF is the best-fitting model for star-forming clouds.

% Papers I-III
These rather different views raise the need for a statistical approach
to understand N-PDFs, covering diffuse and quiescent regions to high-mass
regions.  We thus started a series of papers, of which the first one
(Paper I, Schneider, Ossenkopf, Csengeri et al. 2015a) investigates 
how line-of-sight contamination affects N-PDFs. The second one (Paper
II, Schneider, Klessen, Csengeri et al. 2015b) studies N-PDFs of
massive IRDCs and shows by using complementary molecular line data
that the power law distribution of the N-PDF can be explained by local
and global infall of gas. And finally, the third study, Paper III
(Schneider, Bontemps, Motte et al. 2016), discusses the problems 
of N-PDFs constructed from molecular line observations.

% Objectives of this paper
The objective of this paper is to present N-PDFs for a significant
number of molecular clouds with varying SF activity, using
dust column density maps derived from {\sl Herschel} imaging
only. Though there are methods that combine data from {\sl Herschel},
extinction maps and Planck data
\citep[e.g.,][]{lombardi2014,butler2014,zari2016,abreu2017,pokhrel2020},
we prefer to employ only {\sl Herschel} maps at 18$''$ angular
  resolution, in particular because we do not study the extended
cloud environment. Such analyses would involve Planck and extinction maps, 
and we do not want to introduce systematic effects by using several data 
sets that require a cross-calibration and could introduce a bias. 
The high angular resolution of our maps enables us to better resolve the 
high column density part of the N-PDF that is constituted by molecular clumps 
and cores on a parsec and sub-parsec scale.   
We study the  variation of the N-PDF shapes for diffuse, quiescent, low-, intermediate-,
and high-mass SF regions.  We also establish a well-defined data set
of molecular cloud parameters that can be used for further studies
such as linking the density structure with the dynamics of the gas, 
the SF rate and efficiency, the magnetic field and the
core mass function. Our main goals are: \\
\noindent Quantifying the average column density, total mass, and
LOS confusion for Galactic molecular clouds;\\
\noindent Providing the characteristics of the N-PDF such as PLT slope(s),
widths of the log-normal part(s), the first deviation point (DP1) from
the log-normal to PLT distribution and the second deviation point (DP2) from the first PLT
to the second for this set of molecular clouds; \\
\noindent Investigating how cloud morphology  (for instance filamentary vs. 
spherical) and stellar feedback (such as expanding \HII\ region bubbles) influences the N-PDF shape; \\ 
\noindent Calculating the $\Delta$-variance spectrum \citep{ossk2008a} to characterize the
structural variation in the column density map;\\
\noindent Assessing if there are (column) density
thresholds that signify a change in the dominant physical process or chemistry, such as the
transition from turbulence to gravity or the transition from atomic to molecular hydrogen.

The current paper is organized the following way:
Section~\ref{sec:obs} briefly describes how we derived the {\sl
  Herschel} column density maps (Sect.~\ref{sec:herschel}), chose the
sample of molecular clouds (Sect.~\ref{sec:regions}), estimated LOS
confusion (Sect.~\ref{sec:los}), and determined the N-PDFs and the
$\Delta$-variance (Sect.~\ref{sec:stat}). Section~\ref{sec:results}
presents the column density maps and the resulting cloud parameters
(density, mass, etc.), the N-PDFs, and the $\Delta$-variance of the
observed clouds. Section~\ref{sec:discuss} assesses the value of
N-PDFs as an analysis tool and describes what they tell us about the
column density structure of molecular
clouds. Section~\ref{sec:summary} summarizes the main findings of this
paper.

%__________________________________________________________________  

\section{Observations and data analysis} \label{sec:obs}  
 
%********************************************************* Table 1   *******************************************
%
%***************************************************************************************************************

\begin{table*}  
  \caption{Overview of the molecular cloud sample, ordered by cloud type and name. } \label{table:summary1}   
  \begin{center}  
\begin{tabular}{lcc|c|c|l}  
\hline \hline   
%                &                        &                                        &           &          &   \\ 
Cloud            & $\alpha_{J2000}$       &  $\delta_{J2000}$                      & Distance  & Geometry & References  \\  
                 & [$^{h}$:$^{m}$:$^{s}$] & [$^{\circ}$:$^{\prime}$:$^{\prime\prime}$] &  [kpc]    &           & \\
\hline
\multicolumn{3}{l}{ {\bf High-mass SF regions}}  &  & \\   
\hline
Cygnus North     & 20:37:54 &  41:44:57 &  1.40      &  ridge+filaments    & \citet{hennemann2012,schneider2016} \\ 
Cygnus South     & 20:35:08 &  39:41:50 &  1.40      &  filaments+pillars  & \citet{schneider2016} \\ 
M16              & 18:19:40 & -13:47:34 &  2.00      &  filaments+pillars   & \citet{hill2012,tremblin2013,tremblin2014}\\ 
M17              & 18:18:36 & -16:34:39 &  2.20      &  clumps             & this paper \\     
NGC~6334         & 17:20:58 & -35:51:45 &  1.35      &  massive ridge      & \citet{russeil2013,tige2017} \\  
NGC~6357         & 17:25:06 & -34:25:57 &  1.75      &  dispersed clumps   & \citet{russeil2019}  \\ 
NGC~7538         & 23:14:02 &  61:26:48 &  2.80      &  evolved bubble     & \citet{fallscheer2013} \\ 
Rosette          & 06:33:32 &  04:15:23 &  1.46$^a$  &  \HII\ bubble/      & \citet{motte2010,hennemann2010} \\ % NEW
                 &          &           &            &  ridge              & \citet{schneider2010,james2010}  \\  
                 &          &           &            &                     & \citet{schneider2012,cambresy2013}  \\  
                 &          &           &            &                     & \citet{tremblin2013,tremblin2014}  \\  
Vela~C           & 09:00:37 & -43:56:50 &  0.70      &  ridge          &  \citet{hill2011,giannini2012} \\ %NEW
                 &          &           &            &                 &  \citet{minier2013,tremblin2014}  \\  
\hline   
\multicolumn{3}{l}{ {\bf Intermediate-mass SF regions}} &  &  \\   
\hline
Aquila           & 18:29:43 & -02:46:49 &  0.436$^b$ & bipolar filament& \citet{koenyves2010,koenyves2015,andre2010}  \\ %NEW     
                 &          &           &            &                 & \citet{bontemps2010a,bontemps2010b,schneider2013}  \\  
Mon~R2           & 06:06:45 & -06:17:01 &  0.862     &  hub-filament   & \citet{didelon2015,pokhrel2016}\\ %NEW  
                 &          &           &            &                 & \citet{rayner2017}\\ %NEW  
Mon~OB1          & 06:32:00 &  10:30:00 &  0.80      &  massive clump  & this paper \\ %NEW  
NGC~2264         & 06:40:24 &  09:25:52 &  0.719$^c$  & ridge+filaments & \citet{nony2021} \\ %NEW  
Orion~B          & 05:48:54 &  00:48:08 &  0.40      &  filaments      &  \citet{schneider2013,koenyves2020} \\ %NEW
Serpens          & 18:34:59 &  00:00:00 &  0.436$^b$ & filaments+clumps  &  \citet{roca2015,fiorellino2021}  \\  
\hline   
\multicolumn{3}{l}{ {\bf Low-mass SF regions}} & & \\  
\hline
Cham~I           & 10:55:42 & -77:07:31 &  0.192$^d$ & ridge            &  \citet{catarina2014} \\  
Cham~II          & 12:38:45 & -78:29:35 &  0.198$^d$ & clumps           &  \citet{catarina2014} \\  
IC5146           & 21:48:48 &  47:29:16 &  0.813$^b$ & filament         &  \citet{doris2011,roy2015} \\ %NEW  
Lupus~I          & 15:41:03 & -34:05:34 &  0.182     & filaments+clumps  & \citet{rygl2013,benedettini2015,benedettini2018} \\
Lupus~III        & 16:10:03 & -39:04:40 &  0.162$^c$ & filaments+clumps  &  \citet{rygl2013,benedettini2015,benedettini2018} \\ 
Lupus~VI         & 16:04:36 & -42:04:24 &  0.204     & filament+clumps & \citet{rygl2013,benedettini2015,benedettini2018} \\  
Perseus          & 03:35:41 & 31:31:56  &  0.235$^b$ & filament+clumps  &  \citet{sadavoy2012,sadavoy2014} \\  
                 &          &           &            &                  &  \citet{pezzuto2012,pezzuto2021} \\  
Pipe             & 17:23:08 & -26:24:07 &  0.145     & filament+clumps  &  \citet{peretto2012,roy2014,roy2015} \\ %NEW   
$\rho$Oph        & 16:27:31 & -24:12:40 &  0.140$^b$ & clumps           &  \citet{roy2014,lad2020} \\ %NEW   
Taurus           & 04:21:00 &  27:46:45 &  0.14      & filaments        &  \citet{kirk2013,marsh2014,marsh2016a}\\
                 &          &           &            &                  &  \citet{pedro2013} \\
\hline  
\multicolumn{3}{l}{ {\bf Quiescent regions }} &   \\  
\hline
Cham~III         & 12:38:45 &-78:29:35  &  0.16      & clumps           & \citet{catarina2014}\\  
Musca            & 12:27:36 & -71:38:53 &  0.15      & filament         & \citet{cox2016,bonne2020} \\  %NEW  
Polaris          & 01:50:35 & 88:21:10  &  0.489$^a$ & network filaments& \citet{sascha2010,schneider2013} \\  
                 &          &           &            &                  & \citet{miville2010}  \\  
                 &          &           &            &                  & \citet{derek2010,andre2010}  \\  
\hline  
\multicolumn{3}{l}{ {\bf Diffuse/atomic regions }} &   \\  
\hline  
Draco            & 16:47:57 & 61:45:16  &  0.60      &  clumps          & \citet{miville2017}, this paper \\  
\hline  
\end{tabular}  
\end{center}  
\vskip0.1cm \tablefoot{The coordinates specify the center of the {\sl Herschel} maps. The distances are the ones used in the {\sl
    Herschel} papers that are listed in the last column. Updated distances by recent GAIA or maser parallax publications are
  available for a some sources. (a) Yan et al. 2019. (b) Ortiz et al. 2018.  We note that a carefully selected sample of Gaia eDR3 stars yield a distance of 502 pc \citep{comeron2022}. (c) Dzib et al. 2018. (d) Apellaniz et al. 2019. \citet{zucker2020} present a large overview on distances.}
\end{table*}   

%********************************************************* Table 2  *******************************************
%
%***************************************************************************************************************

\begin{table*}  
  \caption{Global molecular cloud parameters from {\sl Herschel} column density and temperature maps, ordered by cloud type and name. } \label{table:summary2}   
\begin{center}  
\begin{tabular}{l|c|c|c|c|c|c}  
\hline \hline   
%Cloud        & LOS d & <N(H$_2$)>                 & Area(\av$>$1)   &  r [pc] & <n(H$_2$)>    &  M(\av$>$1)          
              &       &                            &                 &     &                     &                   \\ 
Cloud         & LOS   & $\langle N({\rm H_2})\rangle$    & A$_1$           &  $R$  & $\langle n \rangle$ &   $M_1$    \\           
              & dust  &                            & {\tiny \av$>$1} &     & {\tiny \av$>$1} &  {\tiny \av$>$1}       \\  
              & [mag] & [10$^{21}$cm$^{-2}]$        & [pc$^2$]        & [pc] & [cm$^{-3}$]     &   [10$^3$ M$_{\odot}$]  \\  
              &  (1)  & (2)                        & (3)             & (4)  & (5)            & (6)                    \\   
\hline       
\multicolumn{3}{l}{ {\bf High-mass SF regions}}  &   \\   
\hline
Cygnus North  & 5.0   &    5.06                     &   3417          & 33.0   & 24.9        &  324.46               \\ 
Cygnus South  & 5.0   &    5.47                     &   3731          & 34.5   & 25.7        &  383.45               \\ 
M16           & 7.8   &    4.18                     &   1057          & 18.3   & 36.9        &   87.12               \\ 
M17           & 6.6   &    9.99                     &   2478          & 28.1   & 57.7        &  483.76               \\    
NGC 6334      & 8.7   &   13.98                     &    882          & 16.8   & 135.1       &  238.60               \\  
NGC 6357      & 4.2   &    7.22                     &   1004          & 17.9   & 65.4        &  145.14               \\ 
NGC 7538      & 3.3   &    4.22                     &   1524          & 22.0   & 31.0        &  130.39               \\ 
Rosette       & 1.1   &    4.10                     &    881          & 16.8   & 39.7        &   70.64               \\ 
Vela C        &  2.0  &    6.85                     &    450          & 12.0   & 92.7        &    58.00              \\  
\hline   
\multicolumn{3}{l}{ {\bf Intermediate-mass SF regions}} &  \\   
\hline
Aquila        &  2.5  &    2.74                     &    445          & 11.9   & 37.3        &   25.94              \\      
Mon R2        &  1.6  &    2.00                     &     82          &  5.1   & 63.3        &    4.60              \\ 
Mon OB1       &  2.2  &    2.31                     &     59          &  4.4   & 86.1        &    4.87              \\   
NGC 2264      &  1.6  &    2.60                     &    169          &  7.3   & 57.5        &   11.12              \\  
Orion B       &  0.9  &    2.06                     &    683          & 14.8   & 22.7        &    29.69             \\     
Serpens       & 1.6   &    1.46                     &    266          & 10.8   & 21.9        &    14.15             \\   
\hline   
\multicolumn{3}{l}{ {\bf Low-mass SF regions}} &  \\  
\hline
Cham I        & 0.25  &    1.07                     &     18          &  2.4  & 72.5         &    0.767              \\  
Cham II       & 0.23  &    1.00                     &     12          &  2.0  & 82.2         &    0.507              \\   
IC5146        & 0.37  &    1.07                     &     121         &  6.2  & 28.0         &    5.03               \\   
Lupus I       & 0.17  &    0.82                     &     6.6         &  1.4  & 91.7         &    0.364              \\     
Lupus III     & 0.44  &    1.16                     &     5.0         &  1.3  & 148.9        &    0.180              \\      
Lupus VI      & 0.46  &    1.00                     &     7.3         &  1.5  & 106.0        &    0.398              \\    
Perseus       & 0.66  &    1.62                     &    174          &  7.4  & 35.3         &    4.350              \\  
Pipe          & 0.59  &    1.72                     &     40          &  3.6  & 77.8         &    1.452              \\   
$\rho$Oph     & 1.25  &    1.57                     &     26          &  2.9  & 89.0         &    1.244              \\   
Taurus        & 0.4   &    1.73                     &     97          &  5.5  & 50.7         &    4.03               \\   
\hline  
\multicolumn{3}{l}{ {\bf Quiescent regions }} &  \\  
\hline
Cham III      & 0.16  &    0.81                     &     9.4         &  1.7  & 75.4         &    0.317              \\  
Musca         & 0.37  &    0.75                     &     3.2         &  1.0  & 121.2        &    0.110              \\      
Polaris       & 0.31  &    1.73                     &    11.3         &  1.9  & 31.9         &    1.208              \\        
\hline   
\multicolumn{3}{l}{ {\bf Diffuse/atomic regions }} & \\  
\hline   
Draco$^a$     &       &    0.60                     &   1500          & 21.9  &  4.4        &    8.04               \\  
\hline   
\end{tabular}  
\end{center}  
\vskip0.1cm  
\tablefoot{(1) Line-of-sight contamination in visual extinction determined from polygons in the {\sl Herschel} dust column density map. \\ 
\noindent (2) Average H$_2$ column density of the cloud above a visual extinction of \av$>$1. \\ 
\noindent (3) Cloud area in square-parsec above a visual extinction of \av$>$1. \\
\noindent (4) Equivalent radius $R$ in parsec from area with $R=($area$/\pi)^{0.5}$. \\
\noindent (5) Average density $n$ in cm$^{-3}$ above a visual extinction of \av$>$1. The density is calculated by $n({\rm H_2})= N({\rm H_2})/(2\,R)$ with the equivalent radius $R$ of the cloud. \\
\noindent (6) Mass (M $\propto$ $N({\rm H_2}) \,\pi \, R^2$ with N(H$_2$)=0.94$\times$10$^{21}$ \av\, cm$^{-2}$ mag$^{-1}$)   
of the complex determined above \av=1.  \\
\noindent $^a$ Values for area, mass, density, etc. are given for a visual extinction of \av$>$0.} 
\end{table*}   

\subsection{Column density maps from Herschel}\label{sec:herschel}  

% Data from Herschel
For this study, we use the cloud sample from {\sl Herschel} key
programs, the {\sl Herschel} Gould Belt survey (HGBS,
\citet{andre2010}) and the {\sl Herschel} imaging survey of OB Young
Stellar objects (HOBYS, \citet{motte2010}), as well as data from open
time programs such as the {\sl Herschel} Infrared GALactic plane
survey (Hi-GAL, \citet{molinari2010}) and individual PI programs.
Most of the column density maps\footnote{See
  http://gouldbelt-herschel.cea.fr/archives for HGBS data.} used in
this paper were already published (see references in
Table~\ref{table:summary1} for {\sl Herschel} imaging observations for
each region), either at an angular resolution of 18$''$ or 36$''$.

%SED fit 
All column density maps were determined from a pixel-to-pixel graybody
fit to the red wavelength of PACS \citep{poglitsch2010} at 160 $\mu$m
(13.5$''$ angular resolution) and all SPIRE \citep{griffin2010}
wavelengths (250 $\mu$m, 350 $\mu$m, 500 $\mu$m at 18.2$''$, 24.9$''$,
and 36.3$''$ resolution, respectively).  For the SPIRE data reduction,
we used the HIPE pipeline (versions 10 to 13), including the destriper
task for SPIRE, and HIPE and scanamorphos \citep{roussel2013} for
PACS.  The SPIRE maps were calibrated for extended emission.  All maps
have an absolute flux calibration, either by using offset values
determined as described in \citet{bernard2010} for the sources of the
Gould Belt and HOBYS program, or using the {\sc zeroPointCorrection}
task in HIPE for SPIRE and IRAS maps for PACS for the remaining
clouds. For the SED fit, we fixed the specific dust opacity per unit
mass (dust+gas) approximated by the power law
$\kappa_\nu$=0.1$\,(\nu/1000 {\rm GHz})^{\beta_d}$ cm$^{2}$/g with
$\beta_d$=2, and left the dust temperature and column density as free
parameters \citep[see][for details]{hill2011,russeil2013,roy2013}.
The procedure underpinning how high angular resolution maps were
obtained is described in detail in Appendix A of
\citet{pedro2013}. The concept is to employ a multiscale
decomposition of the flux maps and assume a constant LOS 
temperature. The final map at 18$''$ resolution is constructed from
the difference maps of the convolved SPIRE maps (at 500 $\mu$m, 350
$\mu$m, and 250 $\mu$m) and the temperature information from the color
temperature derived from the 160 $\mu$m to 250 $\mu$m ratio.

% dust 
The Draco region has very weak emission so that we used
the classical fitting technique (SED fit to 160 $\mu$m to 500 $\mu$m)
to determine column density maps at 36$''$ resolution to obtain the
best signal-to-noise ratios. In addition, data points at each
wavelength were weighted with a calibration uncertainty of 10\% and
20\% for SPIRE and PACS, respectively.
For the star-forming and quiescent molecular clouds, we used a value
of 3.4$\times$10$^{-25}$~cm$^{-2}$/H for the coefficent $\kappa_\nu$, 
which is in the range of typical values \citep{ossk1994} 
from 1.75$\times$10$^{-25}$~cm$^{-2}$/H for compact grains in diffuse
interstellar clouds to 5.0$\times$10$^{-25}$~cm$^{-2}$/H for fluffy
grains with ice mantles in dense molecular cores.  For Draco, 
we expect rather diffuse cloud conditions without much ice
accretion or dust coagulation.  Based on Planck observations,
\citet{juvela2015} derived
$\kappa_\nu$=2.16$\times$10$^{-25}$~cm$^{-2}$/H for such regions,
following the standard interstellar reddening behavior, so we use
this value for Draco.  For the diffuse cloud Draco, which is mostly atomic, 
we calculated the total hydrogen column density using $N$=\av $\times$1.87$\times$10$^{21}$
cm$^{-2}$ mag$^{-1}$ \citep{bohlin1978}.  For all other clouds, which 
are mostly molecular, we transformed H$_2$ column density into visual
extinction, using the conversion formula $N({\rm H_2})$=\av
$\times$0.94$\times$10$^{21}$ cm$^{-2}$ mag$^{-1}$.

% Uncertainty 
The uncertainty in the {\sl Herschel} column density maps arise from
the uncertainty in the assumed form of the opacity law, including
variations of dust content and dust properties across the clouds and
possible temperature gradients along the LOS. The total
uncertainty is estimated to be around $\sim$30--50\% \citep[see above
  and, e.g.,][for a discussion]{russeil2013}. By comparing an
extinction map and the {\sl Herschel} column density map of Rosette,
\citet{cambresy2013} argued that the optical depth from dust emission
close to heating sources like massive clusters might be overestimated.
Their extinction map, however, suffers from saturation at values above
\av $\sim$20 (only 2MASS) and $\sim$35 (2MASS combined with other
near-IR or mid-IR data), respectively. This limitation makes the study
of very dense regions such as the centers of high-mass SF clouds 
impossible using extinction maps. The multi-temperature column density
mapping procedure PPMAP \citep{marsh2016b} produces differential
column density maps, using {\sl Herschel} flux maps, in a number of
temperature intervals. The PPMAP method, however, includes the 70
$\mu$m data in the SED fit in addition to the 160-500 $\mu$m
wavelength data, but the 70 $\mu$m is mostly tracing  hot dust from
cloud surfaces and not the cool bulk of the atomic and molecular gas in which 
we are interested.

% Error 
Apart from the overall uncertainty of the column density maps, there
is observational noise in the maps, arising from the SPIRE and PACS
instrumental noise. We estimate the noise level in the final column
density maps, using the full N-PDF for regions that are hardly
affected by LOS-contamination and that are sufficiently extended. As
was shown in \citet{ossenkopf2016}, noise produces excess in the low
column density part of the N-PDF and increases the width of the
log-normal part.  When the noise amplitude is less than 40\% of the
peak column density, the excess in the N-PDF at low column densities
is linear. As we see later (Sec.~\ref{pdf}), the N-PDFs with the
highest dynamical range at low column densities indeed show this
linear tail. These N-PDFs go down to values below an \av\ of 0.1
(e.g., Chamaeleon~I-III, Lupus~I, Musca, Polaris, Draco). We perform a
fit including an error tail, a log-normal part, and possbile PLTs,
following \citet{ossenkopf2016} and described in the next section, and
derive as extreme values an error level of \av\ of $\sim$0.02 for
Draco and $\sim$0.1 for Polaris.  Because all sources were observed in
the same way (scanning speed, instrumental setup, etc.), we assume to
first order that all maps, including those of the star-forming clouds,
have a similar low noise level. \\ If there is too much LOS-confusion
or the maps are not extended enough to cover areas without cloud
emission, the noise cannot be estimated this way.  We thus take the
maximum noise level for Polaris (\av=0.1) as a standard value for all
maps of star-forming clouds and conclude that the observational noise
is low enough to resolve a major fraction of the low column density part with \av $<$1 
of the N-PDF, at least for the low-mass and quiescent clouds.  
It should be noted that noise also shifts the peak of the N-PDF 
toward higher column densities \citep{ossenkopf2016}.

\subsection{The molecular cloud sample }\label{sec:regions}  
  
A total of 29 cloud complexes were selected for our study, and their
coordinates and distances are listed in Table~\ref{table:summary1},
together with references for {\sl Herschel} publications. For the
distances, we use values from the literature and update with
  recent results from GAIA. A large overview on distance estimates
  based on a combination of stellar photometric data with GaiaDR2
  parallax measurement is given in \citet{zucker2020}. However, they
  give multiple distance estimates across a single cloud with
  sometimes large differences, so that we prefer to keep the typical
  values from the literature. For the N-PDFs shape, the accurate
  distances are not relevant, they only play a role in the mass
  determination.  Complementary to other N-PDF studies
\citep{kai2009,froebrich2010,lombardi2008,alves2017}, we include more
distant and massive clouds that form intermediate- to high-mass stars,
and quiescent clouds with apparently no SF, and employ
only {\sl Herschel} data.

Generally, throughout the paper, we use the following nomenclature
\citep[e.g.]{bergin2007}: Low-mass regions are molecular clouds
with a mass of 10$^3$--10$^4$ M$_{\odot}$, and a size of up to a few
tens of parsecs that typically form stars of low mass (examples are Taurus or
Perseus). High-mass regions are giant molecular clouds (GMCs)
with a mass of 10$^5$--10$^6$ M$_{\odot}$, a size of up to a $\sim$100
pc, and observational signatures of high-mass SF and cluster formation 
(such as Cygnus). GMCs in addition sometimes contain regions defined as ridges 
\citep{schneider2010,hennemann2012,quang2011,quang2013,didelon2015,motte2018}
that are massive, gravitationally unstable filamentary structures of
high column density (typically N$_{\rm H_2}>$10$^{23}$ cm$^{-2}$) with
high-mass SF.  Some clouds fall in between these
categories as they have masses in the range of 10$^4$--10$^5$
M$_{\odot}$ and form mainly low- and intermediate-mass stars but also
some high-mass stars (such as Orion B). For simplicity, we classify
these as intermediate-mass regions. Quiescent clouds are
those that show very little SF activity (no or only very few
protostars or prestellar cores. Finally, diffuse clouds 
are mostly atomic.
  
\subsection{Line-of-sight contamination}\label{sec:los}  

% LOS contamination intro 
Column density maps from {\sl Herschel} can be affected by 
LOS confusion, in particular in the Galactic plane and
along spiral arms. Unrelated dust emission from LOS clouds can add to
the observed flux in the different wavelength ranges and thus the
column density determined from {\sl Herschel} can be overestimated. In
Paper I, we studied in detail the influence of such confusion on the
maps and the N-PDFs and introduced a simple correction method to
determine the typical background and foreground contribution from the maps
in regions outside the bulk emission of the target. This approach was
then further investigated and justified in
\citet{ossenkopf2016}. Summarizing, it was shown that contamination by
foreground and background emission can be safely removed as a
constant screen if the contaminating N-PDF is log-normal (see
Sec.~\ref{sec:pdfs} for the nomenclature), and its width
$\sigma_{\eta,cloud}$ is narrow, typically $<$0.5$\,\sigma_{\eta,cloud}$,
or the column density of the contaminant is sufficiently small.  We
thus applied the same method as in Paper I and measured the
contamination from a rectangular polygon placed out of the molecular
cloud close to the map borders. We used the procedure developed in
\citet{ossenkopf2016} to obtain a separate N-PDF from these pixels
within the polygon and derived the peak value and width of the
contaminating N-PDF. Appendix~\ref{app-a} (Figs. A.1. and A.2) shows
an example of this method applied to the Aquila cloud. For all maps
used here, the peak of this N-PDF corresponds within 10\% to the
values obtained from averaging the pixels in the rectangular
regions. The widths of the contaminating N-PDFs $\sigma_{\eta,cont}$
are small and vary between $\sigma_{\eta,cont}$=0.05 and
$\sigma_{\eta,cont}$=0.19 so that the ratio
$\sigma_{\eta,cont}$/$\sigma_{\eta,cloud}$ varies between 0.09 and
0.42.  The ratio between the contamination column density and cloud
peak column density $N_{cont}$/$N_{peak}$ is mostly below 1, the
smallest value is 0.41, only three maps have a value of around 3.
From Figs. \ref{etalargecont} and \ref{contcorr} in Appendix
\ref{app-a}, we see that even if the column density ratio
$N_{cont}$/$N_{peak}$ is high, the ratio of the contaminating and
cloud N-PDF widths $\sigma_{\eta,cont}$/$\sigma_{\eta,cloud}$ is
always so small that we can conclude that the LOS correction by
removing a constant screen is indeed a valid method for our 
clouds.  The values determined in this way are listed in
Table~\ref{table:summary2}. It is important to note that LOS
contamination of several log-normal N-PDFs does not create multiple
peaks but instead broadens and shifts the column density distribution.

% Values of LOS 
Massive clouds that are not too distant and more isolated, such as 
Rosette, have a low contamination level (\av$\sim$4) while other GMCs
such as M16 have a high level of \av$\sim$8. Extreme cases are IRDCs
that are far away (typically more than several kpc) because the IRDC
is an intrinsic part of the larger-scale molecular cloud.  NGC6334 is
problematic because there is no consistent concensus on the amount of
contamination as discussed in \citet{russeil2013}. From molecular line
data and the {\sl Herschel} image, we derive a value of around \av=6,
while \citet{froebrich2010} deduce a higher extinction (between \av=7
and 14). The reason for this discrepancy is probably a strong spatial
variation of the contamination due to several clouds along the LOS.  
Therefore, the derived parameters for mass and the N-PDF for
NGC6334 should be treated with care.  For the Cygnus X North and South
regions, we independently determined the average extinction of the
"Cygnus Rift", a feature lying in front of Cygnus at distances $\ll$1
kpc, to be \av$\sim$5 \citep{schneider2007}.  We thus use this value
as an approximation for the contamination.

"Over-correction" of LOS-contamination can also lead to unrealistic
features in the N-PDF for high-mass SF clouds but still provides more
reliable estimates for average column densities and masses and the
slope of the PLT than using values without any correction.
Intermediate-mass clouds such as Aquila, Vela, and MonOB1 are also 
affected by LOS-contamination. The absolute values of the
contamination are low, typically around an \av\ of 2, and the width of
the contaminating N-PDF is small.  Low-mass regions show not much LOS
confusion with values of \av$<$2.

% Limits 
All LOS-values derived from the dust maps are upper limits because the
{\sl Herschel} maps are not always extended enough so that the cloud
borders (if something like a "border" exists) are sufficiently
covered. Accordingly, the polygons may still be placed in areas of
extended cloud emission. On the other hand, one must stay close to the
cloud area because a more remote polygon would not trace the same LOS,
thereby risking to calibrate our correction methods on contaminants
that do not affect our column density maps.

\subsection{Statistical analysis tools}  \label{sec:stat} 

\subsubsection {Probability  distribution functions of column density (N-PDFs)} \label{sec:pdfs} 

Probability distribution functions (PDFs) form the basis for modern
theories of SF 
\citep{dib2007,hennebelle2008,hennebelle2009,fed2012,padoan2014,burkhart2018,burkhart2019},
and are frequently used as an analysis tool for simulations and
observations. We determine the PDFs expressed in column density $N$ or
visual extinction \av\, (we note that $N \propto {\rm A}_{\rm v}$) and call it
N-PDF, following \citet{myers2015}.  The probability of finding gas
within a range [{\av}, {\av}+d{\av}] is given by the surface-weighted
N-PDF of the extinction with $\int_{{\rm A}_{\rm v}}^{{\rm A}_{\rm v}+d{\rm A}_{\rm v}}$
$p_{{\rm A}_{\rm v}} ({\rm A}'_{\rm v})\,d{\rm A}'_{\rm v}$, where $p_{{\rm A}_{\rm v}}
({\rm A}_{\rm v})$ corresponds to the PDF of the extinction. We define
\begin{equation}  
\eta\equiv\rm ln\frac{A_{\rm v}}{\langle A_{\rm v}\rangle}, 
\end{equation}  
as the natural logarithm of the visual extinction \av\, divided by the
mean extinction $\langle {\rm A}_{\rm v}\rangle$.  The quantity
$p_\eta(\eta)$ then corresponds to the probability distribution
function of $\eta$, and by definition 
\begin{equation}
  \int_{-\infty}^{+\infty} p_\eta {\rm d \eta} = \int_0^{+\infty} p_{A_{\rm v}}\rm d A_{\rm v}=1.
\end{equation}
In Paper I, we showed that a binsize of 0.1 in $\eta$ provides the
best compromize between resolving small features in the N-PDF and
avoiding low-number pixel statistics.  We tested four methods to
characterize the N-PDF and derive its characteristic properties. 
In the following, we briefly summarize the methods, but we only use 
the values derived with method 4 for the paper. \\

All methods except method 3  fit a log-normal function at the low column
density range with
\begin{equation}  
p_\eta\,{\rm d}\eta=\frac{1}{\sqrt{2\pi\sigma_\eta^2}}{\rm exp}\Big[ -\frac{(\eta-\mu)^2}{2\sigma_\eta^2} \Big]{\rm d}\eta, 
\end{equation}  
where $\sigma_\eta$ is the dispersion and $\mu$ is the mean
logarithmic column density. For the high column density range, a
single or several PLTs are fitted.  There are, however, subtle
differences in these methods:  \\ 

\noindent {\bf Method 1} used in Paper I performs several fits on a
grid of parameters for $\eta$ and $\mu$ and then calculates the
positive and negative residuals. Then, the range of log-normality is
determined under the premise that the difference between the model and
$p_\eta$ is less than three times the statistical noise in $p_\eta$
and we derive the width and peak of the log-normal part of the
N-PDF. We then perform a linear regression fit to determine the
slope(s) $s$ of the PLT(s). The slope values that are fitted start at
the deviation point (DP) where the log-normal N-PDF turns into one or
two power law distribution(s) and stop where the power law is no
longer well defined (at high column densities) due to a low-number
pixel statistics caused by resolution effects. \\ 

\noindent {\bf Method 2} follows \citet{ossenkopf2016} and fits an error slope
at very low column densities, followed by a log-normal distribution
and a single PLT at high column densities. This method includes
numerical error weighting and map size errors. We tested a large
parameter space and obtained the most reliable results for a 10\% map
size error.  \\

%********************************************************* Table 3   *******************************************
% N-PDF parameter
%***************************************************************************************************************

\begin{table*}[!htpb]
  \caption{Parameters from the N-PDFs study from {\sl Herschel} column density  maps, ordered by cloud type and name.} \label{table:summary3}   
\begin{center}  
\begin{tabular}{l|c|c|c|c|c|c|c|c|c}  
\hline \hline   
%Cloud        & Model & Avpeak1        & Avpeak2      & sigma1              & sigma2              & DP1     & DP2     &s1           & s2
%              &               &         &           &                    &             &             &           &      &           & \\ 
\hline        
Cloud         & Model & A$_{\rm V,pk1}$ & A$_{\rm V,pk2}$ & $\sigma_{\rm \eta 1}$ & $\sigma_{\rm \eta 2}$ & DP1   & DP2     & $s_1$ & $s_2$    \\           
              &       & [mag]         & [mag]          &                    &                     & [mag]   & [mag]   &       &          \\  
              &  (1)  & (2)            & (3)           & (4)                & (5)                 & (6)     & (7)     & (8)   & (9) \\   
\hline        
\hline
\multicolumn{4}{l}{ {\bf High-mass SF regions}}  &  \\   
\hline
Cygnus N      & ELL2P & 2.63          &  2.76         & 0.52                &  1.04               &  18.5   & 79.7    &  -1.83      & -1.40 \\  
Cygnus S      & ELL2P & 1.39          &  3.46         & 0.29                &  0.52               &  15.2   & 37.5    &  -2.37      & -2.66 \\  
M16           & ELLP  & 3.17          &  3.55         & 0.33                &  0.79               &   8.1   & -       &  -2.47      & - \\  
M17           & ELL2P & 4.63          & 12.85         & 0.79                &  0.47               &  37.2   & 87.5    &  -2.67      & -1.70 \\  
NGC~6334      & ELL2P & 4.90          & 14.95         & 0.81                &  0.37               &  28.1   & 68.3    &  -2.18      & -1.55 \\  
NGC~6357      & ELL2P & 2.77          &  9.88         & 0.74                &  0.45               &  26.4   & 45.5    &  -2.57      & -3.11 \\  
NGC~7538      & ELL2P & 1.75          &  3.13         & 0.33                &  0.74               &   9.0   & 18.5    &  -1.44      & -2.00 \\  
Rosette       & ELL2P & 3.03          &  9.09         & 0.63                &  0.49               &  17.3   & 35.9    &  -1.95      & -3.82 \\ 
Vela~C        & ELL2P & 4.64          &  3.80         & 0.25                &  1.13               &  33.8   & 65.5    &  -3.68      & -4.59 \\      
\hline
\small{{\bf mean}} &  & \small{{\bf 3.2$\pm$1.2}} & \small{{\bf 7.1$\pm$4.4}} & \small{{\bf 0.52$\pm$0.21}} & \small{{\bf 0.67$\pm$0.26}} &
\small{{\bf 21.5$\pm$9.8}} & \small{{\bf 54.8$\pm$22.5}} & \small{{\bf -2.35$\pm$0.60}} & \small{{\bf -2.60$\pm$1.08}} \\
\small{{\bf median}} &  & \small{{\bf 3.0}} & \small{{\bf 3.8}} & \small{{\bf 0.52 }} & \small{{\bf 0.52}} &
\small{{\bf 18.5}} & \small{{\bf 55.5}} & \small{{\bf -2.37}} & \small{{\bf -2.33}} \\
\hline   
\multicolumn{3}{l}{ {\bf Intermediate-mass SF regions}}   &  \\   
\hline
Aquila        & ELL2P & 1.42          &  2.77          &  0.57             &  0.35                &  4.6    &  19.1    &  -2.10     & -2.38 \\      
Mon~R2        & ELL2P & 0.84          &  1.71          &  0.36             &  0.52                &  3.1    &  14.6    &  -1.79     & -1.15 \\      
Mon~OB1       & ELL2P & 0.21          &  1.75          &  0.49             &  0.90                &  4.8    &  10.9    &  -1.03     & -3.48 \\      
NGC~2264      & ELL2P & 1.64          &  1.62          &  0.47             &  1.07                &  6.2    &  16.2    &  -2.00     & -1.27 \\      
Orion~B       & ELL2P & 1.53          &  1.66          &  0.10             &  0.46                &  3.6    &  43.3    &  -1.99     & -3.64 \\      
Serpens       & ELL2P & 0.66          &  1.43          &  0.67             &  0.53                &  5.1    &  30.1    &  -1.76     & -2.00 \\      
\hline
\small{{\bf mean}} &  & \small{{\bf 1.1$\pm$0.5}} & \small{{\bf 1.8$\pm$0.4}} & \small{{\bf 0.44$\pm$0.18}} & \small{{\bf 0.64$\pm$0.26}} &
\small{{\bf 4.6$\pm$1.0}} & \small{{\bf 22.4$\pm$11.1}} & \small{{\bf -1.77$\pm$0.35}} & \small{{\bf -2.32$\pm$0.97}} \\
\small{{\bf median}} &  & \small{{\bf 1.1}} & \small{{\bf 1.68 }} & \small{{\bf 0.47}} & \small{{\bf 0.52}} &
\small{{\bf 4.7}} & \small{{\bf 17.7}} & \small{{\bf -1.88}} & \small{{\bf -2.19}} \\
\hline   
\multicolumn{3}{l}{ {\bf Low-mass SF regions}} & \\  
\hline
Cham~I        & EL2P  & 0.47          &  -             &  0.85              &  -                  &  2.35   &  6.0     &  -0.90     & -3.57 \\      
Cham~II       & ELL2P & 0.54          &  0.72          &  0.43              &  1.06               &  2.7    & 26.9     &  -2.36     & -4.27 \\      
IC5146        & ELL2P & 0.71          &  0.82          &  0.37              &  0.79               &  3.4    &  9.6     &  -1.47     & -2.76 \\      
Lupus~I       & ELL2P & 0.44          &  0.63          &  0.29              &  0.76               &  1.6    &  4.4     &  -1.39     & -2.77 \\      
Lupus~III     & ELL2P & 0.91          &  0.93          &  0.36              &  0.84               &  4.8    &  5.9     &  -3.49     & -2.41 \\      
Lupus~IV      & ELL2P & 0.72          &  1.12          &  0.24              &  0.40               &  2.1    &  9.5     &  -2.14     & -2.88 \\      
Perseus       & L2P   & 1.04          &  -             &  0.35              &   -                 &  1.4    &  3.9     &  -1.55     & -2.03 \\      
Pipe          & ELL2P & 0.99          &  1.92          &  0.29              &  0.34               &  5.2    &  13.4    &  -2.91     & -1.72 \\      
$\rho$Oph     & LL2P  & 0.32          &  1.04          &  0.26              &  0.48               &  2.3    &  16.2    &  -1.27     & -2.37 \\      
Taurus        & LL2P  & 0.64          &  1.51          &  0.28              &  0.51               &  2.7    &  18.8    &  -2.27     & -4.40 \\      
\hline
\small{{\bf mean}} &  & \small{{\bf 0.7$\pm$0.2}} & \small{{\bf 1.1$\pm$0.4}} & \small{{\bf 0.37$\pm$0.17}} & \small{{\bf 0.65$\pm$0.24}} &
\small{{\bf 2.9$\pm$1.2}} & \small{{\bf 11.5$\pm$7.0}} & \small{{\bf -2.00$\pm$0.77}} & \small{{\bf -2.92$\pm$0.85}} \\
\small{{\bf median}} &  & \small{{\bf 0.7}} & \small{{\bf 0.99}} & \small{{\bf 0.32}} & \small{{\bf 0.64}} &
\small{{\bf 2.5}} & \small{{\bf 9.5}} & \small{{\bf -1.85}} & \small{{\bf -2.76}} \\
\hline  
\multicolumn{3}{l}{ {\bf Quiescent regions }} &   \\  
\hline
Cham~III      & ELL2P & 0.53          &  0.64          &  0.43              &  0.80               &  3.5    & 12.1     &  -3.83     & -2.02 \\      
Musca         & LL2P  & 0.45          &  0.70          &  0.23              &  0.20               &  0.9    &  4.1     &  -1.73     & -5.05 \\      
Polaris       & ELL2P & 0.45          &  0.62          &  0.38              &  0.61               &  2.35   &  3.4     &  -4.08     & -2.34 \\      
\hline
\small{{\bf mean}} &  & \small{{\bf 0.48$\pm$0.04}} & \small{{\bf 0.65$\pm$0.03}} & \small{{\bf 0.35$\pm$0.08}} & \small{{\bf 0.54$\pm$0.25}} &
\small{{\bf 2.3$\pm$1.1}} & \small{{\bf 6.6$\pm$4.0}} & \small{{\bf -3.21$\pm$1.05}} & \small{{\bf -3.14$\pm$1.36}} \\
\small{{\bf median}} &  & \small{{\bf 0.45}} & \small{{\bf 0.64}} & \small{{\bf 0.38}} & \small{{\bf 0.61}} &
\small{{\bf 2.35}} & \small{{\bf 4.1}} & \small{{\bf -3.83}} & \small{{\bf -2.34}} \\
\hline   
%\hline   
\multicolumn{3}{l}{ {\bf Diffuse/atomic regions }} &   \\  
\hline
Draco         & LL    & 0.13          & 0.40           & 0.32               &  0.34               &  -      &    -     &  -         &  - \\  
\end{tabular}  
\end{center}  
\vskip0.1cm  
\tablefoot{\noindent (1) Best fitting model. E=low column density error slope, L=log-normal, P=power law tail \\ 
\noindent (2,3) Peaks of log-normal parts of the N-PDF in \av.  \\ 
\noindent (4,5) Widths of log-normal parts of the N-PDF in units of $\eta$. \\
\noindent (6) Deviation point in \av\ where the N-PDF changes from log-normal into a PLT.  \\ 
\noindent (7) Deviation point in \av\ for the change in slope from a first to a second PLT.  \\ 
\noindent (8,9) Slopes of PLTs. }
\end{table*}   

%********************************************************* Table 4   *******************************************
%  DELTA-Variance parameters
%***************************************************************************************************************

\begin{table}[!htpb]
  \caption{Parameters from the $\Delta$-variance study from {\sl Herschel} column density  maps,
      ordered by cloud type and name.} \label{table:summary4}   
\begin{center}  
\begin{tabular}{l|c|c|c|c}  
\hline \hline   
%Cloud         &  beta1     & P1   & beta2     & P2 
\hline        
Cloud         &  $\beta_1$ & P1   & $\beta_2$ & P2   \\           
              &            & [pc] &           & [pc] \\  
              &  (1)       & (2)  & (3)       & (4)  \\   
\hline        
\hline
\multicolumn{3}{l}{ {\bf High-mass SF regions}}&  \\   
\hline
Cygnus N      &   2.51     & 0.59  &    -     & - \\ 
Cygnus S      &   2.17     & 1.42  &    -     & - \\ 
M16           &   2.17     & 2.62  &    -     & - \\ 
M17           &   2.22     & 2.57  &    -     & - \\     
NGC 6334      &   2.41     & 1.10  &    -     & - \\  
NGC 6357      &   2.02     & 2.53  &    -     & - \\ 
NGC 7538      &   2.93     & 2.10  &    -     & - \\ 
Rosette       &   2.42     & 4.76  &    -     & - \\ 
Vela C        &   2.30     & 1.83  &    -     & - \\
\hline
\tiny{{\bf mean}}   & \tiny{{\bf 2.35$\pm$0.25}} & \tiny{{\bf 2.17$\pm$1.13}} &  -   &\\ 
\tiny{{\bf median}} & \tiny{{\bf 2.30}}          & \tiny{{\bf 2.10}}          & -    &  - \\ 
\hline   
\multicolumn{3}{l}{ {\bf Intermediate-mass SF regions}}   &  \\   
\hline
Aquila        &   2.28     & 0.32  & 2.62     & 3.90 \\      
Mon R2        &   2.20     & 0.37  & 3.04     & 1.37 \\   
Mon OB1       &   3.38     & 0.76  & 2.50     & 4.36 \\   
NGC 2264      &   2.80     & 0.98  &  -       & - \\   
Orion B       &   2.18     & 1.39  & 2.36     & 4.90 \\     
Serpens       &   2.36     & 0.79  &  -       & - \\  
\hline
\tiny{{\bf mean}}   & \tiny{{\bf 2.53$\pm$0.43}} & \tiny{{\bf 0.92$\pm$0.50}} & \tiny{{\bf 2.63$\pm$0.25}} & \tiny{{\bf 3.63$\pm$2.35}} \\ 
\tiny{{\bf median}} & \tiny{{\bf 2.32}}          & \tiny{{\bf 0.79}}          & \tiny{{\bf 2.56}}          & \tiny{{\bf 4.13}} \\ 
\hline   
\multicolumn{3}{l}{ {\bf Low-mass SF regions}} & \\  
\hline
Cham I        &   2.95     & 0.21  & 2.78     & 2.61 \\  
Cham II       &   2.64     & 0.61  & 2.90     & 2.56 \\  
IC5146        &   2.50     & 0.68  &  -       & - \\   
Lupus I       &   2.77     & 0.11  & 2.45     & 0.82 \\ 
Lupus III     &   3.10     & 0.08  & 2.66     & 1.18 \\   
Lupus VI      &   2.99     & 0.79  & 2.63     & 6.53 \\  
Perseus       &   2.13     & 2.45  &  -       & - \\  
Pipe          &   3.04     & 0.25  &  -       & - \\    
$\rho$Oph     &   2.17     & 0.17  & 2.62     & 1.39 \\   
Taurus        &   3.30     & 0.30  &  -       & - \\  
\hline
\tiny{{\bf mean}}   & \tiny{{\bf 2.76$\pm$0.37}} & \tiny{{\bf 0.55$\pm$0.67}} & \tiny{{\bf 2.67$\pm$0.14}} & \tiny{{\bf 2.51$\pm$1.91}} \\ 
\tiny{{\bf median}} & \tiny{{\bf 2.86}} & \tiny{{\bf 0.28}} & \tiny{{\bf 2.64}} & \tiny{{\bf 1.96}} \\ 
\hline  
\multicolumn{3}{l}{ {\bf Quiescent regions }} &   \\  
\hline
Cham III      &   2.73    & 0.45   & -        & - \\  
Musca         &   3.42    & 0.11   & 2.55     & 0.89 \\    
Polaris       &   2.40    & 0.15   &  -       & - \\  
\hline
\tiny{{\bf mean}}   & \tiny{{\bf 2.85$\pm$0.42}} & \tiny{{\bf 0.24$\pm$0.15}} & \tiny{{\bf 2.55}} & \tiny{{\bf 0.89}} \\
\tiny{{\bf median}} & \tiny{{\bf 2.73}}  & \tiny{{\bf 0.15}} & \tiny{{\bf 2.55}}  & \tiny{{\bf 0.89}} \\ 
\hline   
\multicolumn{3}{l}{ {\bf Diffuse/atomic regions }} &   \\  
\hline
Draco         &  2.27    & 5.95    &  -       & - \\  
\end{tabular}
\tablefoot{Columns (1,3): Exponents $\beta_1$ ($\beta_2$), derived from the $\Delta$-variance calculation. Columns (2,4):
    First and second peak or turnover point in the $\Delta$-variance spectrum in parsec.}
\end{center}  
\end{table}   

\noindent {\bf Method 3}, the adapted BPlfit \citep{veltchev2019}, 
calculates the slope of a power law part of an arbitrary distribution,
without any assumption about the functional form of other parts of
this distribution and constant binning.  The slope and the DP 
are then derived simultaneously as averaged values as the number
of bins is varied.  The method was elaborated further for detection of
a second PLT \citep{marinkova2021} - this technique was also used to
get the results in this paper. \\

\noindent {\bf Method 4} fits models that are a combination of: error slopes at 
low column densities (E), log-normals (L), pairs of log-normals 
(LL), power laws (P) and double power laws (PP). In total 8 models 
are considered: ELP, ELLP, EL2P, ELL2P, LP, LLP, L2P, LL2P. The 
number of parameters for each model ranges from 4 to 11. Error slopes 
contribute 2 parameters: the $\eta$ below which the error slope is fit, 
and the error slope itself. The log-normal contributes 2 parameters: 
the log-normal mean value and its width. Pairs of log-normals contribute 
5 parameters: a mean and width for each log-normal and the ratio of 
amplitudes of the two log-normals. Each power law contributes 2 parameters: 
the $\eta$ value above which the power law is fit, and the power law slope. 
The fitting of each model is done using a Monte-Carlo Markov Chain (MCMC\footnote{Python
    library emcee, https://emcee.readthedocs.io/en/stable/}) to determine 
    the maximum likelihood parameters of each model. The MCMC
  is performed using 500 walkers, each with 100,000 steps. Visual
  inspection of the walker's paths reveal that this is sufficient to
  sample to likelihood distribution and find the maximum. 

To determine the best fitting model out of the 8 models
  considered, we use the Bayesian Information Criterion (BIC):
\begin{equation}
	\begin{aligned}
		\mathrm{BIC}(k) &=  k \ln(n) - 2\ln(\mathcal{L})	\label{eq:BIC}
	\end{aligned}
\end{equation}
where $n$ is the number of data points in the N-PDF, $k$ is number of
model parameters, and $\mathcal{L}$ is the maximum likelihood found
via the MCMC. The model with the minimum BIC is considered the best
fitting model. Further, we use the BIC-weights to illustrate the
evidence of one model over another. If the best-fitting model's
BIC-weight is greater than 10 times the weight of the next most likely
model (an evidence ratio of greater than 10), we consider it to
firmly be the best model; otherwise, we cannot exclude the second
best-fitting model entirely. The BIC and BIC-weights of each model for
each cloud considered can be found in Appendix B. 

\subsubsection {$\Delta$-variance} \label{sec:delta} 

The $\Delta$-variance \citep{stutzki1998,ossk2008a,ossk2008b} is a
method to quantify the relative amount of structural variation in a 2D
map as a function of the size scale.  It measures the amount of
structure on a given scale $L$ in a map $S$, which is a 2D scalar function 
for our column density maps, by filtering the map
with a spherically symmetric wavelet $\bigcirc_{L}$:
$$\sigma^{2}_{\Delta}(L)= \langle \left( S \otimes \bigcirc
_{L}\right) ^{2}\rangle_{x,y}, $$ where $\left\langle \hdots
\right\rangle_{x,y}$ is the ensemble average over coordinates $x$ and
$y$ in the column density map, and $\otimes$ is the convolution
operator.  The $\Delta$-variance probes the variation of the intensity
$S$ over a length $L$ (called "lag") and thus measures the amount of
structural variation on that scale. We use as a filter function the
"Mexican hat filter" with an annulus-to-core-diameter ratio of about
1.5 since it provides the best results for a clear detection of
pronounced scales \citep{ossk2008a,ossk2008b}.  Weighting the image
with the inverse noise function (1/$\sigma_{rms}$) allows us to
distinguish variable noise from real small-scale structure
\citep{bensch2001}. Our {\sl Herschel} column density maps, however,
have such a high dynamic range and very low noise level that there is
no need to include a noise map.

The $\Delta$-variance and power spectra are closely linked.  For any
2D image with a power spectrum $P(k) \propto \vert k \vert^{-\beta}$,
in which $k$ is the spatial frequency, the 2D $\Delta$-variance is
related to the lag by a power law with $\sigma^2_{\Delta} \propto
L^{\beta -2}$ for 0 $< \beta <$ 6.  Practically, we determine the
slope $\alpha$ of the $\Delta$-variance (see below) and derive thus
$\beta$=$\alpha$+2. $\beta$-values range typically between 2 and 3
where numbers at the lower end indicate more structure on smaller
scales and accordingly, high values imply more structure on large
scales.
%
%A horizontal line, i.e., slopes=0, implies that there is no structural
%variation over a certain range of scales.
%
For many regions, the $\Delta$-variance spectrum does not follow a
single power law distribution but shows typically two peaks.  We thus
always start our fit at the resolution limit, defined by the beam size
of 18$''$, until the first peak or turnover point in the spectrum to
obtain a slope value for this first part and thus $\beta_1$.  We
perform a second fit, deriving $\beta_2$, only in cases where there is
a another visible power law behavior in the $\Delta$-variance
spectrum with another peak and turnover point. For the column density maps
discussed here, the values of $\beta_1$ and $\beta_2$, as well as the
peak and turnover point in parsec, are given in Table~\ref{table:summary3}.

On the smallest scales, the $\Delta$-variance spectrum is limited by
the beam size and radiometric noise and on the largest scales, it can
be limited by the map size. The error bars shown in the lower right
panels in Appendix~\ref{app-c} are from the Poisson statistics of each
bin. The $\Delta$-variance performs much faster on rectangular maps
without empty regions. Therefore, we rotated the maps, which were
observed in the coordinate system of right ascension and declination
(J2000), and slightly cut the edges to obtain clean borders. Due to
these rotations, we display the column density maps only using offsets
from the central position (Table~\ref{table:summary1}) in arcmin.  The
calculation of the $\Delta$-variance spectrum and the fit are
performed in IDL, using widget-based routines introduced in
\citet{ossk2008a}\footnote{https://hera.ph1.uni-koeln.de/ossk/Myself/deltavariance.html}.

\begin{figure*}
\centering
\includegraphics[width=9cm, angle=0]{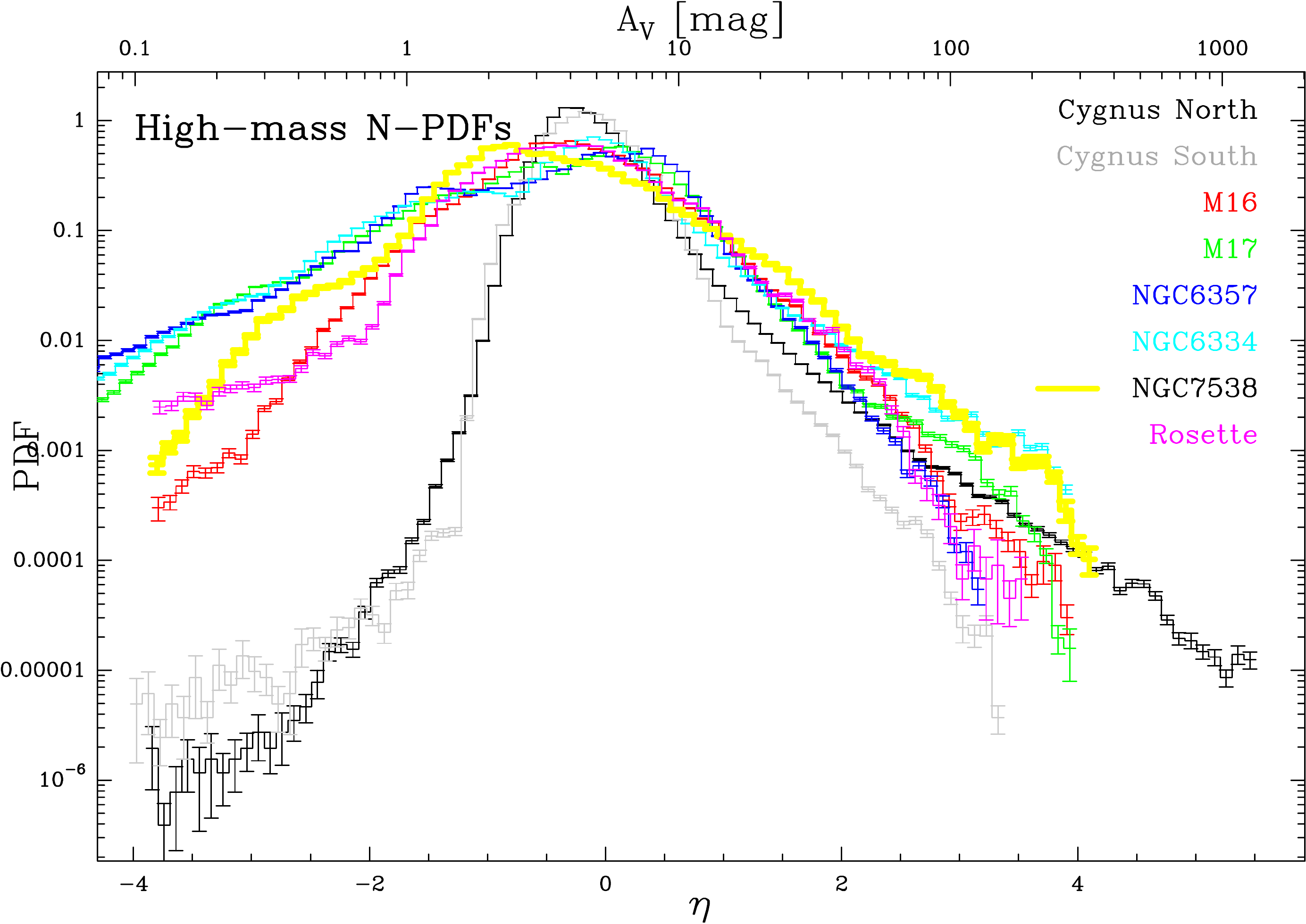}
\includegraphics[width=9cm, angle=0]{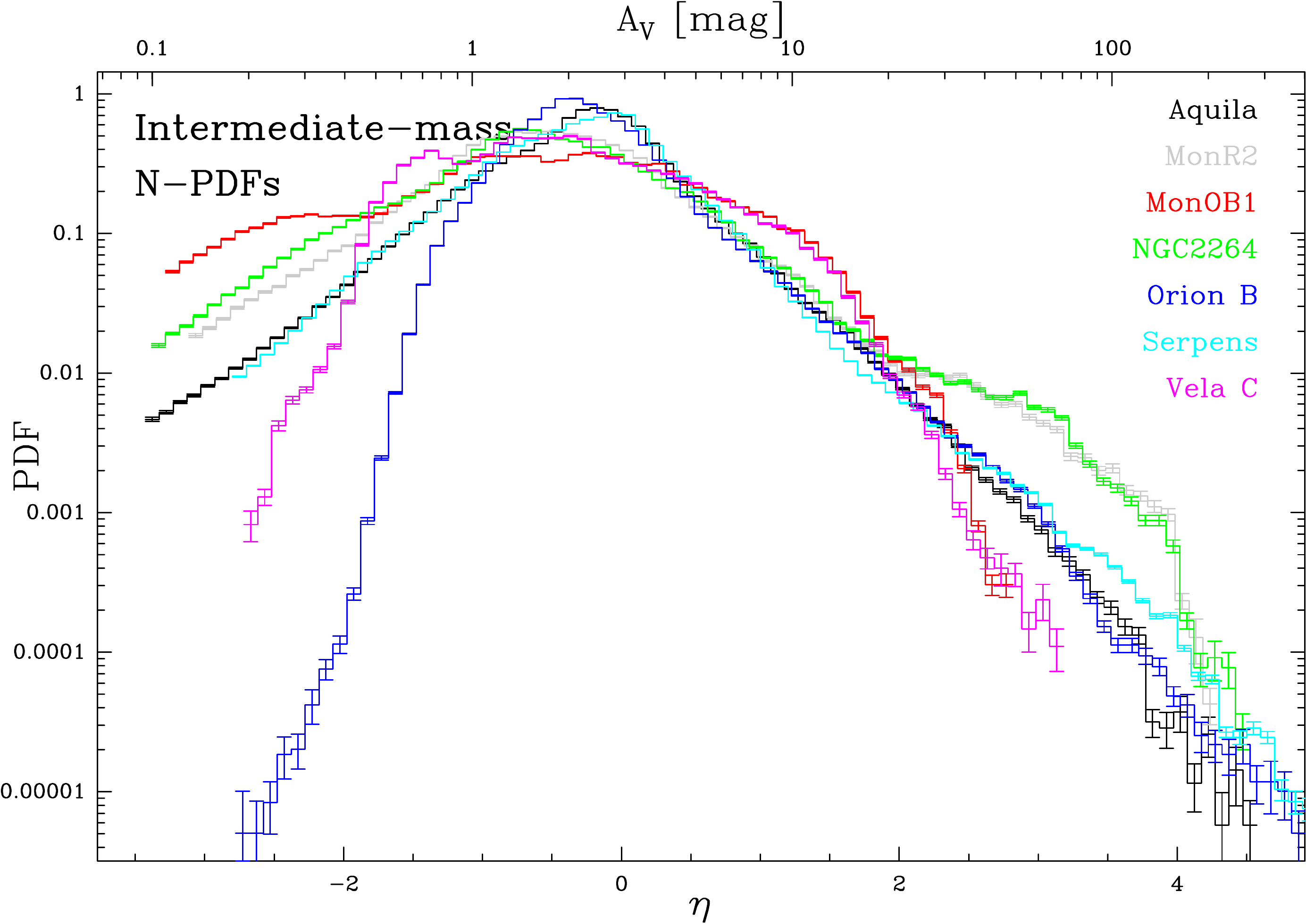}
\includegraphics[width=9cm, angle=0]{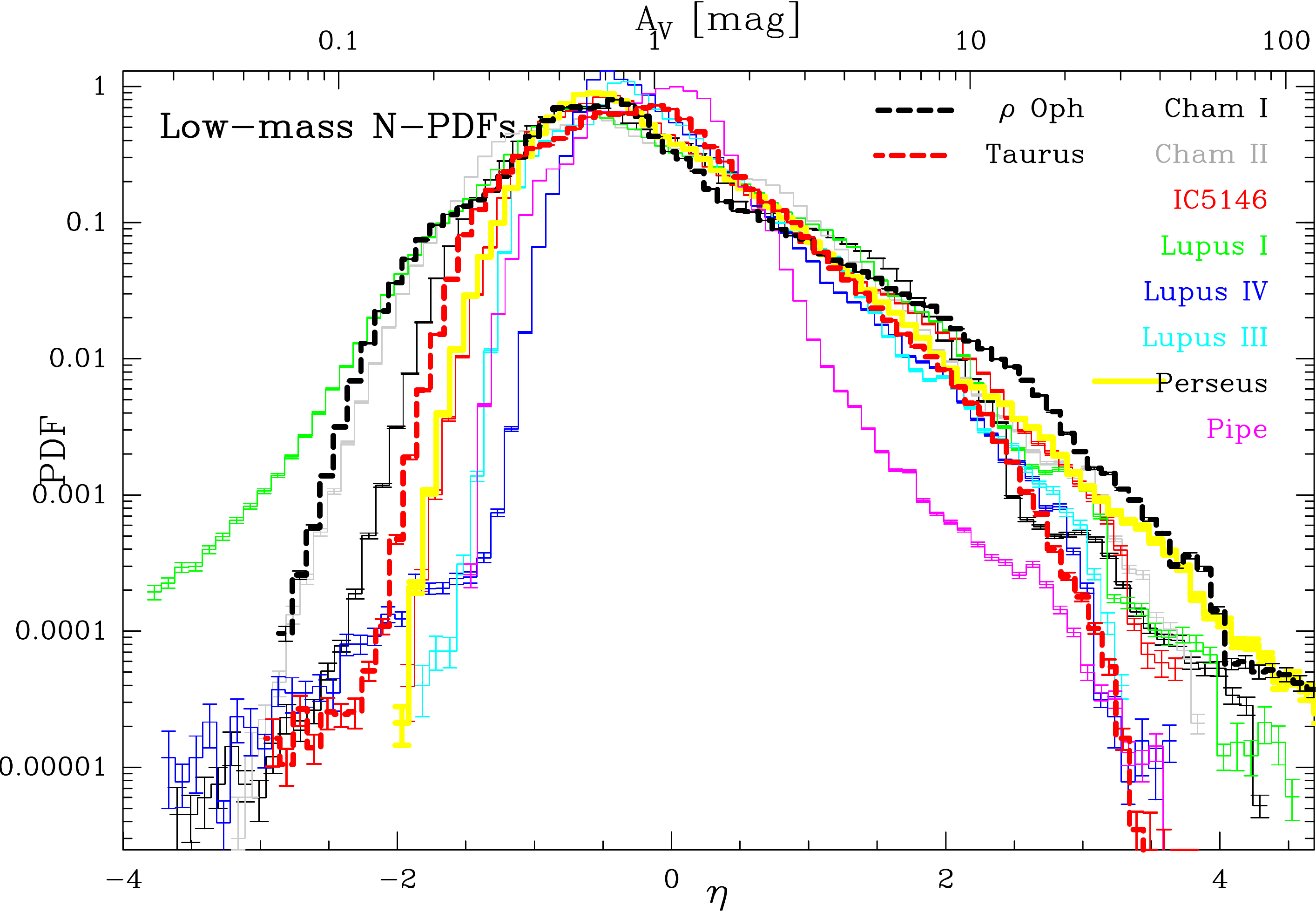}
\includegraphics[width=9cm, angle=0]{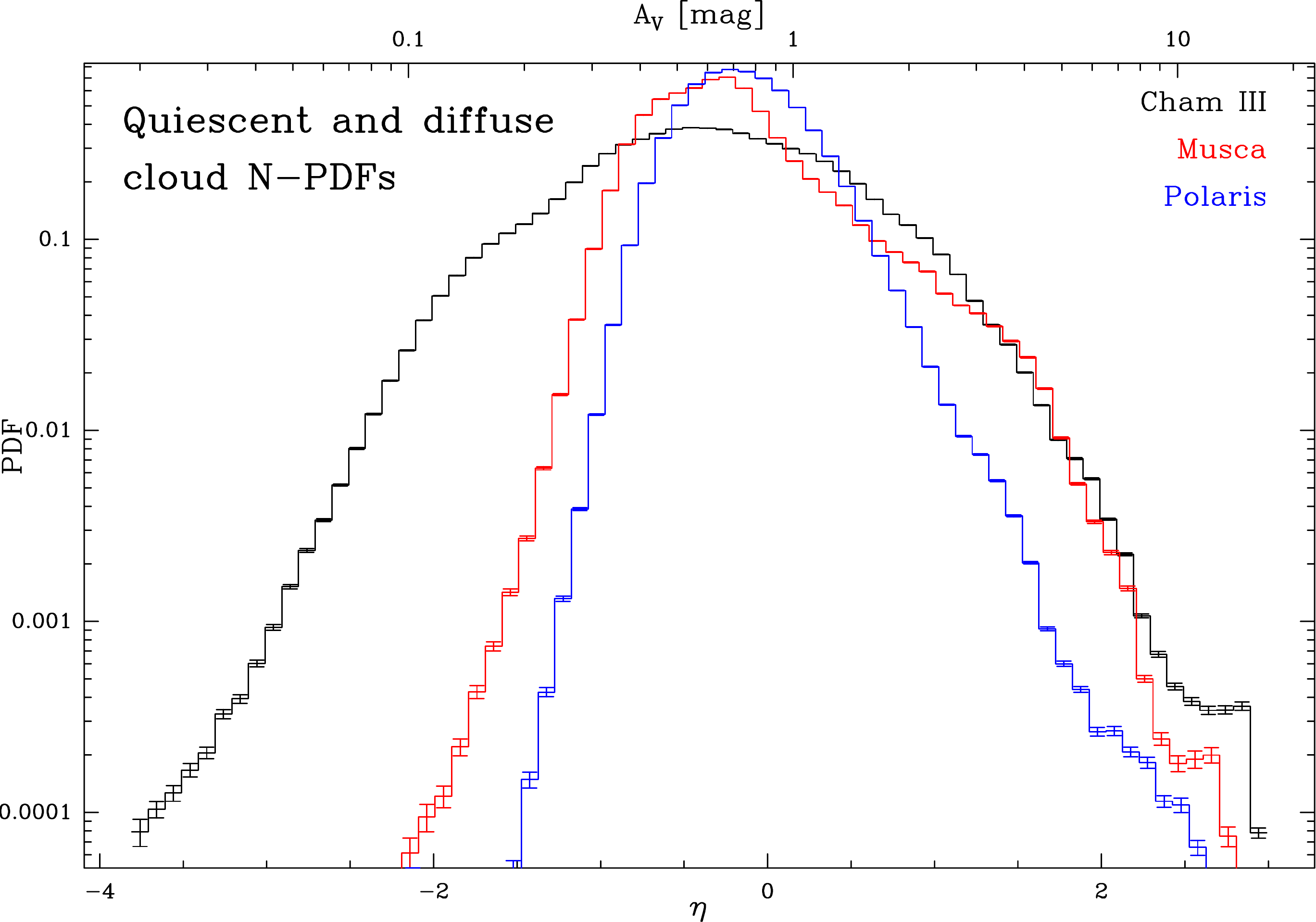}
\caption{N-PDFs of all clouds ordered by cloud type, from high-mass and intermediate mass  
SF regions (top panels) to low-mass and quiescent regions (bottom panels). Each panel shows the 
N-PDFs for clouds in different colors. The column density is expressed in visual extinction 
(upper x-axis) and in $\eta$ (lower x-axis). Error bars are calculated using Poisson statistics.  
For better visibility, we reduced the error bars by a factor of 2 (see Figs. in Appendix~\ref{app-c} 
for plots with the full error bars). We note that recently, a more sophisticated method was proposed 
by \citet{jaupart2022} to derive the statistical error bars of PDFs employing the autocovariance function. } 
\label{pdfs-all}
\end{figure*}

\section{Results and Analysis} \label{sec:results}  

\subsection{Molecular cloud parameters} \label{coldens}   

In Figs.~\ref{dr21}-\ref{draco} in Appendix~\ref{app-c}, we show the
column density maps of each cloud -- expressed in visual extinction --
together with its respective N-PDF and $\Delta$-variance
spectrum. Table \ref{table:summary2} gives cloud parameters such as
average column density $\langle N({\rm H_2}) \rangle$, as well as mass
($M$) above \av=1 and mean density $n$.  Table \ref{table:summary3}
displays the properties of the N-PDF (peak, width, DPs 
from log-normal to PLT and PLT to PLT, and the slopes of the PLTs) and
the $\beta$-values of the $\Delta$-variance.  For a better comparison
to other studies \citep{lada2010}, we use the common threshold of
\av\,=1 for the mass determination for all clouds except Draco, which 
is a diffuse region with a very low overall column density and we do
not apply any threshold.

The masses given in Table~\ref{table:summary2} justify the
classification of the clouds into the categories of high-,
intermediate-, and low-mass. The high-mass clouds cover a range
between $\sim$7$\times$10$^4$ M$_{\odot}$ (Rosette) up to
$\sim$5$\times$10$^5$ M$_\odot$ (M17) and the intermediate-mass ones
cover a range between $\sim$5$\times$10$^3$ M$_\odot$ (MonR2, MonOB1)
and $\sim$6$\times$10$^4$ M$_\odot$ (Vela C), respectively.  The
low-mass clouds comprise rather different types of cloud. For example,
the small Lupus regions have only a few hundred M$_\odot$ and show
little SF activity while Taurus and Perseus are extended (more than
100 pc$^2$) and more massive ($\sim$5$\times$10$^3$
M$_\odot$). Because only low-mass stars are forming in the latter
clouds, we classify them in the low-mass cloud category.  The diffuse
and quiescent clouds have low masses, except for Draco, which is a very
extended region (around 1500 pc$^2$). We do not compare our values of
molecular cloud parameters to the ones published elsewhere (see
Table~\ref{table:summary1}) because we applied a LOS-correction and
thus derive possibly lower values for certain clouds, and we determine
the cloud parameters above \av=1.
  
\subsection{N-PDFs of molecular clouds} \label{pdf}   
 
\subsubsection{Shapes of the N-PDFs} \label{pdfgeneral}   
 
Though N-PDFs from {\sl Herschel} studies have been already presented
in various previous publications, we show all N-PDFs from the cloud
sample in this paper to present a homogeneous data set.
We exclude all bins with low probability ($\sim$10$^{-4}$-10$^{-5}$)
at the high column density range for fitting because otherwise, the
fit would suffer from low pixel number statistics. We note that the maps
are sampled on a finer grid (typically 4$''$) while the angular
resolution is 18$''$ (36$''$ for Draco). The gridding, however, has no
significant influence on the N-PDF, as was shown in
Appendix~\ref{app-a} in Paper I, but it can lead to some bumps in the
N-PDF at high column densities. Some N-PDFs exhibit a sharp drop at
the very last high column density bins, which is a resolution effect.
For the regions where we correct for LOS contamination, we show in
Appendix~\ref{app-c} the original N-PDF (in blue) and the corrected
N-PDF (in black), of which the latter is used for determination of the
N-PDF parameters. In Appendix~\ref{app-d}, we display the N-PDF
  with the best fitting model and the residuals. The LOS-correction
leads to a pronounced tail in the low-column density range
\citep{schneider2015a,ossenkopf2016}.  For clarity of display, we cut
all other N-PDFs at the \av\,=0.1 level, which we consider to be
approximately the noise level (see Sec. \ref{sec:herschel}).

For a first overview, Fig.\ref{pdfs-all} shows all N-PDFs for each
cloud type in one figure.  The shapes of the N-PDFs are very complex
and do not reflect the perfect examples of N-PDFs often found from
simulations, typically a simple log-normal part and a PLT.  Moreover, 
we note that the classification in log-normal parts and PLTs is only 
a simple analytic expression which tries to approximate the N-PDF shape.  
In reality, N-PDFs are probably many overlapping log-normals and PLTs, 
describing different areas and physical processes with deviations from 
log-normal due to intermittency in the turbulent fields \citep{fed2010}. The maps are 
so large with sufficient resolution that the underlying complexity is 
evident in the plots and the errors so small that it is clear that the 
models are only approximate. Nevertheless, we will try to limit the possible 
models and give explanations for the shapes that are physically motivated. 
We first order the N-PDF shapes by increasing complexity:
Single/double log-normal: the diffuse region Draco;
Single log-normal and double PLT with error slope: Cham~I;
Double log-normal and double PLT without error slope: Musca, Pipe, $\rho$Oph, Taurus;  
Double log-normal and double PLT with error slope and $|s_1| > |s_2|$: Cygnus North, M17, Mon~R2, NGC~2264, NGC~6334,
  Cham~III, Lupus~III, Polaris, Pipe; 
Double log-normal and double PLT with error slope and $|s_1| < |s_2|$: Aquila, Cygnus South, Mon~OB1, NGC~6357, NGC~7538,
  Vela C, Cham~II, IC5146, Lupus~I, Lupus~IV, Orion~B, $\rho$Oph, Serpens, Taurus

The combination of two log-normal and two PLTs is the most frequent one. We confirm 
the detection of a flatter second PLT than the first one ($|s_1| > |s_2|$) for Mon~R2 and NGC~6334 
\citep{schneider2015c}, and find more examples (see above), for all cloud types.  
Furthermore, a new class of N-PDFs was detected where the second PLT is steeper than the 
first one ($|s_1| < |s_2|$), and this category contains clouds with low-, intermediate-, and high 
mass. There is thus no striking correlation between cloud type and slope(s) of the PLTs. 
In particular the second, flatter PLTS is not limited to massive clouds but also occurs in low-mass and 
quiescent clouds. 

For all N-PDFs, the best fitting model is the one that contains two log-normal distributions for the 
lower column density range.  We propose two possible explanations, depending on cloud type: in quiescent clouds 
and regions of low-mass SF, the two log-normal parts may represent the N-PDFs of atomic hydrogen (lowest column 
density range) and molecular hydrogen\footnote{We recall that all column density maps obtained from {\sl Herschel} 
dust observations contain hydrogen in atomic and molecular form.}. For massive and intermediate mass clouds, 
we suggest that both peaks arise from the fully molecular gas and that the peak or bump at higher column densities 
is caused by stellar feedback when gas is compressed by expanding \HII-region or stellar wind. We come 
back to this point in the next section.
  
\subsubsection{N-PDFs of high-mass star forming clouds} \label{high-mass} 
 
Figure~\ref{pdfs-all}, Figs.~\ref{dr21}-\ref{vela} in Appendix \ref{app-c}, and Figs.~\ref{dr21+dr15-npdf}-\ref{rosette+vela-npdf} 
in Appendix~\ref{app-d} display the LOS-contamination corrected column density maps and N-PDFs of massive clouds. For all clouds 
except of M16, the best fitting model was the one of two log-normals and two PLTs. \\

% double peak LL
\noindent {\bf Log-normal distribution(s)}\\
The peaks of the first and second log-normal are at \av=3.0 and \av=3.8 (median values), 
respectively, and the corresponding widths are $\sigma_{\rm \eta 1}$ = $\sigma_{\rm \eta 2}$ = 0.52.  
A double peak or a broadening of the N-PDF (see also Sec.~\ref{inter-mass}) is
frequently observed in regions with stellar feedback. For Rosette,
NGC6334, and M16, we confirm with our high-resolution maps what was
found using {\sl Herschel} low-resolution maps
\citep{schneider2012,russeil2013,tremblin2014}. In addition, these
types of N-PDFs were reported for W3 \citep{alana2013}, RCW36, and
RCW120 \citep{tremblin2014}.  The second, higher column density peak
is interpreted as a gas layer compressed by an expanding \hii-region
\citep{schneider2012,tremblin2014}.  As shown in hydrodynamic
simulations including radiation
\citep{tremblin2012a,tremblin2012b,tremblin2014}, the presence of a
double-peak in the N-PDF depends on the turbulent state of the cloud, 
it is only visible at low Mach numbers and when the cloud is
dominated by ionized-gas pressure.  Therefore, the double-peak is not
a general feature of regions with stellar feedback. \\

% PLTS
\noindent{\bf Power law tail(s)}\\
From the nine clouds in our sample, only one (M16) shows a single 
PLT, all others have two PLTs from which three sources have a flatter  
second slope and five a steeper one. The LOS correction has a 
strong influence on the first slope, making it flatter than in the
original N-PDF (see Figures \ref{dr21}-\ref{vela}), but has nearly no 
impact on the second PLT.  The resulting slope(s) of the PLT(s) vary 
between -1.4 and -3.7 for the first PLT, with a median of -2.37, 
and between -1.4 and -4.6 for the second PLT, with a median of -2.33.  
The higher statistics compared to \citet{schneider2015c} shows that there 
is no systematic trend for high-mass SF regions that the second PLT is 
flatter than the first one. Both PLT slopes are thus consistent with that anticipated for
the gravitational collapse of an isothermal spherical density ($\rho$)
distribution of equivalent radius $R$
\citep{larson1969,penston1969,shu1977,whitworth1985,foster1993} with
$\rho \propto r^{-\alpha}$ and $\alpha$=2. The exponent $\alpha$ and
the slope $s$ are linked via $\alpha$=($-2/{\rm s}$)+1
\citep{fed2013,girichidis2014,veltchev2019}. \\

% DPs
\noindent{\bf Deviation point(s) and structure}\\
The DP from the log-normal part to the first PLT (DP1) and the DP from 
the first to the second PLT (DP2) show a very large spread, with \av(DP1)$\sim$8-37 
and \av(DP2)$\sim$19-88, respectively. The high value of DP1 is partly due to 
the fact that the LOS correction may still underestimate the emission along the 
LOS and that the maps are not extended enough. The regions characterized by high 
column densities (above DP2) are outlined in the plots of Appendix~\ref{app-c} with a black contour.
Interestingly, there is a direct link to the $\Delta$-variance spectra
that are shown in the lower right panels of Appendix \ref{app-c},
including the values of the $\beta$-exponent(s). First, we observe
that the largest variation in structure occurs at small scales because
the exponent $\beta_1$ is small, typically between 2.0 and 2.5, with a
median of 2.3 (Table~\ref{table:summary3}). Second, the extent of the
area defined in the column density map by the contour at DP2
corresponds approximately to the peak or turnover point of the
$\Delta$-variance spectrum. For example, DP2 for M17 lies at
  \av=88 and the northern clump outlined in the column density map by
  the black contour at that value has a linear scale of $\sim$2-3 pc
  and the peak of the $\Delta$-variance spectrum lies at 2.57 pc
  (Table~\ref{table:summary3}).  On the other hand, the prominent
peak in the $\Delta$-variance spectrum for NGC7538 ($\beta_1$=2.93)
occurs at $\sim$2-3 pc, which translates into a physical size of the
structure\footnote{As explained in \citet{ark2016} and
  \citet{ossenkopf2019}, the peak in the $\Delta$-variance spectrum
  occurs at 1.7$\times$FWHM size of the structure.} of 1.2-1.8 pc.
This characteristic size can either be caused by the dominating bubble
in this source (at offset 15$'$,10$'$ in Fig.~\ref{ngc7538}) or by the
high-density clumps in the southeast of the map.  Summarizing, the
$\Delta$-variance thus points toward a scenario where the structure
in massive clouds is dominated by sub-parsec scale clumps and not long
filaments or ridges \citep{dib2020}.  From the column density maps, it is obvious that
these clumps are located inside the most massive regions,
preferentially where several filaments merge
\citep{myers2011,schneider2012}. 

%%%%%%%%%%%%%%%%%%%%%%%%%%%%%%  

\subsubsection{N-PDFs of intermediate-mass star forming clouds} \label{inter-mass}  
 
The N-PDFs for intermediate-mass SF regions (Fig.~\ref{pdfs-all}, Figs.~\ref{aquila}-\ref{serpens} in Appendix
\ref{app-c} and Figs.~\ref{aquila+monr2-npdf}-\ref{orionb+serpens-npdf} in Appendix
\ref{app-d}) show a very complex shape (in particular MonOB1, Vela, and NGC2264), similar to that of 
high-mass SF clouds.\\

% double peak LL
\noindent {\bf Log-normal distribution(s)}\\
The peaks of the first and second log-normal are at \av=1.10 and \av=1.68 (median values), respectively, and 
the corresponding widths are $\sigma_{\rm \eta 1}$ = 0.47 and $\sigma_{\rm \eta 2}$ = 0.52. 
The N-PDFs of Mon OB1 and NGC2264 are broader than the others, which can be explained by external compression.  As shown in
\citet{schneider2013}, and conforming with numerical models \citep{tremblin2012a,tremblin2012b}, external compression mainly due
to radiative effects caused by close-by \hii\,-regions leads to a broadening of the N-PDF.  Observationally, this influence becomes also
obvious in cuts of column density profiles \citep{peretto2012,schneider2013,tremblin2013}. \\

% PLTs
\noindent{\bf Power law tail(s)}\\
All clouds in the sample have two PLTs from which only two regions (MonR2, NGC2264) have a flatter second slope.  
For the first PLT, typical values for the slope scatter around -2 (the median is $s_1$=-1.88). For the second PLT, 
the variation is large, the median of all sources is $s_2$=-2.19. These values are again consistent with what is 
expected for gravitational collapse. \\ 

% DPs
\noindent{\bf Deviation point(s) and structure}\\
The first DP shows a small scatter with a median of 4.7, while DP2 varies more, with a median of 17.7. \\
% MORPHOLOGY
There seems to be no clear correlation between cloud morphology and
N-PDF shape.  The two sources with the clearest flatter second PLT are
MonR2 with a dominant hub-filament geometry and NGC2264 with a
dominant ridge structure. And the two sources with a steeper second
PLT are MonOB1, which is basically a large clump, and Vela C, which has
also a dominant ridge structure. The morphology for the areas
constituting the densest gas (above DP2), however, is always clumpy,
over scales from sub-parsec sizes up to a few parsecs.

%DELTA
The $\Delta$-variance spectra are more complex than those for
high-mass SF regions. We typically observe an increase in structure
until a first peak (or turnover into a flat spectrum) around 0.3 pc to
1.8 pc (Table~\ref{table:summary3}) with a median value of
$\beta_1$=2.32, followed by a second increase of the spectrum with a
median $\beta_2$=2.56 and a peak around 4 pc. Similar to high-mass SF
regions, small $\beta$ values indicate the largest structure variation
on small scales.  These are then possibly the sub-parsec- to
parsec-scale dense clumps, filaments and cores that are embedded in
the molecular cloud. The questions arises how the $\Delta$-variance spectrum now links 
to the N-PDF. One correlation is seen in MonR2. The dense, central
clump, in which a whole cluster is forming, has a size scale of around
1-2 pc (Fig.~\ref{monr2}), which is also the size derived
from the peak in MonR2's $\Delta$-variance spectrum (peak at $\sim$2
pc, corresponding to a size of 1.2 pc).  The N-PDF, on the other hand,
shows a slope change (from a steep into a flat PLT) at an \av\ around
15. This level of emission corresponds in the column density map (left
panel in Fig~\ref{monr2}) exactly to the central clump, visible 
where the color changes from green to yellow. Another good 
example is Mon OB1 (Fig~\ref{monob1}), where the N-PDF
PLT slope change occurs at \av$\sim$10, which corresponds to regions
with a size scale of around 1 pc.

%%%%%%%%%%%%%%%%%%%%%%%%%%%%%%  

\subsubsection{N-PDFs of low-mass regions} \label{low-mass}   
  
The N-PDFs of low-mass star-forming regions are displayed in
Fig.~\ref{pdfs-all}, Figs.~\ref{chamI}-\ref{taurus} in
Appendix~\ref{app-c} and Figs.~\ref{chamI+chamII-npdf}-\ref{rhooph+taurus-npdf} in
Appendix~\ref{app-d}.  There are some clouds that have N-PDFs with a
well-defined shape, defined by a rather clear log-normal part at lower
column densities and a PLT at higher column densities (IC5146, Lupus
III, Lupus IV, Perseus, $\rho$Oph, Taurus). Others, however, exhibit
N-PDFs with a bumpier shape (Cham I, Cham II, Cham III, Lupus I,
Pipe). \\

% double peak LL
\noindent {\bf Log-normal distribution(s)}\\
All clouds except of Perseus are best fitted with two log-normals of which the first 
one has values between \av=0.3 and 1 for the first peak (the median is \av = 0.7) 
and the second one a median \av\ of $\sim$1. The widths of the log-normals are $\sigma_{\rm \eta 1}$ 
= 0.32 and $\sigma_{\rm \eta 2}$ = 0.64. In contrast to high-mass SF regions, where the 
two log-normals can both be attributed to purely molecular gas and the second bump to the 
effect of stellar feedback, we are here in a regime where there can be a contribution 
from atomic hydrogen \citep{mandal2020}. The peak of the first log-normal N-PDF always lies below A$_{\rm V}$ = 1 
(in most of the cases significantly lower, only Perseus and Pipe have a value of 
A$_{\rm V,pk1}\sim$1). The \HI-to-H$_2$ transition depends on many parameters such as 
the external radiation field, the density, and turbulence, and is predicted to happen 
between \av$\sim$0.1-0.4 \citep{roellig2007,glover2010,wolfire2010,bialy2017,bisbas2019}. 
The extinction when CO arises from a fully molecular phase ("CO-bright H$_2$") is around A$_{\rm v} \sim$1 
\citep{roellig2007,visser2009,sternberg2014}.  We emphasize that these A$_{\rm v}$ values given in the 
literature are local values, expressed as A$_{\rm v,3D}$ \citep{seifried2020}. The observational 
visual extinction is derived by averaging along the LOS dubbed as A$_{\rm v,2D}$, and is a factor of 
a few larger (see discussion in \citet{seifried2020}. Considering this fact, it is thus reasonable 
that the peak of the first log-normal part of the N-PDF arises from the atomic gas.  
Reducing the observational median A$_{\rm v,2D}$ of 0.7 and 0.45 for low-mass and quiescent regions, 
respectively, by a factor of three for example, would imply a A$_{\rm v,3D}$ of 0.23 and 0.15, 
which is well in the range of the \HI-to-H$_2$ transition. In Sec.~\ref{diffuse}, we discuss the 
N-PDFs of Draco that shows only two log-normal parts of the N-PDF, where we present additional 
evidence for our proposition. But the low-mass clouds we present here can be in an early evolutionary 
state and the atomic envelope \citep{imara2016} may be more prominent. The atomic and the molecular 
gas then have both log-normal N-PDFs (caused by turbulent mixing) that overlap \citep{mandal2020}.  \\
  
% PLTS
\noindent{\bf Power law tail(s)}\\
All clouds are again fitted with two PLTs, but interestingly, the majority (8 out of 10) have a 
steeper second PLT compared to the first one. The median value for the first PLT is $s_1$ = -1.85 and 
for the second PLT $s_2$ = -2.76. This is a new feature in N-PDFs and discovered most likely thanks 
to the higher angular resolution of the maps. While the first PLT slope is consistent with that 
expected for gravitational collapse at early stages 
\citep{larson1969,penston1969,shu1977,whitworth1985,fed2013,girichidis2014}, it is unclear what could cause 
the steeper second PLT. It can be related to magnetic fields, which we discuss in more detail 
in Sect.~\ref{sec:discuss}. \\

% DPs
\noindent{\bf Deviation point(s) and structure}\\
The DP1 is overall at lower values (median \av(DP1) = 2.5) but there is a large scatter in the values 
between 1.4 and 5.2. \\
There are interesting differences in the shapes of the
$\Delta$-variance spectra. Nearly all sources (the best example,
however, is Perseus) show little variation in structure (flat
spectrum) between $\sim$0.3 pc and 1 pc. Below $\sim$0.3-0.5 pc (the
first peak, P1 in the $\Delta$-variance spectrum, see
Table~\ref{table:summary3}), there is the largest structural variation
where the median of the exponent $\beta_1$ is 2.86.  Most of the
sources show an increase in the $\Delta$-variance spectrum after the
flat range with a second peak typically at 1-3 pc (only Lupus IV shows
a peak at 6.5 pc). The median of the exponent $\beta_2$ is 2.64.  We
observe a similar correlation between the slope change of the N-PDF at
DP2 and the structural change visible in the $\Delta$-variance
spectrum (at scale P1) as was detected for high-mass and
intermediate-mass SF regions. We, however, do not distinguish here
between a change into a flatter or steeper PLT.  In the clouds Cham I,
Cham II, Pipe, Taurus, the extent of the higher density regions with a
flatter or steeper PLT outlined by the black contour in the column
density map corresponds approximately to the first characteristic
scale (P1) in the $\Delta$-variance spectrum.  The values of P1,
however, are not always the same and there is a trend that clouds with
higher DP2 show smaller values for P1. For example for Cham I,
\av(DP2)=6.0 and P1=0.21 pc while for Cham II, \av(DP2)=26.9 and
P1=0.61 pc. We come back to this point in the next section.  In
some sources (IC5146, Lupus I, Perseus), there is no clear correlation
between DP2 and P1, or it is less obvious ($\rho$Oph). There is
another trend that the sources with the flattest $\Delta$-variance
spectra (such as Perseus) have the best defined single PLT. This behavior is
consistent with numerical experiments where small-scale fluctuations
increase as the medium becomes full of shock compressed high-density
clumps and filamentary structures, which shape the high-density end of
the N-PDF.

%****************************************************

% Draco 

\begin{figure}
\centering
\includegraphics[width=9cm, angle=0]{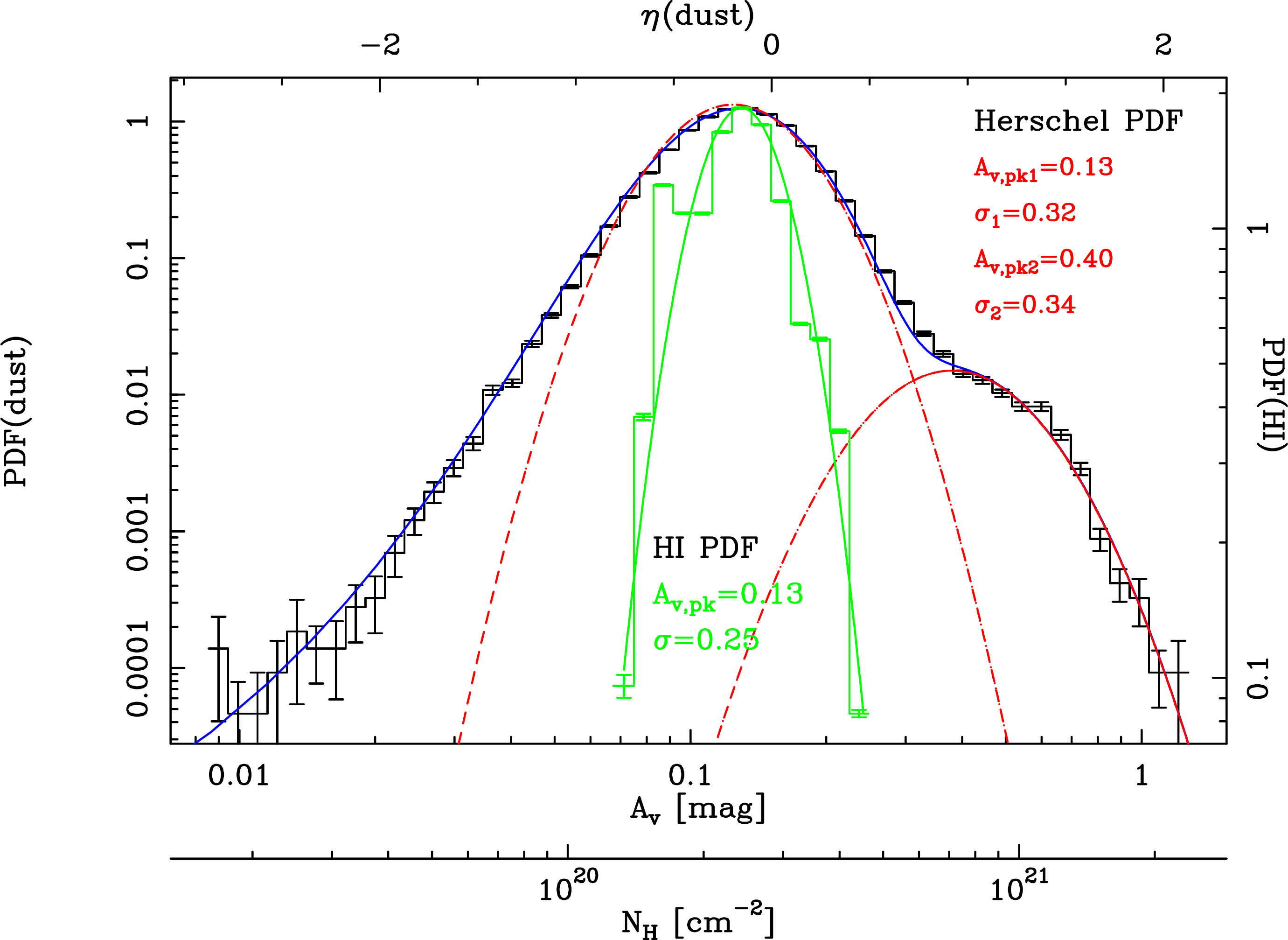}
\caption{N-PDF of the Draco region. 
 The black histogram shows the N-PDF obtained from {\sl 
  Herschel} data and the blue line its analytic description. The red
  line is the result from fitting two log-normal PDFs and considering
  the noise contribution that leads to a linear behavior at low column
  densities \citep{ossenkopf2016}.  The green histogram displays the
  N$_{\rm HI}$-PDF of the \HI\ data and the continuous line a single log-normal fit.
  The fitted peak positions of the PDFs and the widths ($\sigma$ in
  units of $\eta$=ln(N/$\langle N \rangle$)) are given in the panel.  Error bars are
  based on Poisson statistics. The left y-axis gives the probability
  density for the {\sl Herschel} map and the right y-axis for the \HI\
  map. }
\label{draco-spider-pdf}
\end{figure}

\subsubsection{N-PDFs of quiescent regions} \label{quiet}   

The sample we have for quiescent clouds that are not actively forming
stars is small (Fig.~\ref{pdfs-all}).  Only three clouds are included,
namely Cham III (Fig.~\ref{chamIII} and Fig.~\ref{chamIII+polaris-npdf}), Musca (Fig.~\ref{musca} and
Fig.~\ref{musca-npdf}), and Polaris (Fig.~\ref{polaris} and Fig.~\ref{chamIII+polaris-npdf}). 
All sources are fitted with two log-normals and two PLTs. \\

% double peak LL
\noindent {\bf Log-normal distribution(s)}\\
The two log-normals fitted for quiescent clouds both have their peak at 
low column densities (median \av = 0.45 and \av = 0.64, respectively) so that we here 
also attribute the first log-normal to a mostly atomic gas distribution and second one to a mostly molecular one. \\

% PLTS
\noindent{\bf Power law tail(s), Deviation point)s) and structure}\\
The full Polaris region has a first PLT with a steep slope ($s_1$=-4.08) and a 
second flatter one ($s_2$=-2.34) though \citet{schneider2013} noted that the 
N-PDFs of quiescent subregions in Polaris are better described by a single 
log-normal part. If the log-normality is caused by turbulence, then there is a 
direct link to the structure, which should show a self-similar behavior \citep{stutzki1998,schneider2011}. 
Indeed, the $\Delta$-variance spectrum of Polaris displays such a self-similar 
behavior over more than an order of magnitude in size between 0.02 pc and 0.6 pc. 
We come back to the slopes of the PLTs in Sect~\ref{sec:discuss}.

Musca displays a rather unusual N-PDF compared to other quiescent or
low-mass clouds because it shows two PLTs that separate at
\av(DP2)$\sim$4. This \av\ is approximately that defined by
\citet{cox2016} as the "high-density filament crest" of Musca, in
contrast to the lower density surrounding filamentary structures
called "striations" \citep{pedro2013}. This change in behavior also
becomes obvious in the column density map, shown in Fig.~\ref{musca},
where the crest stands out as a prominent skeleton (red areas) within
the whole Musca cloud.
           
The $\Delta$-variance spectrum of Musca shows a characteristic first
scale (P1) at around 0.1 pc, which is even smaller than the width of
the crest. Interestingly, the \av=4 contour, where the slope change
occurs, outlines clumps of $\sim$0.1 pc size. In any case, the slope
change of the PLT from a value $s_1$=-1.7 into a much steeper one of
$s_2$=-5.0 when entering the crest indicates a change of the dominant
process governing the column density distribution. While the first
slope is consistent with self-gravity, the much steeper second slope
could be explained with the influence of the magnetic field.
Observationally, \citet{soler2019} showed that slopes of the N-PDF are
steepest in regions where the magnetic field B and the column density
distribution are close to perpendicular.  This configuration is the
case for Musca, as it was shown in \citet{cox2016}. \citet{auddy2019}
argue that clouds with a strong magnetic field with a subcritical
mass-to-flux ratio and small amplitude initial perturbations develop a
steep PLT in the PDF. They reason that gravitationally driven ambipolar 
diffusion leads to shallower core density profiles than in a hydrodynamic collapse.

%%%%%%%%%%%%%%%%%%%%%%%%%%%%%%  

\subsubsection{N-PDFs of a diffuse region} \label{diffuse}   

With the Draco region \citep{mebold1985,herbst1993,miville2017}, 
we include an example of a diffuse cloud that is probably only at the verge of becoming
molecular and does not show star-forming activity.  Draco is an 
intermediate velocity cloud (IVC, velocity around --20 km s$^{-1}$)
that most likely originates from a Galactic fountain process in which
disk material is lifted above the plane and falls back to the disk at
high velocities (\citep{lenz2015} and references therein), or infall
of extragalactic gas. Figure~\ref{draco} shows the
column density map, the 250 $\mu$m {\sl Herschel} map, the N-PDF
and the $\Delta$-variance spectrum for this regions. We include here
the 250 $\mu$m map and derive the $\Delta$-variance spectrum from this
map because of the higher angular resolution of 18$''$ (the column
density map is at 36$''$). In addition, we display in 
Fig.~\ref{draco-spider-pdf} the N-PDF again, this time together with
the N$_{\rm HI}$-PDF from atomic hydrogen\footnote{For constructing
  the N$_{\rm HI}$-PDF, we use the all-sky \HI\ data from the
  Effelsberg-Bonn \HI\ survey (EBHIS) \citep{winkel2016} at an angular
  resolution of $\sim$10$'$.  We assume that the \HI\ line is
  optically thin \citep{herbst1993}, and calculate the \HI\ column
  density $N_{\rm HI}$ using $N_{\rm HI}$ [cm$^{-2}$]= 1.82 10$^{18}$
  W(HI) with the line integrated intensity W(HI) in [K km s$^{-1}$].}.

We show in Fig.~\ref{draco-spider-pdf} the N-PDF of the
pixel distribution over a wide column density range covering the
noise and the log-normal parts (and potentially PLTs).  
The N-PDF shows a tail at very low column densities, followed by a more
complex shape for column densities above $\sim$10$^{20}$ cm$^{-2}$.
The best model fitting the distribtion is the ELL one, two log-normals and an error PLT.  
The model fit is shown as a blue line and the two individual log-normal parts of the N-PDF as
red dashed lines.  The dust N-PDF parts have widths of $\sigma$=0.32
and 0.34, maxima at \av(peak)=0.13 and 0.40 (N=2.4 and
7.5$\times$10$^{20}$ cm$^{-2}$), respectively.  The location where the
two N-PDF parts have the same contribution is at A$_{\rm V}$=0.33
(N=6.2$\times$10$^{20}$ cm$^{-2}$). 

The N$_{\rm HI}$-PDF determined from the \HI\ data is shown as a green
histogram in Fig.~\ref{draco-spider-pdf}. A log-normal fit 
(green line) is the simplest approach
with only three parameters. Indeed, log-normal shapes for N$_{\rm
  HI}$-PDFs were obtained for other \HI\ observations
\citep{berk2008,burkhart2015b,imara2016}.  For Draco's log-normal fit,
the peak is found at A$_{\rm V}$(peak$_{\rm HI}$)=0.13 and the width
is $\sigma_{\rm HI}$=0.25.  Given the low angular resolution of the
\HI\ map, however, we may blur small-scale structure in the
\HI\ distribution and thus possibly underestimate the width of the
N$_{\rm HI}$-PDF.  We assume that the N$_{\rm HI}$-PDF consists mostly
of CNM (cold neutral medium) gas and not WNM (warm neutral medium),
similar to observations reported by \citet{burkhart2015b} and
\citet{stani2014}. The peak of the N$_{\rm HI}$-PDF at A$_{\rm V}$(peak$_{\rm HI}$)=0.13
corresponds remarkably well to the left log-normal low column density
{\sl Herschel} dust N-PDF so that the most simple and straightforward
explanation is that this feature reflects cold \HI\ gas, i.e., the
atomic CNM phase. The transition between the two log-normals occurs at
N$\sim$6.2$\times$10$^{20}$ cm$^{-2}$ (A$_{\rm V}\!\sim$0.33), and
the peak of the second N-PDF is at N$\sim$7.5$\times$10$^{20}$
cm$^{-2}$ (A$_{\rm V}$=0.40) and has no counterpart in the N$_{\rm
  HI}$-PDF. We thus attribute this second feature as arising mostly
from the molecular H$_2$ phase. The transition between \HI\ to H$_2$
around an \av\ of 0.33 is consistent with typical values found for
IVCs and diffuse clouds
\citep{federman1979,reach1994,lagache1998,lockman2005,gillmon2006,roehser2014}.
Furthermore, PDR models \citep{roellig2007} and cloud simulations
including radiative transfer
\citep{glover2010,bisbas2019} determine the transition
to be around \av=0.3, but can be slightly higher
  \citep{wolfire2010}, depending on incident UV field and density.
Integrating over the N-PDF, we determine that 89\% of mass is in
low-column density gas and 11\% at high column densities.  Given that
the absolute values for the column density are below $\sim$10$^{21}$
cm$^{-2}$, and thus well below the limit of significant CO formation
\citep{lee1996,visser2009}, we suspect that a significant part of
  the H$_2$ N-PDF is made up out of CO-dark gas. For further details, see the
summaries given in \citet{klessen2016}, \citet{clark2012} and
\citet{smith2016} for numerical simulations.

Summarizing, this is the first time that such purely bimodal
log-normal dust N-PDFs without a high column density PLT are
observed. In particular the higher column density log-normal part of
the Draco dust N-PDF is well resolved and sampled, and attributed to
CO-bright and CO-dark H$_2$.  This finding is consistent with current
analytic theories of SF as both clouds are in a very early
stage of their evolution where turbulence dominates over self-gravity,
so that a high column density PLT is not expected.

\section{Discussion} \label{sec:discuss}   
\subsection{General remarks}   
Probability distribution functions derived from visual extinction maps
and {\sl Herschel} column density maps are useful tools for the
analysis of the density structure of a molecular cloud but they have
their drawbacks.  Extinction maps are affected by LOS-confusion,
limited angular resolution that leads to small number statistics in
the high-density pixel regime, and limited map sensitivity. In
addition, the absolute scale depends on the conversion factor
\av\ into N(H$_2$), which can be controversial. A completely different
approach to obtain column density maps is to perform pixel-to-pixel
SED-fitting to the FIR-data from {\sl Herschel}. Other errors are
introduced with these maps (opacity, assumption of isothermal dust
distribution, etc.), but the cut-off in the high column density regime
is much higher (up to a few hundred \av), also because the angular
resolution is much higher. The {\sl Planck} all-sky survey also
provides the ability to obtain column density maps by SED fits, but
the angular resolution is much lower (around 5$'$). Combining {\sl
  Herschel} and {\sl Planck} dust emission observations
\citep[e.g.,][]{lombardi2014,zari2016,abreu2017} is a way to cover
larger cloud areas, but does not solve the angular resolution
limitation of {\sl Planck}.

In any case, correcting for LOS confusion, as we have done in this
study, is an important improvement because it affects N-PDF parameters
such as width and PLTs. Without the correction these parameters show a
larger spread, as shown in many other studies (the widths become
narrower and the slopes steeper).

\subsection{Discussion of N-PDFs parameters}   
  
In the following, we discuss the properties of the N-PDF parameters alone 
and their possible correlation for which we plot and discuss in Appendix~\ref{app-e} 
key parameters such as width of log-normal, PLT slopes etc. against the mass. 
This is a purely qualitative comparison since our sample is still too small to 
perform a more quantitative analysis. \\
  
\noindent {\bf The width of the log-normal part of the N-PDF} \\
\noindent Different physical processes are responsible for shaping the
N-PDF \citep{nordlund1999,fed2008,fed2010}. As shown in the
simulations presented in \citet{fed2008,fed2010,fed2012,molina2012},
the width and the peak position of density and column density PDFs
depend on the Mach-number, the forcing (compressive or solenoidal
driving), and the ratio between thermal and magnetic energy.  For
example, compressive modes cause a broader log-normal part of the
N-PDF with the peak shifted to lower densities. \\ 
In our study,  we mostly fit two log-normals to the low-column density range, and 
we attribute the first one for high-, and intermediate SF regions to turbulently 
mixed molecular gas, and the second one to compression by stellar feedback. 
Indeed, $\sigma_{\eta2}$ is broader compared to  $\sigma_{\eta1}$ in clouds exposed to external compression from
  expanding \hii-regions and stellar winds (CygnusX N and S, M16,
  NGC7538, Vela, Orion B, Mon OB1, NGC2264).  On the other hand, it is
  not obvious why the massive GMCs M17, Rosette, and NGC6334, which
  are strongly exposed to radiative feedback, have a narrower 
  second log-normal N-PDF. However, as shown in \citet{tremblin2014},
  a double-peak or generally "bumpiness" in the N-PDF is only visible
  when the Mach number is low and the cloud is dominated by
  ionized-gas pressure.  There, the case of Rosette is well modeled
  with a Mach 2 turbulent cloud with ionization.  \\ For low-mass and
  quiescent regions, $\sigma_{\eta1}$ of the N-PDF, we attribute to
  turbulently mixed \HI\ gas, is clearly lower with a mean of 0.32 and
  0.38 with respect to more massive regions.  $\sigma_{\eta2}$ has
  high values, 0.64 and 0.61, respectively, and characterizes the
  width of turbulently mixed molecular gas and corresponds in our
  interpretation to $\sigma_{\eta1}$ of intermediate and high-mass SF
  regions.  \\

\noindent {\bf The deviation point(s) of the N-PDF} \\ 
\noindent
Our sample of clouds of different masses shows that there is no  
common value for DP1, but a trend that the group of quiescent, low-mass 
and intermediate mass clouds has values between \av(DP1)=2-5.  
High-mass clouds have a median value of \av(DP1)$\sim$18.5, which can partly 
be attributed to the LOS-contamination correction that may not be perfect. 
The question now is whether these values reflect a change in the dominant physical
process within these cloud types or a threshold in (column) density
for core- or star-formation.  While \citet{kai2011} explained their
rather constant \av(DP1)-value of 2--4 as due to a phase transition
between lower-density interclump gas and pressure-confined clumps,
\citet{froebrich2010} proposed that there is a universal threshold of
\av\,=6$\pm$1.5 where gravity dominates over turbulence. 
If gravity starts to play a significant role during molecular cloud
formation, an increasing fraction of gas will be above a certain
threshold of column density/extinction and form stars. Such extinction
thresholds were identified by \citet{lombardi2008} for
Ophiuchus or \citet{roman2010} for the Pipe nebula. Furthermore,
\citet{heider2010} claim to have found a "star-formation threshold" 
around \av$\sim$8 (corresponding to a surface density of $\sim$130
M$_{\odot}$ pc$^{-2}$) that was defined as a steep increase of the
SF rate surface density $\Sigma_{\rm {SFR}}$ over gas
surface density $\Sigma_{\rm {gas}}$. Studies using {\sl Herschel} or
other continuum data do not give a common picture.  For regions such
as Aquila \citep{andre2014,koenyves2015}, Orion B
\citep{koenyves2020}, and Taurus \citep{marsh2016a}, the majority of
pre- and protostellar cores is found in gravitationally collapsing
filaments above an \av\ threshold of around 6--7. On the other hand,
in regions with low overall column density like the Lupus clouds
\citep{benedettini2018}, prestellar cores are detected above a
background of only \av=2. \citet{pokhrel2020} studied the relation between the stellar mass
surface density and the mass surface density of a subsample of our
clouds and deduce that there is no gas column density threshold for
SF.  Summarizing, we conclude that each of these studies
has its own biases and no clear threshold value for the formation of
self-gravitating cores has emerged.

The circumstance that different values of the threshold are reported
in these studies and in the N-PDFs are an indication that the SF 
column density threshold, if it exists, might depend on the
local properties of the host cloud such as the strength of its
magnetic field, the local radiation field, and its nonthermal
velocity dispersion. These properties control the mechanisms that can
provide support against gravitational collapse, such as the magnetic
pressure mediated by collisions between neutrals and ions and the
turbulent motions that supply nonthermal pressure support
\citep{klessen2016}. Recently, \citet{jaupart2020} showed in
  their analytic formalism that the threshold density (accordingly
  also the column density DP1) evolves with time on the same timescale
  as the global, average properties of the cloud and is thus not
  constant.

The fact that we nevertheless observed a clustering of DP1 around
\av(DP1)$\sim$2--5 could be due a chemical transition. For a given UV
radiation field, there is a minimum column density necessary to
self-shield CO and to maintain significant molecular abundances. This
transition occurs typically at \av=1.5 if we take into account the
effect that the typical \av\ that a cloud element experiences toward 
the external radiation field is lower by a factor of 3 relative to the
total column an observer sees \citep{glover2010}.  Alternatively, it
may represent a change of the dust properties for cold material
leading to ice mantles and dust grain growth as the gas temperature at
\av=4-5 falls below some condensation threshold.

Concerning high-mass SF, \citet{krumholz2008} proposed on the basis of
theoretical considerations a threshold of 1 g cm$^{-2}$ (equivalent to
an \av\ of $\sim$300), for high-mass stars to form.  Observationally,
no clear picture emerges. Our study of N-PDFs does not reveal a
characteristic value of the second deviation point (DP2) of
\av=300. Instead, there is a large variation of \av(DP2) for different
regions (Table~\ref{table:summary3}  and Appendix~\ref{app-e}).\\

\noindent {\bf Power law tail slopes} \\ 
\noindent
The overall median slope values for the first and second PLT are 
  $s_1$=-2.125 and $s_2$=-2.34, respectively, which correspond to an
exponent $\alpha_1$=1.94 and $\alpha_2$=1.85, for an isothermal
spherical density ($\rho$) distribution of equivalent radius $R$ with
$\rho \propto R^{-\alpha}$ (see Sec.~\ref{high-mass}).  The
gravitational collapse of an isothermal sphere has been studied for a
long while \citep{larson1969, penston1969,shu1977,whitworth1985}, and
though all models start with different initial conditions, they arrive
at the same $\alpha$=2 for early stages and $\alpha$=1.5 after a
singularity forms at the center of the sphere. It is thus possible that the PLTs then stem
only from local core collapse. However, in most of the clouds, this 
explanation is unlikely because cores constitute only a small mass
fraction of the total gas mass (e.g., 15\% of dense gas in Aquila,
\citet{andre2014}). In addition, there are clear observational
signatures for gravitational collapse on much larger scales, for example as
observed in the DR21 ridge \citep{schneider2010}, the Serpens filament
\citep{kirk2013}, W33A \citep{galvan2010}, and IRDCs (Paper II).
Gravitational fragmentation of dense filaments into prestellar cores,
possibly fed by accretion via filaments oriented orthogonal to the
main filament, called "striations" \citep{pedro2013}, is proposed
as the main process to form solar-type protostars \citep{andre2014}.
Mass accretion by larger subfilaments is considered further as an
important process to build up the large mass reservoir to form massive
star(s). Observational examples are  found in  \citet{schneider2010,schneider2012,galvan2010,quang2011,hennemann2012,kirk2013,peretto2014,motte2018}, 
simulations in \citet{heitsch2001,smith2011,smith2013}.  
The PLT of the N-PDF is thus not only due to
local core collapse, but can also arise from the aforementioned
processes. We thus interpret the PLTs as due to gas that is controlled
by gravity on all scales (global collapse and accretion, core
collapse).

Special attention is devoted to the second PLT, which can be
  steeper or flatter than the first PLT. One would expect that a
second flatter PLT appears after the first PLT has developed and was
explained by thermodynamic effects, radiative feedback, or small-scale
convergent flows \citep{schneider2015c}. It is also possible that
  a flatter second PLT can have different reasons, depending on cloud
  type. For massive, evolved GMCs, \citet{tremblin2014} put forward 
stellar feedback as an explanation because they show that the PLT of
the N-PDF becomes flatter going from the cloud center toward the
interaction zone between an \HII\ region and the cloud. This shift in
slope implies that compression of gas takes place and that
self-gravity then takes over in the densest regions to form cores and
finally stars.  \citet{fed2013} showed in their models how the whole
slope of the N-PDF flattens with increasing star-formation efficiency 
and more stellar feedback. Recently, two other theoretical
  explanations were given for the occurence of a flatter second PLT in
  less evolved molecular cloud without stellar feedback.
  \citet{jaupart2020} developed an analytical theory of a
  nonstationary density PDF including gravity with a first PLT with
  slope $s_1\lesssim -$4, which reaches an asymptotic value of
  $s_2\lesssim -$2 in freefall collapsing regions. These values fit
  very well to the observed slopes of the quiescent regions Cham~III
  and Polaris (Table~\ref{table:summary3}), but not with the Musca
  filament. However, the highest density part (above \av$\sim$10) of
  the N-PDF of Musca is not well resolved due to limited angular
  resolution and may hide a flatter PLT. In this case, there would
  indeed be a succession of a steep PLT and a flatter
  one. \citet{donkov2021} on the other hand discuss a model in which
  cores of an averaged representative of a whole class of molecular
  clouds are considered. They propose that the thermodynamic state of
  the gas (only turbulence and gravity included) changes from
  isothermal on large scales to a polytropic equation of state of the
  gas $p \propto \rho^\Gamma$ with pressure $p$ and density $\rho$ and
  an exponent $\Gamma$ larger than 1 on the sub-parsec proto-stellar
  core scale. A density profile $\rho(r) \propto r^{-p}$ with $p$ = -3
  and an exponent $\Gamma$ = 4/3 then produces a flatter second PLT. 

As a counterpoint, there are also N-PDFs with a steeper second PLT
from all cloud types, for instance Rosette for high-mass clouds, Mon OB1 for
intermediate-mass clouds, and Taurus for low-mass clouds.  It is
unclear what physical cause may be behind such behavior.
Observationally, resolution effects may also play a role, but we emphasize 
that we excluded all pixels at the highest-column density range
(typically above \av=100 for massive/intermediate mass SF regions and
\av=30 for low-mass and quiescent regions) so that we can be sure
about the significance of a second flatter or steeper PLT. 
There is also a link between the extinction value, \av(DP2), where the
slope turn occurs, and the $\Delta$-variance, suggesting that there is
indeed a change in the column density structure or the dominant
physical process. In addition, for most of the clouds the second
steeper PLT is well pronounced and not a small feature (see Vela C in
Fig.~\ref{vela}).  A possible explanation for a steeper PLT is the
magnetic field orientation. \citet{soler2019} investigated the
relative orientation between the magnetic field $B$ projected onto the
plane of sky derived from polarized thermal emission of dust observed
by Planck, and the distribution of dust column density, obtained from
{\sl Herschel}\footnote{The column density maps were obtained by
  cross-correlating {\sl Herschel} and Planck data and have an angular
  resolution of 36$''$ \citep{abreu2017}.}, for a number of nearby
low-mass and quiescent molecular clouds. They derived that the slope
of the N-PDF is steepest in clouds or regions within a cloud where the
magnetic field and the column density distribution are close to
perpendicular. We note, however, that such a correlation had not yet been
observed in simulations \citep{seifried2020}.  In the study of Soler,
the PLT fit was restricted to a single slope, without the more
detailed distinction into one or two PLTs we perform here.
Nevertheless, we find that the steeper, high-column density PLTs seen
from N-PDFs in the low-mass SF or quiescent regions ChamI, ChamII,
IC5146, LupusI, Orion B, Perseus, Taurus, and Musca indeed correlates
with the presence of a magnetic field that is oriented perpendicular
to the higher column density regions (see \cite{soler2019} for
magnetic field observations in these clouds). For the intermediate-
and high-mass SF regions that show a second steeper PLT (Rosette, Mon
OB1, Vela), there are no high angular resolution maps available. The
Rosette cloud was studied in \citet{planckXXXIV} and Fig.~3 in this
paper shows that the high column density regions are partly aligned
with the magnetic field and partly perpendicular. Given the limited
sample, we thus only tentatively propose that the second steeper PLT
for some cloud regions is a result of magnetic field orientation.

Numerical simulations show different density and column density PDF
shapes, depending on evolutionary stage, but a steeper PLT is normally
only found when the simulation reaches its resolution limit. 
Since we mostly find PLTs (first or second) with a slope around -2 that indicates that
high-density gas in all clouds is collapsing in free-fall. The
remaining variations in slope values are then most likely caused by
processes that are only partly considered in simulations and analytic
descriptions such as stellar feedback or magnetic fields.

\subsection{Tracing the HI-to-H$_2$ transition with N-PDFs} \label{transition}   

\citet{alves2017} argued that the log-normal portion of N-PDFs
(typically \av$<$1) cannot be safely traced if the map is not
complete, which means that the last closed contour must be sufficiently low and
above the noise of the map to represent the
PDF. \citet{ossenkopf2016,chen2018,koertgen2019} investigated this
problem in more detail and concur that observational limitations such
as noise, LOS effects, and incompleteness can have an impact
on the N-PDF. Paper I and \citet{ossenkopf2016}, however, showed that
there are efficient methods to correct for noise and contamination,
and \citet{chen2018,koertgen2019} deduced that a model with a
log-normal part and a PLT for the N-PDF gives the best
fitting model for star-forming clouds.

As outlined in Sec.~\ref{diffuse}, we find for Draco a double
log-normal dust N-PDF, attributed to the atomic and to the molecular
CNM, respectively, both turbulently mixed. The higher-column density
molecular part of the Draco N-PDF can be fitted by a log-normal
distribution between N$\sim$3-4$\times$10$^{20}$ cm$^{-2}$ and
N$\sim$2$\times$10$^{21}$ cm$^{-2}$, which is well above the noise
level of around 0.03$\times$10$^{21}$ cm$^{-2}$ (\av$\sim$0.016).  The
data points constituting the molecular part of the N-PDF
(Fig.~\ref{draco-spider-pdf}) are defined by pixels comprising larger
clumps and filaments (green and red colors in Fig.~\ref{draco}, size
scale 1-6 pc), or from very small-scale structures,  molecular
gas that is intimately mixed with the lower column density material
lower or at the resolution limit. A distribution of resolved small
clumps was identified by \citet{miville2017} in their {\sl Herschel}
Draco study. They determined the clump size to be $\sim$0.1-0.2 pc
with an average density of $\sim$10$^{3}$ cm$^{-3}$.  Our observations
thus support a scenario that is put forward by many authors
\citep[e.g.,][]{heitsch2005,hartmann2001,glover2007,valdivia2016,seifried2017}
in which H$_2$ rapidly forms in dense clumps and then diffuses into
lower density gas. We speculate that these small clumps may constitute
the major reservoir of CO-dark gas because they are already
molecular but not yet realistically detectable in CO
\citep{pringle2001,koyama2000,smith2014}. 

In general, while it is possible to construct a last closed contour
for the extended structures in the column density map, the turbulent
nature of the gas naturally prevents the  construction of a clearly defined
closed contour for the small-scale structures.  \citet{koertgen2019}
arrived at the same conclusion using magneto-hydrodynamic simulations
of colliding \HI\ flows with and without self-gravity and investigated
in detail the issue of the last closed contour in a turbulent
environment. They point out that ``in a fully turbulent medium, there
will essentially be no closed contour anymore for a sufficiently low
column-density threshold because this is just natural for a turbulent
medium''.

Obviously, to perform reliable studies of N-PDFs in molecular clouds,
the total area studied should be sufficiently large and the map should
have a high dynamic range to sample the N-PDF well, conforming to what
is stated by \citet{schneider2015a,ossenkopf2016,alves2017,chen2018}.
In contrast to \citet{alves2017}, however, we propose that we have
found in Draco an observational example of a cloud whose N-PDFs indeed
shows a log-normal part for the low column density molecular range,
similar to what is seen in early time steps of simulations or analytic
descriptions of molecular cloud formation
\citep{vaz1994,ball2011,collins2012,fed2012,fed2013,koertgen2019,jaupart2020}.
This log-normal part of the N-PDF is thus consistent with analytic
models of star formation that are based upon a log-normal
(column)-density PDF of turbulent gas
\citep{padoan2002,hennebelle2008,hennebelle2009,elmegreen2011,donkov2012,parravano2012,hopkins2012,fed2012}.  
Later in the cloud's evolution, an additional PLT develops
\citep{klessen2000,ball2011,kritsuk2011,fed2013,valdivia2016}, which is
mostly attributed to self-gravity
\citep{kritsuk2011,girichidis2014,jaupart2020}.

For Draco, we propose that we found an observational example of a non-biased 
log-normal N-PDF with an atomic and molecular part, the question is now to which 
extent this finding applies to more evolved clouds.  
\citet{burkhart2017} suggest that the log-normal part of
N-PDFs of star-forming regions can be attributed to \HI\ and the
PLT tail to H$_2$, and developed an analytic model for
determining the transition point between log-normal and 
components.  The model is based on the typical coincidence of a common
density threshold for H$_2$ formation and the onset of gravitational
instability. For a high-latitude cloud such as Draco, however, this
scenario does not apply as we observe molecular gas, even CO, but no
signatures of gravitationally bound structures.

For star-forming clouds, the \HI\ contribution mostly stems from
extended \HI\ envelopes around the molecular cloud
\citep{motte2014,imara2016,kabanovic2022} and most of the H$_2$ is either already
locked in larger, dense clumps and filaments that are dominated by
self-gravity or is still in very small structures - like in Draco -
mixed with the CNM, but with a low volume and mass filling
factor. This picture would be consistent with simulations
\citep[e.g.,][]{ball1999,ward2014}) where the log-normal N-PDF part is
always present, also after the development of a  tail.  
All N-PDFs in quiescent and low-mass SF regions are fit by a double-log-normal 
plus PLTs, and the peak of the first log-normal is always lower than A$_{\rm v,3D}<$1, 
which translates into an A$_{\rm v,2D}\lesssim$0.3 \citep{seifried2020,glover2010,mandal2020} 
and thus indicates the transitional \av\ for H$_2$ formation. We thus suggest 
that we may observe in these N-PDFs the distinct contributions of \HI\ and H$_2$. 
This finding is similar to the one already put forward by \citet{burkhart2015b,burkhart2017,imara2016,chen2018}  
where the authors found that the log-normal part of the N-PDF is built up by both \HI\ and 
H$_2$, while the PLT is composed of molecular material that is self-gravitating.

\section{Summary} \label{sec:summary}   

We present dust column density maps derived from {\sl Herschel}
imaging for 29 Galactic cloud complexes, covering diffuse gas regions
and quiescent (mostly non-star-forming) clouds as well as low-,
intermediate-, and high-mass star-forming clouds. The maps have an
angular resolution of 18$''$ and are presented in visual extinction
\av.  Line-of-sight contamination is considered for high-mass and
  intermediate-mass clouds by subtracting a constant value. From
these maps, we then determined column density probability distribution
functions (N-PDFs).  Different methods for fitting a log-normal and
power law tails (PLTs) to the N-PDF are discussed. In addition, we
investigate the cloud structure using the $\Delta$-variance and
discuss the spectrum and its exponent $\beta$. The characteristic
properties of the N-PDFs are presented and explored as a function of
cloud type. Summarizing our main results, we find:
\begin{itemize}  
\item The shapes of the N-PDFs are complex, but can generally be
  described with two log-normal low-column density parts and one or two
  PLTs. Massive clouds such as Cygnus North, M17, NGC6334, MonR2, and NGC2264
  often show a second PLT that is flatter than the first one, confirming
  an earlier study of \citet{schneider2015c}. Two low-mass SF regions (Pipe and
  Lupus III) also show this behavior. The reason for this
  accumulation of high (column) density is unclear but may be related
  to radiative feedback effects. A steeper second PLT is found for
  clouds of all masses (for example Vela C, Rosette, Taurus, Musca)
  and is thus not an intrinsic feature of a
  certain cloud type. Its origin is also unclear but may be related to
  the magnetic field orientation, as proposed by \citet{soler2019}. He
  found steeper PLTs in regions where the magnetic field is oriented
  perpendicular to the column density distribution, similar to what we
  anticipate in our study.
\item The first deviation point between log-normal and PLT (DP1) is
  not constant, but varies between \av(DP1)$\sim$1 and
  \av(DP1)$\sim$18.5 with a clustering around \av(DP1)$\sim$2--5.  We
  thus do not find a correlation between the DP1 and the proposed
  threshold of \av=8 ($\sim$130 M$_{\odot}$ pc$^{-2}$) in nearby
  clouds \citep{heider2010,lada2010,andre2014} above which 
  dense cores and YSOs are found. Moreover, the value of
  \av(DP1)$\sim$2--5 could signify the
  minimum column density necessary to self-shield H$_2$ and CO 
  to build and maintain significant molecular abundances. The
  change of dust properties may also play a role since ice mantles and
  particles are expected to grow as the gas temperature at
  \av=4-5 falls below some condensation threshold.
\item The diffuse cloud Draco has a well-resolved and sampled N-PDF
  with two log-normal distributions peaking at \av(peak)=0.13 and
  0.40, respectively. We interpret the low column density part as
  arising from the cold neutral medium and the higher column density
  part originating mostly from H$_2$.  The \HI-to-H$_2$ transition is
  defined where the two log-normal dust N-PDFs have equal
  contributions and takes place at A$_{\rm V}\!\sim$0.33
  (N$\sim$6.2$\times$10$^{20}$ cm$^{-2}$). This is the first time that
  such a bimodal log-normal dust N-PDF without a high column density
  PLT is observed. We also find that all quiescent and
    low-mass SF regions show a double-log-normal part at low column
    densities and propose that we observe the N-PDFs of the atomic and
    molecular gas. This finding is consistent with current analytic
  theories of star formation, where a log-normal density PDF is a key
  feature. It challenges the proposal of \citet{alves2017} that all
  clouds, including non-star-forming ones, have N-PDFs described by a
  PLT.
\item Most of the $\Delta$-variance spectra of the observed clouds
  show two peaks. As a result, we fitted the power law exponent of the
  $\Delta$-variance in two intervals and derived from that the
  exponent $\beta$.  $\beta_1$ was determined starting at the
  resolution limit until the first peak and $\beta_2$ until the second
  peak. We find that $\beta_1$ decreases with increasing cloud mass,
  while $\beta_2$ is rather constant for all cloud types. For high-
  and intermediate mass clouds, the largest structural variation
  happens on small scales, $\beta_1$ is typically between 2.0 and 2.5
  with a median value of 2.3.  Low-mass and quiescent clouds are
  dominated by structural variations on larger scales, the median of
  $\beta_1$ is 2.86 and 2.73, respectively. There is an intriguing
  correlation between the $\Delta$-variance spectrum and the
  N-PDF. The first characteristic size scale detected in the
  $\Delta$-variance spectrum (P1) depends on the cloud type and the
  second deviation point (DP2) of the N-PDF. Quiescent and low-mass SF
  clouds have P1 values below 0.6 pc, which signify filament widths and
  clumps. Intermediate- and high-mass SF regions are dominated by
  structures around 1 pc, possibly the typical size of cluster-forming
  clumps. This structural variation is correlated with the column
  density structure, because the value where the slope of the first
  PLT changes into a flatter or steeper one (DP2) increases with P1.
\end{itemize}

The final interpretation from this study is that atomic and molecular
gas are turbulently mixed at low column densities while the high
column density part of the N-PDF is constituted by molecular gas,
dominated by self-gravity.  The model of log-normal distributions
  at low column densities followed by one or two PLTs is thus the best
  description for molecular clouds.  The gas mass reservoir above an
extinction value \av $\sim$4-5 is strongly affected by self-gravity,
and indeed may be globally contracting in most clouds (whether massive
or not).  A clear separation in the N-PDF between global (such as
filament) collapse, and local core collapse (and other effects like
radiative feedback) awaits further studies that make the link between
the core population (pre- and protostellar) and the N-PDF.

The characteristic parameters of the N-PDF (deviation point from
log-normal, power law tails, existence of a double-peak) depend on
environmental properties and allow a distinction to be made between cloud type
(quiescent, low-, high-mass SF cloud).
Comparing to simulations, we find the best correspondance to the
observed N-PDFs in the case of large-scale turbulence with gravity, 
consistent with the analysis of the velocity structure of observed clouds.  

\begin{acknowledgements}  
This work was supported by the Agence National de Recherche
(ANR/France) and the Deutsche Forschungsgemeinschaft (DFG/Germany)
through the project "GENESIS" (ANR-16-CE92-0035-01/DFG1591/2-1) and by
the German \emph{Deut\-sche For\-schungs\-ge\-mein\-schaft, DFG\/}
project number SFB 956, project ID 184018867.  This research has made use of data from the
Herschel Gould Belt survey project. The HGBS is a Herschel Key Project
jointly carried out by SPIRE Specialist Astronomy Group 3 (SAG3),
scientists of several institutes in the PACS Consortium (CEA Saclay,
INAF-IAPS Rome and INAF-Arcetri, KU Leuven, MPIA Heidelberg), and
scientists of the Herschel Science Center (HSC). \\
T.V. acknowledges support by the German Research Foundation (DFG) under grant 
KL 1358/20-3 and additional funding from the Ministry of Education and Science 
of the Republic of Bulgaria, National RI Roadmap Project DO1-383/18.12.2020. \\
D.E. acknowledges support by the INAF Main-stream Grant “The ultimate exploitation of 
the Hi-GAL archive and ancillary infrared/mm data” (1.05.01.86.09). \\
C.F.~acknowledges funding provided by the Australian Research Council (Future Fellowship 
FT180100495), and the Australia-Germany Joint Research Cooperation Scheme (UA-DAAD).\\
J.D.S. acknowledges funding from the European Research Council under the Horizon 2020 
Framework Program via the ERC Consolidator Grant CSF-648 505. 

\end{acknowledgements}  

%  N.S., S.B., and P.A. acknowledge support by the ANR-11-BS56-010 
%  project ``STARFICH''.  R.S.K. acknowledges subsidies from the 
%  Deutsche Forschungsgemeinschaft, priority program 1573 (``Physics of 
%  the Interstellar Medium'') and the collaborative research project 
%  SFB 881 (``The Milky Way System'', subprojects B1, B2, and B5).  C. 
%  Federrath acknowledges the Australian Research Council for a 
%  Discovery Projects Fellowship (grant No. DP110102191).  T.Cs. 
%  acknowledges financial support for the ERC Advanced Grant GLOSTAR 
%  under contract no. 247078. 

\begin{appendix}      

%\counterwithin{figure}{section}
%\setcounter{figure}{0} \renewcommand{\thefigure}{A.\arabic{figure}}
  
\section{Line-of-sight contamination correction} \label{app-a}

In \citet{ossenkopf2016}, we simulated the effect of LOS contamination
on the column-density N-PDF of a molecular cloud assuming a
contamination by a typical diffuse cloud that has a total column density below
that of the investigated molecular cloud, but a spread in column
densities that may be wider. We found that the underlying N-PDF of the
observed cloud can be approximately restored from the observations by
treating the contamination like a constant screen, systematically
shifting the column densities.  As this is only a first-order 
correction, we also provided estimates for the residual change of the
N-PDF parameters in terms of the peak position and the N-PDF width in case
of a log-normal distribution.

For many clouds in this paper, we are in a somewhat different regime.
The contamination is given by the Galactic structure that provides a
larger column density but is more homogeneous on the scale of the individual
molecular clouds. Hence, we repeat here the computations from
\citet{ossenkopf2016} for an adjusted parameter range. Here, we allow for
contaminations of up to four times the typical column density of the
considered cloud, but the contaminating cloud has a lower spread in its N-PDF
width of at most half that for the considered cloud. This new
regime covers all configurations from this paper.

\begin{figure}[ht] 
\begin{center}
\includegraphics[angle=0,width=8cm]{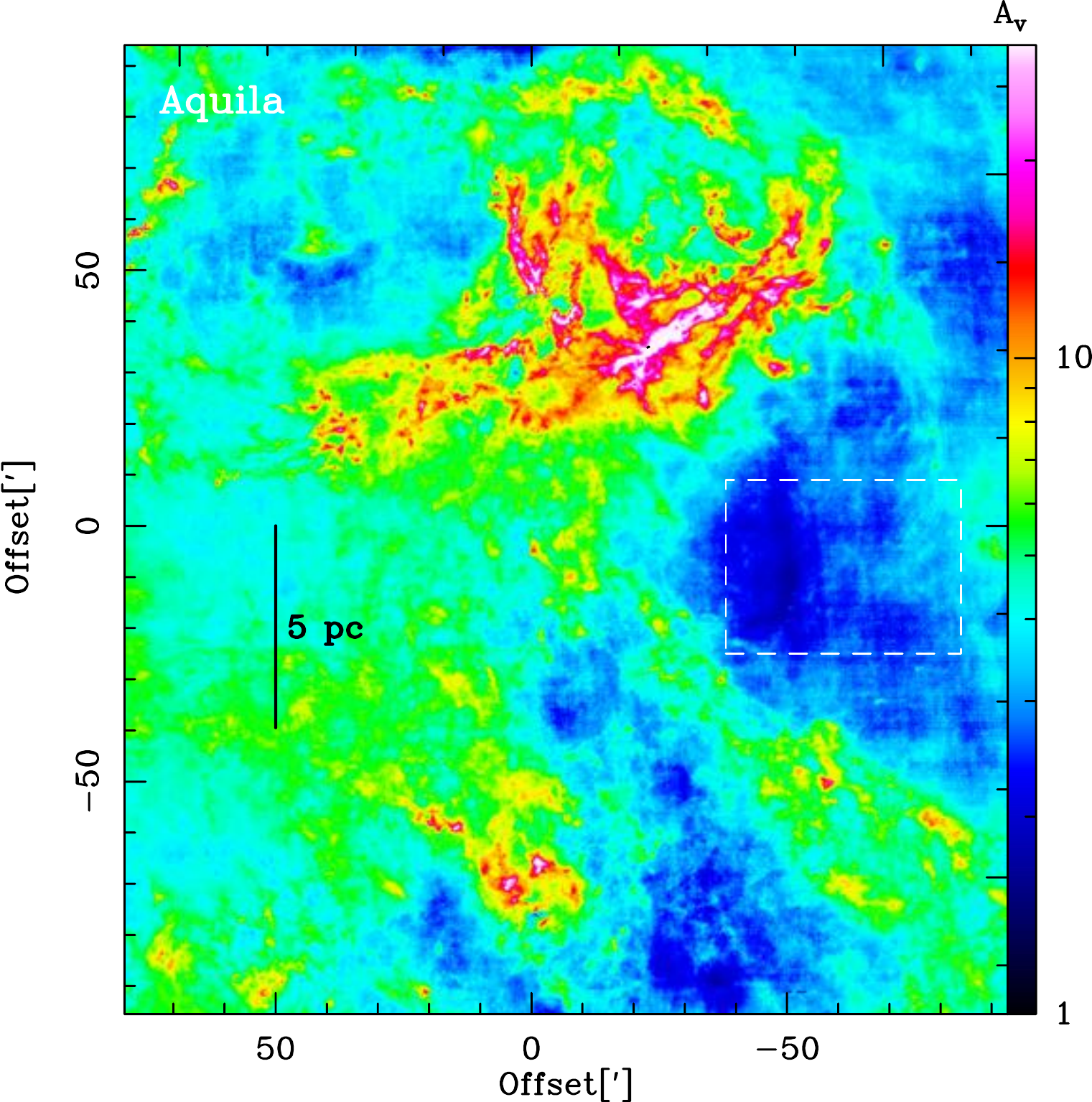} 
\end{center}
\caption[]{LOS-uncorrected column density of the Aquila cloud
  expressed in visual extinction.  The white dashed rectangle
  indicates the region used for evaluating the LOS-contamination by
  (i) taking the mean of all pixels inside the rectangle (\av=2.5)
  and (ii) constructing an N-PDF from these pixels and determining the
  peak of the distribution (\av=2.4, see Figure \ref{etacont}).
}  \label{aquilacont}
\end{figure}

\begin{figure}[ht] 
\begin{center}
\includegraphics[angle=0,width=9cm]{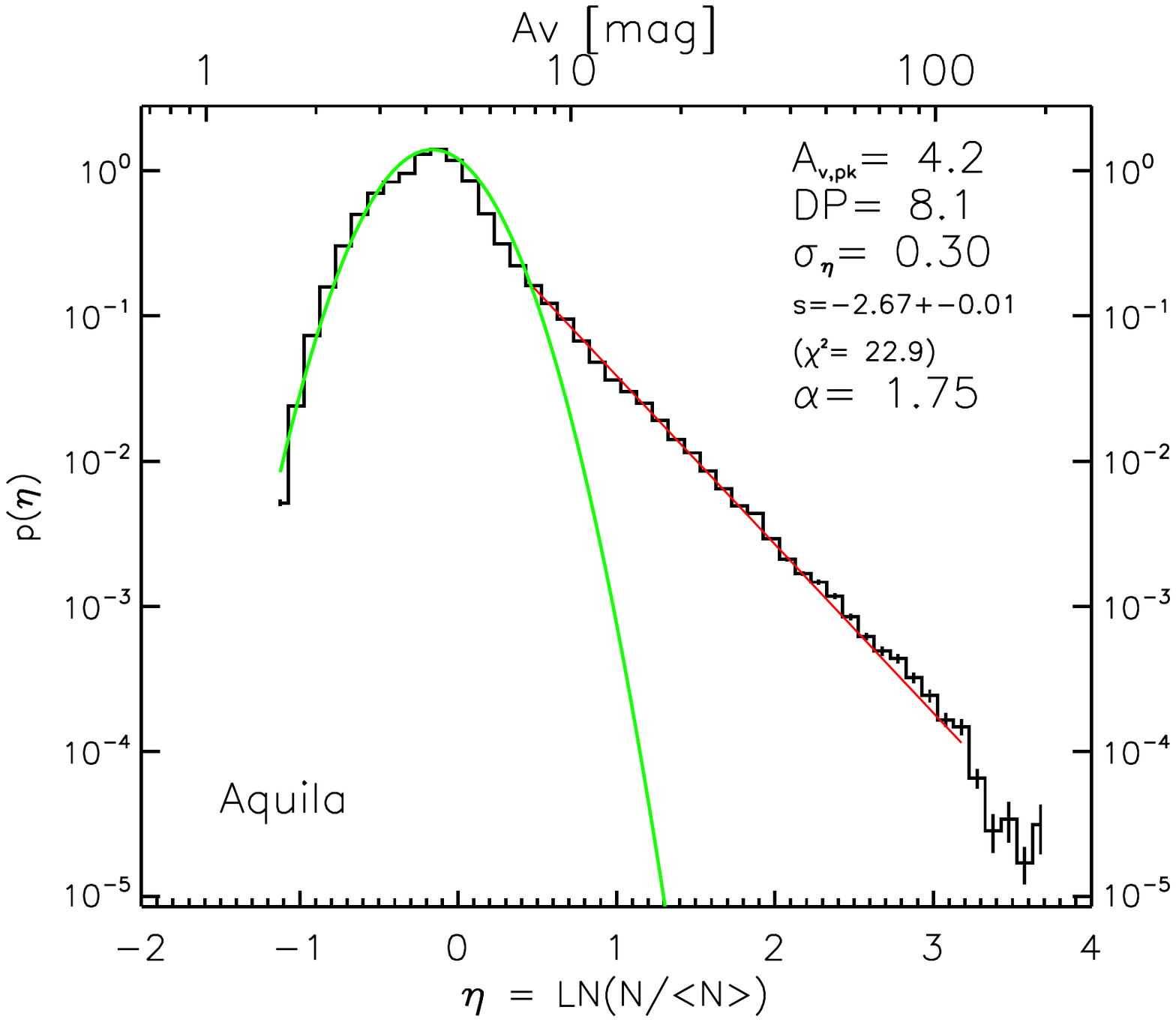}
\vspace{-1cm}
\includegraphics[angle=0,width=9cm]{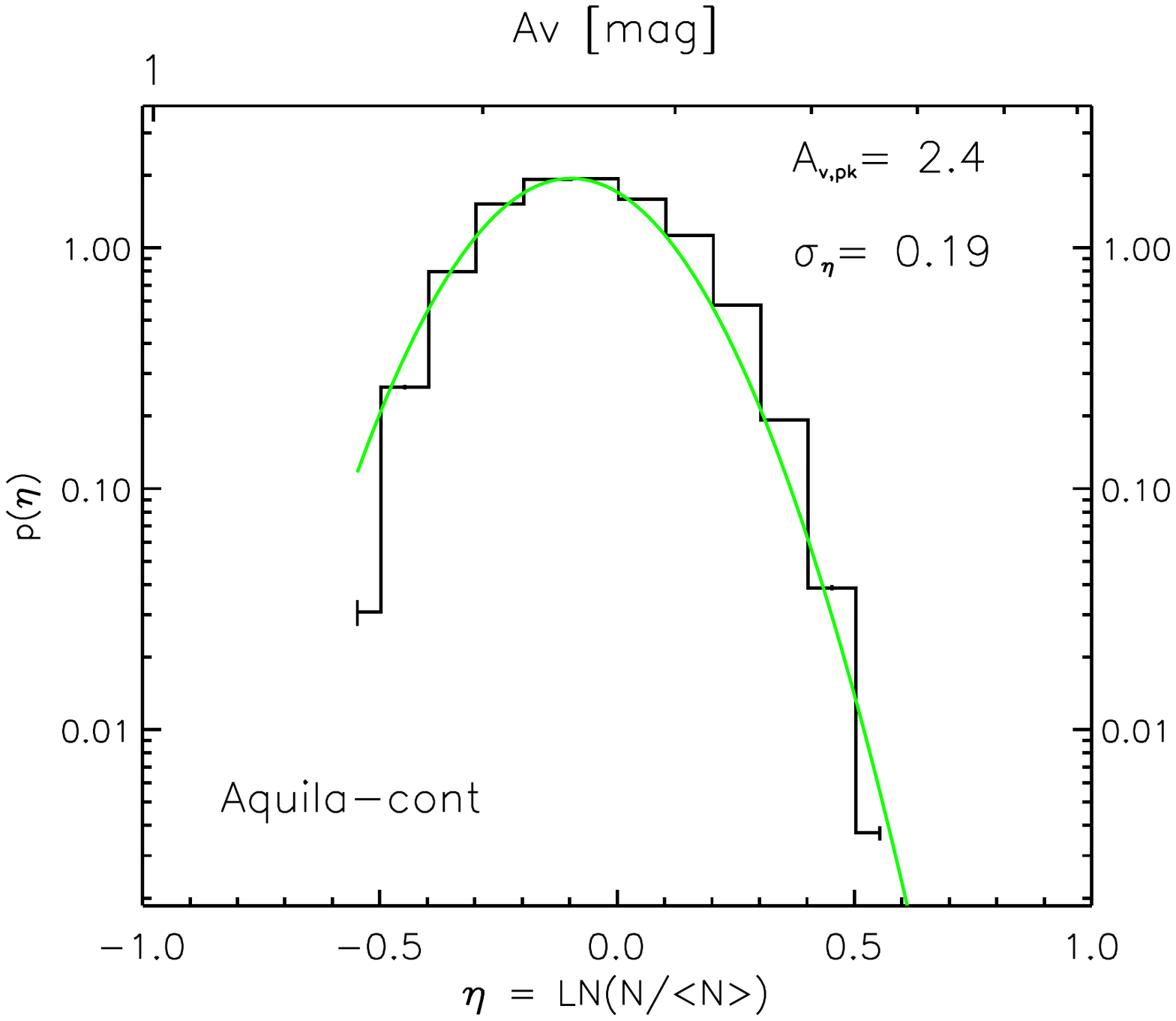} 
\vspace{-1cm}
\includegraphics[angle=0,width=9cm]{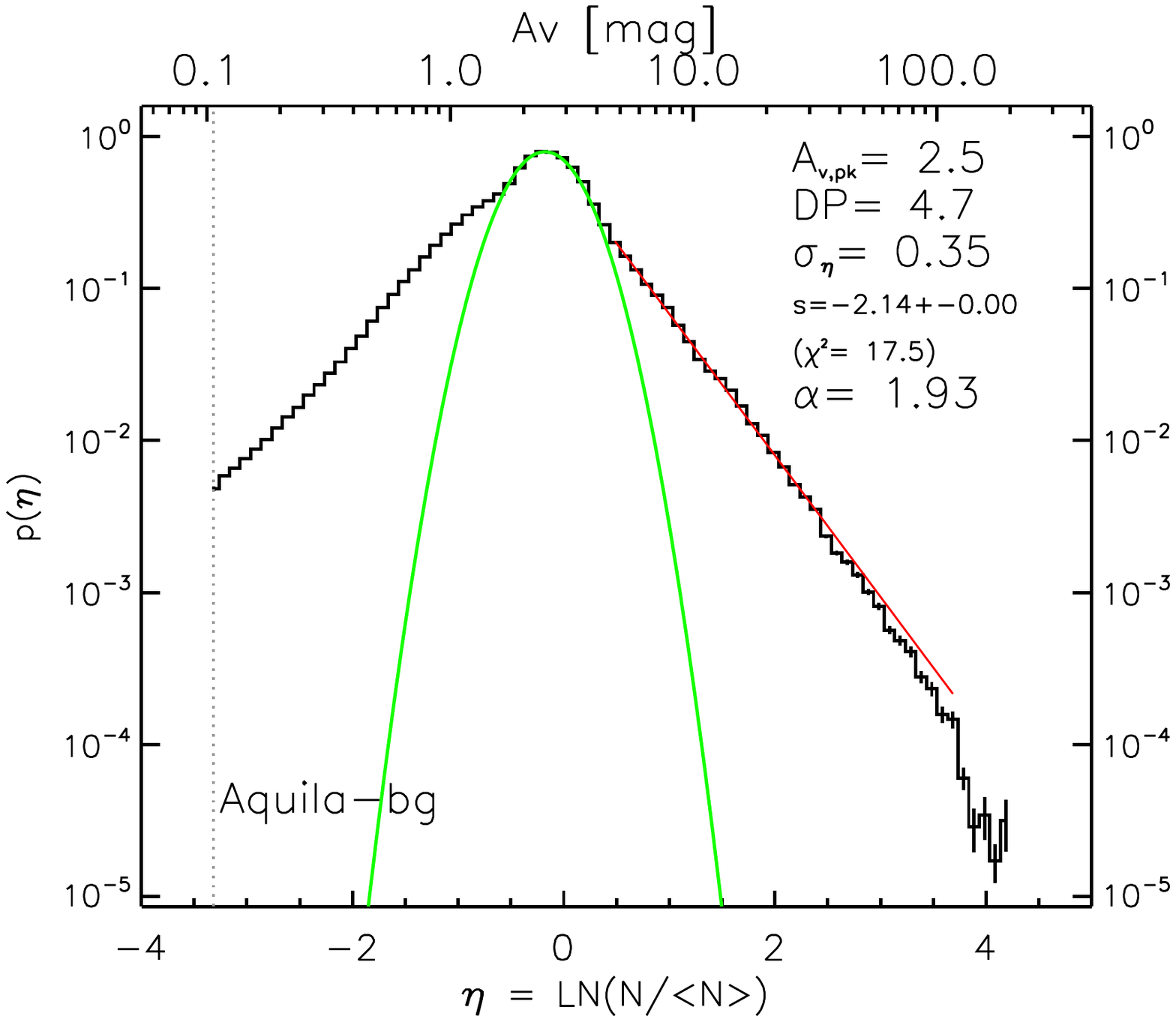} 
\vspace{-1cm}
\end{center}
\caption[]{N-PDFs obtained from the LOS-uncorrected column density map
  (top), from the pixels within the white rectangle from
  Fig. \ref{aquilacont} (middle), and from the LOS-corrected column
  density map (bottom). }  \label{etacont}
\end{figure}

We first illustrate the LOS-contamination correction using the example
of the Aquila cloud and then validate the results. We note that we used Method 1 here 
for the fitting so that the derived values are slightly different to what was obtained 
with Method 4. Figure \ref{aquilacont} shows the uncorrected Aquila column density map and
the rectangular subregion with the lowest level of emission that we
consider to be a measure for the contamination level. The mean of the
pixel values inside this latter area is \av=2.5. From these pixels, we
obtained the N-PDF of the contamination. In the Aquila case, this area 
has 468$\times$350 pixels on a 5.8$''$ grid, which was sufficient to
obtain a reliable N-PDF. For all other clouds, we had similar 
contamination sample sizes. Most of the N-PDFs show a clear log-normal distribution like
Aquila, and only in Serpens we did obtain an additional PLT. Figure
\ref{etacont} (top) displays the N-PDFs of Aquila obtained from the
original (LOS-uncorrected) column density map. The middle panel shows
the N-PDF constructed from the pixels in the rectanglur subregion, and on
the bottom the final N-PDF derived from the LOS-corrected map (using a
constant value of \av=2.5) is displayed. The peak value and width of
the contaminated N-PDF are then used to calculate the ratios
$\sigma_{\eta,cont}$/$\sigma_{\eta,cloud}$ and $N_{contam}$/$N_{peak}$
to assesss if the removal of a constant value is an adequate choice,
following \citet{ossenkopf2016}. The ratio
$\sigma_{\eta,cont}$/$\sigma_{\eta,cloud}$ for Aquila is 0.19, the
highest value for all clouds, and the ratio $N_{contam}$/$N_{peak}$ is
0.82.  We note that the normalization of the N-PDF differs between
Paper I and \citet{ossenkopf2016}. While
Paper I uses the normalization $\eta\equiv\rm
ln\frac{N}{\langle N\rangle}$, \citet{ossenkopf2016} take the
logarithmic peak $N_{peak}$ for log-normal distributions as they center 
them at $\eta$=0. The relation between the two normalizations is
$N_{peak}(N) = {\langle N \rangle} \,\, \exp(-1.5
\,\sigma_{\eta}^2)$.

\begin{figure}[ht] 
\begin{center}
\includegraphics[angle=0,width=8cm]{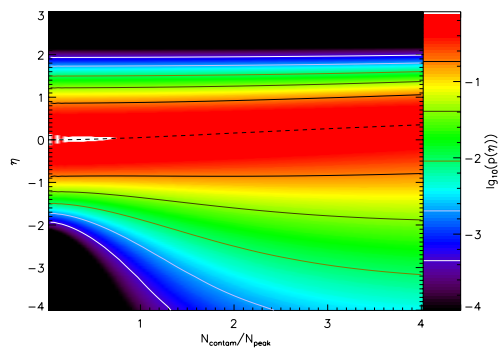} 
\end{center}
\caption[]{Two-dimensional representation of the N-PDFs of
    contaminated clouds after applying the constant screen correction
    as a function of the ratio between the contamination strength
    $N_{contam}$ and the typical cloud column density
    $N_{peak}$. The width of the cloud N-PDF was assumed to be
    $\sigma_{\eta,cloud}=0.5$ and the width of the contamination
    $\sigma_{\eta,cont}=0.15$, using some typical values for the
    clouds in this paper.  The N-PDFs are represented through colors
    showing the logarithm.}  \label{etalargecont}
\end{figure}

\begin{figure}[ht] 
\begin{center}
\includegraphics[angle=0,width=8cm]{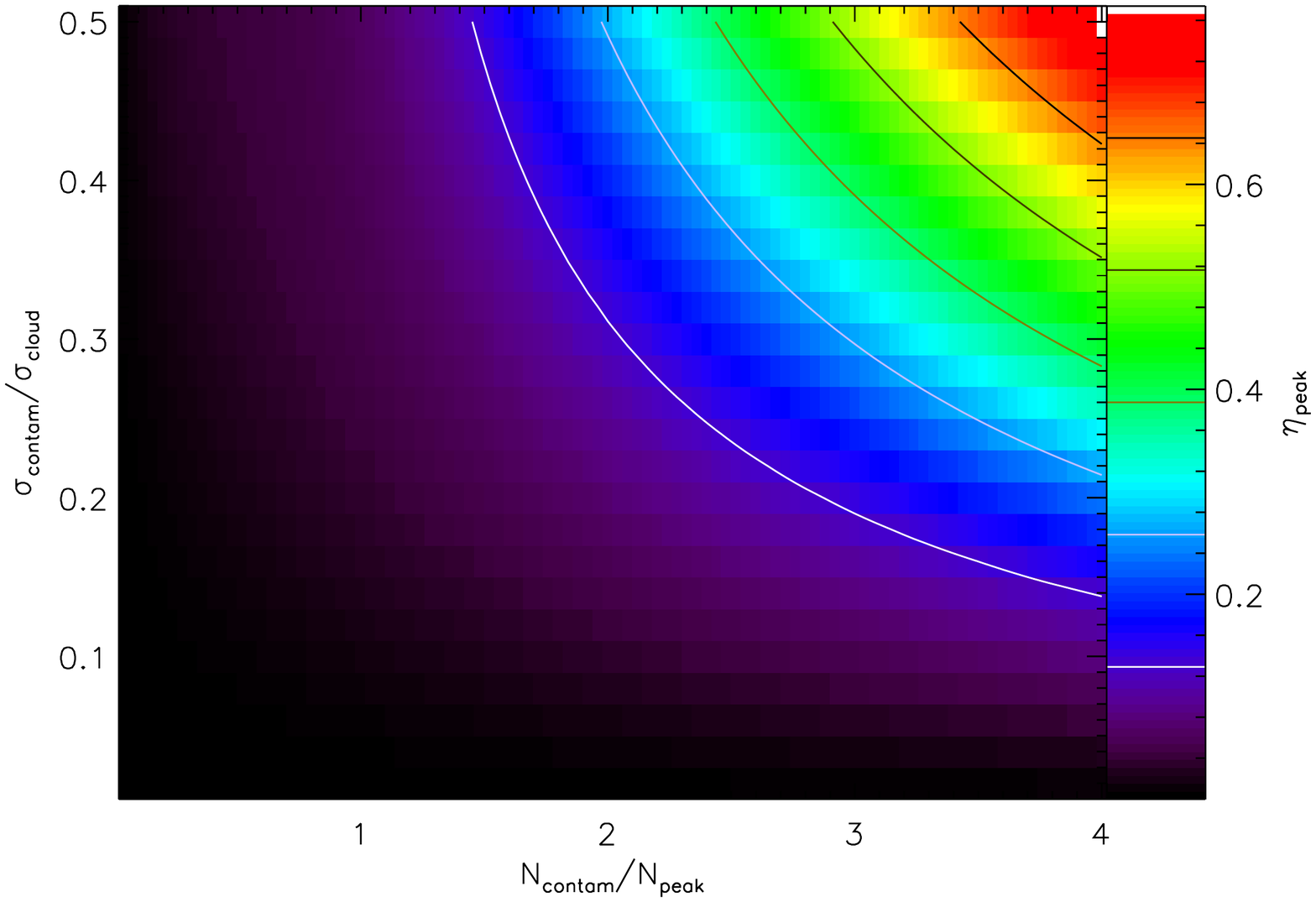} 
\includegraphics[angle=0,width=8cm]{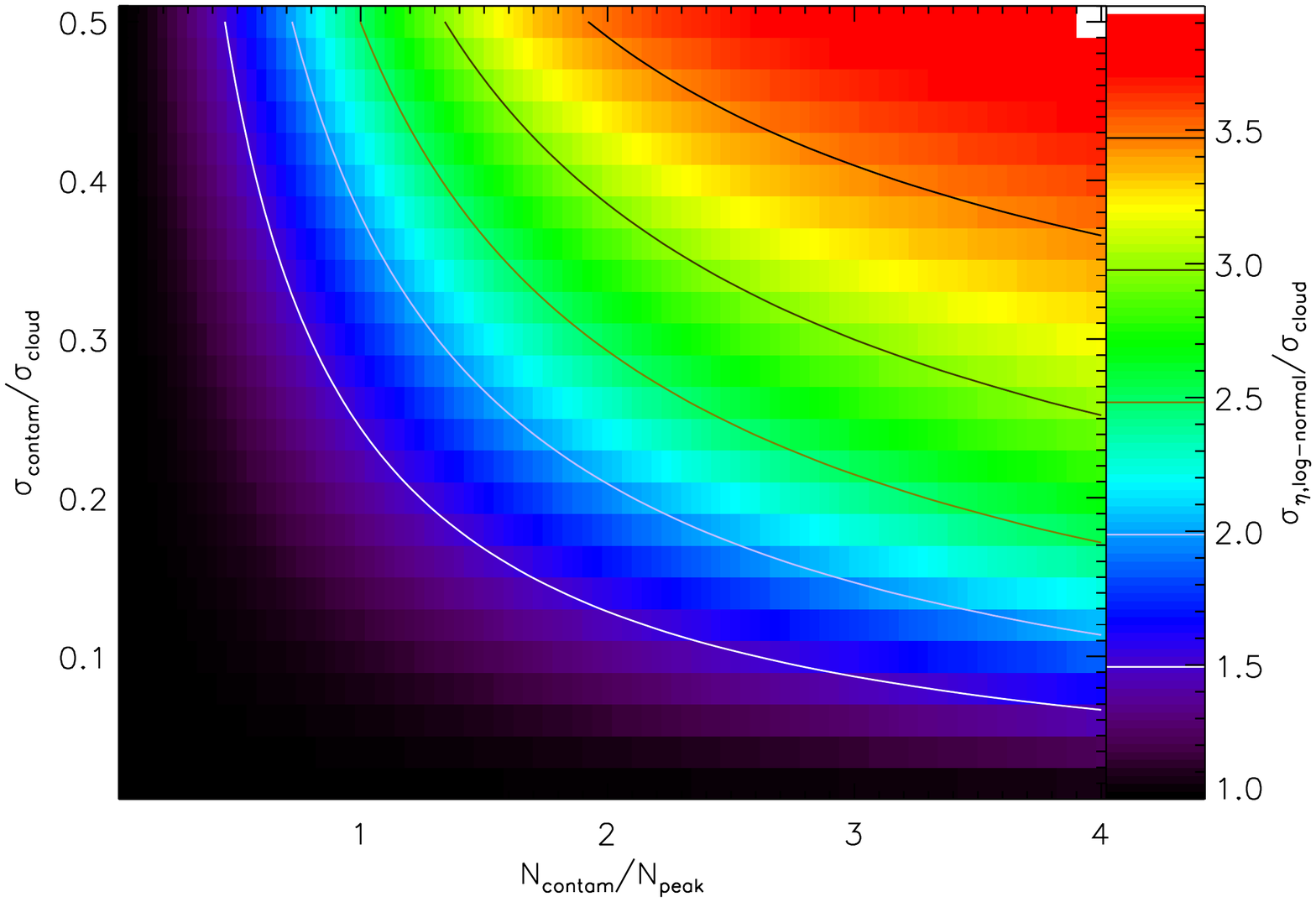} 
\end{center}
\caption[]{Parameters of the N-PDFs of contaminated clouds corrected for
  the contamination through the subtraction of a constant offset given
  by the peak of the N-PDF of the contaminating structure.  The upper
  plot shows the position of the N-PDF peak on the logarithmic column
  density scale $\eta$ relative to the original peak column of the
  cloud. Consequently a value of 0 represents the correct peak
  position; a value of 0.5 indicates a 65\,\% overestimate of the
  column. The lower plot shows the width of the corrected cloud N-PDF
  relative to the original cloud N-PDF. On the horizontal axis, we varied
  the amplitude of the contamination, while on the vertical axis its width
  relative to the cloud N-PDF width. }  \label{contcorr}
\end{figure}

Figures~\ref{etalargecont} and \ref{contcorr} show the results
equivalent to Figs.~13 and 14 from \citet{ossenkopf2016},
respectively.  Figure~~\ref{etalargecont} displays the distribution of
resulting N-PDFs obtained when convolving a log-normal cloud N-PDF
with a second log-normal ``contaminating'' N-PDF and subsequently
correcting this contamination by subtracting a constant $N_{contam}$
contamination column density. Each vertical line in the plot
represents one reconstructed N-PDF. The horizontal axis gives the
dependence on the ratio between the contamination and the cloud column
densities $N_{contam}/N_{peak}$. Here, the cloud and contamination
column densities, $N_{peak}$ and $N_{contam}$, denote the most probable
column density on a logarithmic scale, providing the peaks of the
log-normal N-PDFs. In this example, the width of the cloud N-PDF was
assumed to be $\sigma_{\eta,cloud}=0.5$ and the width of the
contamination $\sigma_{\eta,cont}=0.15$. The distribution at the left
edge represents the original cloud N-PDF because it was computed for a
contamination and correction with $N_{contam}=0$.  Like in
\citet{ossenkopf2016}, we find a good reproduction of the central part
of the original N-PDF for the whole range of contamination amplitudes
but a shift of the N-PDF peak position by up to $\Delta \eta
=0.35$. Moreover, there is a residual broadening of the distribution,
in particular toward lower column densities where the N-PDF becomes
shallower.  The logarithmic scale used in the plot, however, strongly
emphasizes these deviations. They actually occur at levels of less than
1\,\% of the N-PDF peak.

For the description of these residuals after the constant screen
correction, we also extended the parameter scan from
\citet{ossenkopf2016} over the full parameter range
$N_{contam}/N_{peak}$=0$\dots$4 and
$\sigma_{\eta,cont}/\sigma_{eta,cloud}$=0$\dots$0.5.
Figure~\ref{contcorr} shows the results.  The shift of the peak of the
N-PDF, shown in the upper plot, goes up to a factor of 2 in column
density for strong contaminations with a large width.  The lower plot
shows the broadening of the distribution relative to the original
cloud value. With these cases in hand, we can look up the most extreme
cases from our cloud sample.

Summarizing, a LOS-correction has a clear influence on the N-PDF parameters and
needs to be considered before  all further analysis. In the Aquila case,
the PLT slope changes from a rather steep one (-2.67) to a flatter one
(-2.14). Earlier estimates of the slope of the uncorrected
column density also obtained steep PLTs with a slope of -2.59
\citep{schneider2013} and -2.9 \citep{koenyves2015}.  The width of the
N-PDF increases from $\eta$=0.30 to 0.35 and the DP moves
from \av=8.1 to 4.6. 

% APPENDIX B

\section{Significance of fitted models and slope comparison} \label{app-b} 

\subsection{BIC information} \label{app-b1} 

Tables~\ref{tab:BIC} and \ref{tab:BIC-weights} give the Bayesian
information criterion (BIC) values and the weights, respectively, for
the clouds in the study that were analyzed with method 4. The most
likely model is the one with the lowest BIC values and the ratios of
the weights gives a measure of how favored a model is compared to
another. There are some models with a weight of 0 because it is just
so low.

\begin{table*}[htb!]
\centering
\caption{Bayesian information criterion values for all clouds and for all models. The model with the minimum BIC is shown in bold. We note that for Draco, no PLT could be fit, the most likely model is the ELL one with a BIC of -382.2. Other models (EL, L, LL, LP etc.) have higher BICs.}
\label{tab:BIC}
\begin{tabular}{l|c|c|c|c|c|c|c|c}
\hline
\hline
Model  & ELP & ELLP & EL2P & ELL2P & LP & LLP & L2P & LL2P \\
\hline
\multicolumn{3}{l}{ {\bf High-mass SF regions}}  &  & \\   
\hline
Cygnus North  & -503.88   & -871.82  & -719.87  & \bf{-914.78}  &  8671.25  &  41.67   & 8300.31   & -174.01 \\
Cygnus South  & -309.28   & -608.02  & -318.54  & \bf{-617.44}  & 11267.88  &  715.80  & 9802.92   & 707.52  \\
M16           &   23.91   & \bf{-536.85}  & -123.23  & -406.03  & 1210.30   & -307.46  & 1045.74   & -190.63 \\
M17           & 4356.28   & 264.01   & 3212.22  & \bf{98.91}   & 23378.30  & 5528.30  & 19586.52  & 5642.49 \\
NGC6334       & 8063.80   & -236.74  & 7851.92  & \bf{-542.82}  & 61030.20  & 3553.64  & 51465.35  & 3469.84 \\
NGC6357       & 3563.59   & -226.32  & 3541.21  & \bf{-250.39}  & 14412.52  & 6246.08  & 11648.62  & 4234.76 \\
NGC7538       &  483.82   &  71.95   & 25.69    & \bf{-64.33}   & 2687.11   & 607.64   & 1798.76   & 601.15 \\
Rosette       & -92.52    & -518.24  & -416.58  & \bf{-546.29}  & 1293.80   & 167.18   & 1033.42   & 175.69 \\
Vela C        &  575.47   & -105.57  & 116.86   & \bf{-352.96}  & 8364.11   & 1178.53  & 2497.10   & 87.91 \\
\hline
\multicolumn{3}{l}{ {\bf Intermediate-mass SF regions}}  &  & \\   
\hline
Aquila        & -283.75   & -618.65  & -415.31  & \bf{-749.90}  & 15509.10  &  991.36  & 13294.37  & 890.44 \\
MonR2         & 221.51    & -65.93   & -447.38  & \bf{-503.61}  & 8809.96   & -299.52  &  7972.57  & -410.60 \\
MonOB1        & 4339.68   & 116.11   & 4239.46  & \bf{-176.29}  & 10159.28  &  774.17  & 10044.87  & 311.15 \\
NGC2264       & 277.97    & 216.35   & 216.46   & \bf{75.87}    & 10524.29  &  745.90  & 10105.42  & 1106.19 \\
Orion B       & 149.43    & 86.70    & -470.14  & \bf{-542.29}  &  144.98   & 129.05   & -474.24   & -492.82 \\
Serpens       & 1099.54   & 145.28   & 546.30   & \bf{90.14}    & 23029.18  & 2674.87  & 19708.70  & 2617.06 \\
\hline
\multicolumn{3}{l}{ {\bf Low-mass SF regions}}  &  & \\   
\hline
ChamI         & 9394.67   & 372.33   & \bf{134.39}   & 217.27   & 43476.62  & 1435.52  & 3104.40  & 1321.47 \\
ChamII        & 404.72    & -264.60  & 324.52   & \bf{-361.85}  & 9221.02   & -148.55  &    7.01  & -237.61 \\
IC5146        & 2062.21   & 1964.76  & 216.23   & \bf{-365.36}  & 2258.17   & 2058.07  &  339.46  & -265.88\\
Lupus~I       &  117.50   &  75.00   & -23.75   & \bf{-36.72}   & 16701.75  &  423.47  & 1761.06  & 542.80 \\
Lupus~III     & 208.33    & -160.58  & -103.01  & \bf{-178.68}  & 280.77    & -43.10   & -22.35   & -116.40 \\
Lupus~IV      & -113.56   & -366.87  & -320.08  & \bf{-425.10}  &   47.10   & -145.50  &  -46.51  & -350.43\\
Perseus       & 155.23    & -359.51  & -446.61  & -441.65  & 1419.38   & -330.10  & \bf{-454.69}  & -442.44 \\
Pipe          & -152.55   & -311.61  & -320.53  & \bf{-375.61}  & -156.14   & -281.39  & -304.53  & -376.78 \\
$\rho$Oph     & 12934.22  & 12189.37 &  450.90  & -248.50  & 12928.56  & 12182.47 & 443.59   & \bf{-254.23} \\
Taurus        & 2085.79   & 2079.25  & -290.70  & -352.77  & 2314.11   & 2026.09  & -99.35   & \bf{-356.92} \\
\hline
\multicolumn{3}{l}{ {\bf Quiescent regions}}  &  & \\   
\hline
ChamIII       & 109.96    & -470.85  &   5.27   & \bf{-487.08}   &  273.33  & -317.40  &  -3.95   & -333.85 \\
Musca         & 3406.39   & -347.94  & -307.86  & -350.00   & 8338.84  & -180.95  &  -221.75 & \bf{-352.48} \\
Polaris       & -245.60   & -353.62  & -258.46  & \bf{-359.30}   & -194.85  & -315.35  &  -217.47 & -318.71 \\
\hline
\end{tabular}
%\tablefoot{}
\end{table*}

\begin{table*}[htb!]
\centering
\caption{Bayesian information criterion weight values for all clouds and for all models. The most likely model is denoted by its weight being bold. If the second most likely model has a weight within a factor of 10 of the most likely model (i.e. may be an alternative model) it is shown with an italic, bold weight. We note that for Draco, no PLT could be fit, the most likely model is the ELL one with a weight of 
9.826514e-01. Other models (EL, L, LL, LP etc.) have higher weights.}
\label{tab:BIC-weights}
\begin{tabular}{l|c|c|c|c|c|c|c|c}
\hline
\hline
Model  & ELP & ELLP & EL2P & ELL2P & LP & LLP & L2P & LL2P \\
\hline
\multicolumn{3}{l}{ {\bf High-mass SF regions}}  &  & \\   
\hline
Cygnus North  & 5.946e-90  & 4.692e-10 & 4.741e-43  & \bf{1.000e+00} & 0.000e+00 & 2.039e-208  & 0.000e+00 & 1.393e-161 \\
Cygnus South  & 1.202e-67  & 8.924e-03 & 1.233e-65  & \bf{9.911e-01} & 0.000e+00 & 3.067e-290  & 0.000e+00 & 1.926e-288 \\
M16           & 1.708e-122 & \bf{1.000e+00} & 1.526e-90  & 3.916e-29 & 0.000e+00 & 1.544e-50   & 0.000e+00 & 6.596e-76 \\
M17           & 0.000e+00  & 1.409e-36 & 0.000e+00  & \bf{1.000e+00} & 0.000e+00 & 0.000e+00   & 0.000e+00 & 0.000e+00 \\
NGC6334       & 0.000e+00  & 3.432e-67 & 0.000e+00  & \bf{1.000e+00} & 0.000e+00 & 0.000e+00   & 0.000e+00 & 0.000e+00 \\
NGC6357       & 0.000e+00  & 5.933e-06 & 0.000e+00  & \bf{1.000e+00} & 0.000e+00 & 0.000e+00   & 0.000e+00 & 0.000e+00 \\
NGC7538       & 9.348e-120 & 2.554e-30 & 2.834e-20  & \bf{1.000e+00} & 0.000e+00 & 1.212e-146  & 0.000e+00 & 3.111e-145 \\
Rosette       & 2.911e-99  & 8.076e-07 & 6.816e-29  & \bf{9.999e-01} & 0.000e+00 & 1.177e-155  & 0.000e+00 & 1.665e-157 \\ 
Vela C        & 2.477e-202 & 1.905e-54 & 9.547e-103 & \bf{1.000e+00} & 0.000e+00 & 0.000e+00   & 0.000e+00 & 1.846e-96  \\
\hline
\multicolumn{3}{l}{ {\bf Intermediate-mass SF regions}}  &  & \\   
\hline
Aquila        & 5.982e-102 & 3.158e-29 & 2.212e-73  & \bf{1.000e+00} & 0.000e+00 & 0.000e+00   & 0.000e+00 & 0.000e+00 \\
MonR2         & 3.485e-158 & 9.099e-96 & 6.163e-13  & \bf{1.000e+00} & 0.000e+00 & 4.813e-45   & 0.000e+00 & 6.355e-21 \\
MonOB1        & 0.000e+00  & 3.207e-64 & 0.000e+00  & \bf{1.000e+00} & 0.000e+00 & 4.076e-207  & 0.000e+00 & 1.425e-106\\
NGC2264       & 1.302e-44  & 3.127e-31 & 2.960e-31  & \bf{1.000e+00} & 0.000e+00 & 3.198e-146  & 0.000e+00 & 1.857e-224 \\
Orion B       & 6.236e-151 & 2.609e-137 & 2.152e-16 & \bf{1.000e+00} & 5.771e-150 & 1.661e-146 & 1.672e-15 & 1.810e-11 \\
Serpens       & 6.480e-220 & 1.063e-12 & 8.833e-100 & \bf{1.000e+00} & 0.000e+00 & 0.000e+00   & 0.000e+00 & 0.000e+00 \\
\hline
\multicolumn{3}{l}{ {\bf Low-mass SF regions}}  &  & \\   
\hline
ChamI         & 0.000e+00  & 2.148e-52 & \bf{1.000e+00}  & 1.007e-18 & 0.000e+00 & 2.905e-283  & 0.000e+00 & 1.694e-258 \\
ChamII        & 3.479e-167 & 7.628e-22 & 9.050e-150 & \bf{1.000e+00} & 0.000e+00 & 4.814e-47   & 8.000e-81 & 1.051e-27 \\
IC5146        & 0.000e+00  & 0.000e+00 & 5.121e-127 & \bf{1.000e+00} & 0.000e+00 & 0.000e+00   & 8.918e-154 & 2.501e-22\\
Lupus~I       & 3.243e-34  & 5.491e-25 & 1.524e-03  & \bf{9.985e-01} & 0.000e+00 & 1.176e-100  & 0.000e+00 & 1.439e-126  \\
Lupus~III     & 9.158e-85  & 1.174e-04 & 3.702e-17  & \bf{9.999e-01} & 1.705e-100 & 3.623e-30  & 1.131e-34 & 2.992e-14 \\
Lupus~IV      & 2.238e-68  & 2.267e-13 & 1.567e-23  & \bf{1.000e+00} & 2.905e-103 & 1.930e-61  & 6.169e-83 & 6.104e-17\\
Perseus       & 3.530e-133 & 2.099e-21 & 1.725e-02  & 1.444e-03 & 0.000e+00  & 7.782e-26  & \bf{9.792e-01} & 2.134e-03 \\
Pipe          & 1.308e-49  & 4.531e-15 & 3.919e-13  & \textit{\textbf{3.578e-01}} & 7.876e-49 & 1.242e-21   & 1.315e-16 & \bf{6.422e-01}\\
Taurus        & 0.000e+00  & 0.000e+00 & 3.708e-15  & \textit{\textbf{1.116e-01}} & 0.000e+00 & 0.000e+00   & 1.042e-56 & \bf{8.884e-01}\\
\hline
\multicolumn{3}{l}{ {\bf Quiescent regions}}  &  & \\   
\hline
ChamIII       & 2.261e-130 & 2.989e-04 & 1.223e-107 & \bf{9.997e-01} & 7.567e-166 & 1.427e-37  & 1.229e-105 & 5.326e-34 \\
Musca         & 0.000e+00  & 7.418e-02 & 1.469e-10  & \textit{\textbf{2.078e-01}} & 0.000e+00 & 4.063e-38   & 2.941e-29 & \bf{7.180e-01}\\
Polaris       & 1.931e-25  & 5.520e-02 & 1.197e-22  & \bf{9.448e-01} & 1.843e-36 & 2.702e-10   & 1.504e-31 & 1.450e-09 \\
\hline
\end{tabular}
%\tablefoot{}
\end{table*}

\subsection{Comparison between different slope determinations}  \label{app-b2} 
As pointed out in Sec. 2.4.1, the three methods used to fit the N-PDF
differ in their premises. In contrast to Methods 1 and 2, the adapted
BPLFIT technique (Method 3, see \citet{veltchev2019} extracts 
possible PLTs, without any assumption on the rest of the
distribution. Marinkova et al. (2020; in preparation) modified this
technique further -- through introduction of varying lower and upper
density cutoffs -- to allow for extraction of a second PLT.

It is therefore instructive to compare the slopes obtained through
Method 3 with those from the other two.  In general, the
BPLFIT slopes correlate well with their counterparts.  Method 2
extracts only single PLTs but some of them are identified with the
first PLT (with a single exception) obtained through Method 3 (right
panel). A few more significant discrepancies are due to differences
between the estimated DPs. Methods 1 and 3 agree on the existence of two
PLTs in eight  studied regions of all types, with a good agreement between
the obtained slope values (open squares in the left panel). For the
rest of the regions, the first PLTs from Method 1 are typically
identified with single PLTs extracted by Method 3 (filled
triangles). The correlation between the slope values is even better
than in the regions for which both techniques extract two PLTs.

\begin{figure}[ht] 
\begin{center}
\includegraphics[angle=0,width=9cm]{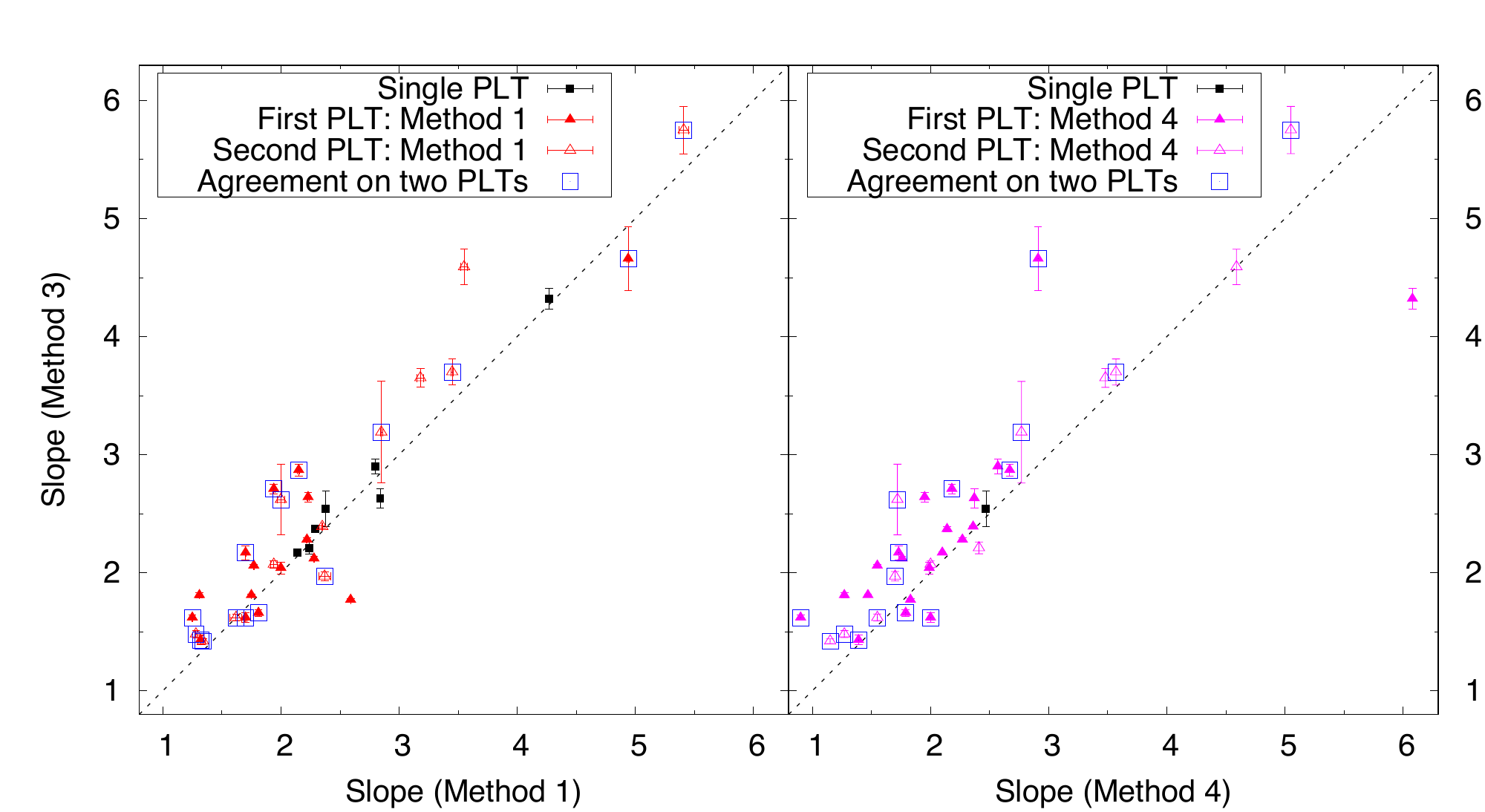} 
\end{center}
\caption[]{Comparison of slope estimations from the three PLTs 
  fitting methods described in Sec.~\ref{sec:pdfs}.  Absolute slope
  values are given and the identity line is plotted as a
  dashed line. Different symbols are used for one or two PLTs. We note 
  that the methods do not always agree on the number of
  PLTs.} \label{methods}
\end{figure}

% APPENDIX C

\section{Column density maps, N-PDFs and $\Delta$-variance spectra} \label{app-c}  

Figures C.1-C.29 display the column density maps expressed in
visual extinction on the left and the $\Delta$-variance and N-PDF on
the right.  The plot range is 0.5 to 200 in \av\ for high-mass SF
molecular clouds. Since all these clouds are affected by LOS contamination, 
we show the corrected column density maps and N-PDFs. The plot
range for intermediate-mass SF clouds is 0.5 to 100 in \av, for
low-mass SF clouds 0.1 to 20 in \av, for quiescent clouds 0.1 to 10 in
\av, and for Draco 0.1 to 2 in \av.  The N-PDFs are
presented as they are without the fits to the log-normal part and
PLT(s). These are shown separately in Appendix~D.  As outlined in Sec. 
\ref{sec:stat}, the values of the slope(s) of the PLT(s), the
DP and the width of the log-normal part are taken from
the fit of method 4.

% *******************************************************************
% HIGH-MASS SF REGIONS 
% *******************************************************************

% Cygnus North (DR21) 
\begin{figure*}[ht] 
\begin{center}
\includegraphics[angle=0,width=17cm,height=9cm,keepaspectratio]{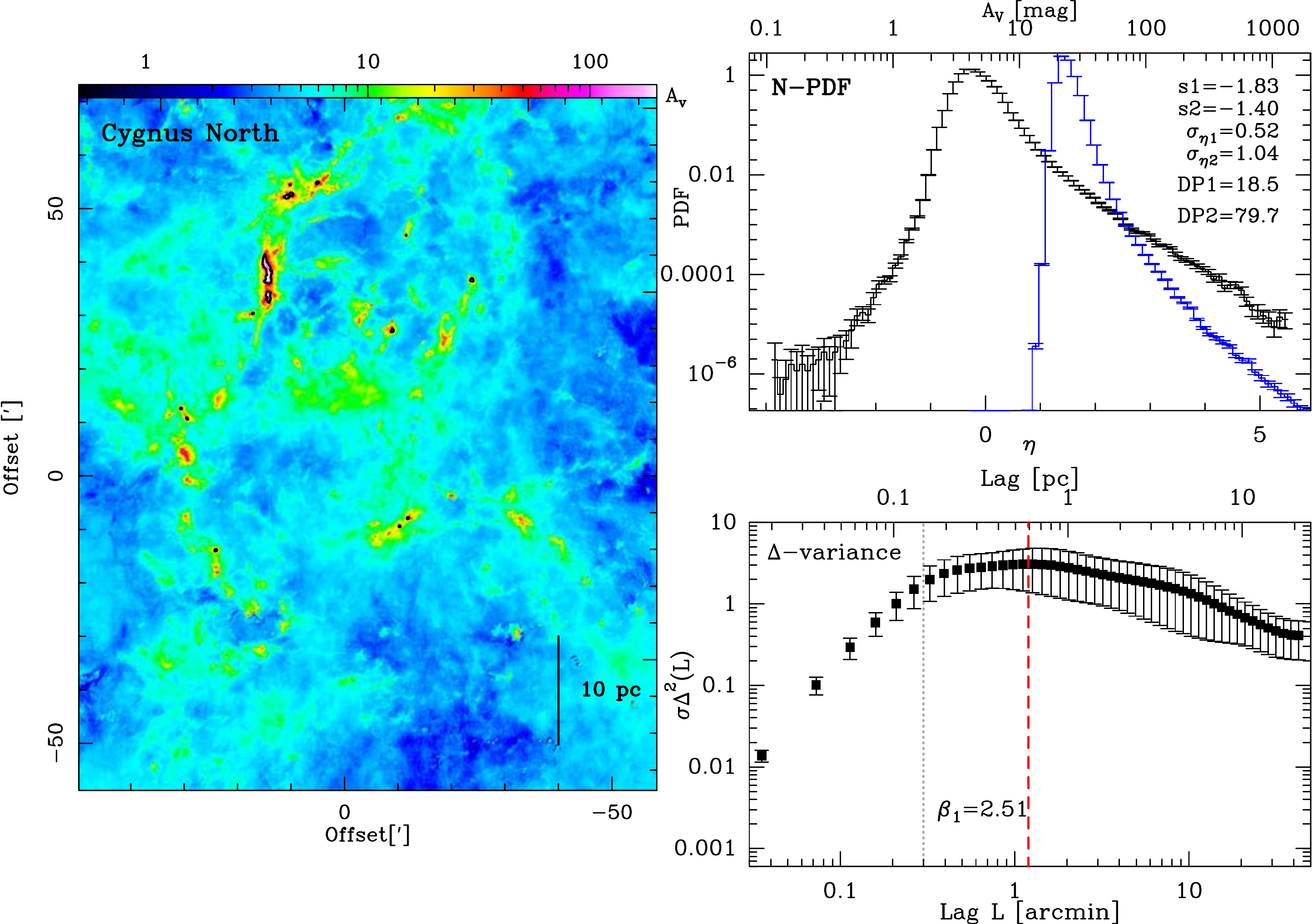} 
\end{center}
\caption[]{CYGNUS X NORTH (DR21): {\bf Left:} LOS corrected
  column density map in visual extinction. The image is rotated and a
  length scale is given in the panel. For the cloud N-PDFs with two PLTs,
  the contour of the second DP is plotted in black. {\bf
    Right (lower panel):} $\Delta$-variance spectrum. X-axis units are
  arcmin (bottom) and parsec (top).  The black dashed line indicates
  the angular resolution (18$''$), the first red dashed line indicates
  the upper limit for the fit of $\beta_1$ (the lower limit is the
  resolution limit) and in case there are two fitting intervalls, two
  other red dashed lines indicate the fit range for $\beta_2$.  The
  values of $\beta_1$ and $\beta_2$ are given in the panel. The errors
  are omitted for better visibility, they are always on the order of
  0.01 to 0.03.  {\bf Right (top panel):} N-PDF of LOS corrected
  column density in black, expressed in visual extinction (upper
  x-axis) and in $\eta$ (lower x-axis). For comparison, the N-PDF of
  the uncorrected map is displayed in blue.  The slope s of the
  PLT, the width $\sigma$ (expressed in $\eta$) of the
  log-normal part of the (corrected) N-PDF, and the deviation point DP
  (expressed in \av) are given in the panel. }  \label{dr21}
\end{figure*}

% Cygnus South (DR15) 
\begin{figure*}[ht] 
\begin{center}
\includegraphics[angle=0,width=18cm,height=10cm,keepaspectratio]{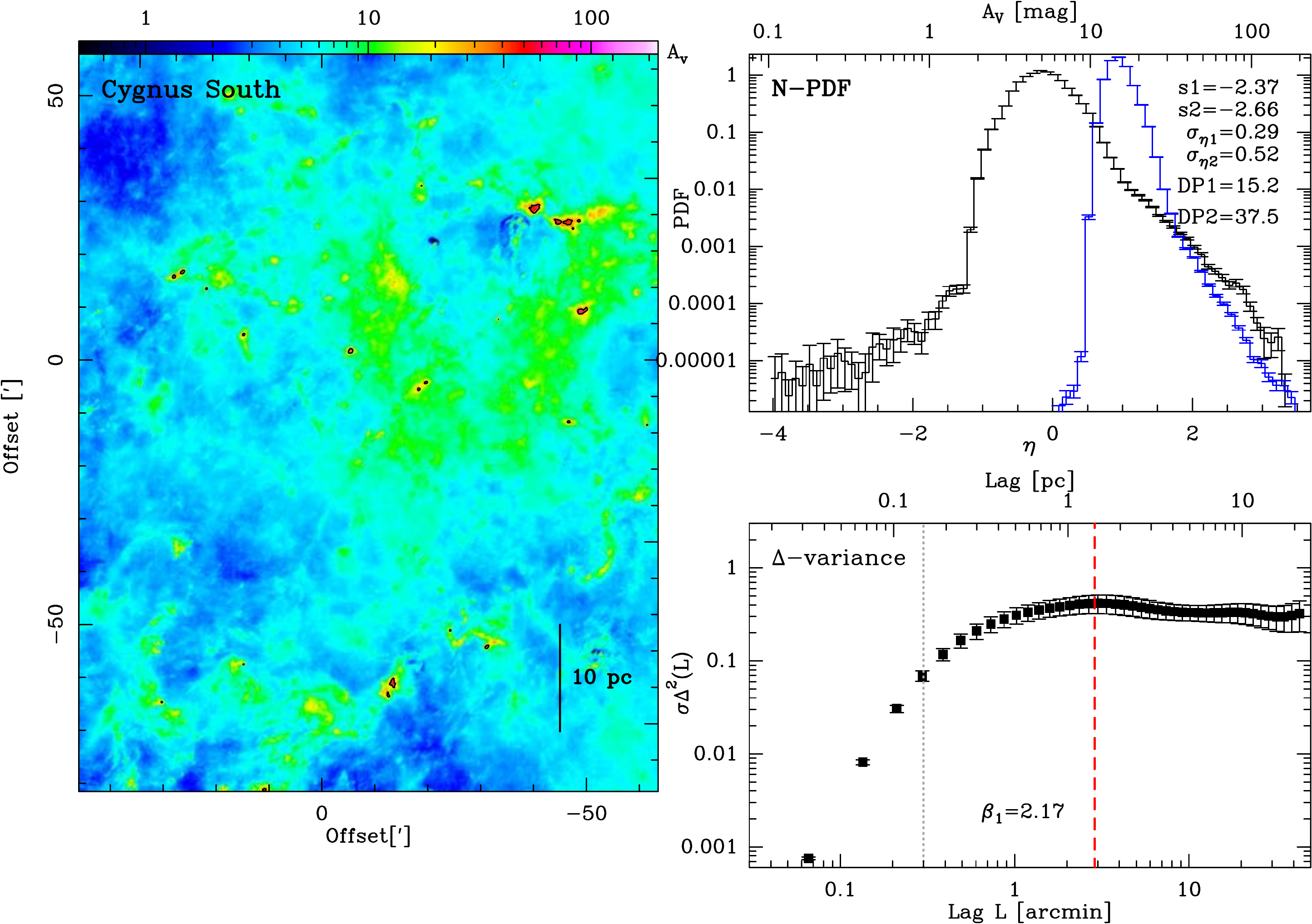} 
\end{center}
\caption[]{CYGNUS X SOUTH (DR15): Fig. caption see Fig. C.1.} \label{dr15} 
\end{figure*} 
%\vspace{-5cm}

\clearpage

% M16
\begin{figure*}[ht]
\begin{center}
\includegraphics[angle=0,width=18cm,height=10cm,keepaspectratio]{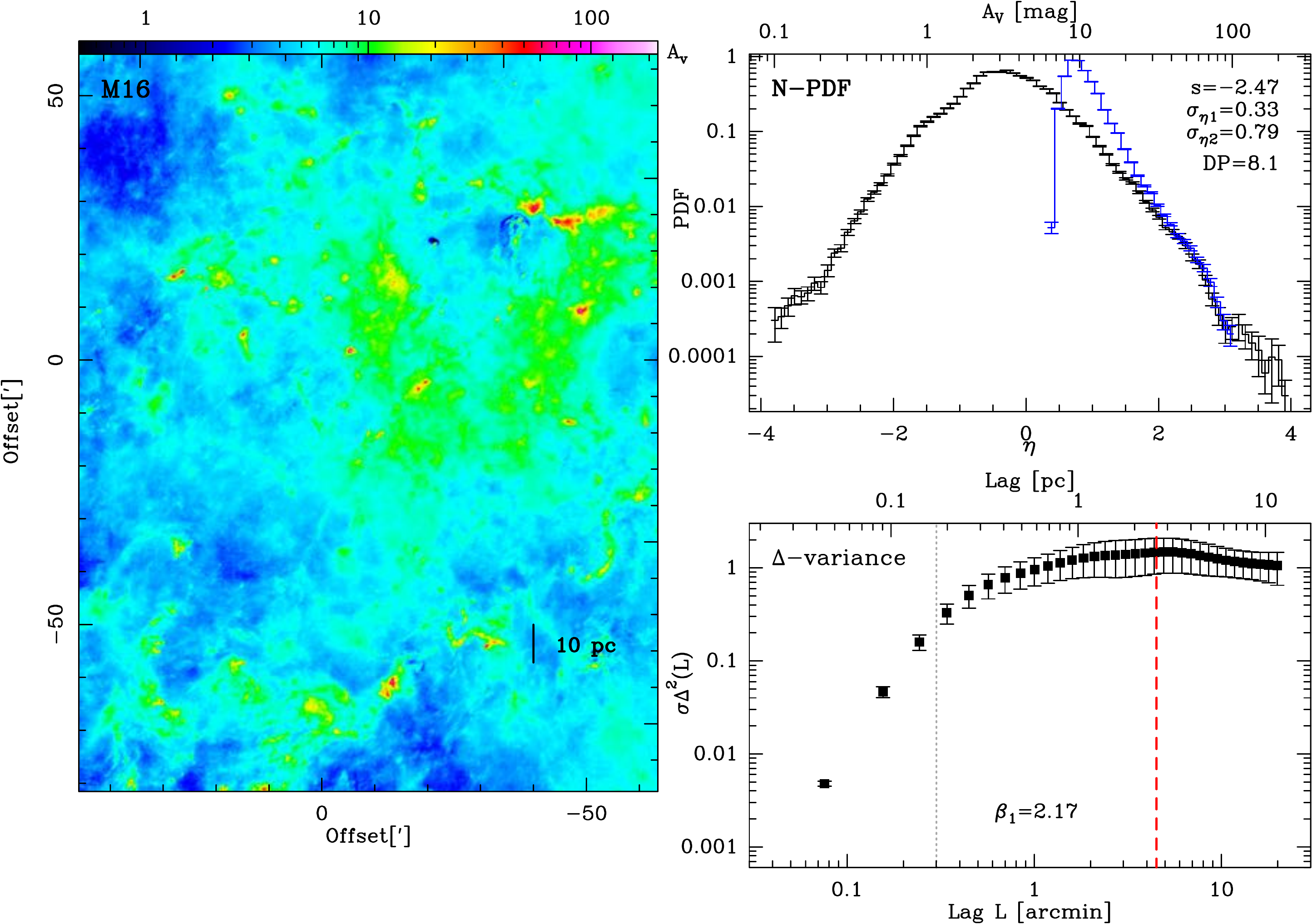} 
\end{center}
\caption[]{M16: Fig. caption see Fig. C.1.}   \label{m16} 
\end{figure*}

% M17
\begin{figure*}[ht] 
\begin{center}
\includegraphics[angle=0,width=18cm,height=10cm,keepaspectratio]{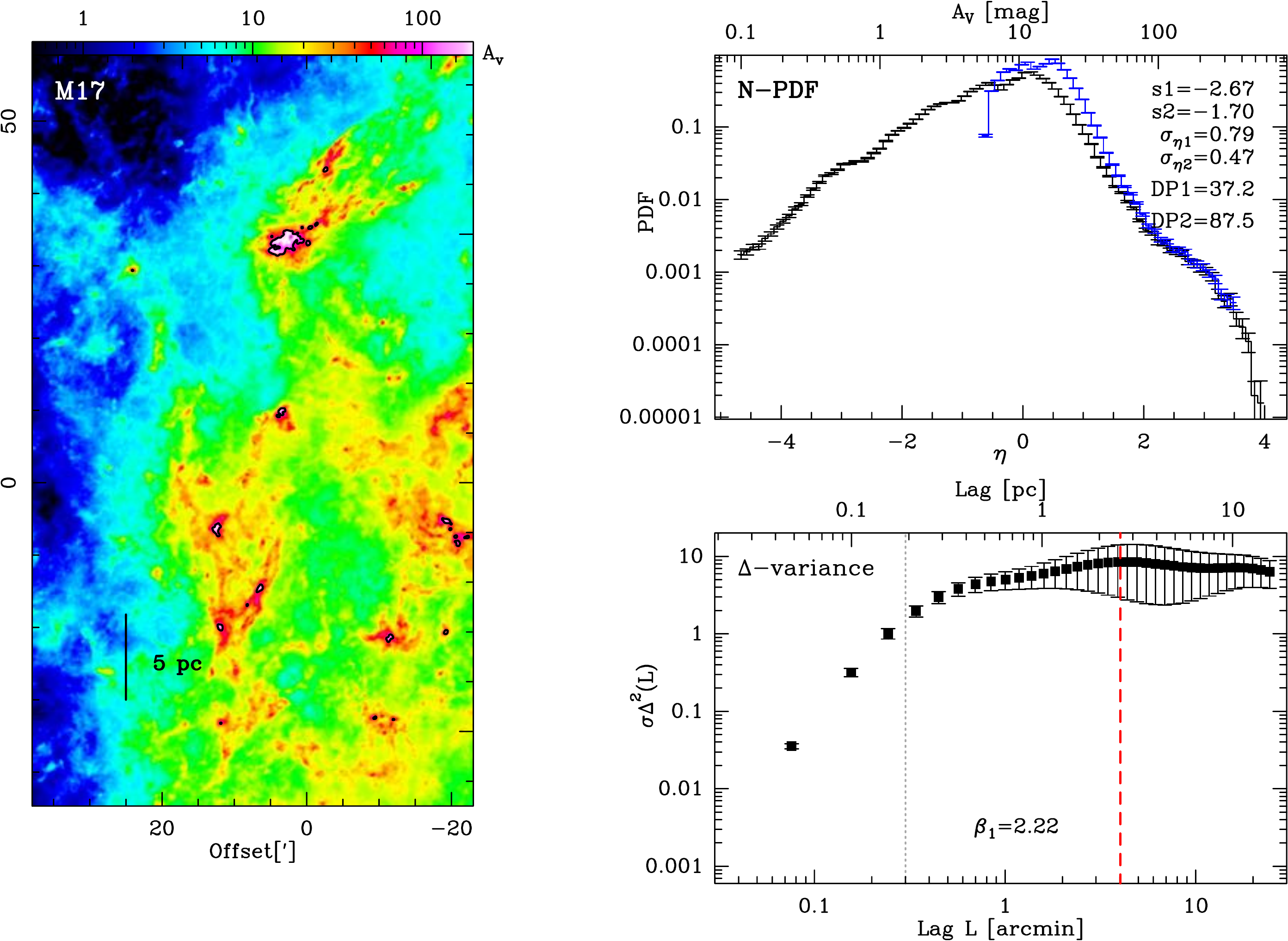} 
\end{center}
\caption[]{M17: Fig. caption see Fig. C.1.}  \label{m17} 
\end{figure*}
%\vspace{-5cm}

% NGC6334
\begin{figure*}[ht] 
\begin{center}
\includegraphics[angle=0,width=18cm,height=10cm,keepaspectratio]{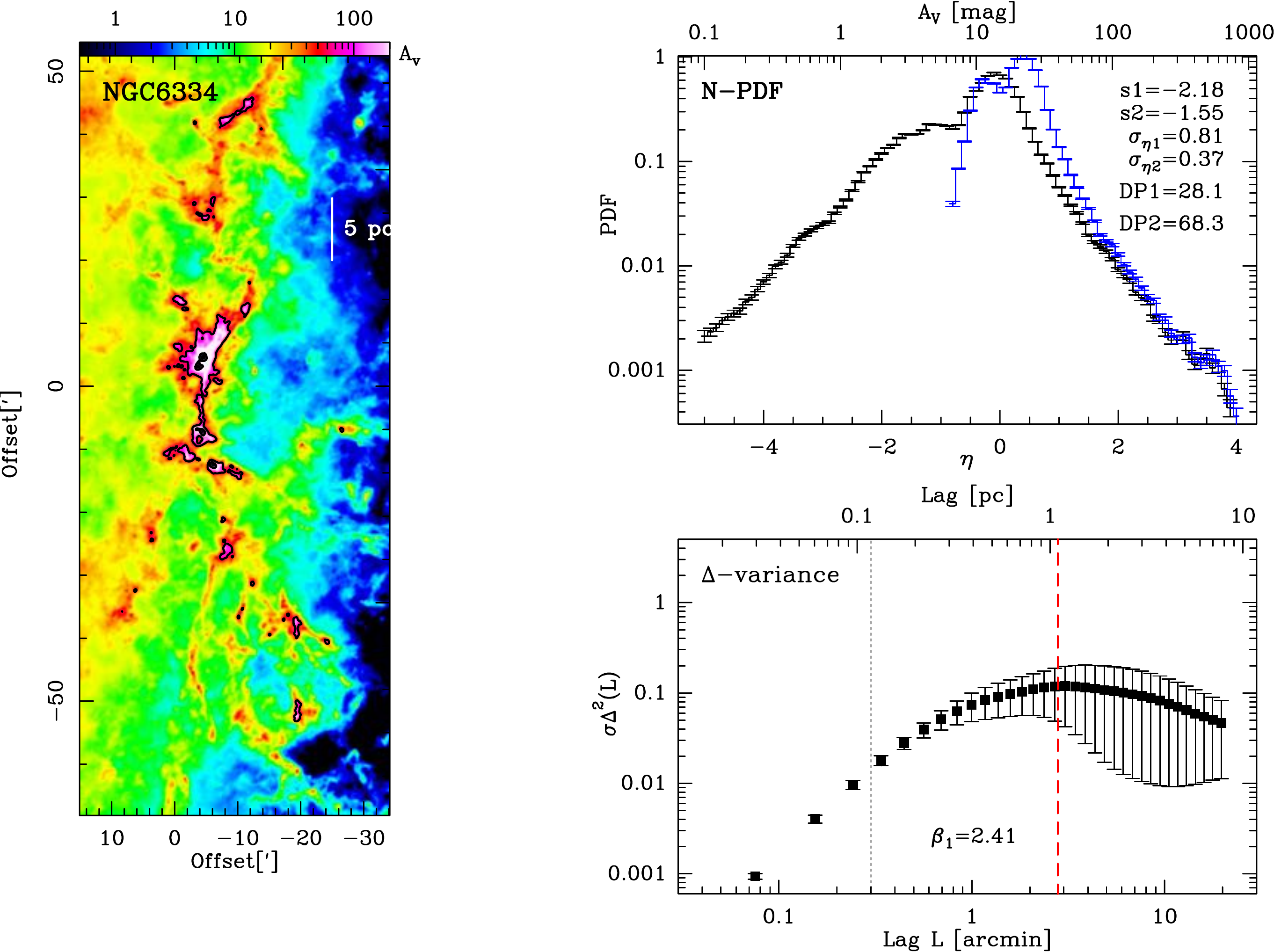} 
\end{center}
\caption[]{NGC6334: Fig. caption see Fig. C.1.} \label{ngc6334} 
\end{figure*} 

% NGC6357
\begin{figure*}[ht] 
\begin{center}
\includegraphics[angle=0,width=18cm,height=10cm,keepaspectratio]{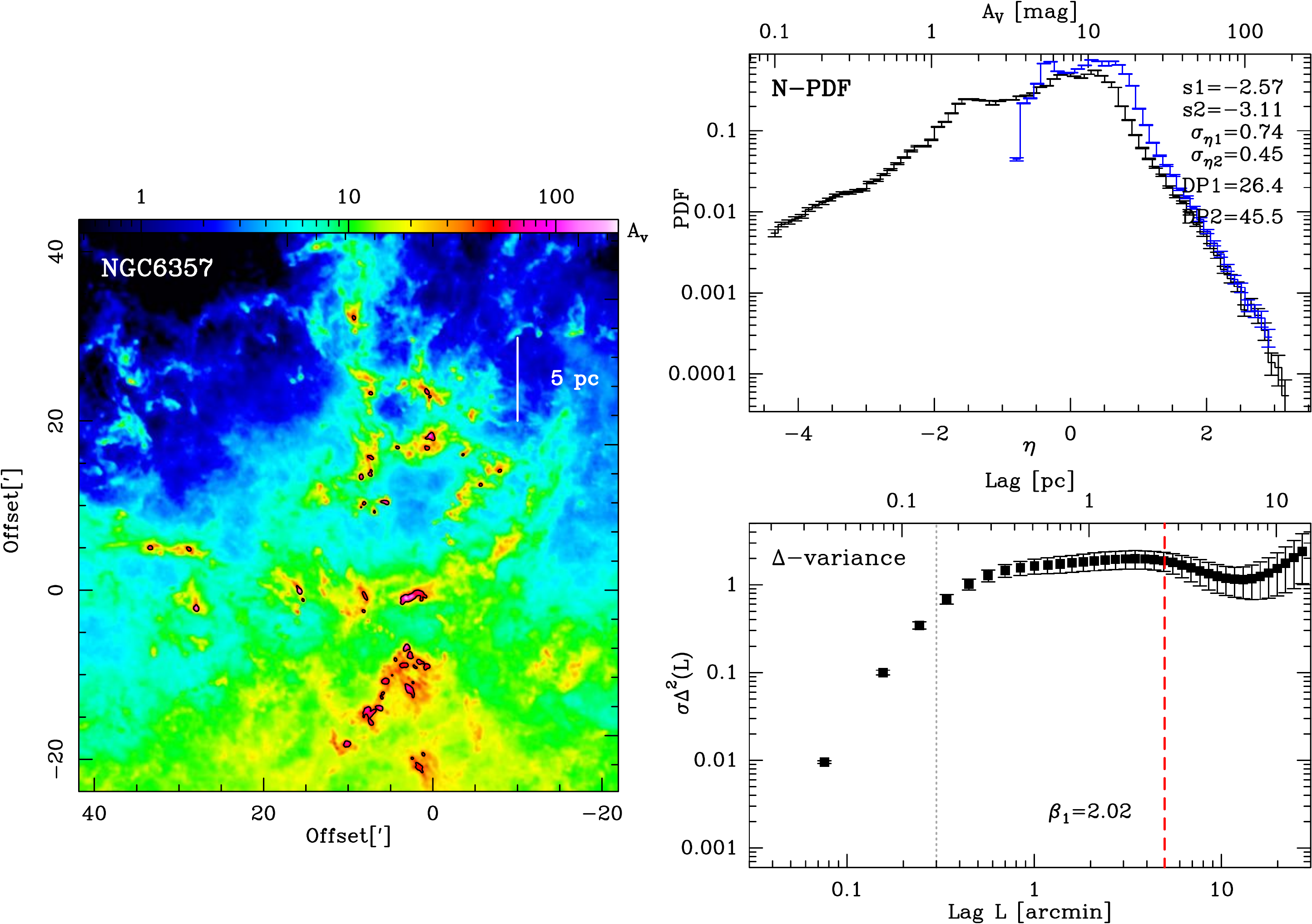} 
\end{center}
\caption[]{NGC6357: Fig. caption see Fig. C.1.} \label{ngc6357} 
\end{figure*}

\clearpage

% NGC7538
\begin{figure*}[ht]
\begin{center}
\includegraphics[angle=0,width=18cm,height=10cm,keepaspectratio]{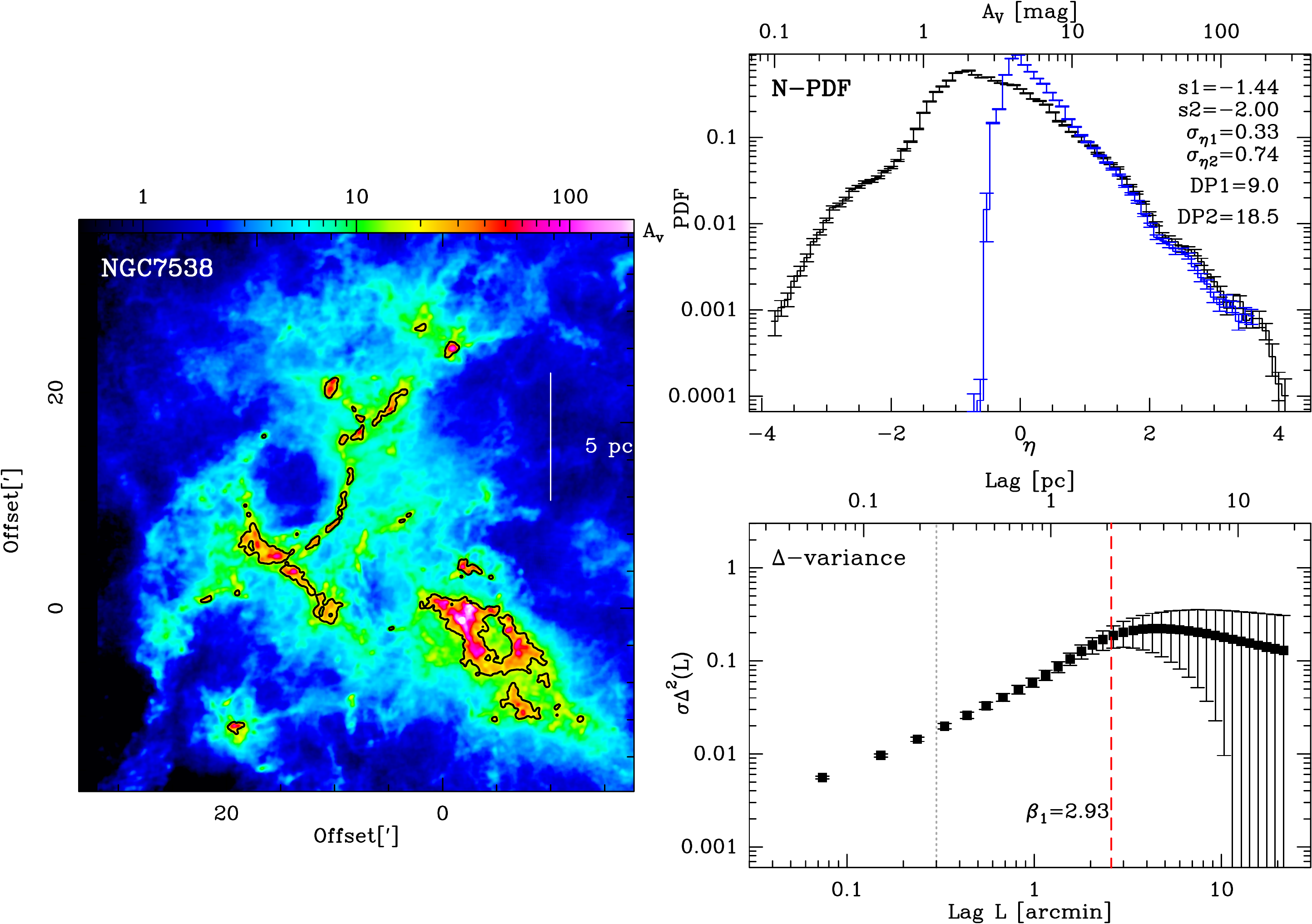} 
\end{center}
\caption[]{NGC7538: Fig. caption see Fig. C.1.}  \label{ngc7538} 
\end{figure*}

% Rosette 
\begin{figure*}[ht] 
\begin{center}
\includegraphics[angle=0,width=18cm,height=10cm,keepaspectratio]{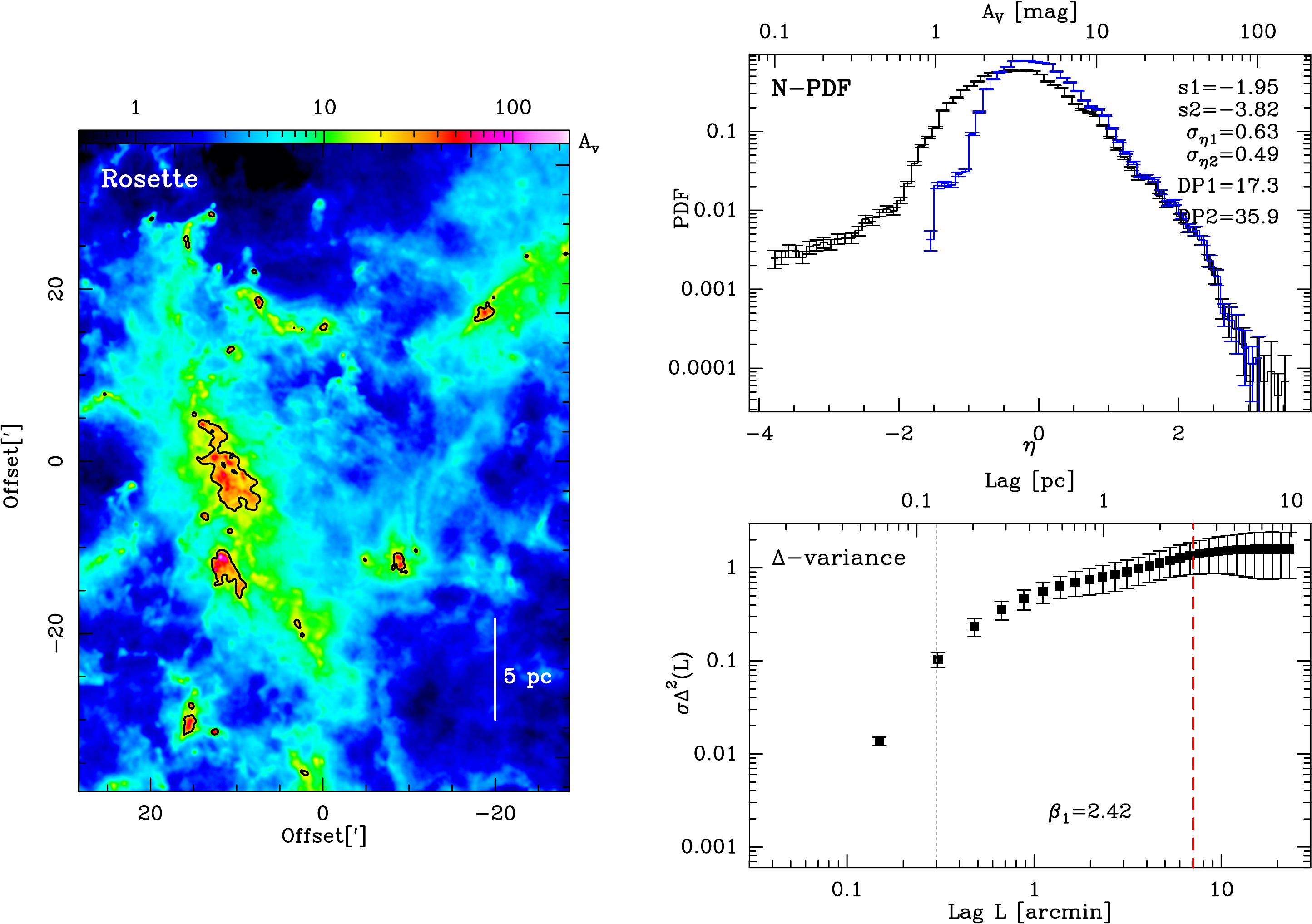} 
\end{center}
\caption[]{ROSETTE: Fig. caption see Fig. C.1.} \label{rosette} 
\end{figure*}

% Vela
\begin{figure*}[ht] 
\begin{center}
\includegraphics[angle=0,width=18cm,height=10cm,keepaspectratio]{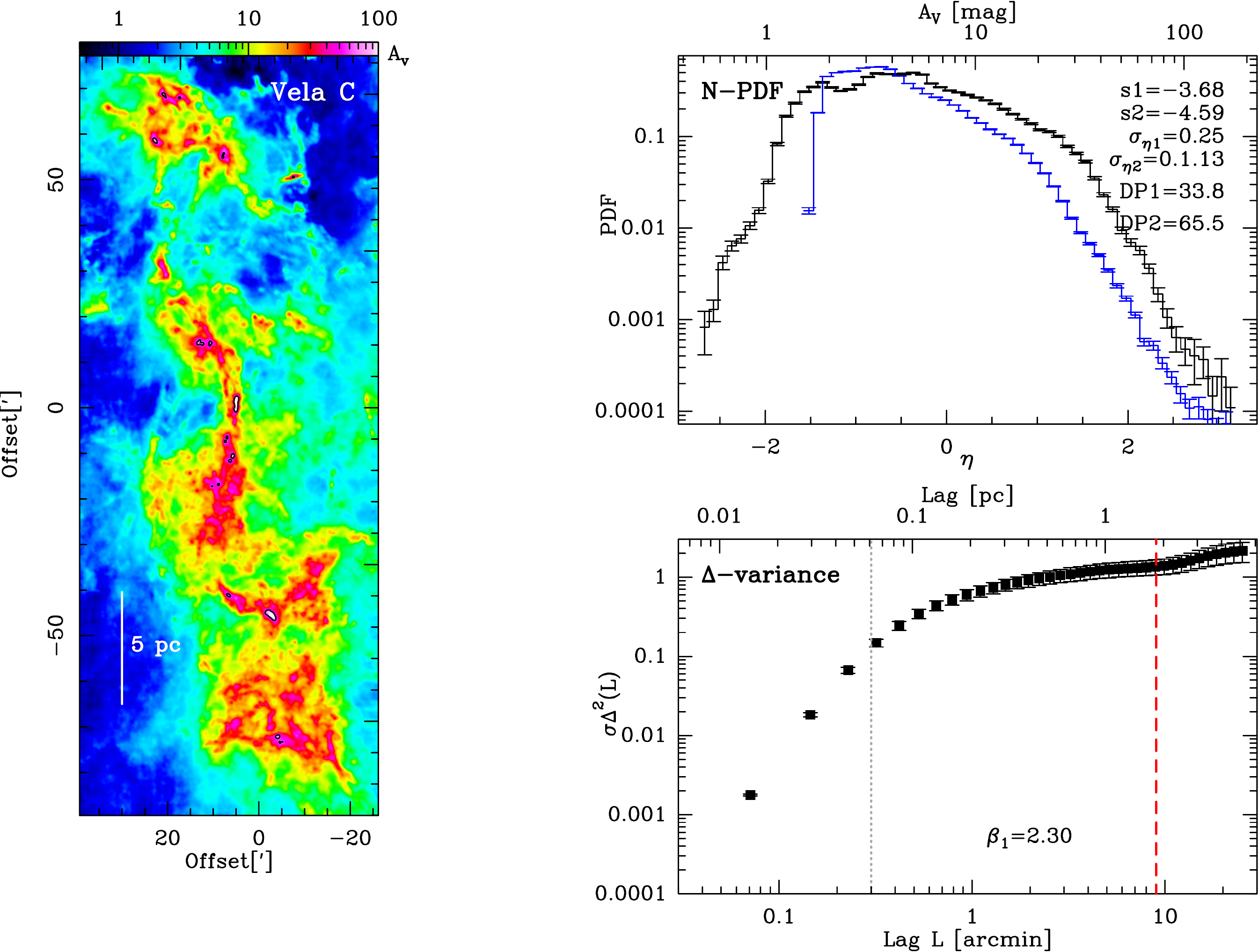} 
\end{center}
\caption[]{VELA C: Fig. caption see Fig. C.1.} \label{vela} 
\end{figure*}
%\vspace{-5cm}

% *******************************************************************
% INTERMEDIATE-MASS SF REGIONS 
% *******************************************************************

% Aquila
\begin{figure*}[ht] 
\begin{center}
\includegraphics[angle=0,width=18cm,height=10cm,keepaspectratio]{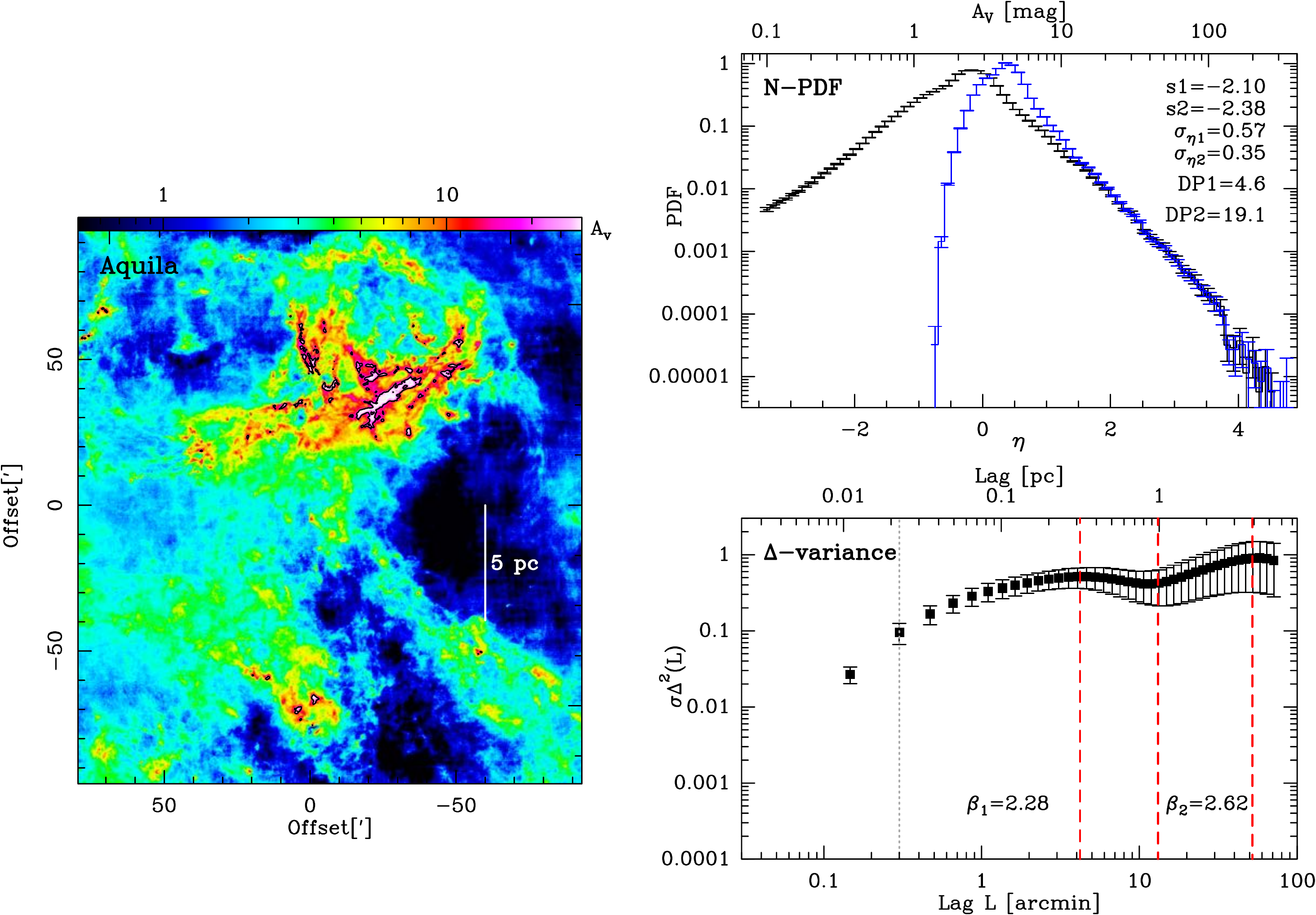} 
\end{center}
\caption[]{AQUILA: Fig. caption see Fig. C.1.} \label{aquila} 
\end{figure*}

% MonR2
\begin{figure*}[ht] 
\begin{center}
\includegraphics[angle=0,width=18cm,height=10cm,keepaspectratio]{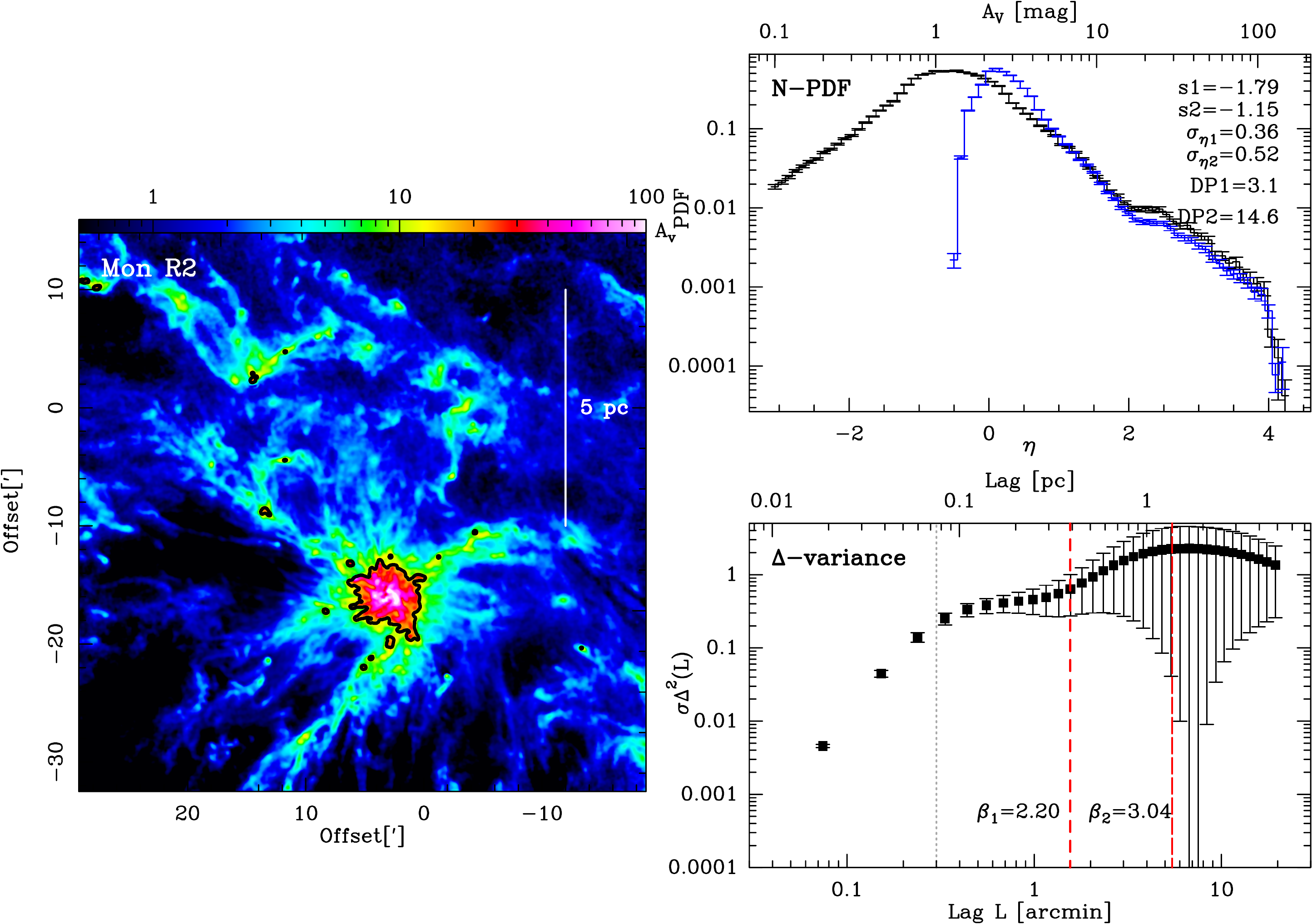} 
\end{center}
\caption[]{MONR2: Fig. caption see Fig. C.1.} \label{monr2} 
\end{figure*}
%\vspace{-5cm}

\clearpage

% MonOB1
\begin{figure*}[ht]
\begin{center}
\includegraphics[angle=0,width=18cm,height=10cm,keepaspectratio]{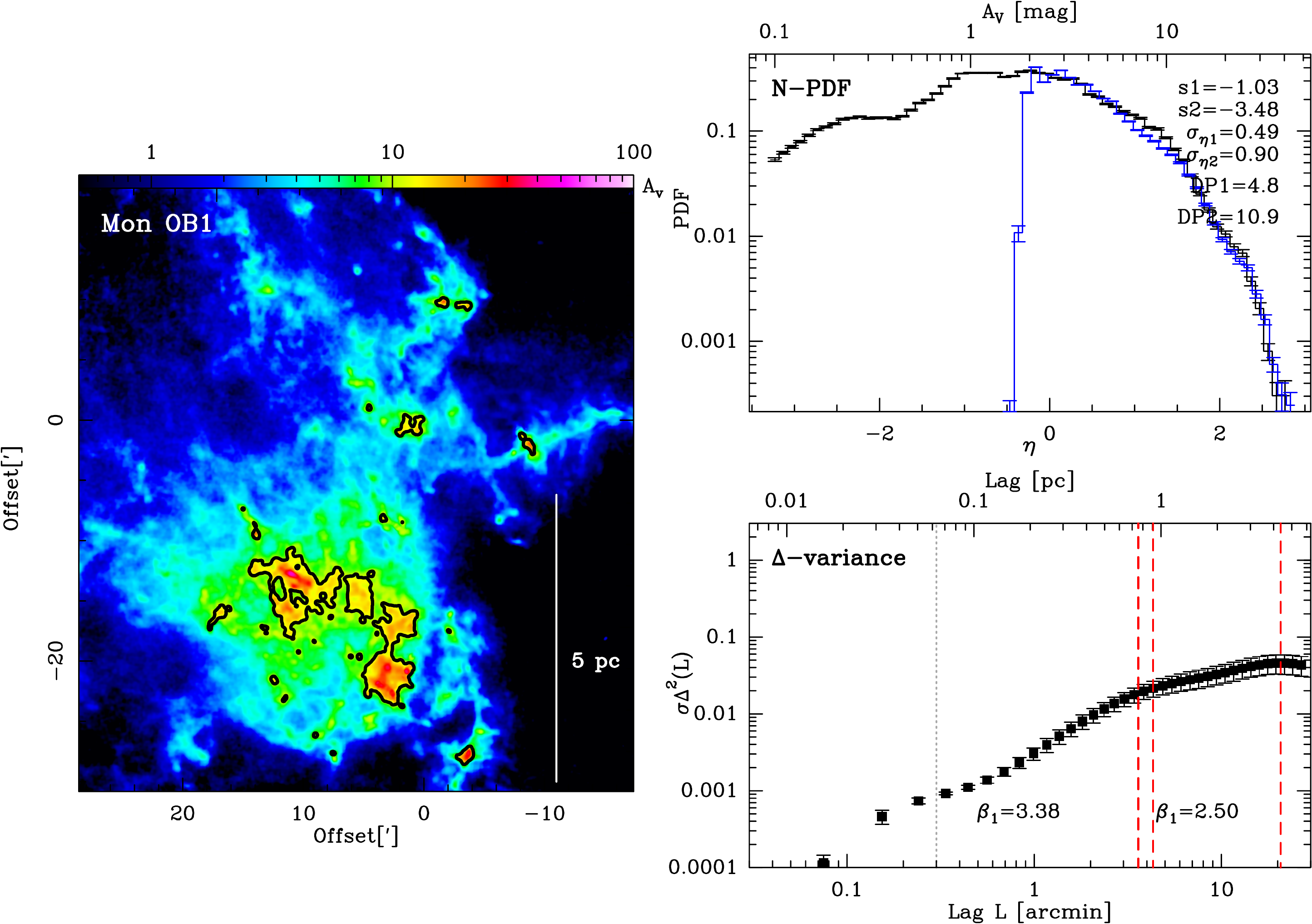} 
\end{center}
\caption[]{MONOB1: Fig. caption see Fig. C.1.} \label{monob1} 
\end{figure*}

% NGC2264
\begin{figure*}[ht] 
\begin{center}
\includegraphics[angle=0,width=18cm,height=10cm,keepaspectratio]{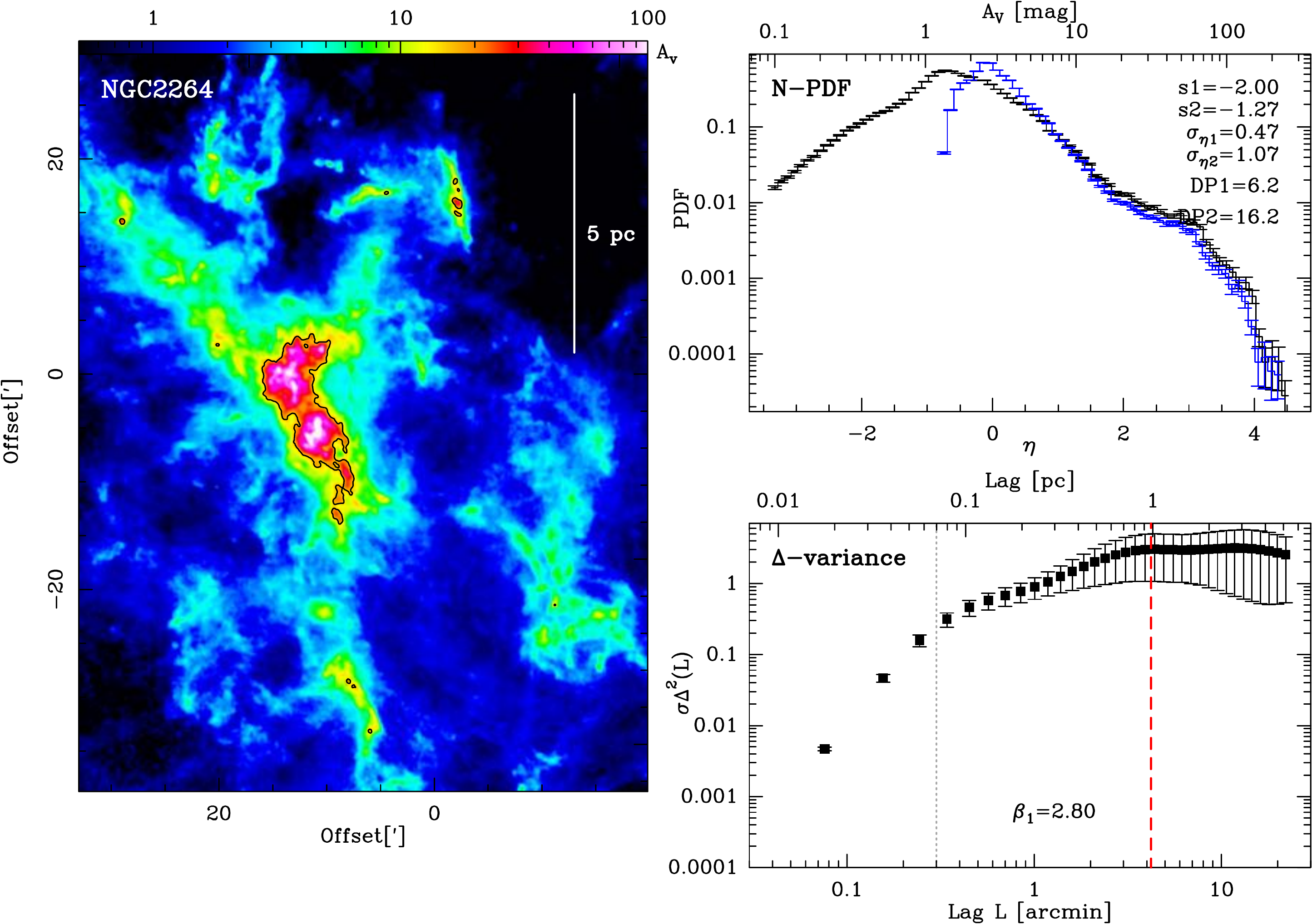} 
\end{center}
\caption[]{NGC2264: Fig. caption see Fig. C.1.} \label{ngc2264} 
\end{figure*}
%\vspace{-5cm}

% Orion B 
\begin{figure*}[ht]
\begin{center}
\includegraphics[angle=0,width=18cm,height=10cm,keepaspectratio]{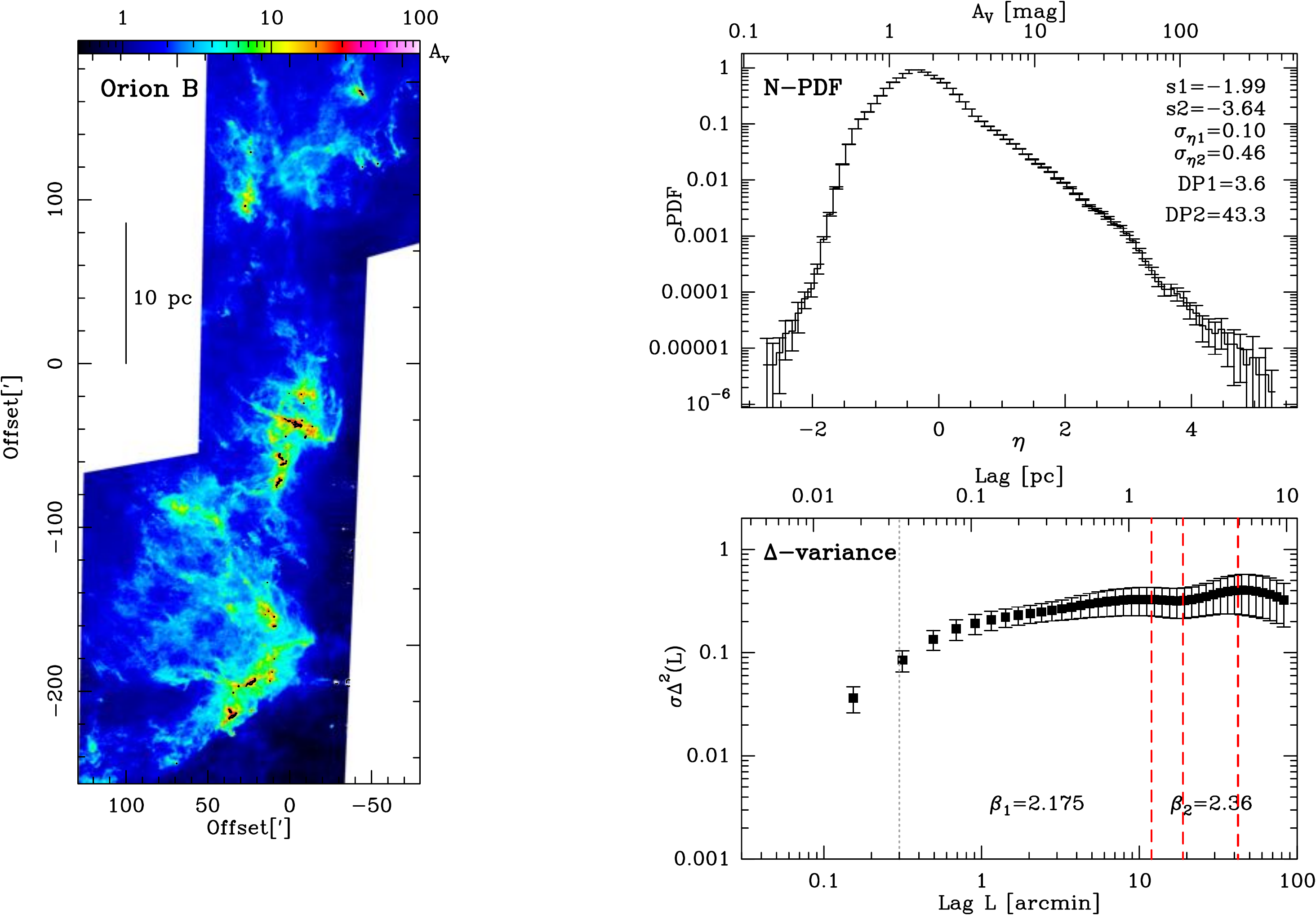} 
\end{center}
\caption[]{ORION-B: Fig. caption see Fig. C.1.}  \label{orionb} 
\end{figure*}

% Serpens
\begin{figure*}[ht] 
\begin{center}
\includegraphics[angle=0,width=18cm,height=10cm,keepaspectratio]{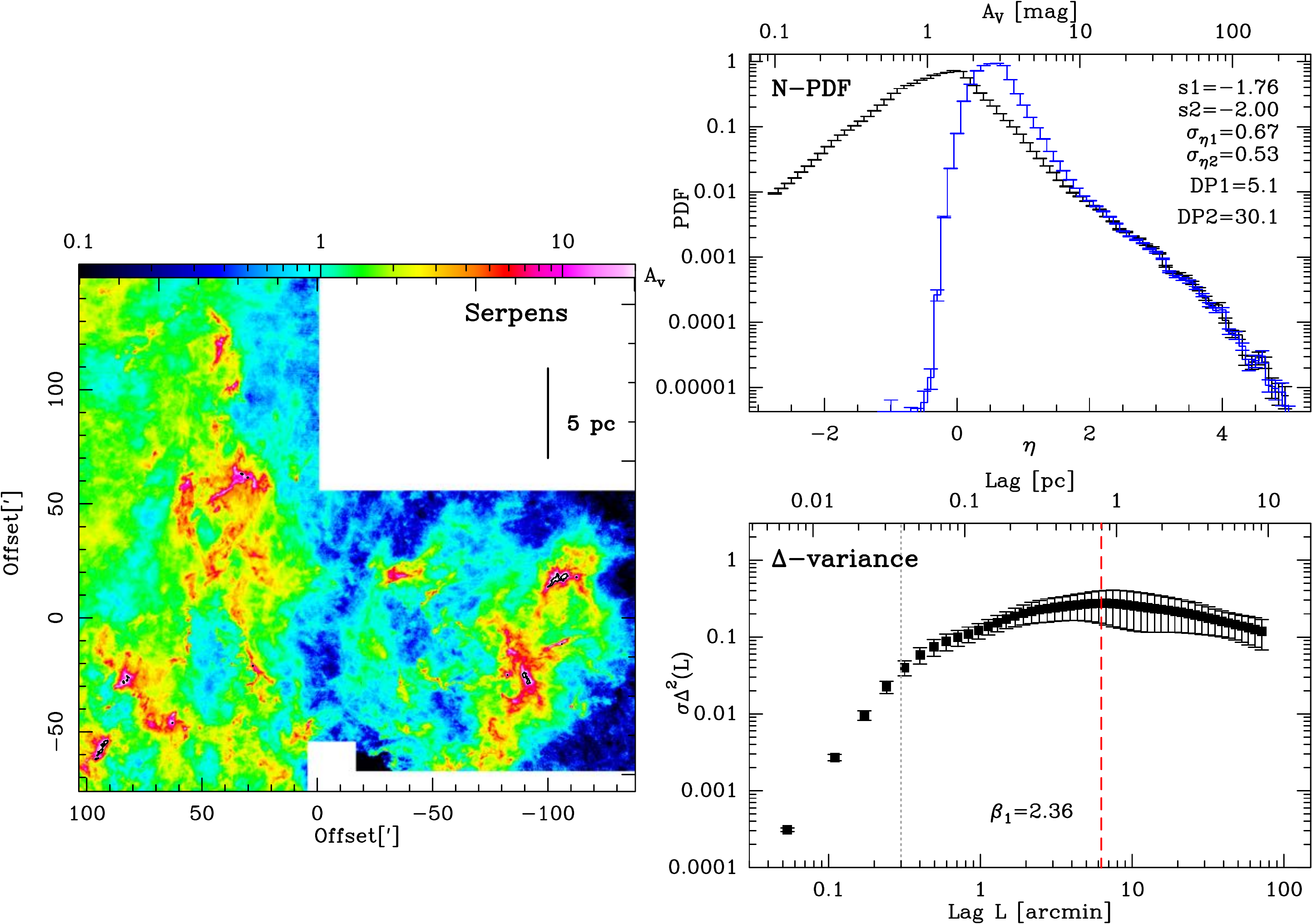} 
\end{center}
\caption[]{SERPENS: Fig. caption see Fig. C.1. For comparison, the
  N-PDF of the uncorrected map is displayed in grey.} \label{serpens} 
\end{figure*}

\clearpage

% *******************************************************************
% LOW MASS SF REGIONS
% *******************************************************************

% Cham I 
\begin{figure*}[ht] 
\begin{center}
\includegraphics[angle=0,width=18cm,height=10cm,keepaspectratio]{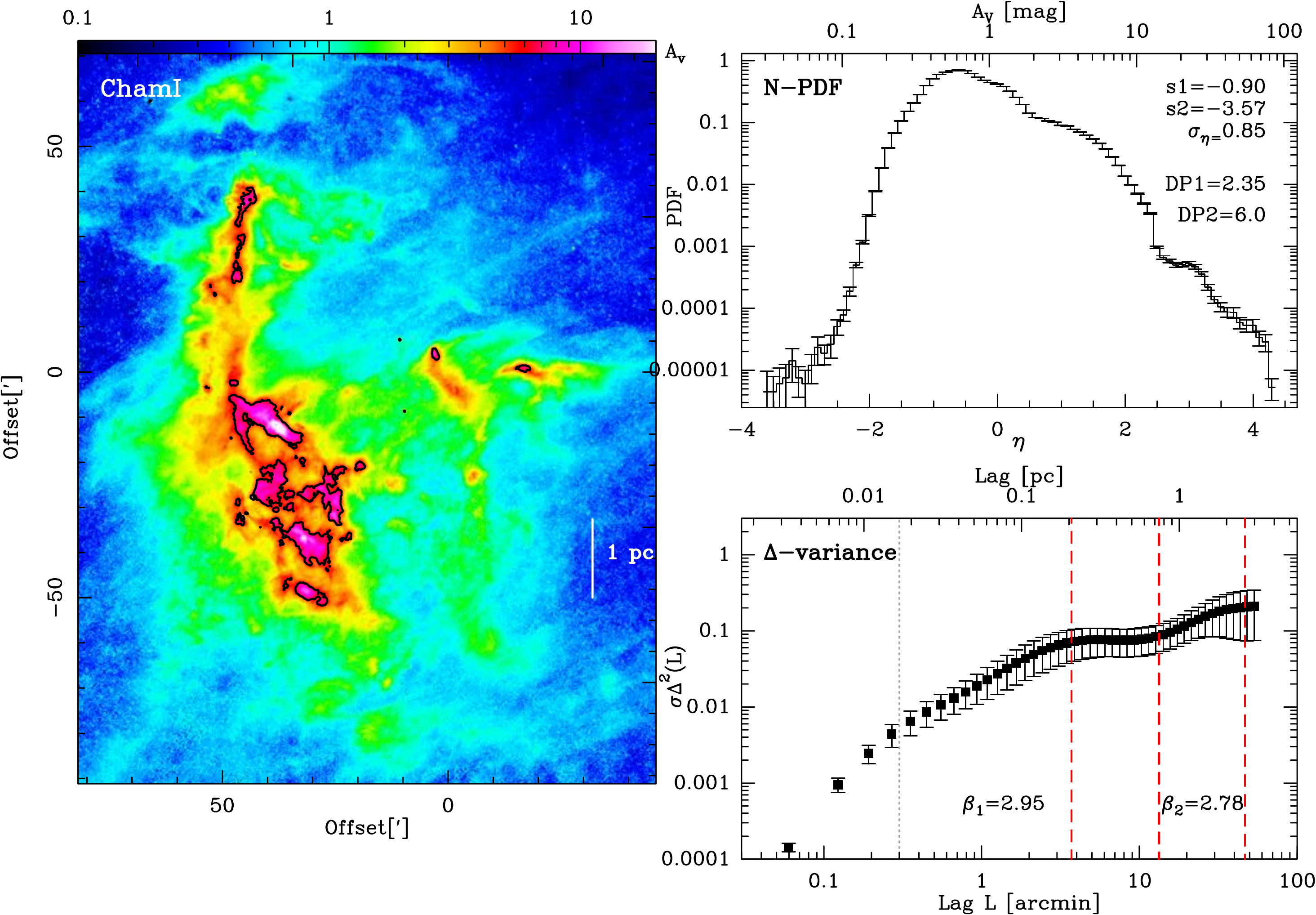} 
\end{center}
\caption[]{Chamaeleon I: Fig. caption see Fig. C.1.} \label{chamI} 
\end{figure*}

% Cham II
\begin{figure*}[ht] 
\begin{center}
\includegraphics[angle=0,width=18cm,height=10cm,keepaspectratio]{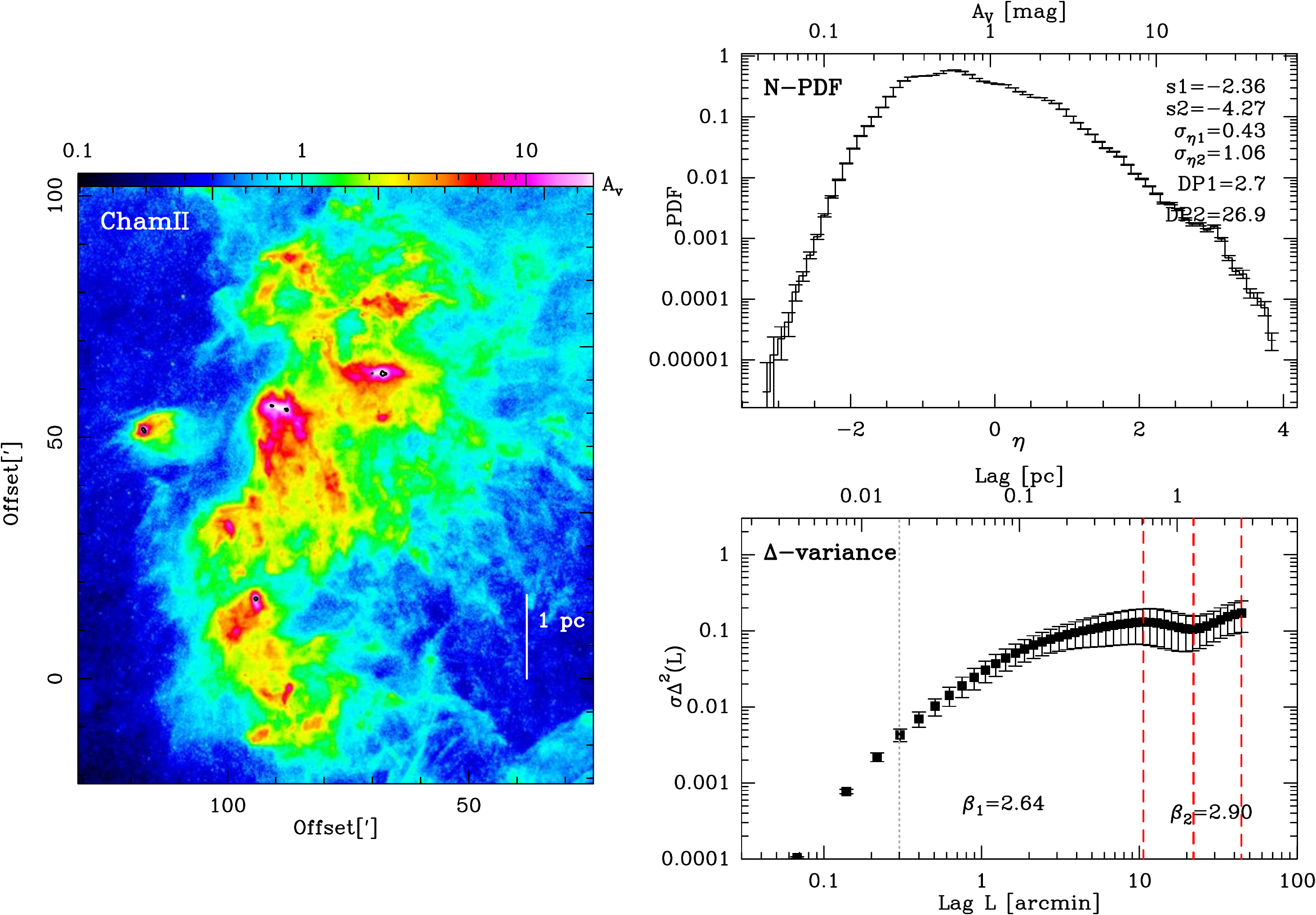} 
\end{center}
\caption[]{Chamaeleon II: Fig. caption see Fig. C.1.} \label{chamII} 
\end{figure*}
%\vspace{-5cm}

% IC5146
\begin{figure*}[ht] 
\begin{center}
\includegraphics[angle=0,width=18cm,height=10cm,keepaspectratio]{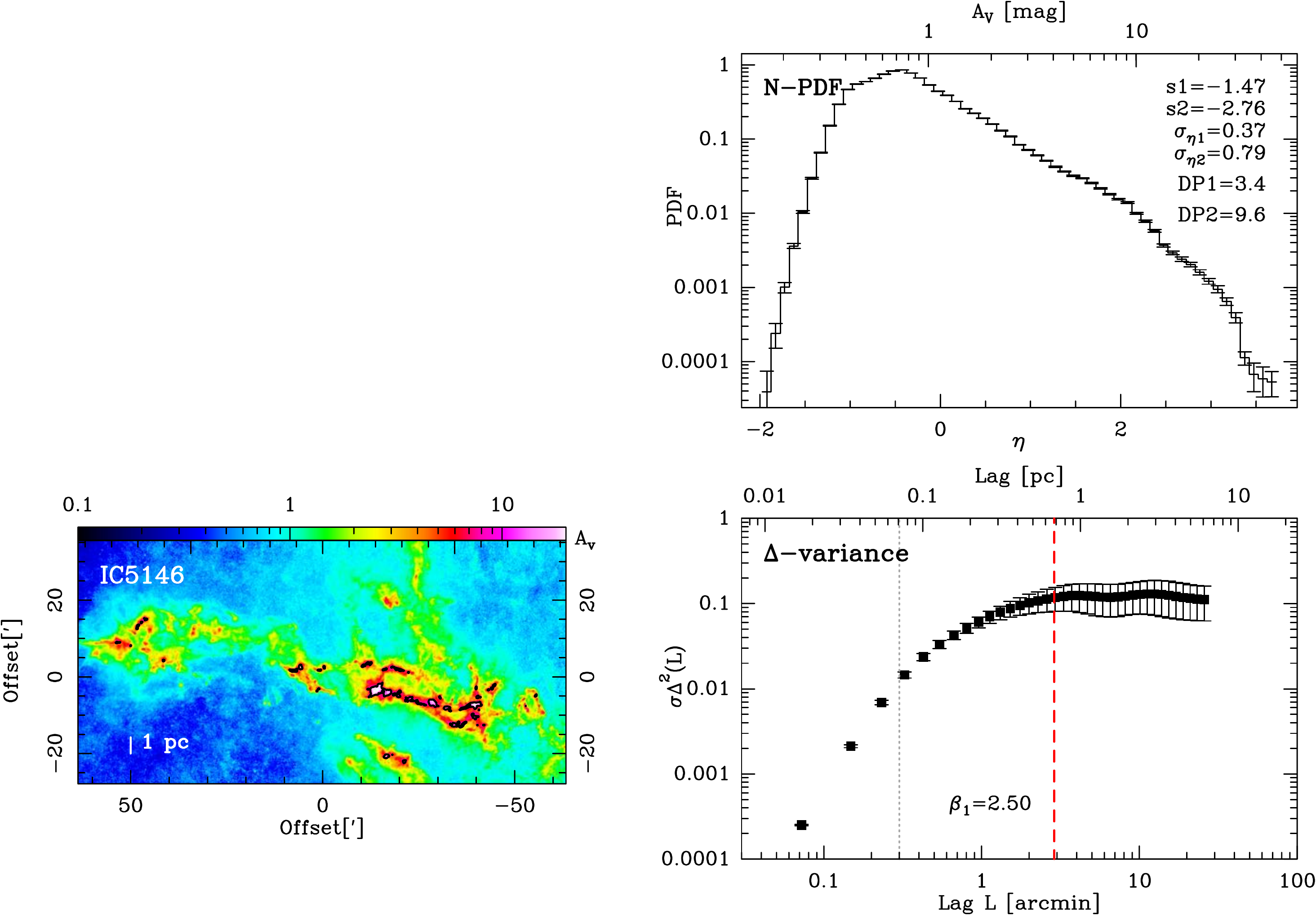} 
\end{center}
\caption[]{IC5146: Fig. caption see Fig. C.1.} \label{ic5146} 
\end{figure*}

% Lupus I
\begin{figure*}[ht] 
\begin{center}
\includegraphics[angle=0,width=18cm,height=10cm,keepaspectratio]{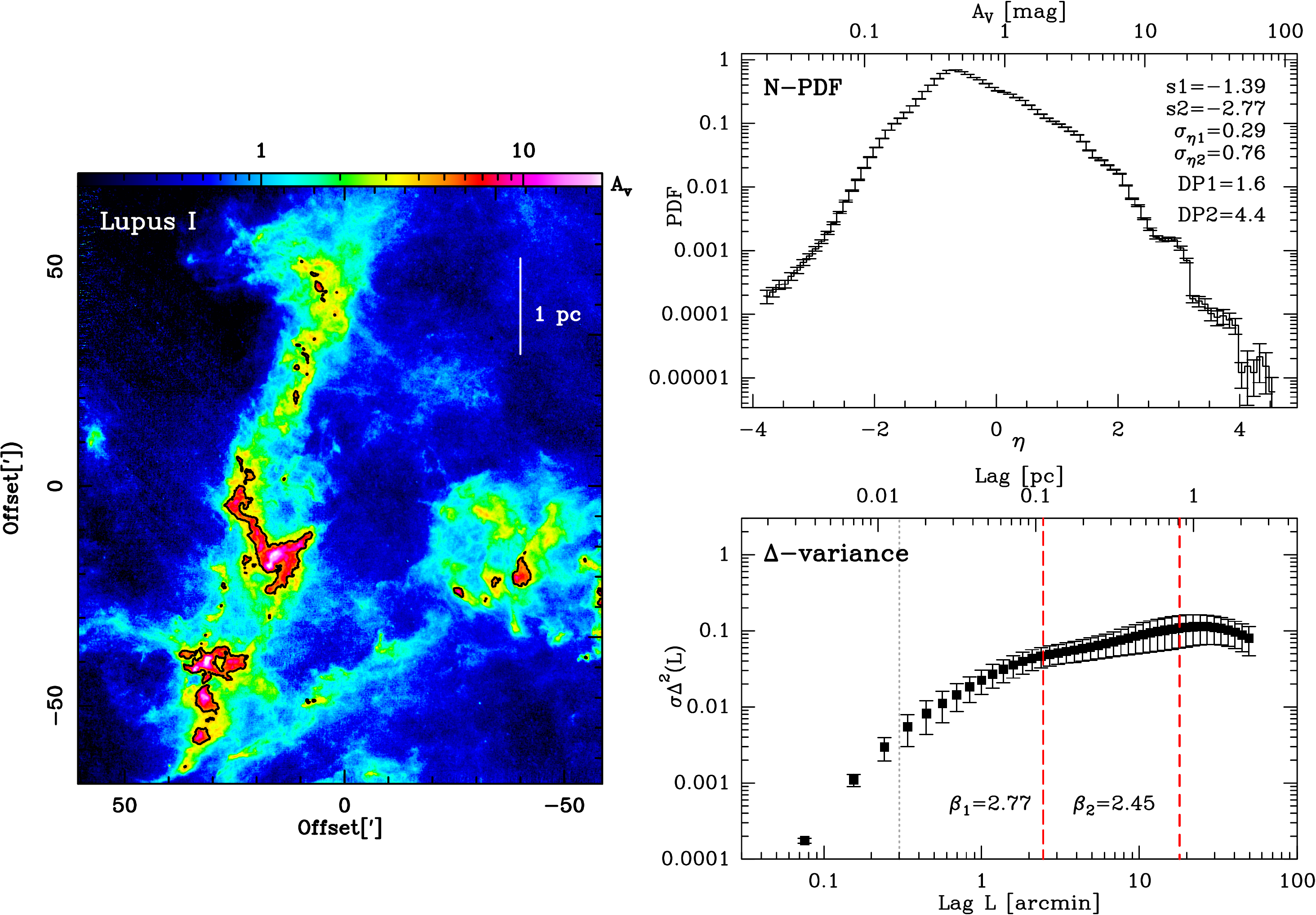} 
\end{center}
\caption[]{LUPUS I: Fig. caption see Fig. C.1. } \label{lupusI} 
\end{figure*}
%\vspace{-5cm}

\clearpage

% Lupus III
\begin{figure*}[ht]
\begin{center}
\includegraphics[angle=0,width=18cm,height=10cm,keepaspectratio]{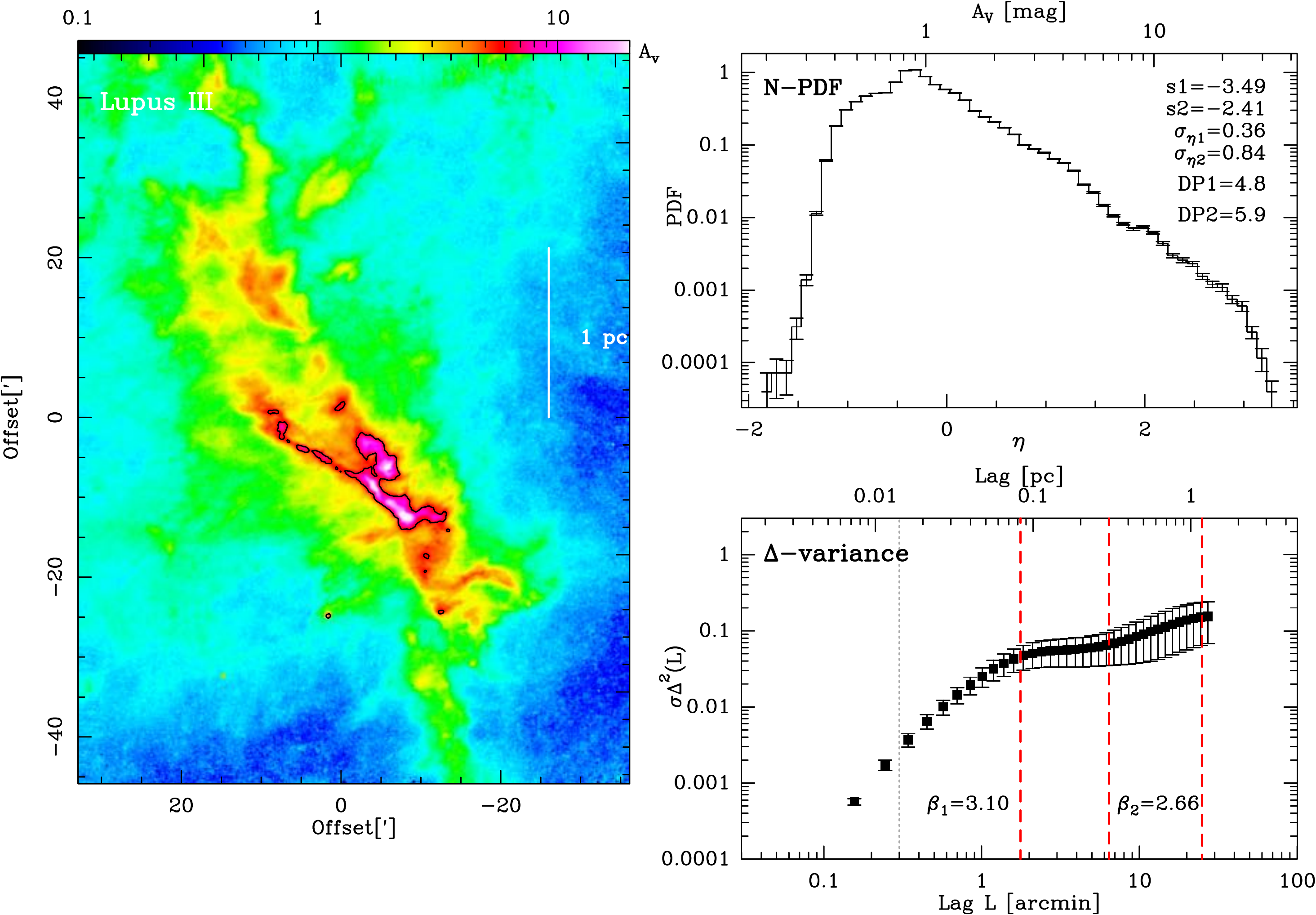} 
\end{center}
\caption[]{LUPUS III: Fig. caption see Fig. C.1.}  \label{lupusIII} 
\end{figure*}
%\vspace{-5cm}

% Lupus IV
\begin{figure*}[ht] 
\begin{center}
\includegraphics[angle=0,width=18cm,height=10cm,keepaspectratio]{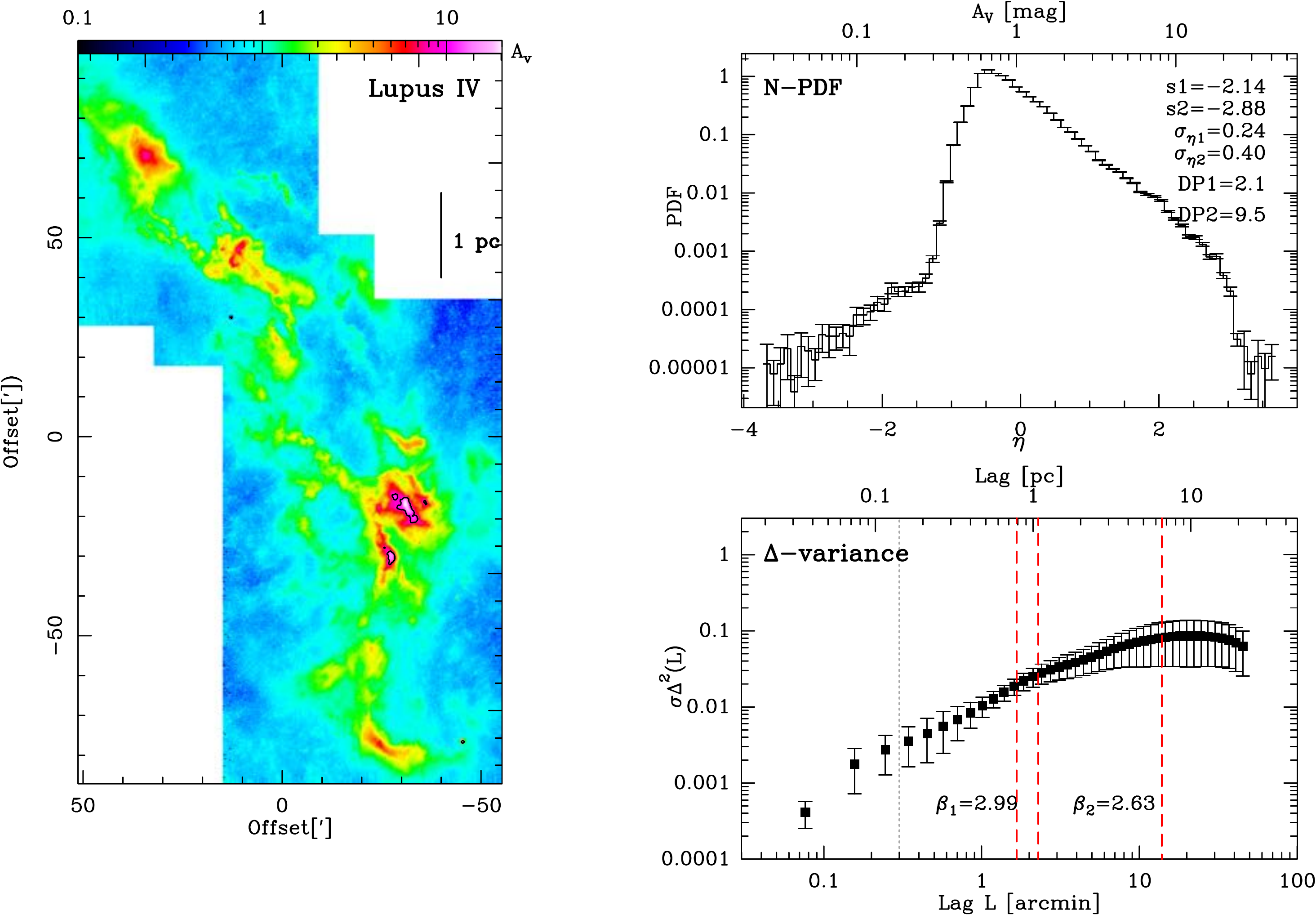} 
\end{center}
\caption[]{LUPUS IV: Fig. caption see Fig. C.1. } \label{lupusIV} 
\end{figure*}

% Perseus
\begin{figure*}[ht] 
\begin{center}
\includegraphics[angle=0,width=18cm,height=10cm,keepaspectratio]{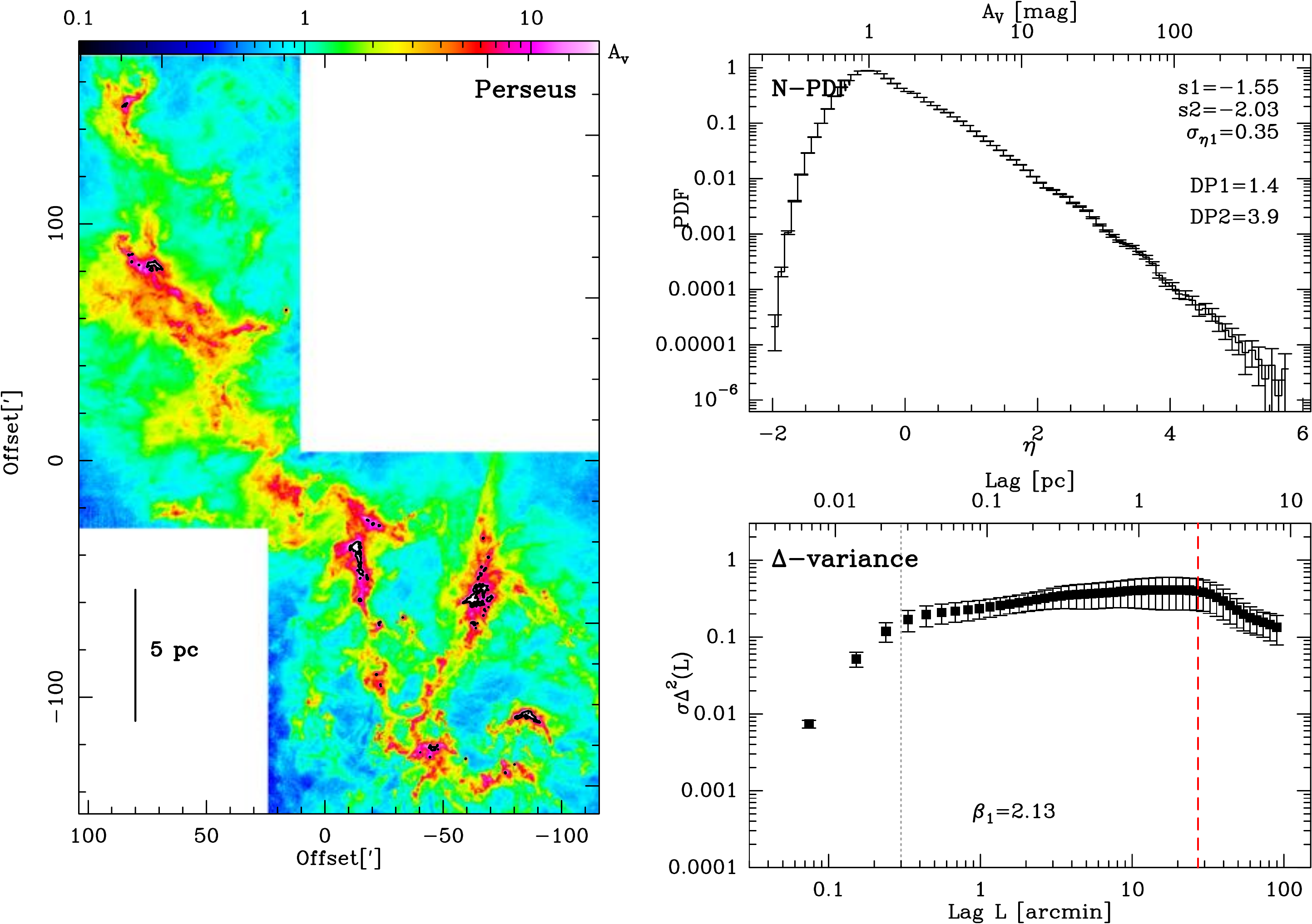} 
\end{center}
\caption[]{PERSEUS: Fig. caption see Fig. C.1. } \label{perseus} 
\end{figure*}

% PIPE
\begin{figure*}[ht]
\begin{center}
\includegraphics[angle=0,width=18cm,height=10cm,keepaspectratio]{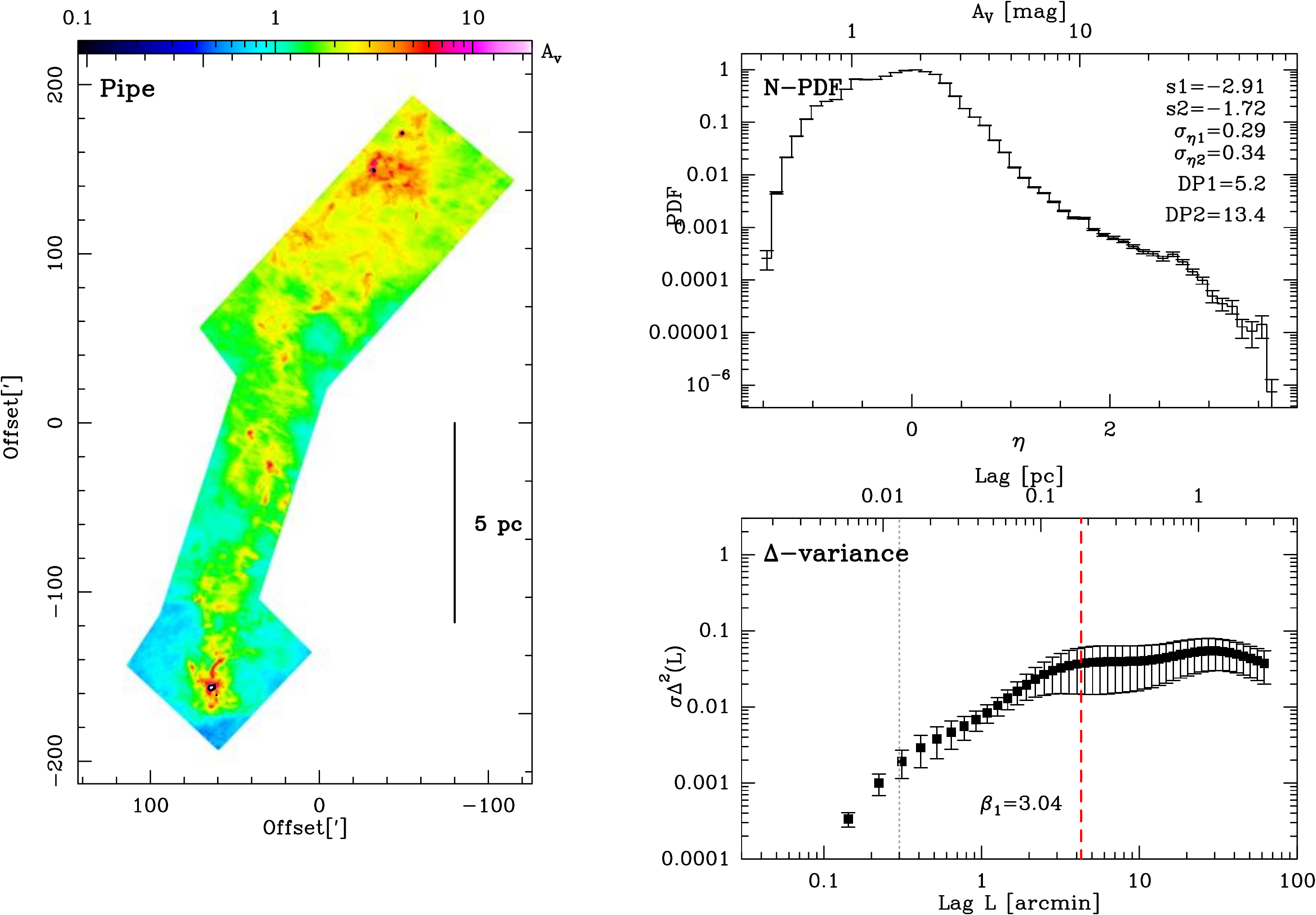} 
\end{center}
\caption[]{PIPE: Fig. caption see Fig. C.1. }  \label{pipe} 
\end{figure*}
%\vspace{-5cm}

\clearpage

% Rhooph
\begin{figure*}[ht]
\begin{center}
\includegraphics[angle=0,width=18cm,height=10cm,keepaspectratio]{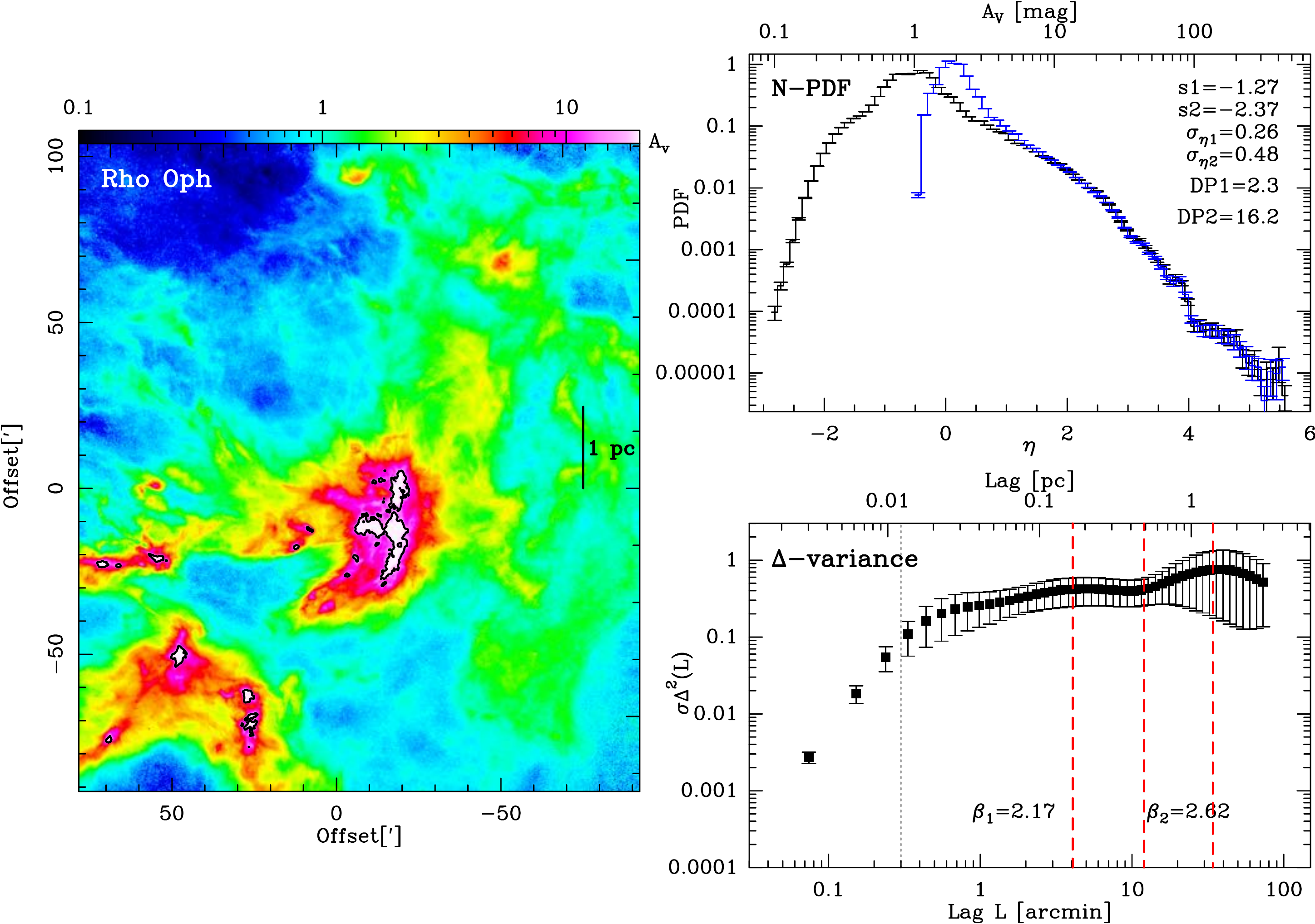} 
\end{center}
\caption[]{RHO OPH: Fig. caption see Fig. C.1. For comparison, the
  N-PDF of the uncorrected map is displayed in grey.}  \label{rhooph} 
\end{figure*}

% Taurus
\begin{figure*}[ht] 
\begin{center}
\includegraphics[angle=0,width=18cm,height=10cm,keepaspectratio]{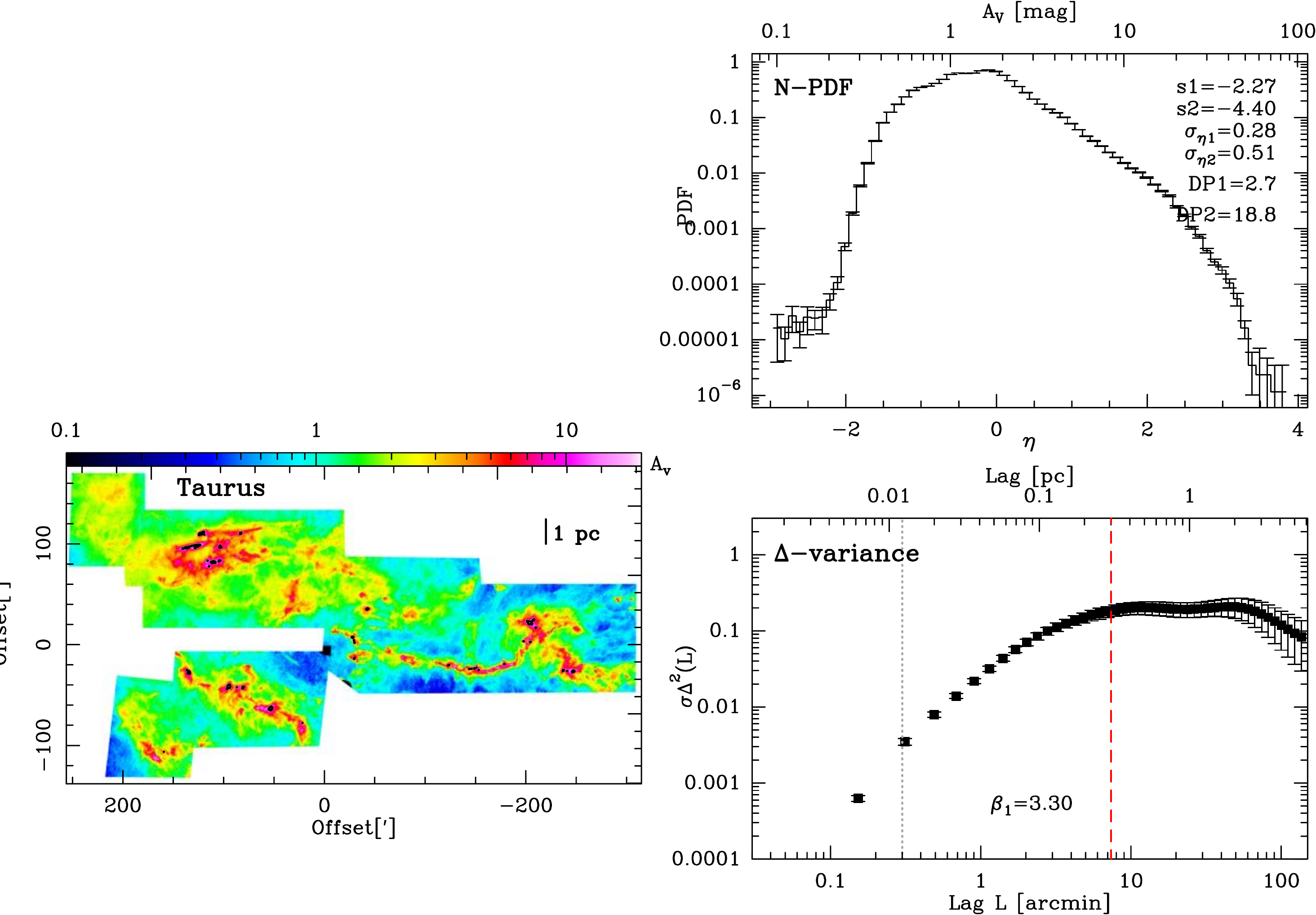} 
\end{center}
\caption[]{TAURUS: Fig. caption see Fig. C.1.} \label{taurus} 
\end{figure*}

% *******************************************************************
% QUIESCENT REGIONS
% *******************************************************************

% Cham III
\begin{figure*}[ht] 
\begin{center}
\includegraphics[angle=0,width=18cm,height=10cm,keepaspectratio]{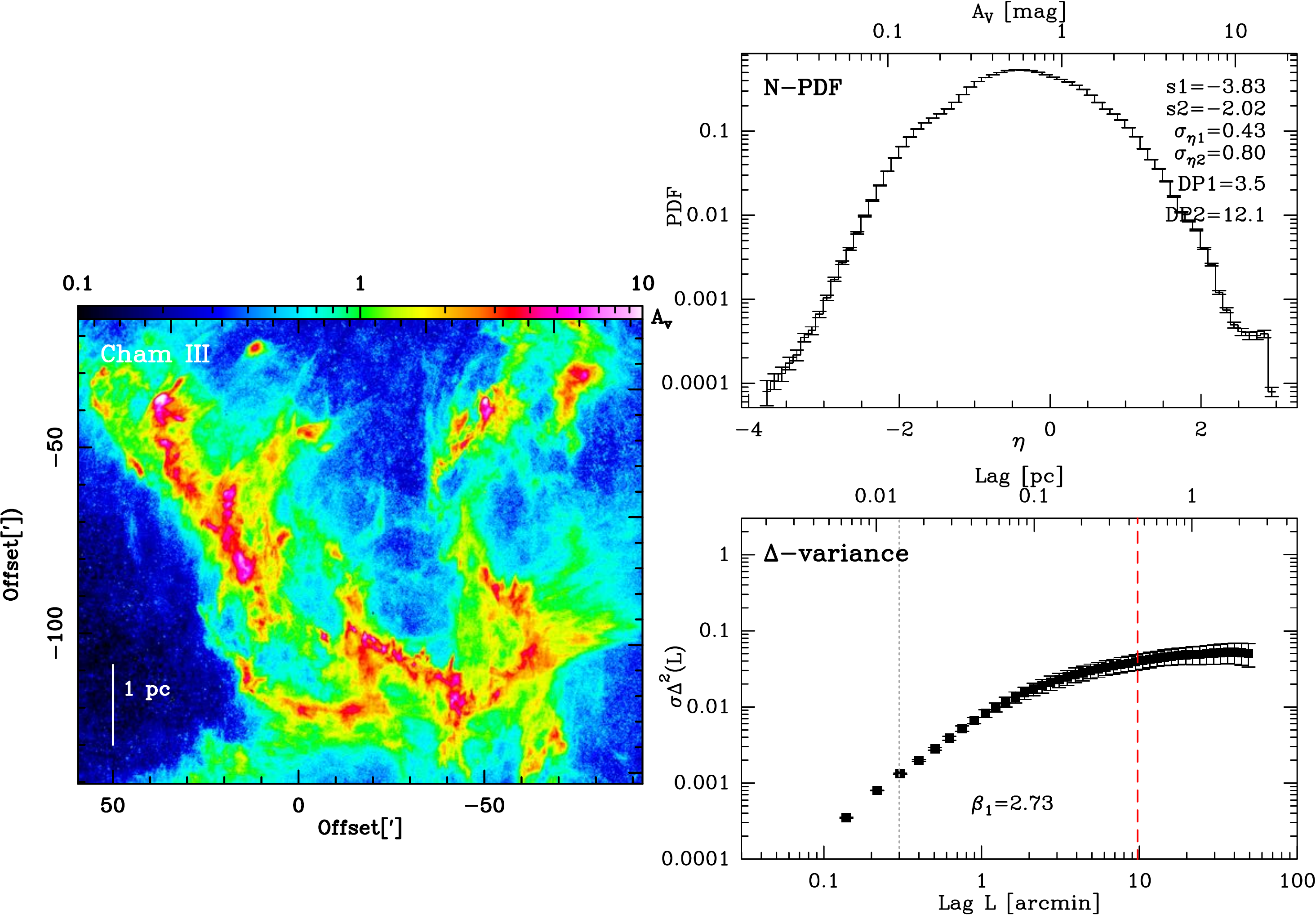} 
\end{center}
\caption[]{Chamaeleon III:  Fig. caption see Fig. C.1.} \label{chamIII} 
\end{figure*}
%\vspace{-5cm}

\clearpage

% Musca
\begin{figure*}[ht] 
\begin{center}
\includegraphics[angle=0,width=18cm,height=10cm,keepaspectratio]{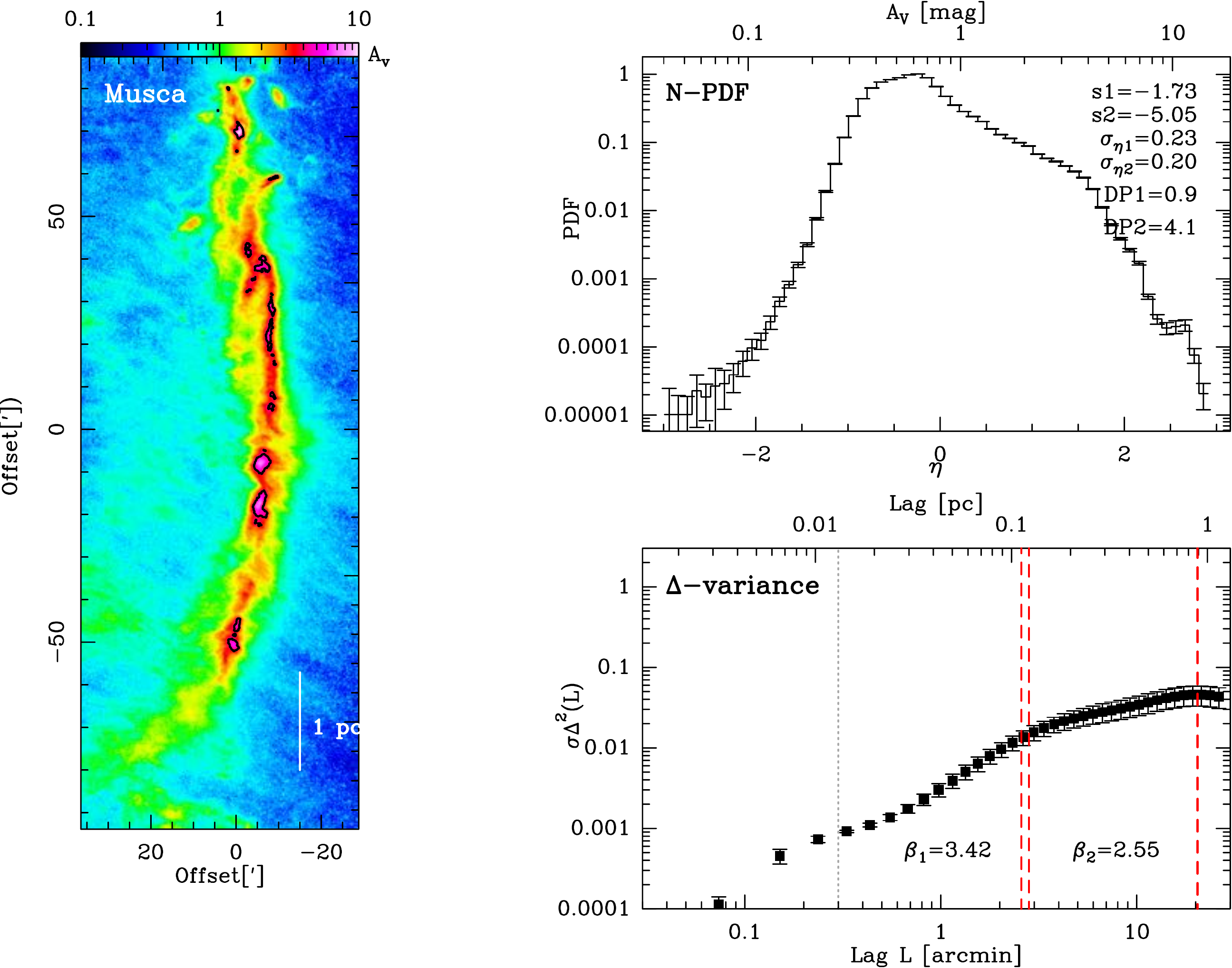} 
\end{center}
\caption[]{MUSCA: Fig. caption see Fig. C.1. } \label{musca} 
\end{figure*}
%\vspace{-5cm}

% Polaris
\begin{figure*}[ht] 
\begin{center}
\includegraphics[angle=0,width=18cm,height=10cm,keepaspectratio]{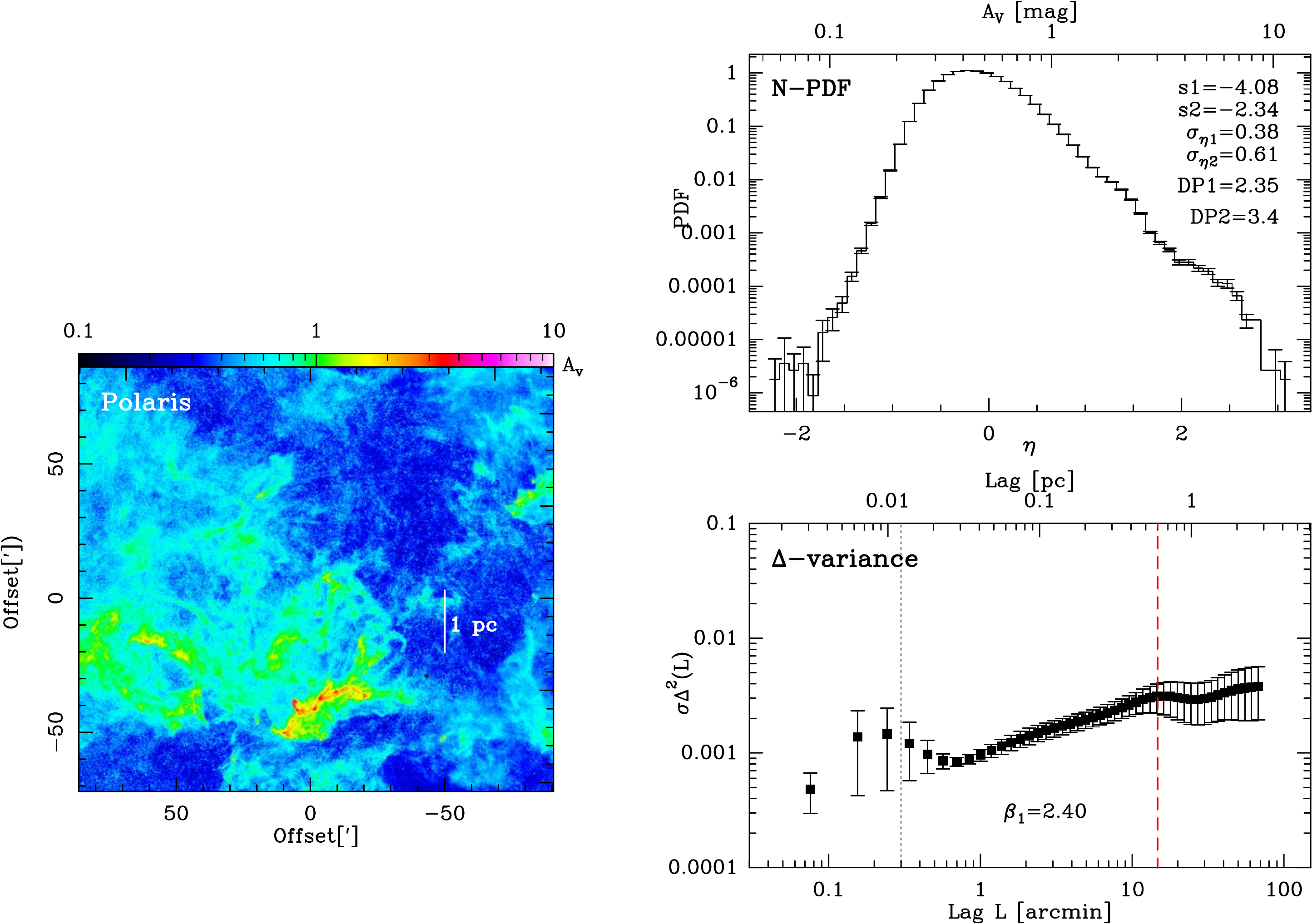} 
\end{center}
\caption[]{POLARIS: Fig. caption see Fig. C.1. } \label{polaris} 
\end{figure*}

% Draco
\begin{figure*}[ht] 
\begin{center}
\includegraphics[angle=0,width=18cm,height=10cm,keepaspectratio]{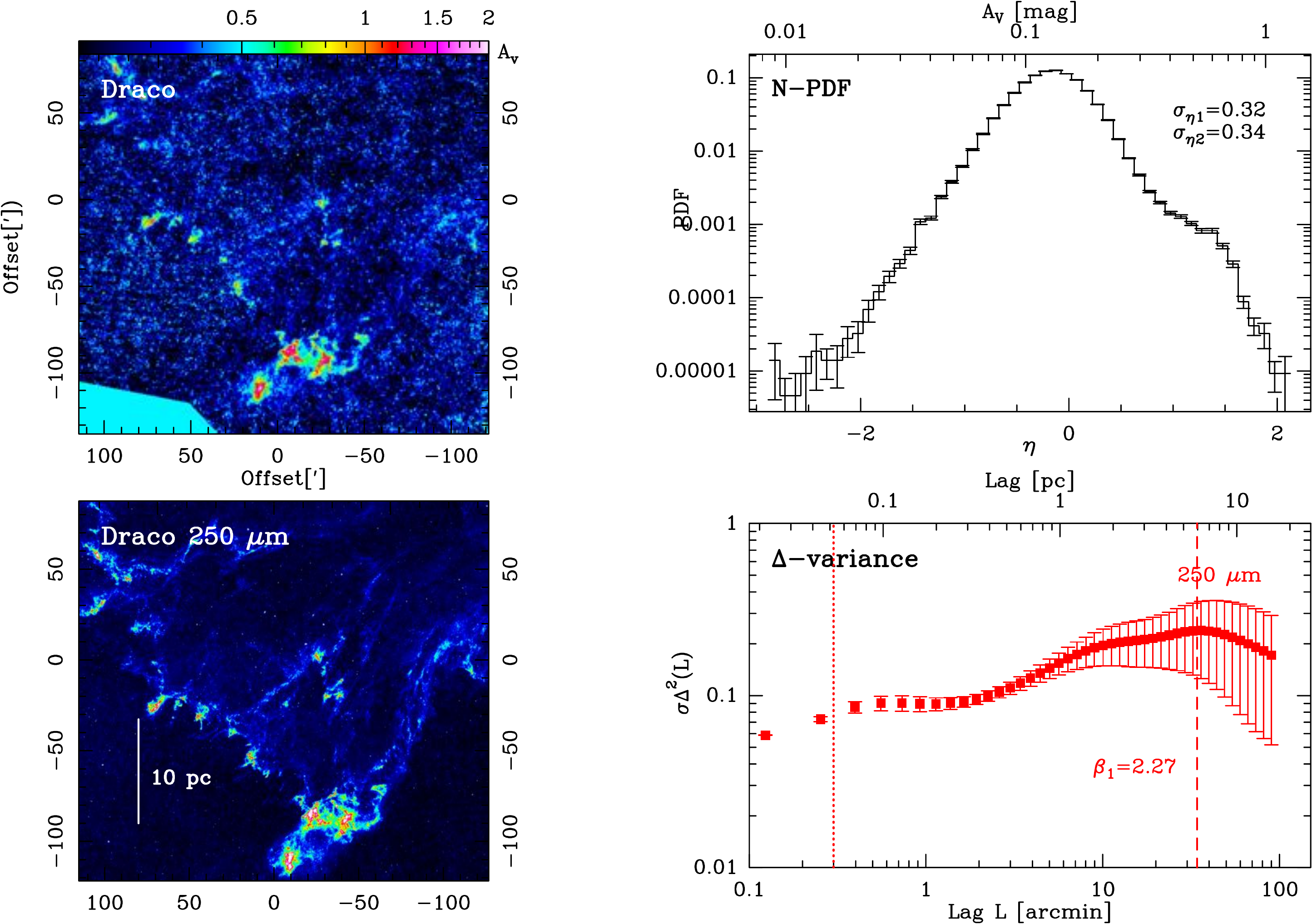} 
\end{center}
\caption[]{DRACO: Fig. caption see Fig. C.1. } \label{draco} 
\end{figure*}

\clearpage

% APPENDIX D

\section{Individual N-PDFs} \label{app-d}  

Figures D.1 to D.29 display the N-PDFs of all clouds and the best
  fitting model. For all regions where we applied a LOS correction, we
  used this column density for performing the N-PDF. The models are
  indicated in the figure caption and follow the syntax explained in
  Sec.~\ref{sec:pdfs}, (1) ELP: a single log-normal (L) convolved
  with a Gaussian noise distribution that creates an error slope (E)
  on the left-hand side and a power law tail (P); (2) ELLP: the same
  as (1) but with two log-normals, (3) LL2P: Two log-normals and two
  PLTs, (4) EL2P: the same as (1) but with two PLTs, (5) ELL2P: the
  same as (2) but with 2 PLTs.
   
% *******************************************************************
% HIGH-MASS SF REGIONS 
% *******************************************************************

\begin{figure*}[!h]
\centering
 \begin{subfigure}[c]{0.40\textwidth}
 \includegraphics[width=1.\textwidth]{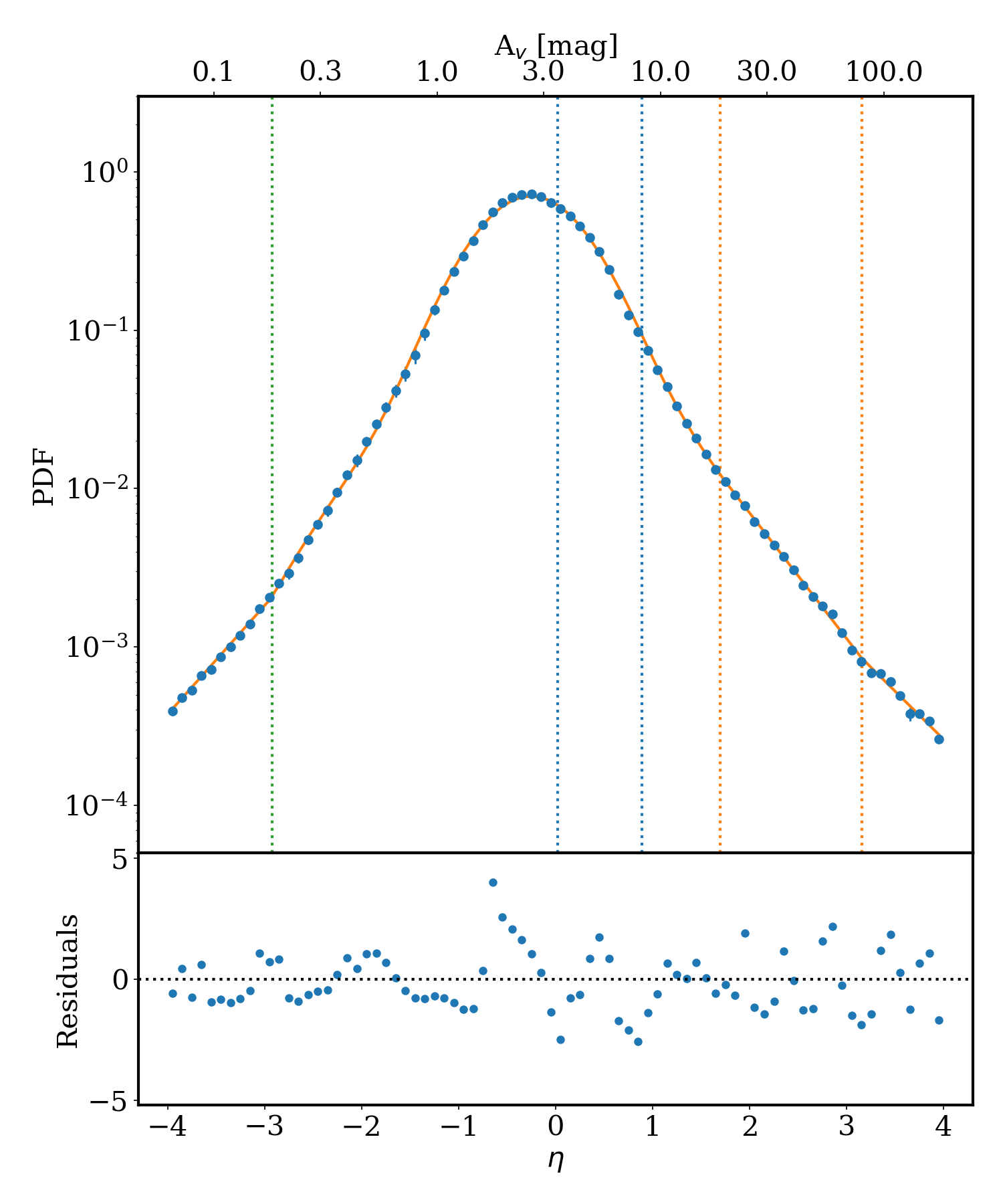} 
\end{subfigure}
\begin{subfigure}[c]{0.40\textwidth}
\includegraphics[width=1.\textwidth]{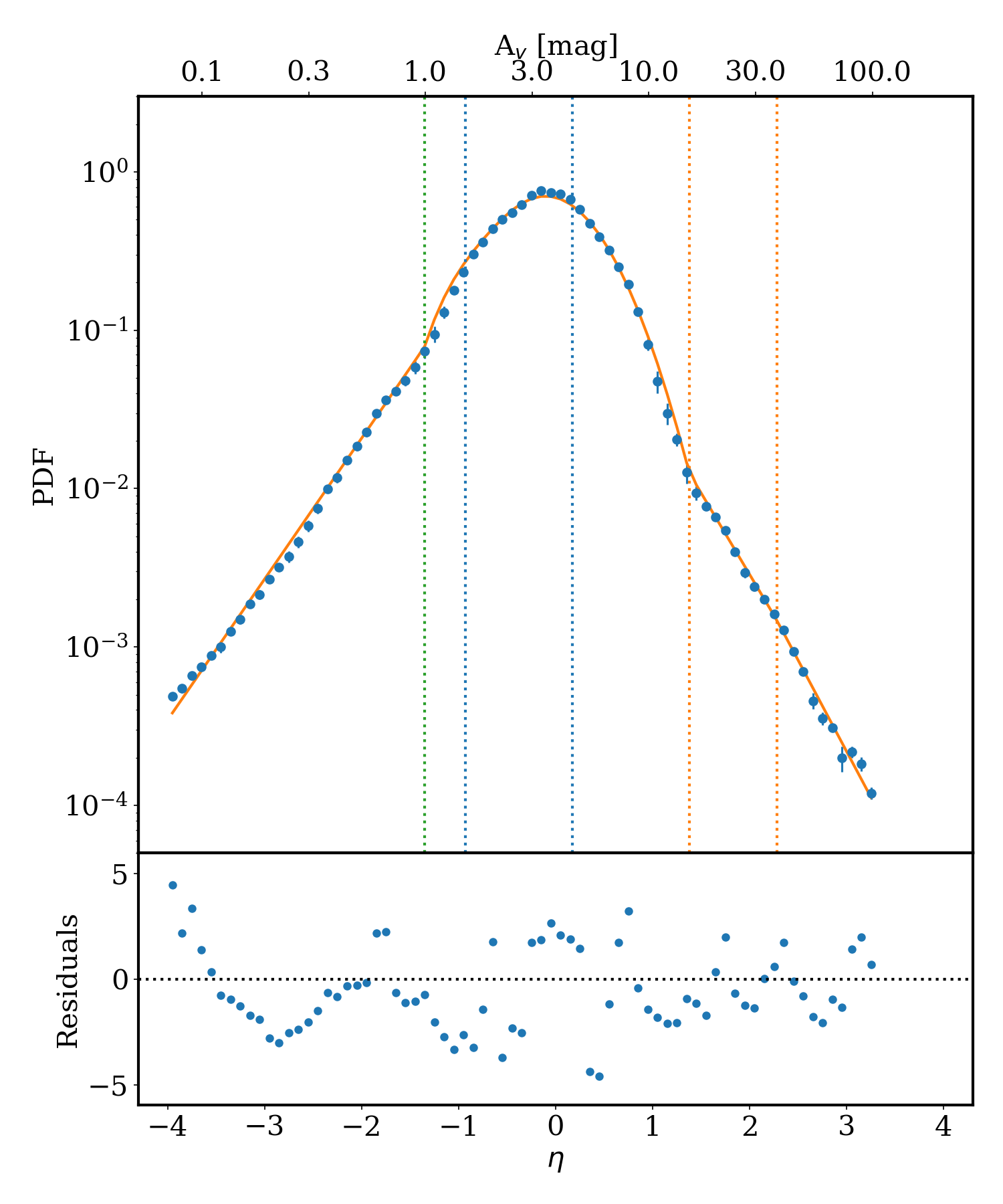}
\end{subfigure}
\caption{N-PDFs of LOS corrected column density (blue points),
    expressed in visual extinction (upper x-axis) and in $\eta$ (lower
    x-axis) of DR21 (left) and DR15 (right).  The left y-axis gives
    the PDF (there can be small differences compared to the
    plots in Appendix~C because for the model fit, we excluded the
    extreme low- and high density ranges, which leads to a slightly
    different normalization). The orange curve indicates the best
    fitting model for the N-PDF (see Table~\ref{tab:BIC} and
    \ref{tab:BIC-weights}). The vertical lines show the peak values of
    the log-normal(s) in blue, the break points for the power laws in
    orange and the error power law break point in green.  Underneath
    is the standardized residuals, a perfect model would give
    numbers with mean of 0 and a variance of 1. \label{dr21+dr15-npdf}}
\end{figure*}

\begin{figure*}[!h]
\centering
 \begin{subfigure}[c]{0.40\textwidth}
 \includegraphics[width=1.\textwidth]{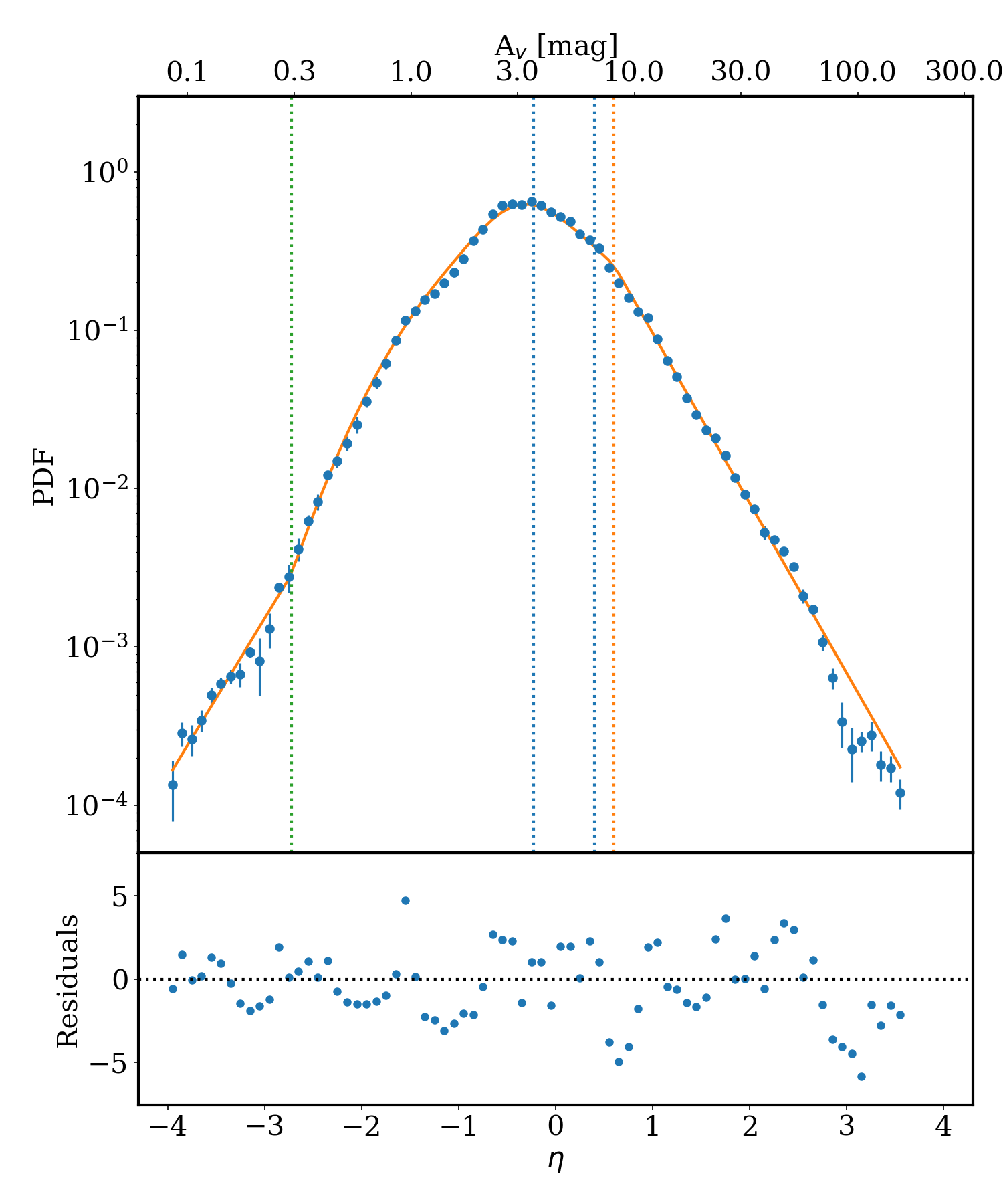} 
\end{subfigure}
\begin{subfigure}[c]{0.40\textwidth}
\includegraphics[width=1.\textwidth]{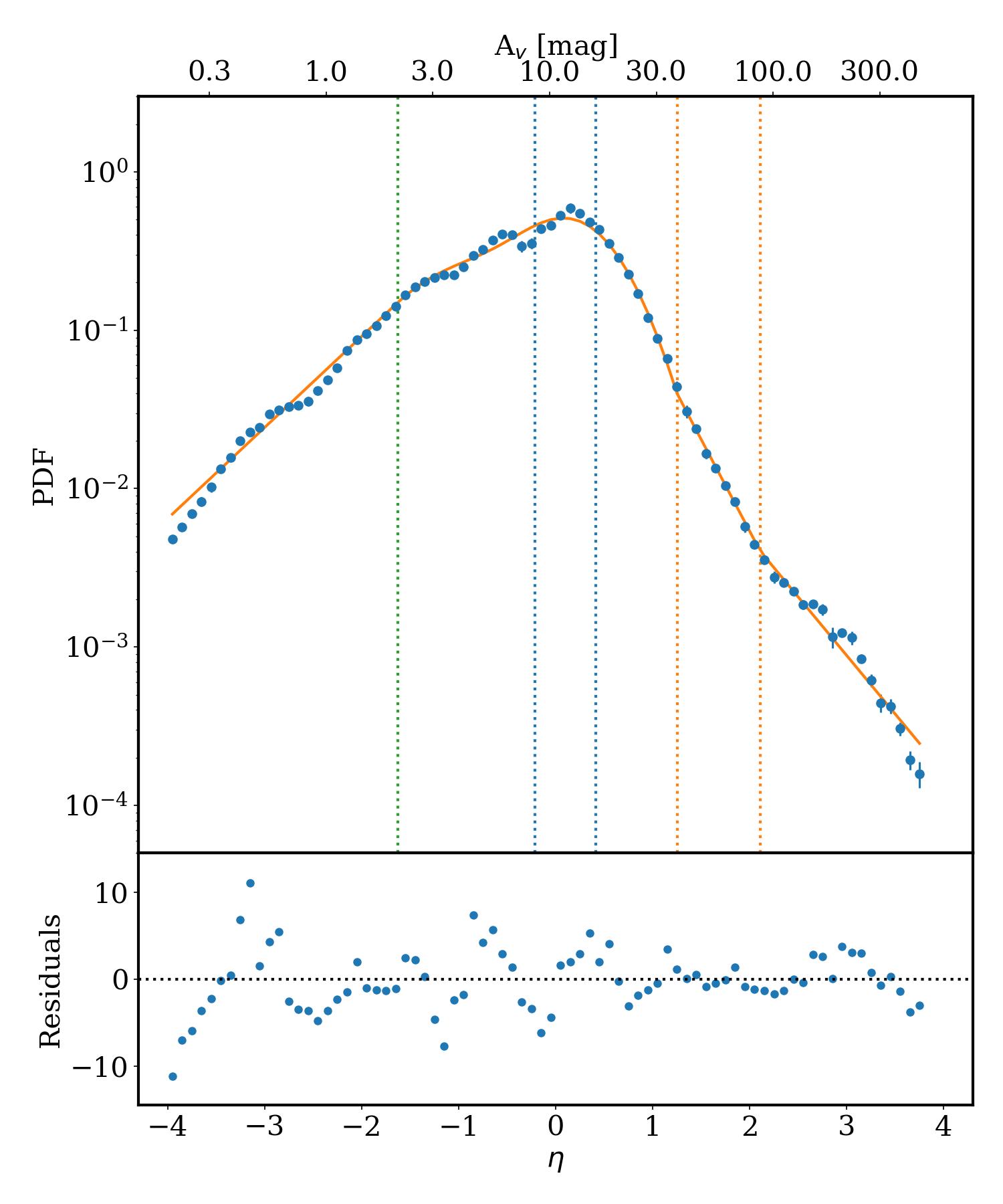}
\end{subfigure}
\caption{M16 (left) and M17 (right). \label{m16+m17-npdf}}
\end{figure*}

\begin{figure*}[!h]
\centering
 \begin{subfigure}[c]{0.40\textwidth}
 \includegraphics[width=1.\textwidth]{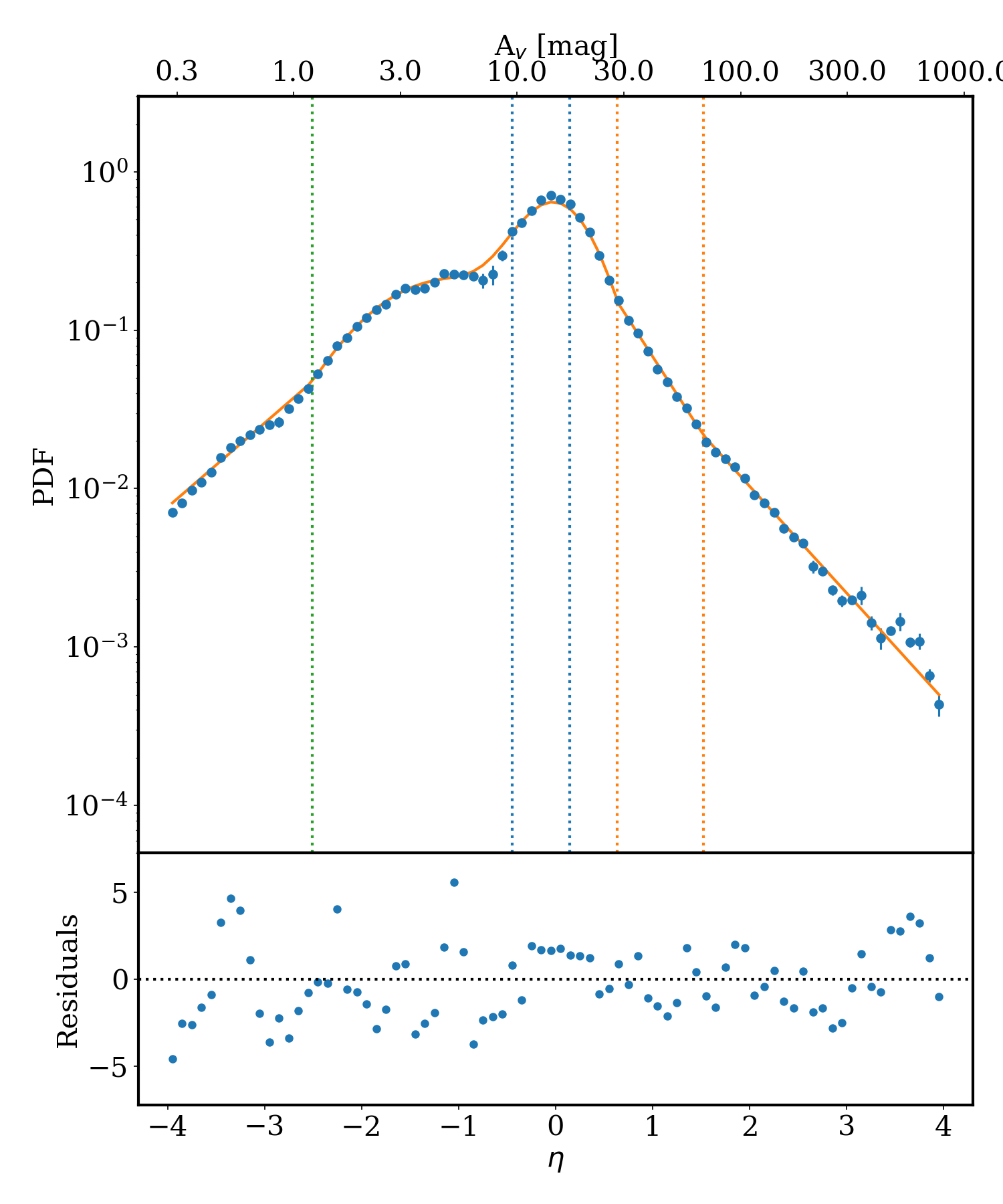} 
\end{subfigure}
\begin{subfigure}[c]{0.40\textwidth}
\includegraphics[width=1.\textwidth]{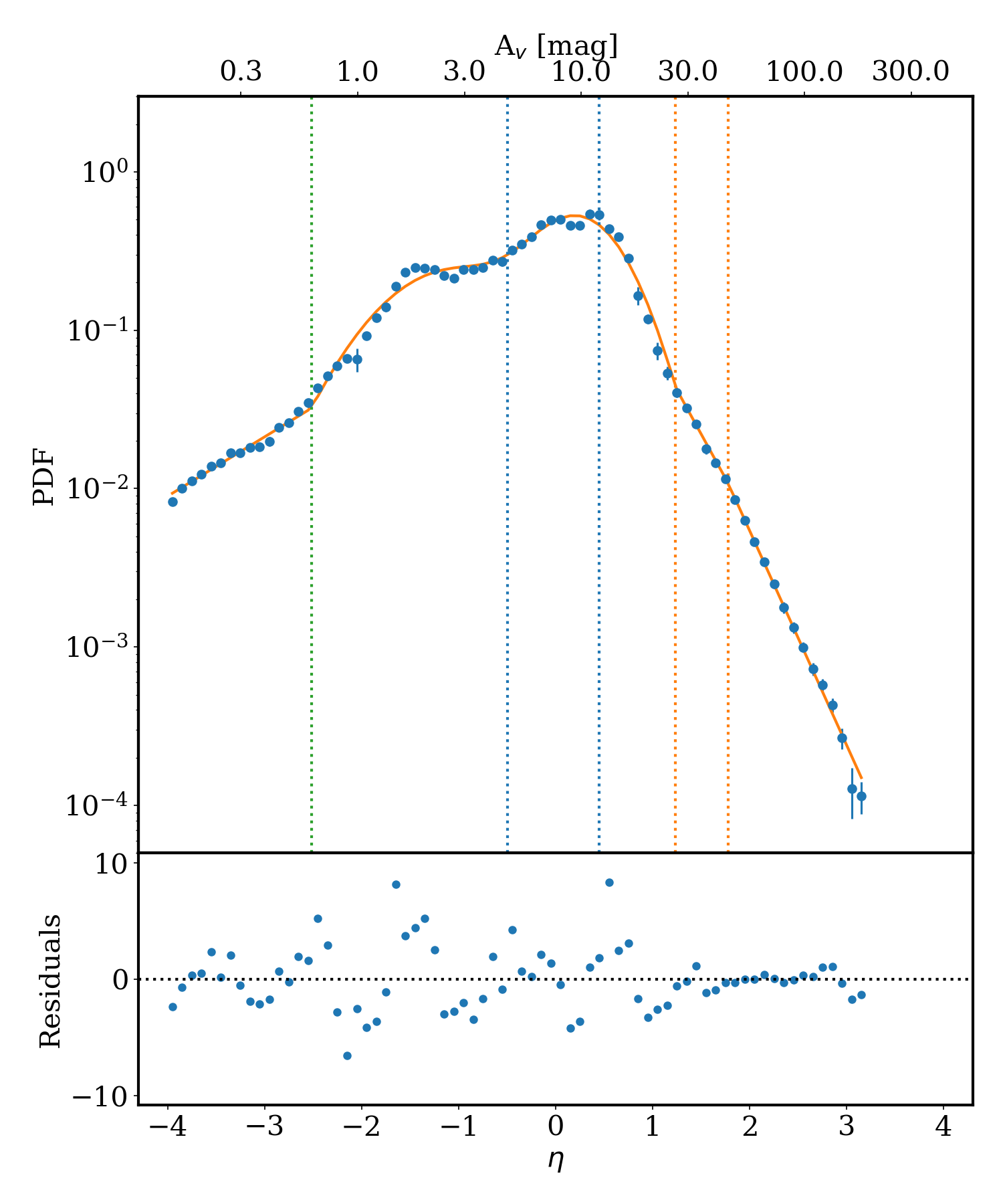}
\end{subfigure}
\caption{NGC6334 (left) and NGC6357 (right). \label{ngc6334+ngc6357-npdf}}
\end{figure*}

\begin{figure*}[!h]
\centering
\begin{subfigure}[c]{0.40\textwidth}
\includegraphics[width=1.\textwidth]{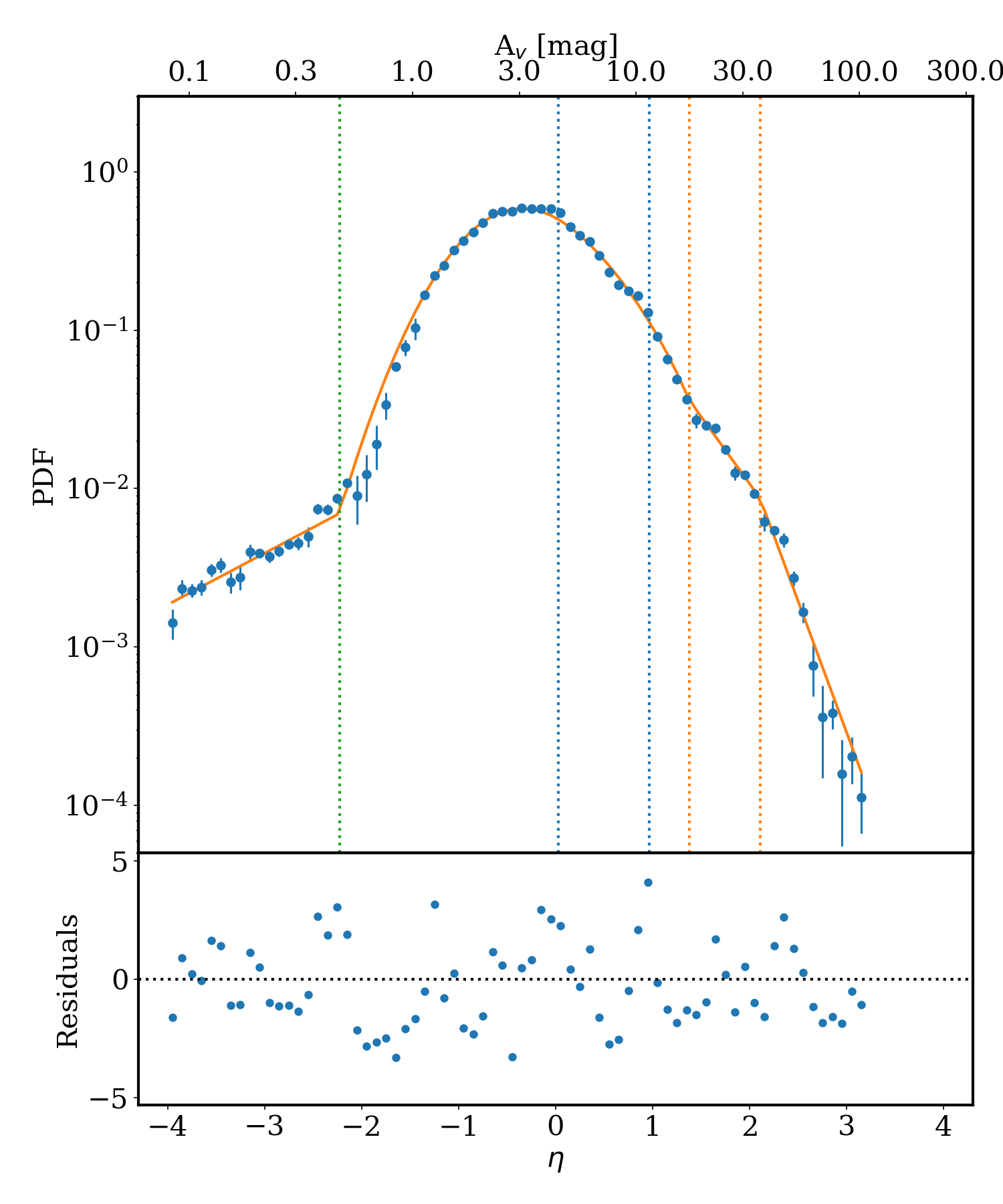}
\end{subfigure}
 \begin{subfigure}[c]{0.40\textwidth}
 \includegraphics[width=1.\textwidth]{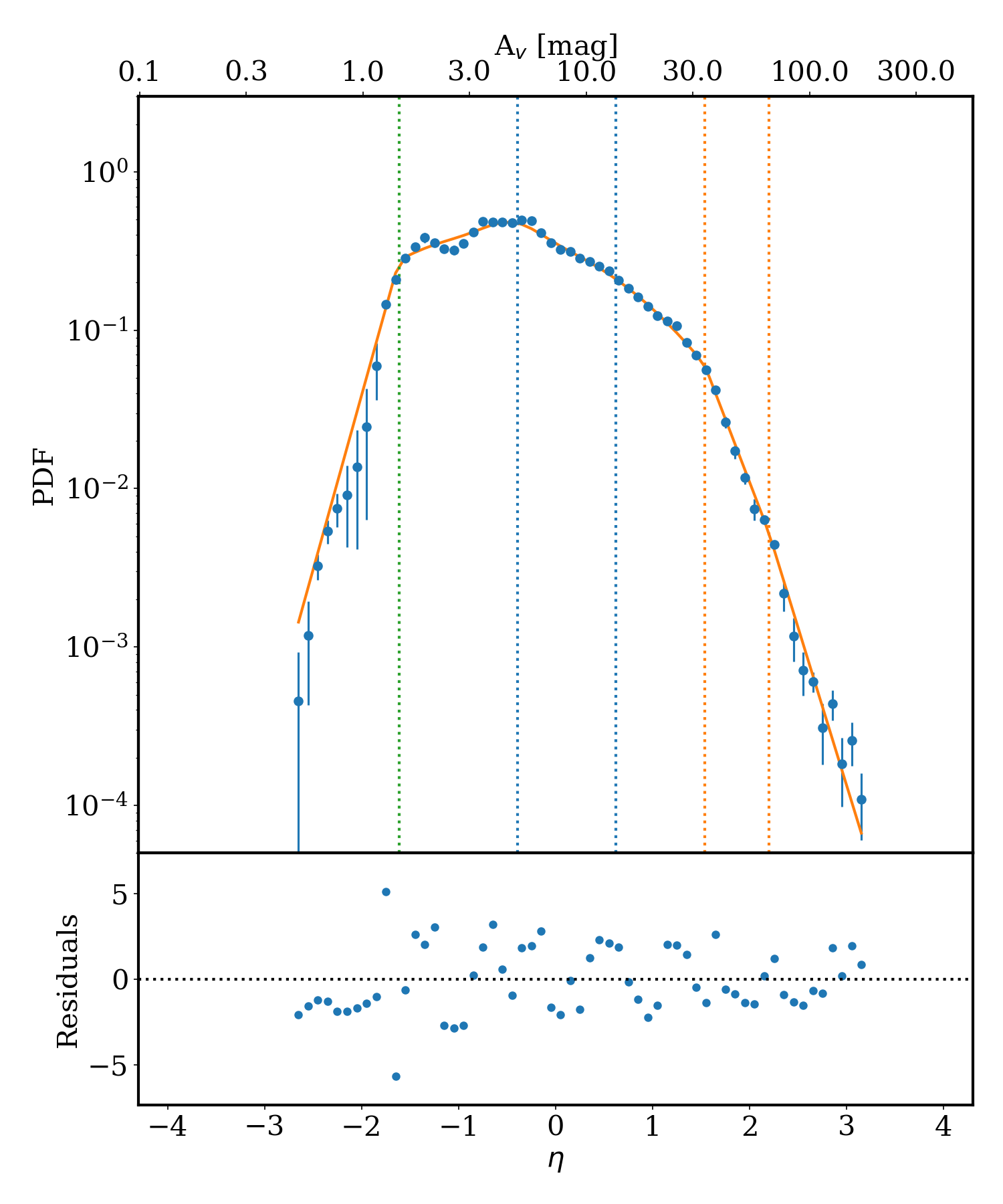} 
\end{subfigure}
\caption{Rosette (left) and Vela (right). \label{rosette+vela-npdf}}
\end{figure*}

\begin{figure*}[!h]
\centering
 \begin{subfigure}[c]{0.40\textwidth}
 \includegraphics[width=1.\textwidth]{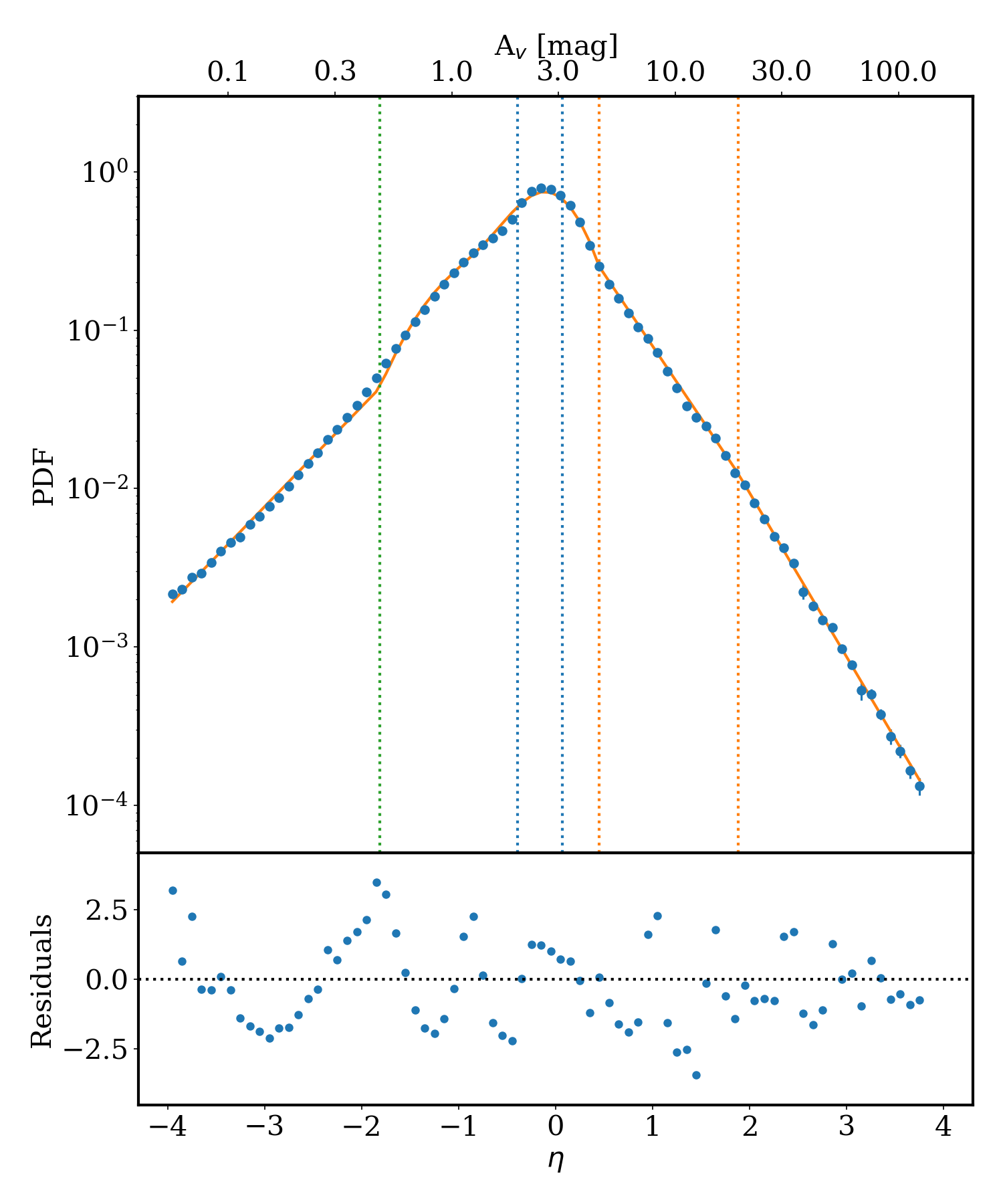} 
\end{subfigure}
\begin{subfigure}[c]{0.40\textwidth}
\includegraphics[width=1.\textwidth]{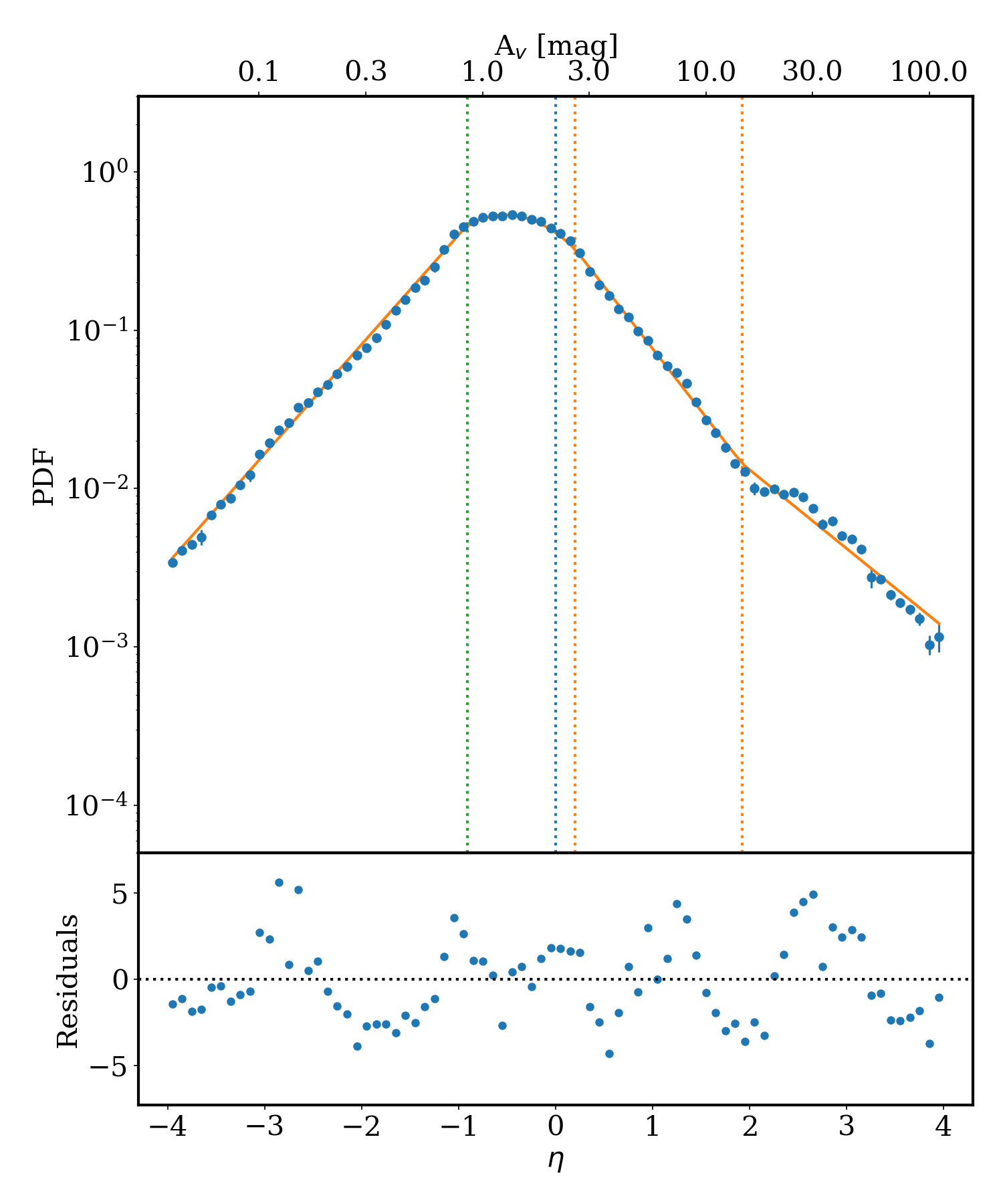}
\end{subfigure}
\caption{Aquila (left) and Mon~R2 (right). \label{aquila+monr2-npdf}}
\end{figure*}

\begin{figure*}[!h]
\centering
 \begin{subfigure}[c]{0.40\textwidth}
 \includegraphics[width=1.\textwidth]{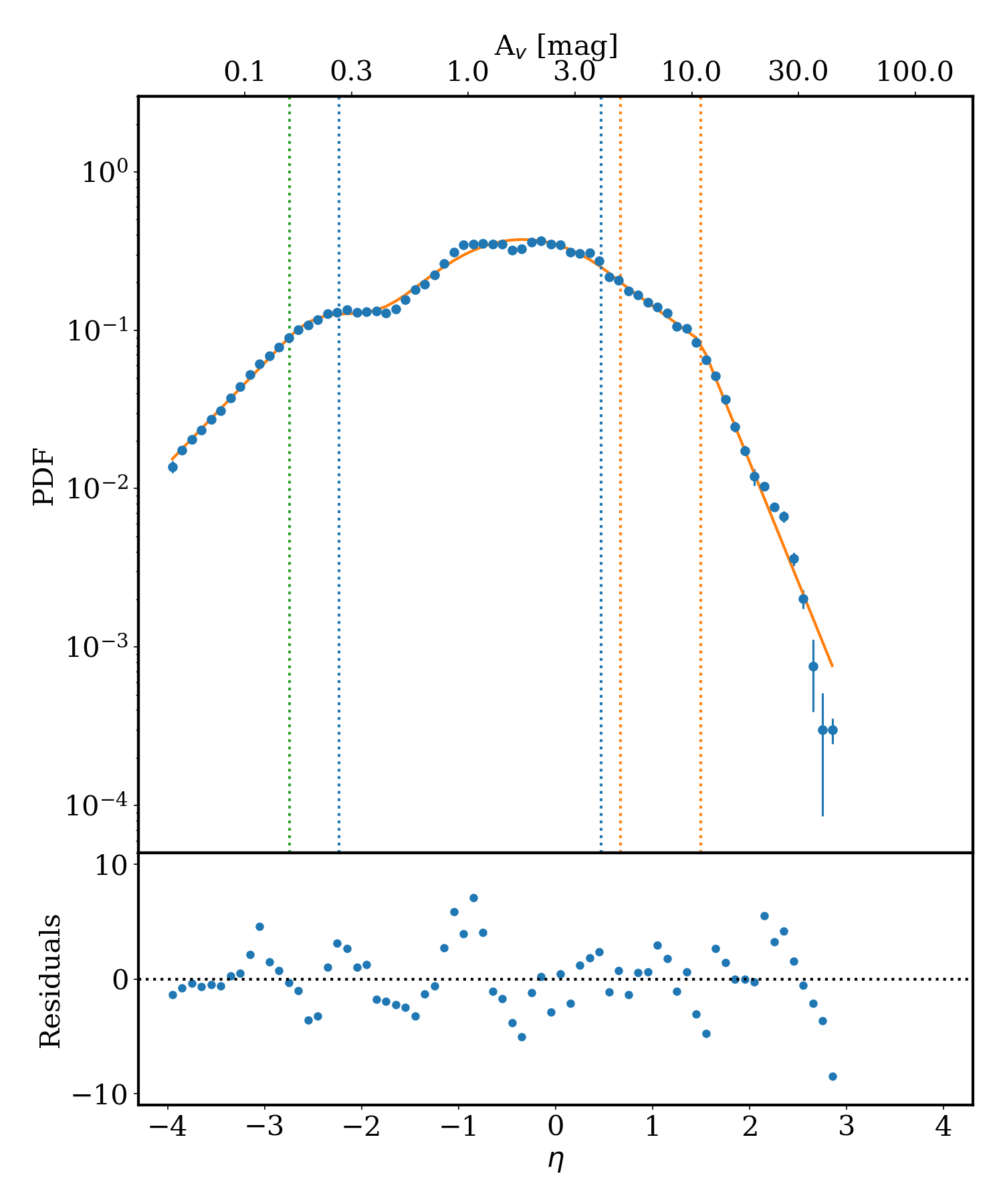} 
\end{subfigure}
\begin{subfigure}[c]{0.40\textwidth}
\includegraphics[width=1.\textwidth]{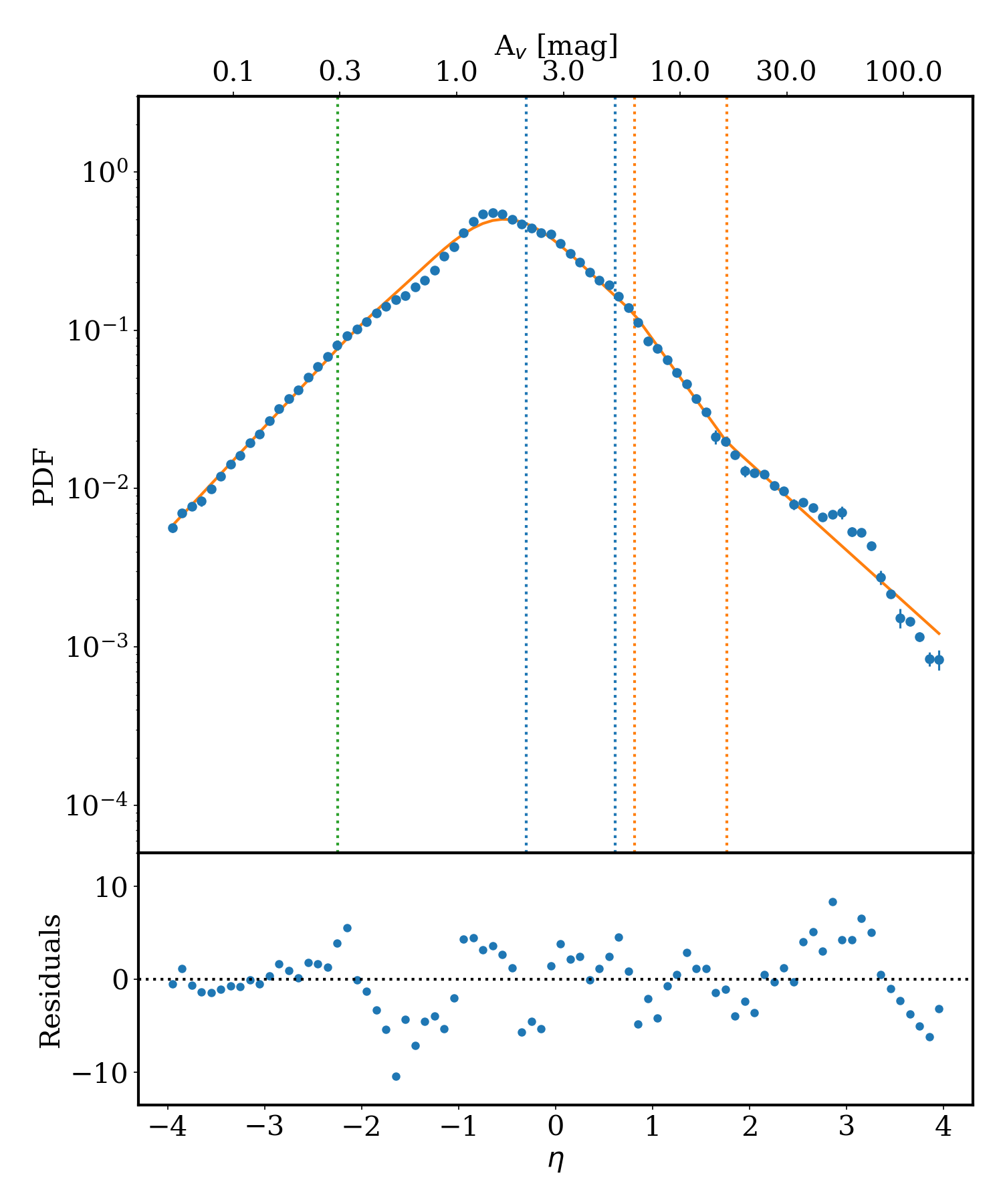}
\end{subfigure}
\caption{MonOB1 (left) and NGC2264 (right). \label{monob1+ngc2264-npdf}}
\end{figure*}

\begin{figure*}[!h]
\centering
 \begin{subfigure}[c]{0.40\textwidth}
 \includegraphics[width=1.\textwidth]{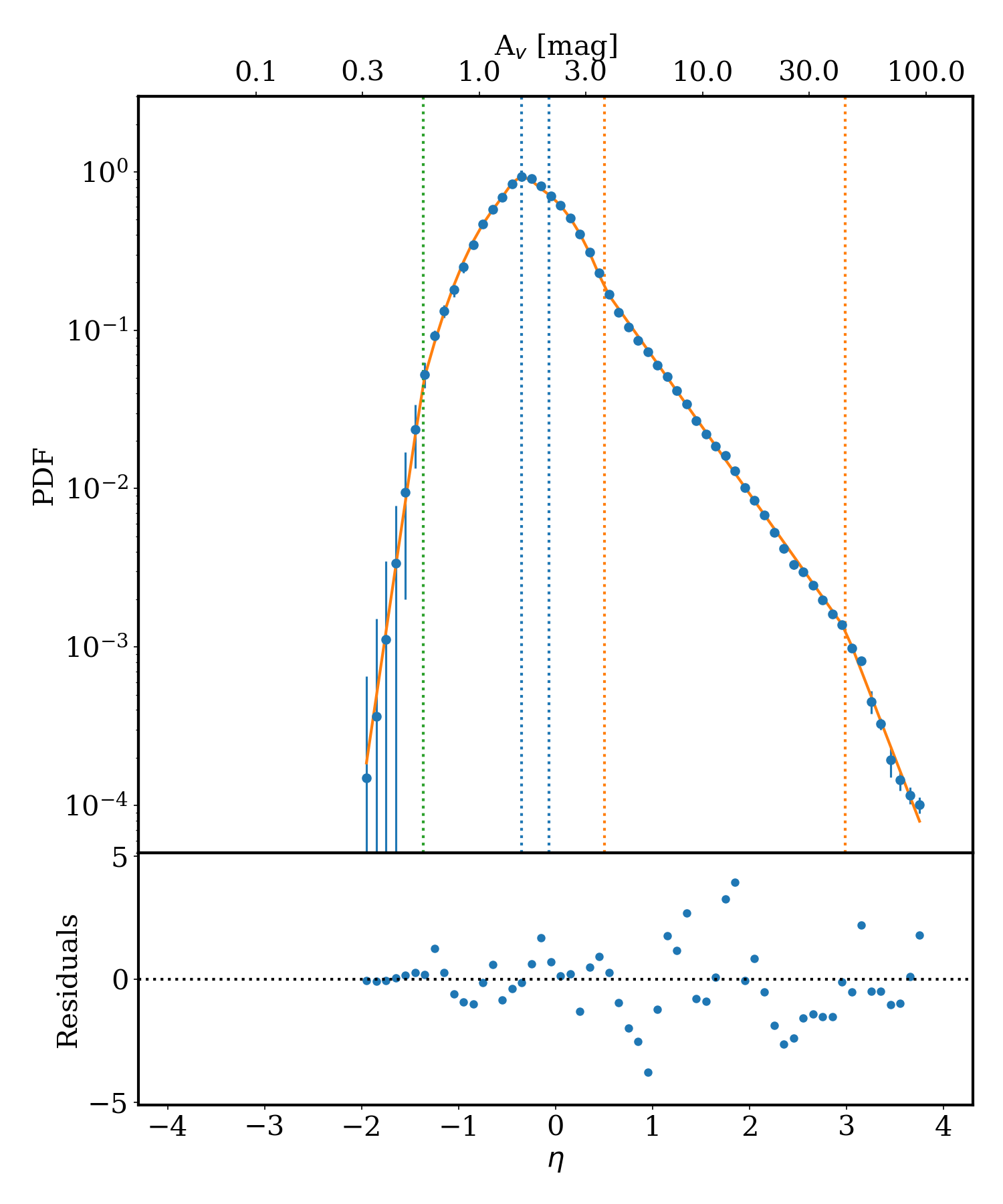} 
\end{subfigure}
\begin{subfigure}[c]{0.40\textwidth}
\includegraphics[width=1.\textwidth]{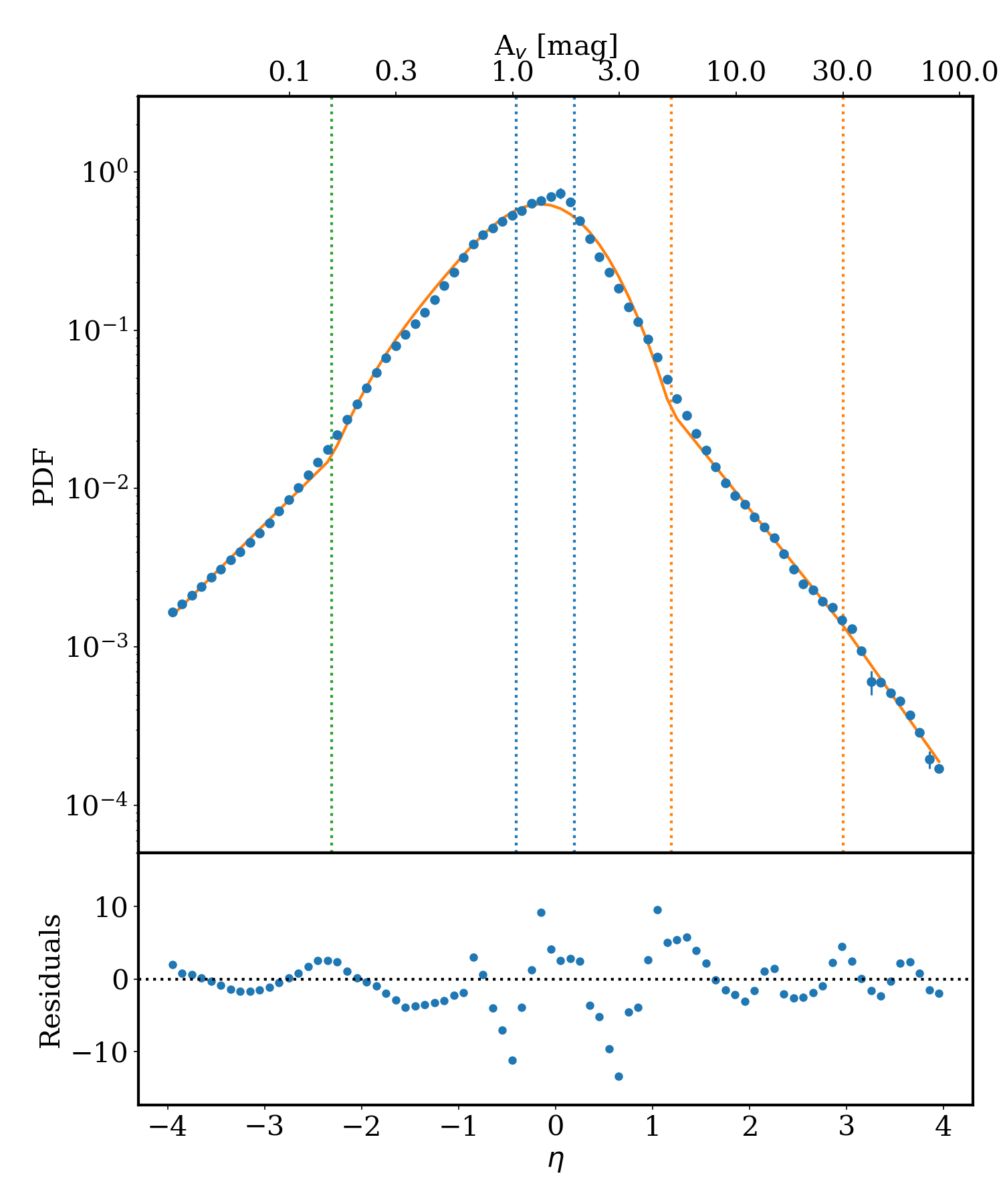}
\end{subfigure}
\caption{Orion~B (left) and Serpens (right). \label{orionb+serpens-npdf}}
\end{figure*}

\begin{figure*}[!h]
\centering
 \begin{subfigure}[c]{0.40\textwidth}
 \includegraphics[width=1.\textwidth]{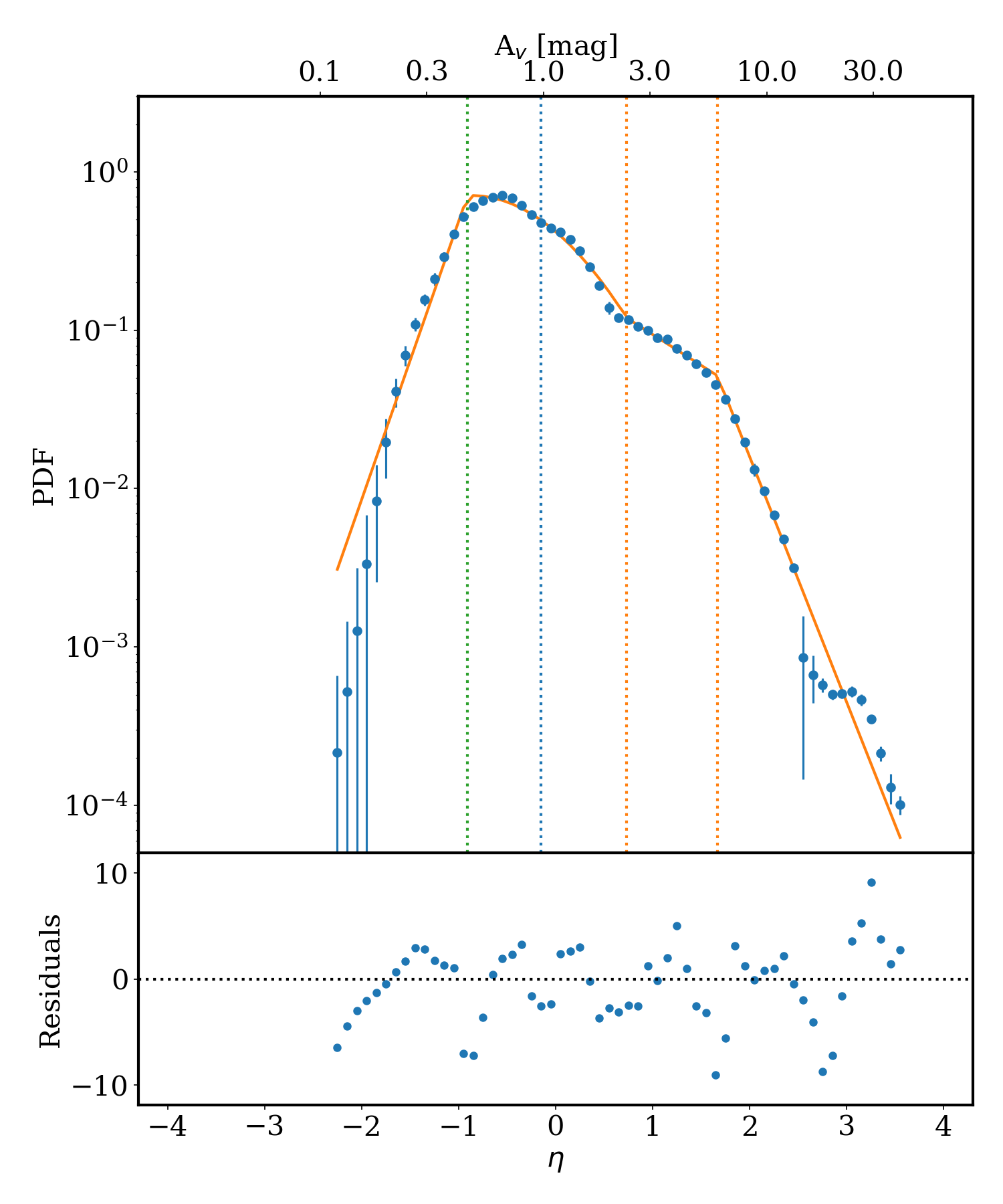} 
\end{subfigure}
\begin{subfigure}[c]{0.40\textwidth}
\includegraphics[width=1.\textwidth]{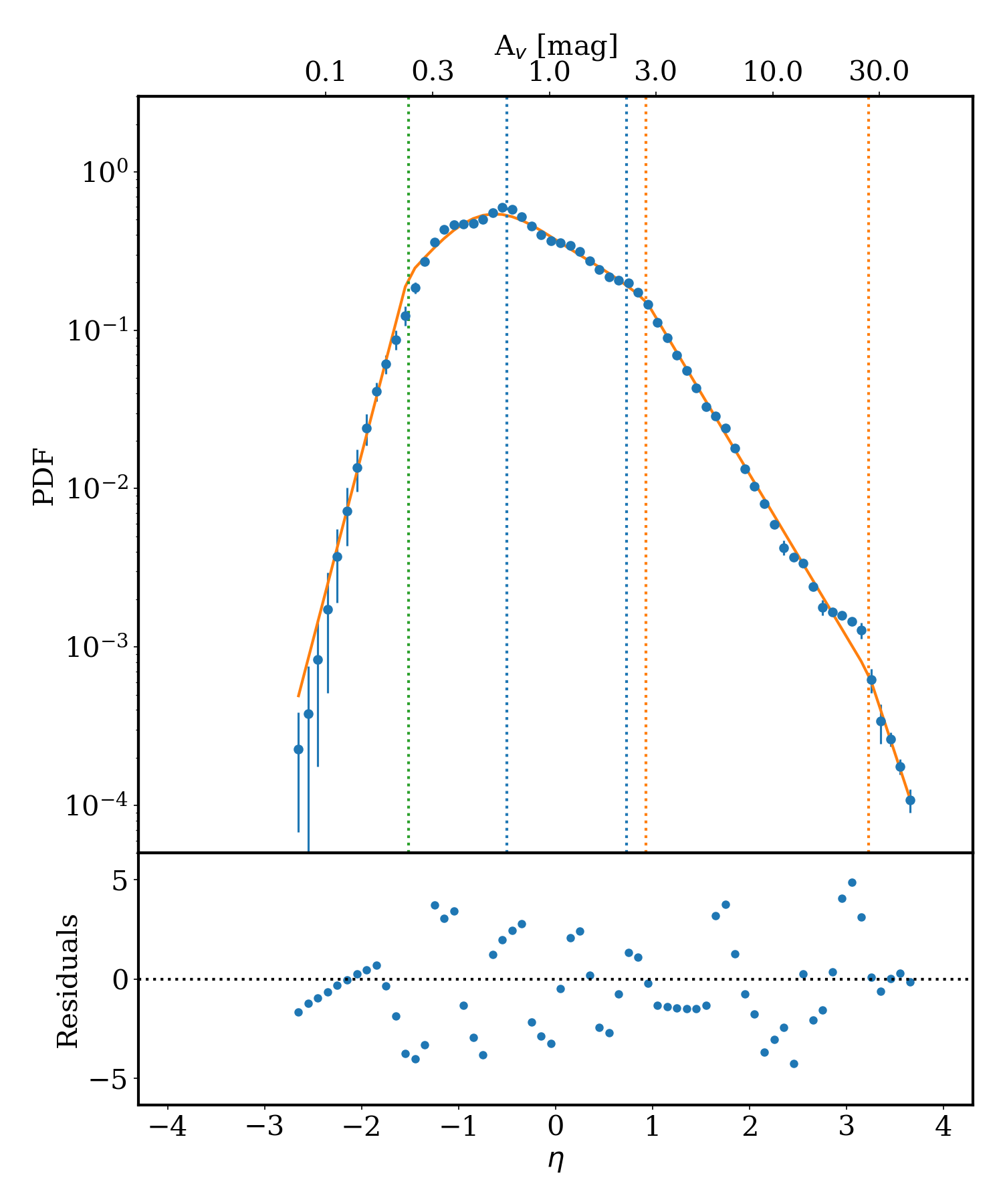}
\end{subfigure}
\caption{ChamI (left) and ChamII (right). \label{chamI+chamII-npdf}}
\end{figure*}

\begin{figure*}[!h]
\centering
 \begin{subfigure}[c]{0.40\textwidth}
 \includegraphics[width=1.\textwidth]{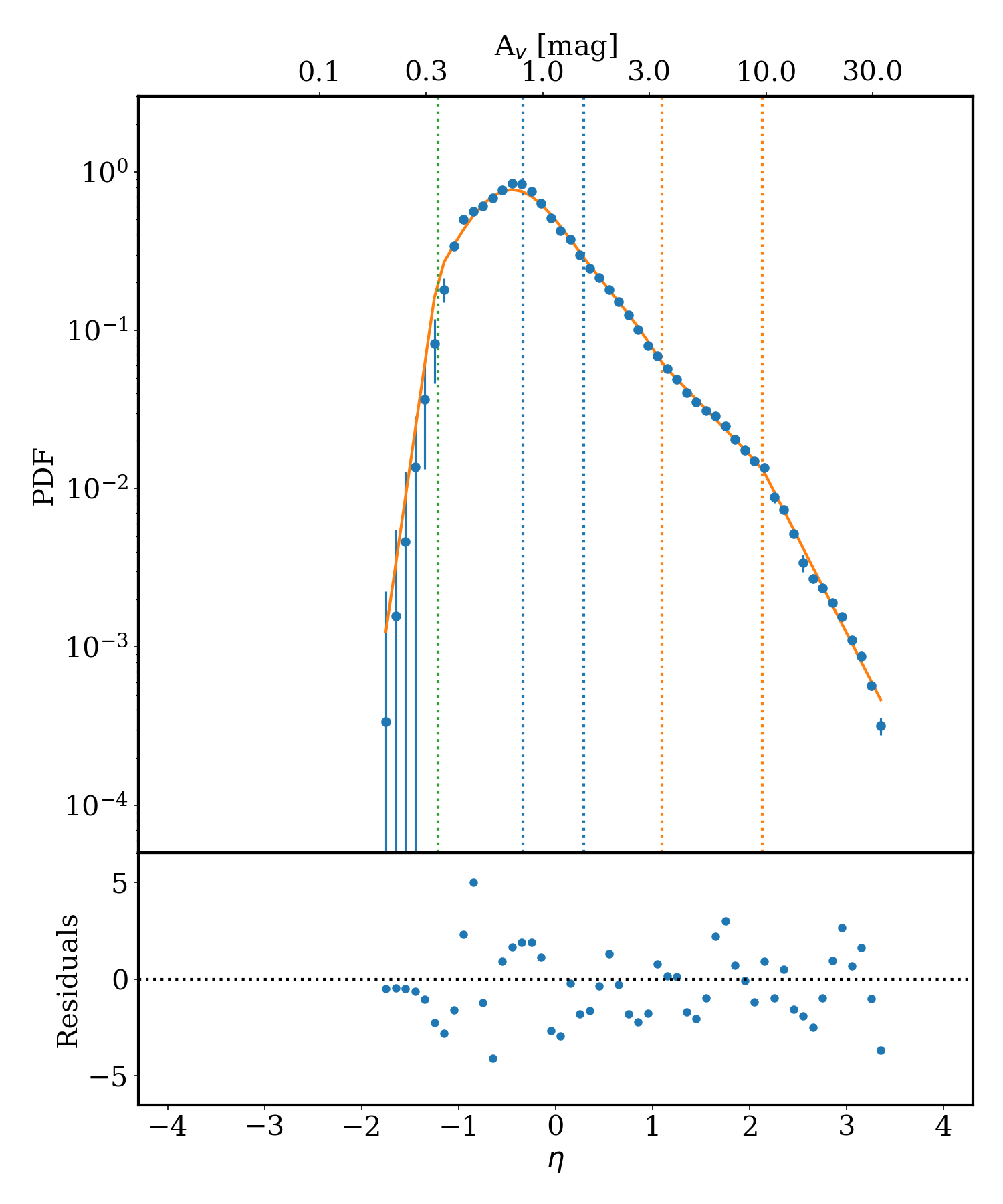} 
\end{subfigure}
\begin{subfigure}[c]{0.40\textwidth}
\includegraphics[width=1.\textwidth]{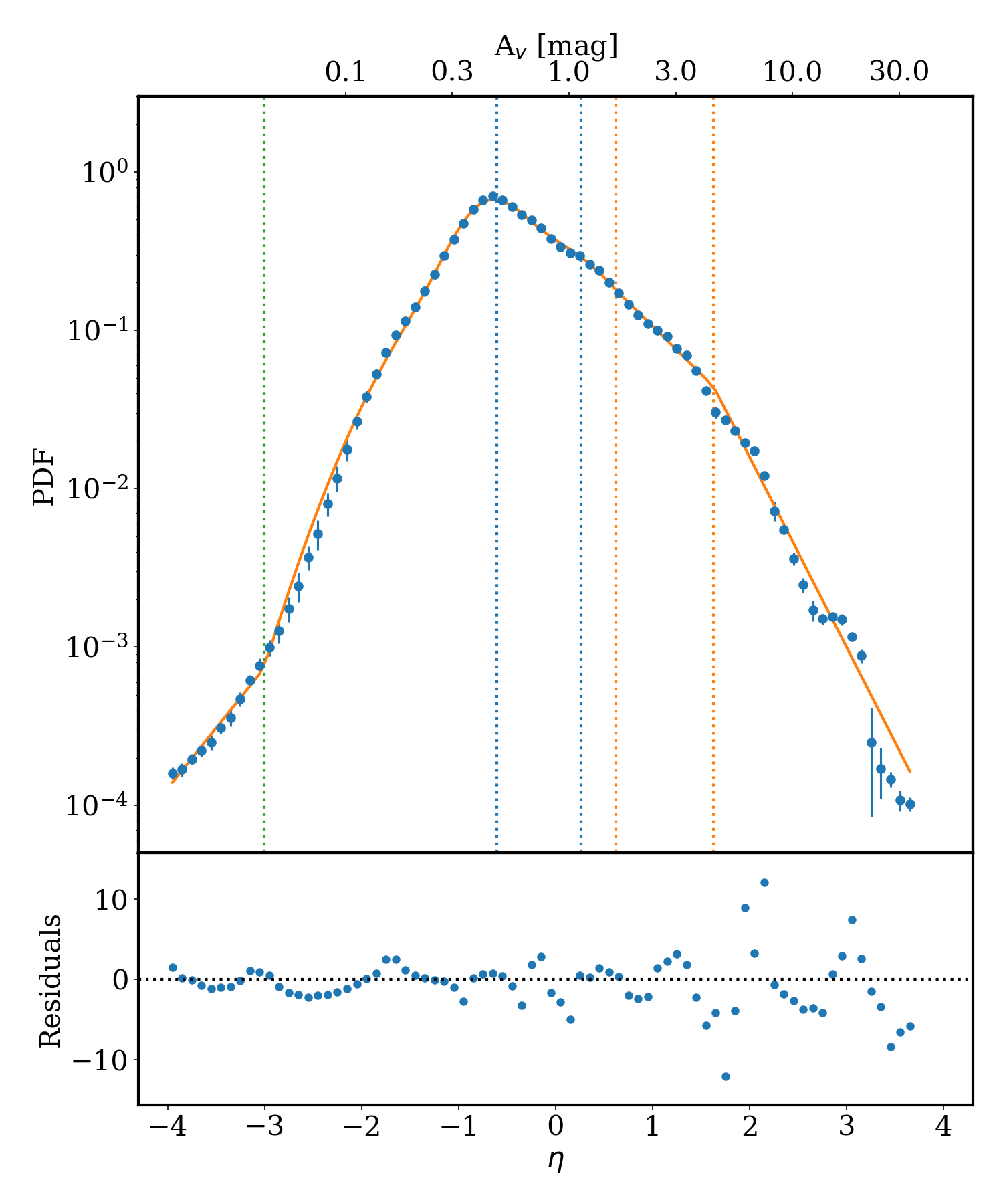}
\end{subfigure}
\caption{IC5146 (left) and Lupus~I (right). \label{ic5146+lupusI-npdf}}
\end{figure*}

\begin{figure*}[!h]
\centering
 \begin{subfigure}[c]{0.40\textwidth}
 \includegraphics[width=1.\textwidth]{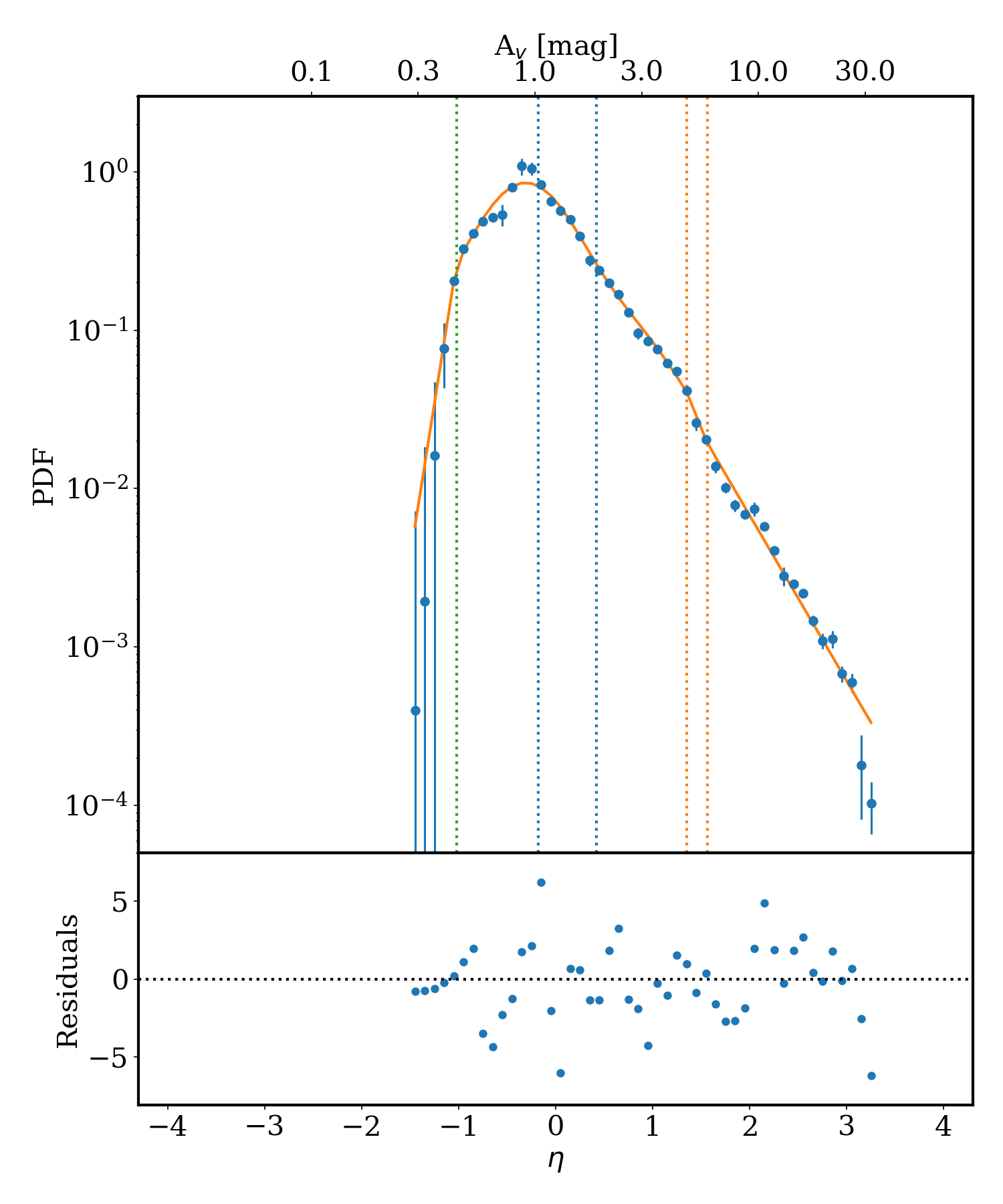} 
\end{subfigure}
\begin{subfigure}[c]{0.40\textwidth}
\includegraphics[width=1.\textwidth]{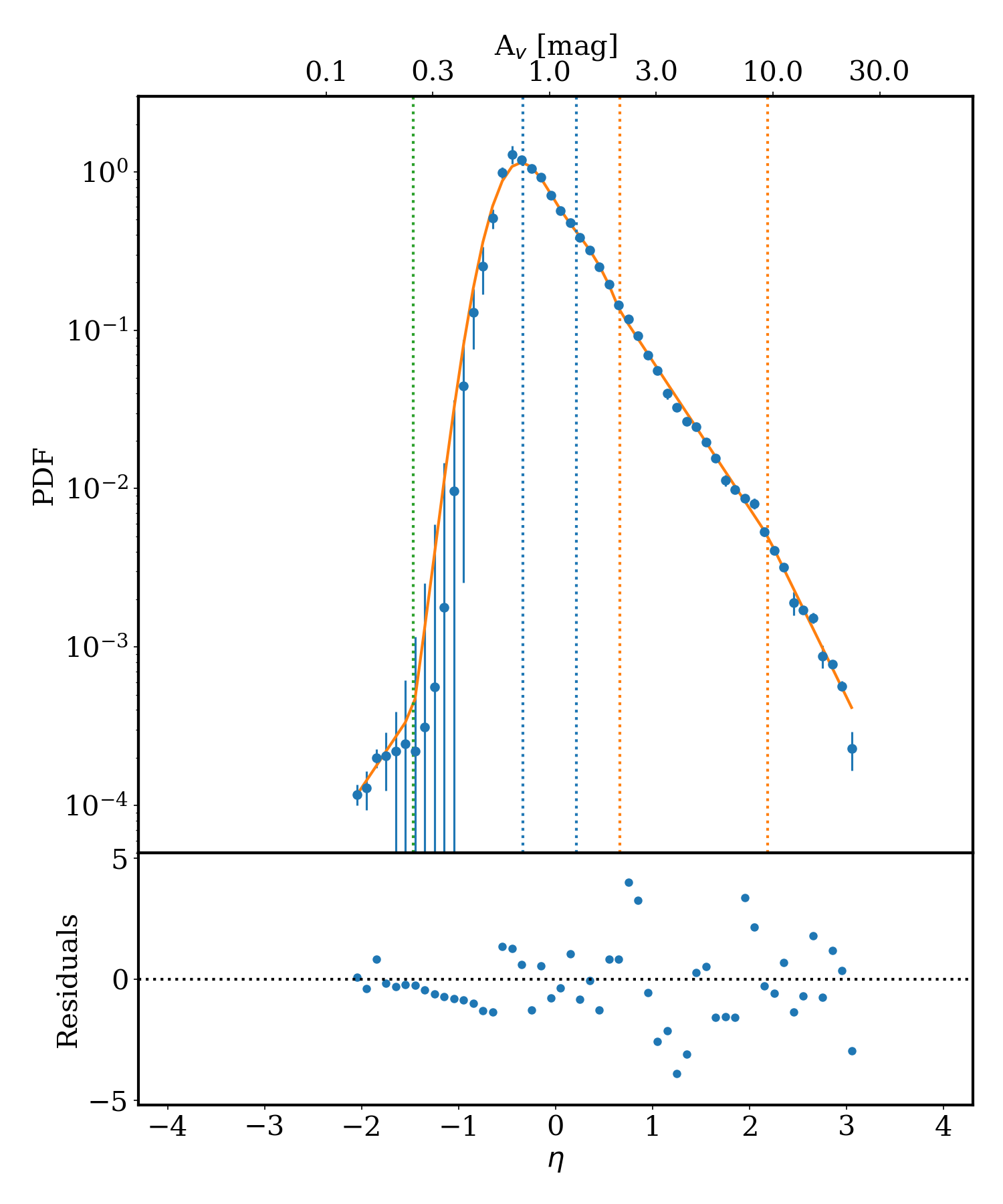}
\end{subfigure}
\caption{Lupus~III (left) and Lupus~IV (right). \label{lupusIII+lupusIV-npdf}}
\end{figure*}

\begin{figure*}[!h]
\centering
 \begin{subfigure}[c]{0.40\textwidth}
 \includegraphics[width=1.\textwidth]{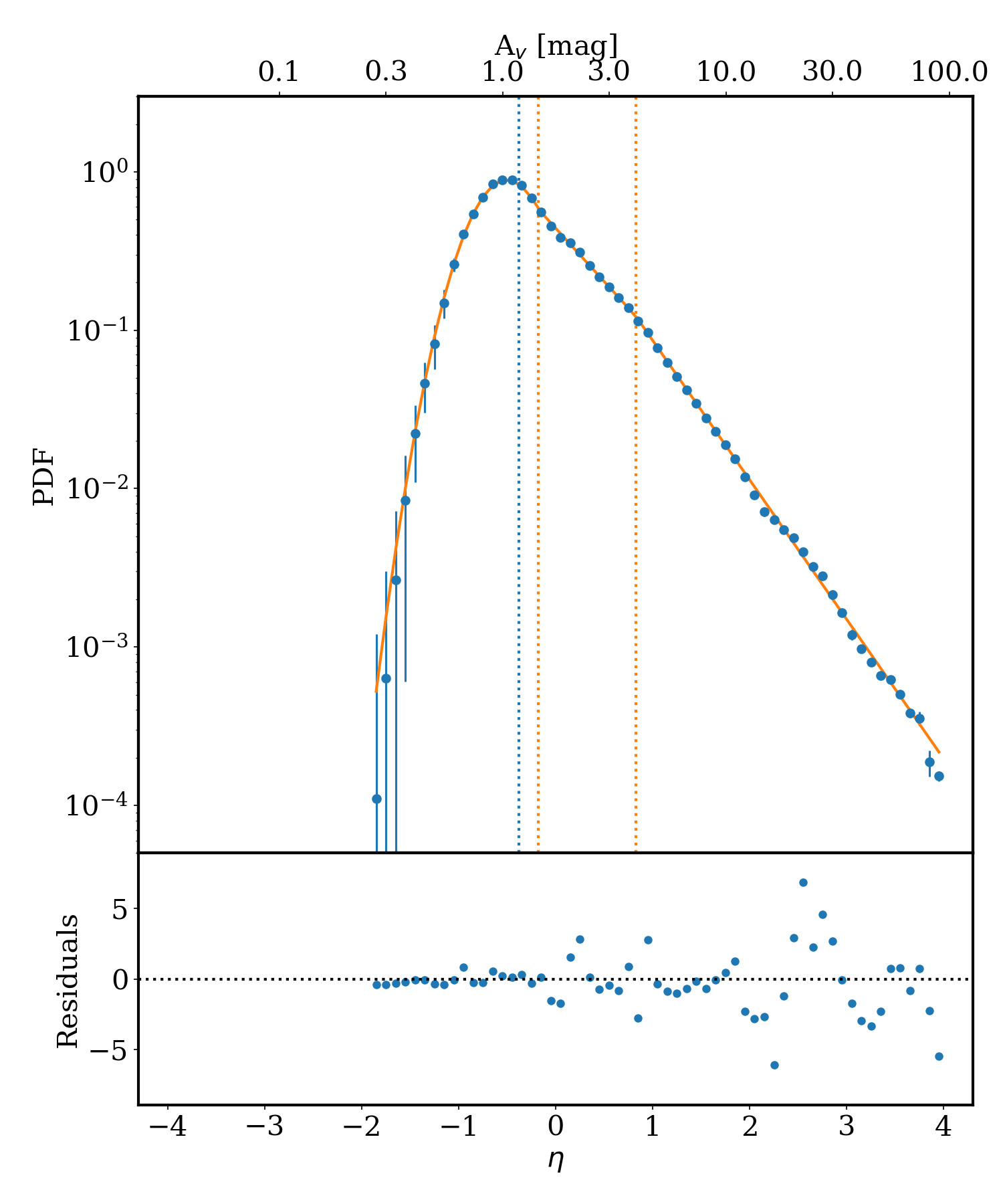} 
\end{subfigure}
\begin{subfigure}[c]{0.40\textwidth}
\includegraphics[width=1.\textwidth]{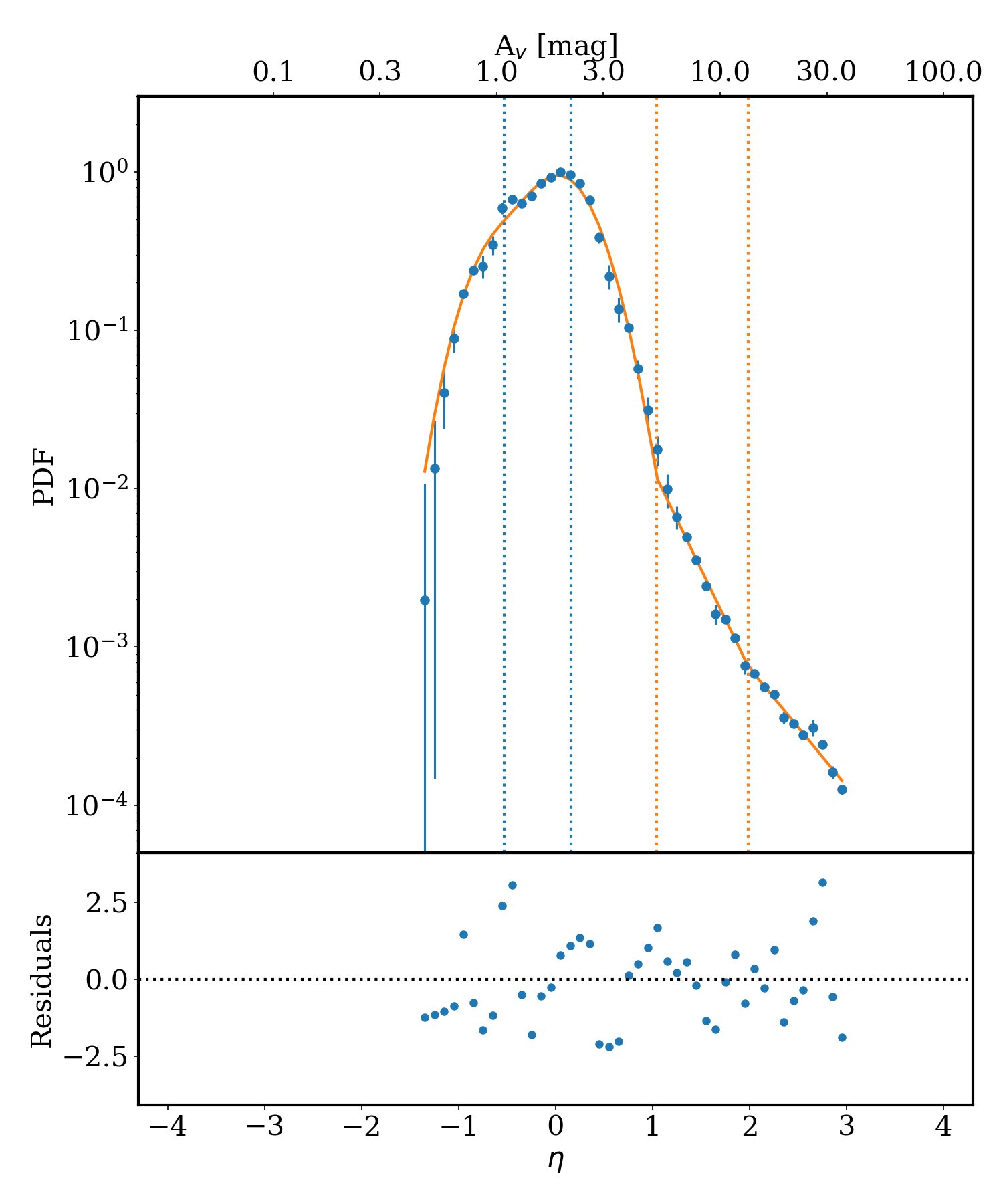}
\end{subfigure}
\caption{Perseus (left) and Pipe (right). \label{perseus+pipe-npdf}}
\end{figure*}

\begin{figure*}[!h]
\centering
 \begin{subfigure}[c]{0.40\textwidth}
 \includegraphics[width=1.\textwidth]{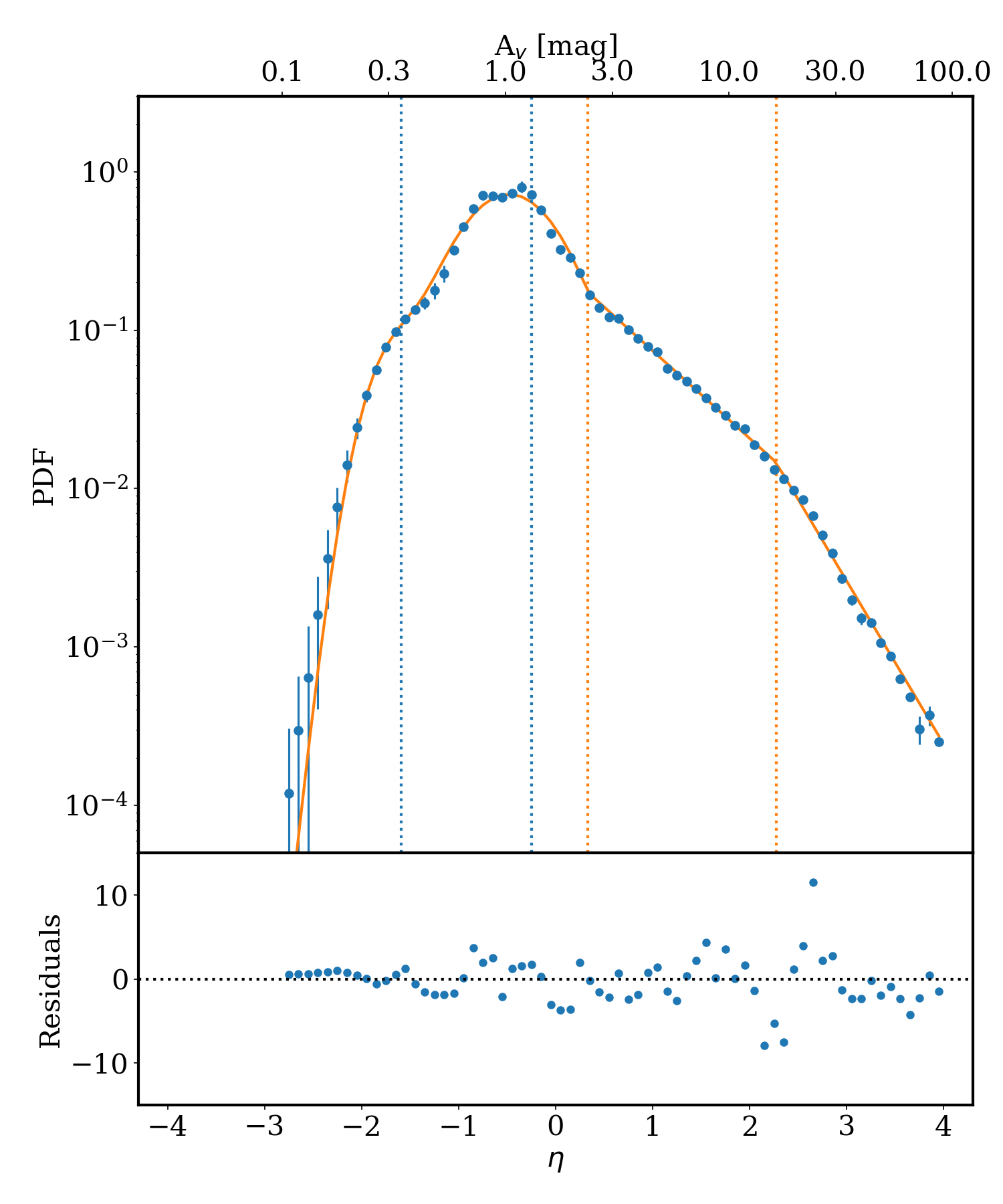} 
\end{subfigure}
\begin{subfigure}[c]{0.40\textwidth}
\includegraphics[width=1.\textwidth]{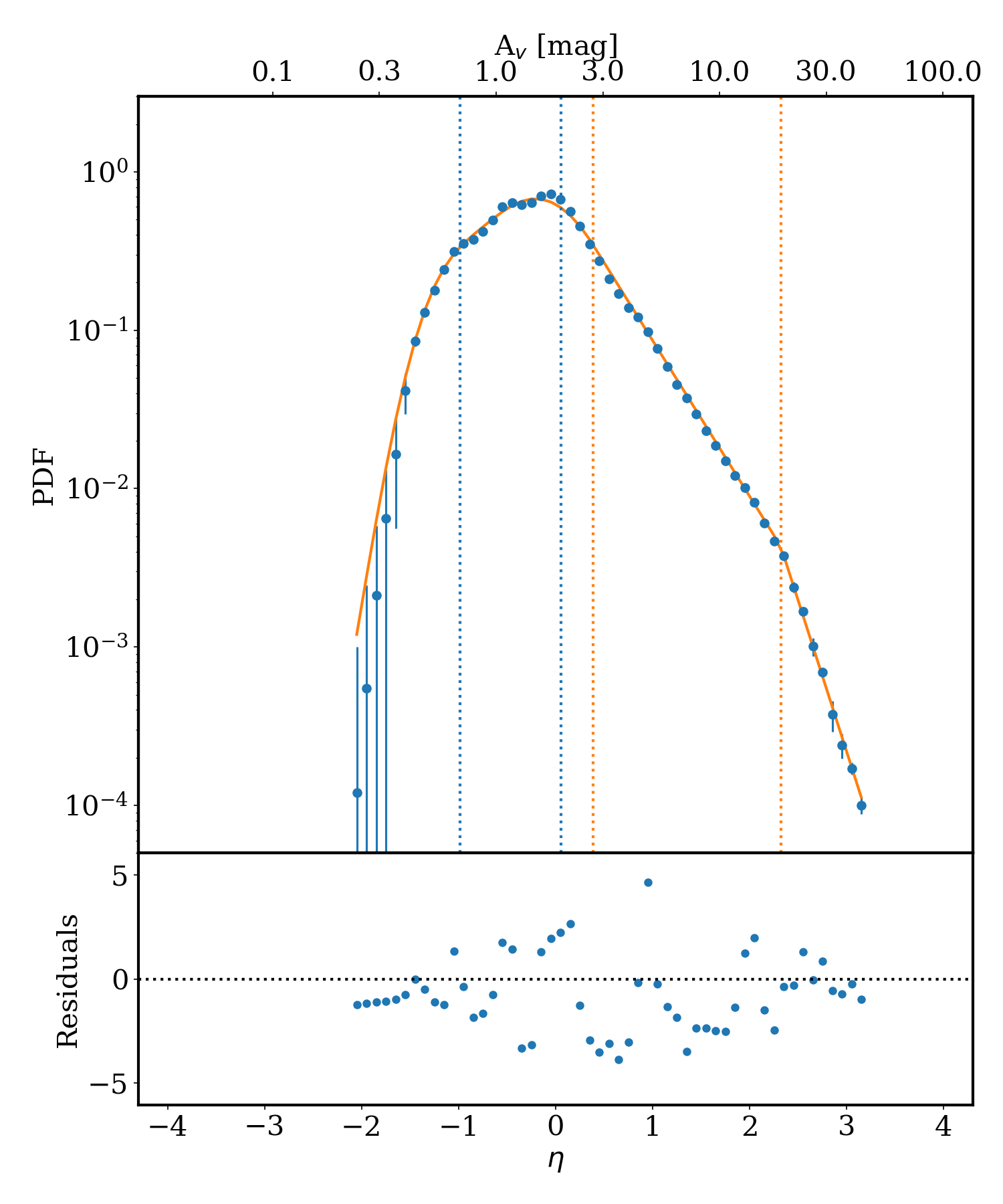}
\end{subfigure}
\caption{Rhooph (left) and Taurus (right). \label{rhooph+taurus-npdf}}
\end{figure*}

\begin{figure*}[!h]
\centering
 \begin{subfigure}[c]{0.40\textwidth}
 \includegraphics[width=1.\textwidth]{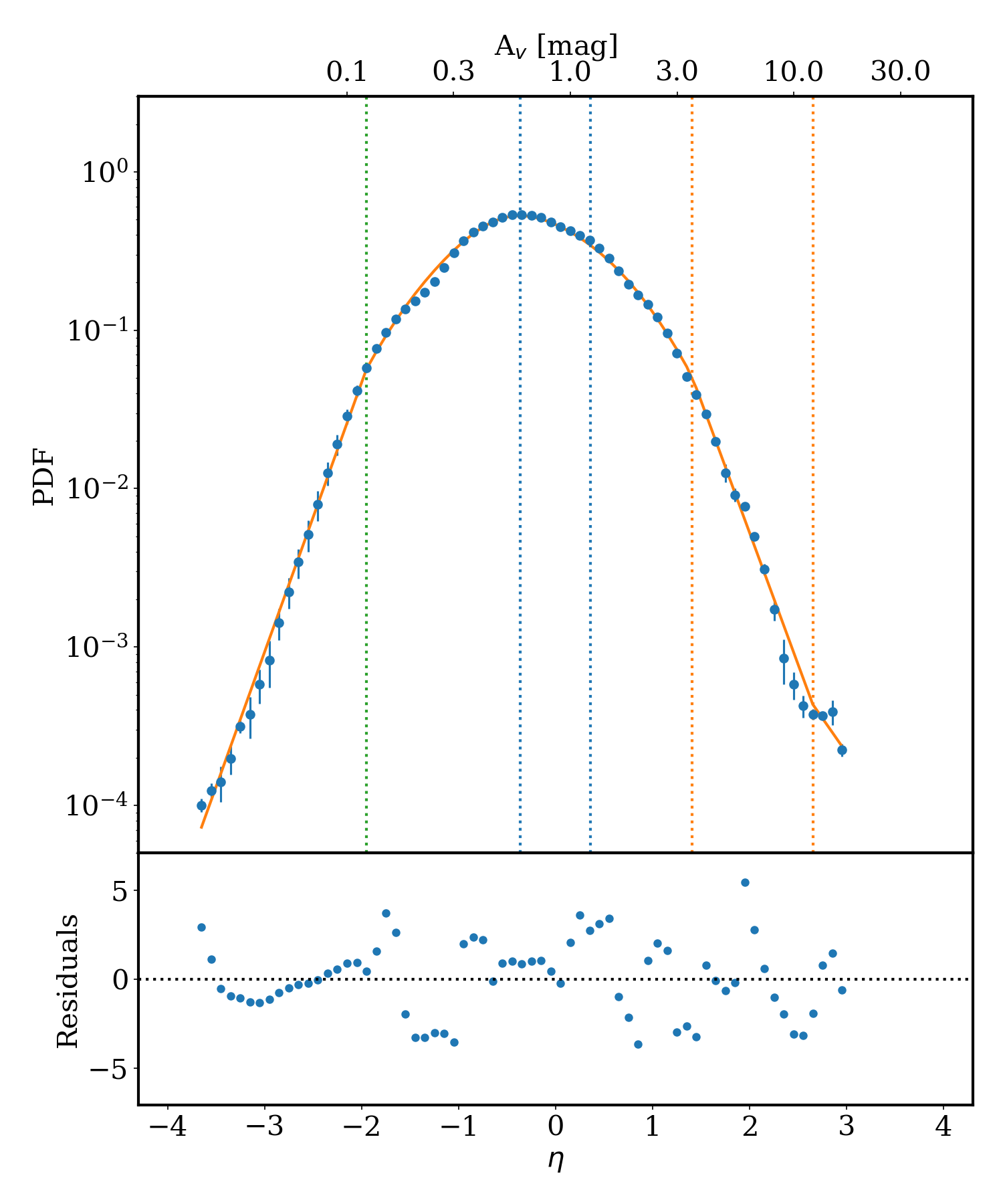} 
\end{subfigure}
\begin{subfigure}[c]{0.40\textwidth}
\includegraphics[width=1.\textwidth]{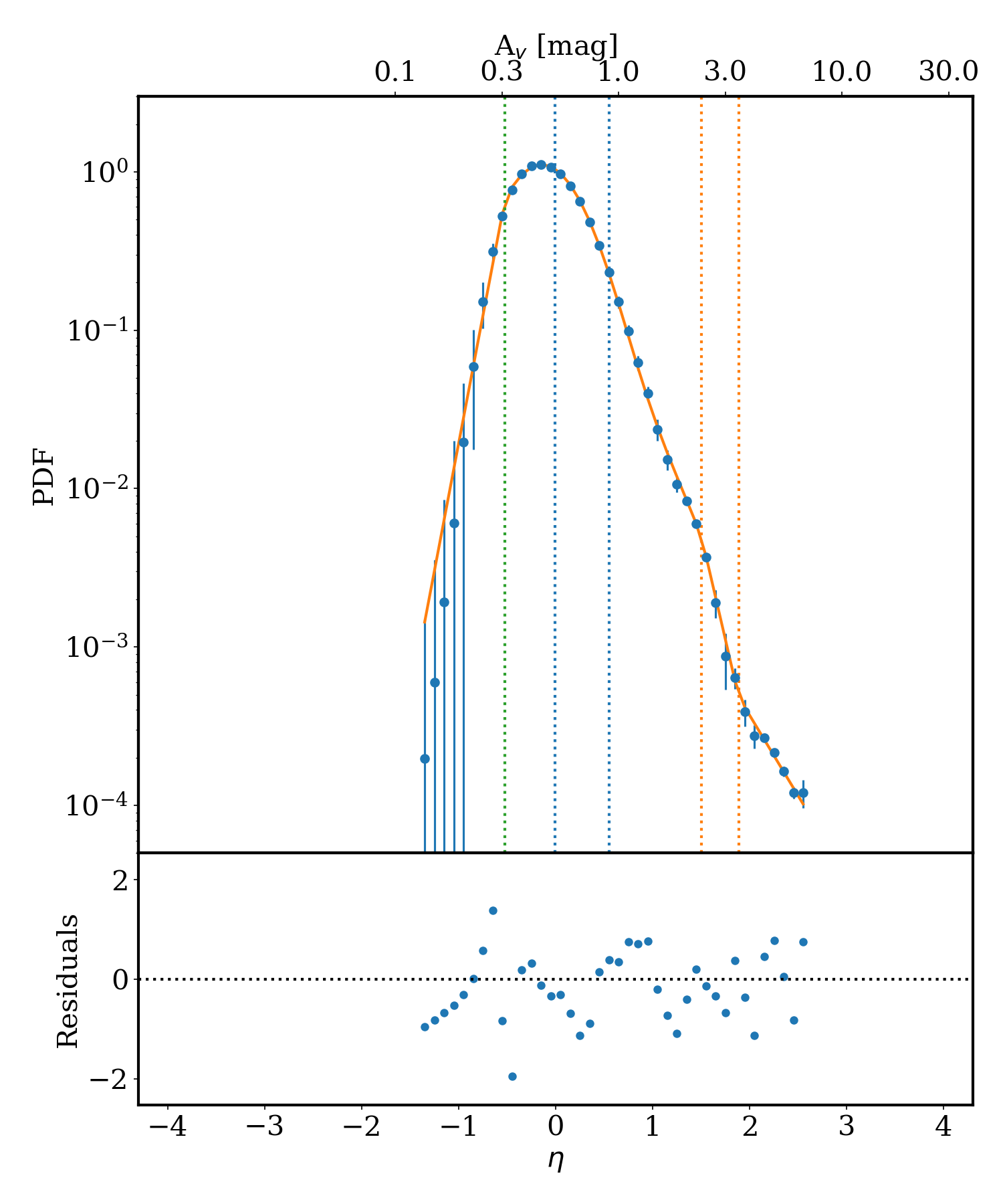}
\end{subfigure}
\caption{Cham~III (left) and Polaris (right). \label{chamIII+polaris-npdf}}
\end{figure*}

\begin{figure*}[!h]
\centering
 \begin{subfigure}[c]{0.40\textwidth}
 \includegraphics[width=1.\textwidth]{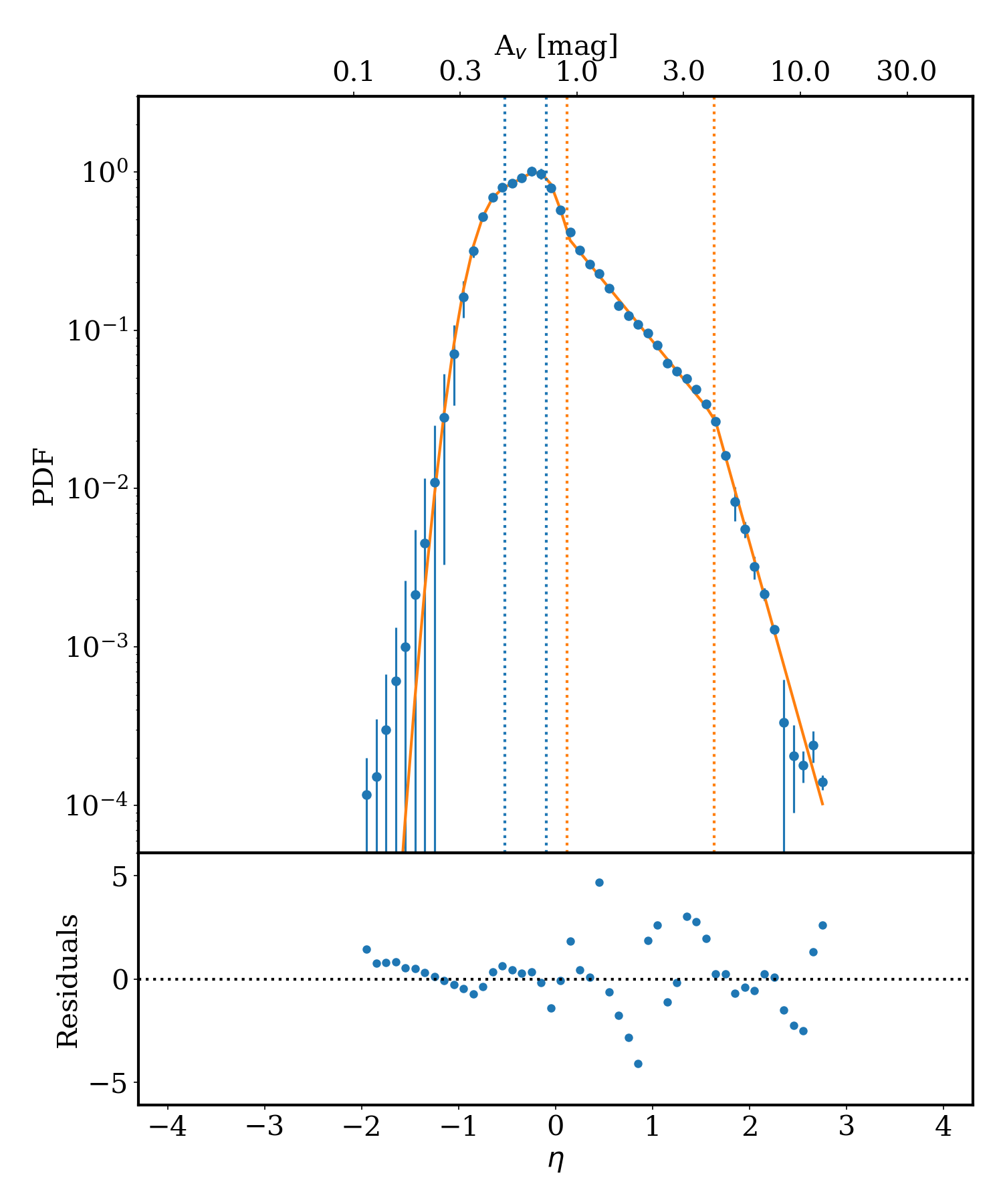} 
\end{subfigure}
\begin{subfigure}[c]{0.40\textwidth} 
 \includegraphics[width=1.\textwidth]{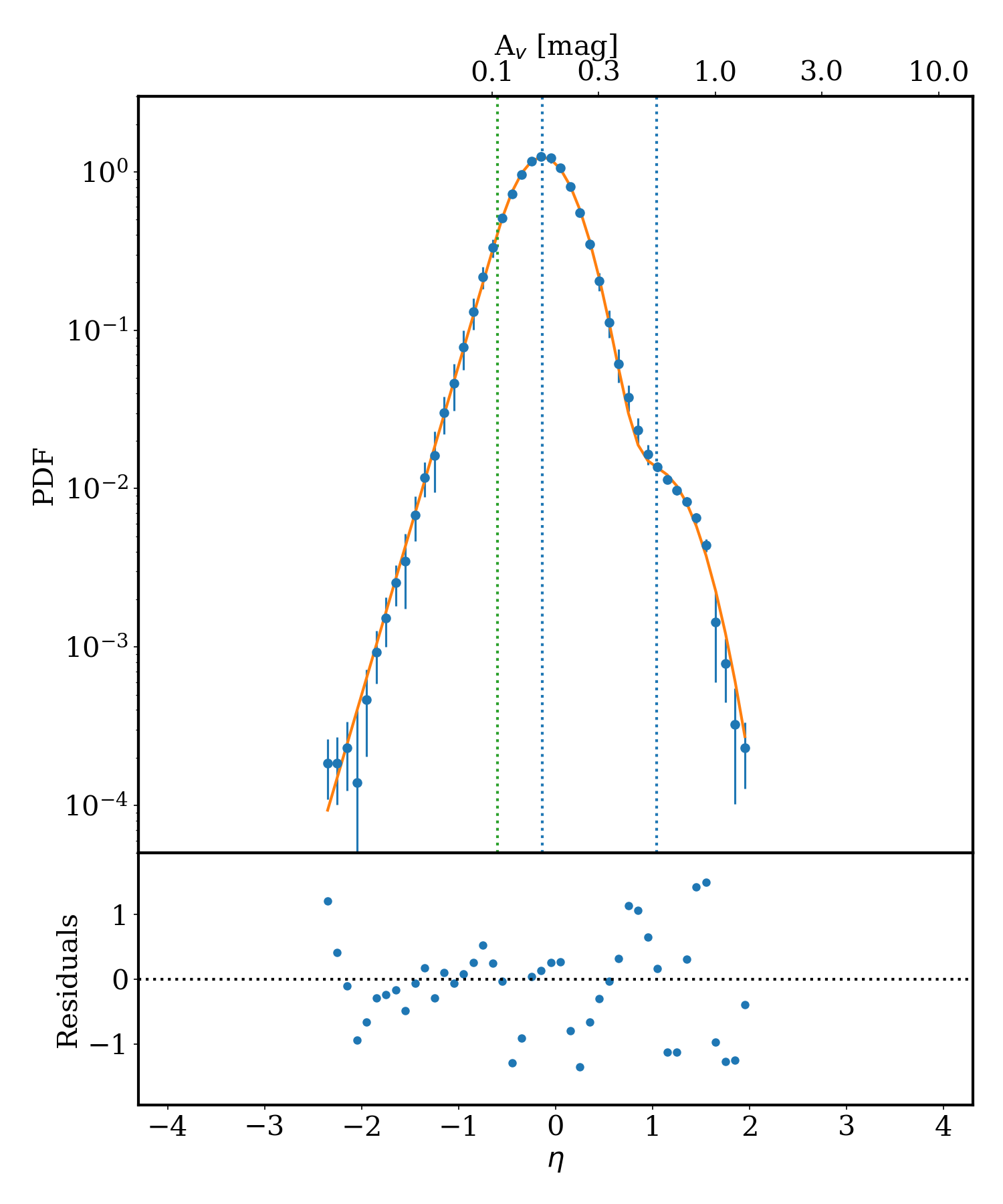} 
\end{subfigure}
\caption{Musca (left) and Draco (right). \label{musca-npdf}}
\end{figure*}

\clearpage

% APPENDIX E

\section{Correlations of N-PDF parameters with cloud type} \label{app-e}  

\begin{figure*}
\centering
\includegraphics[width=8cm, angle=0]{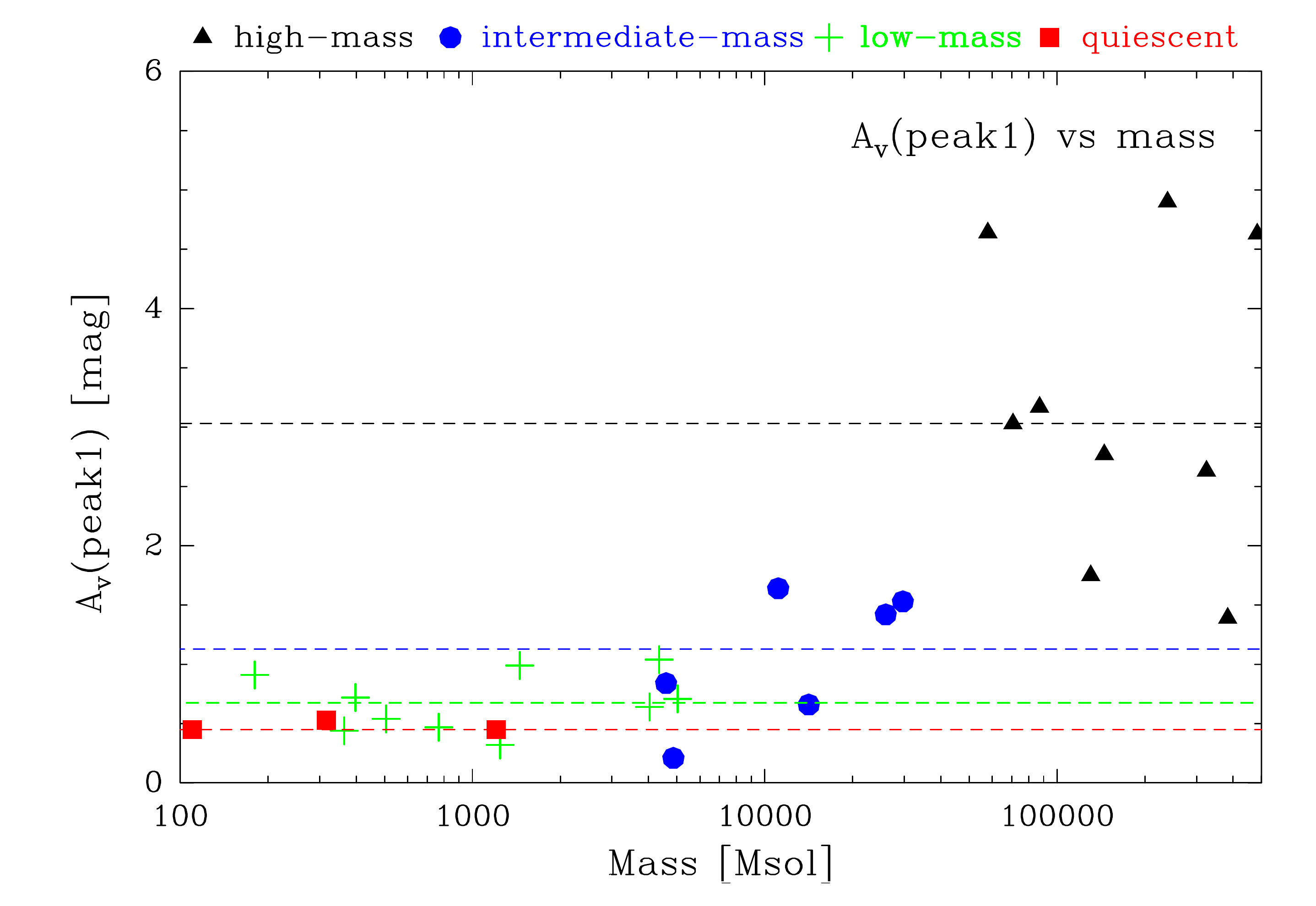}
\includegraphics[width=8cm, angle=0]{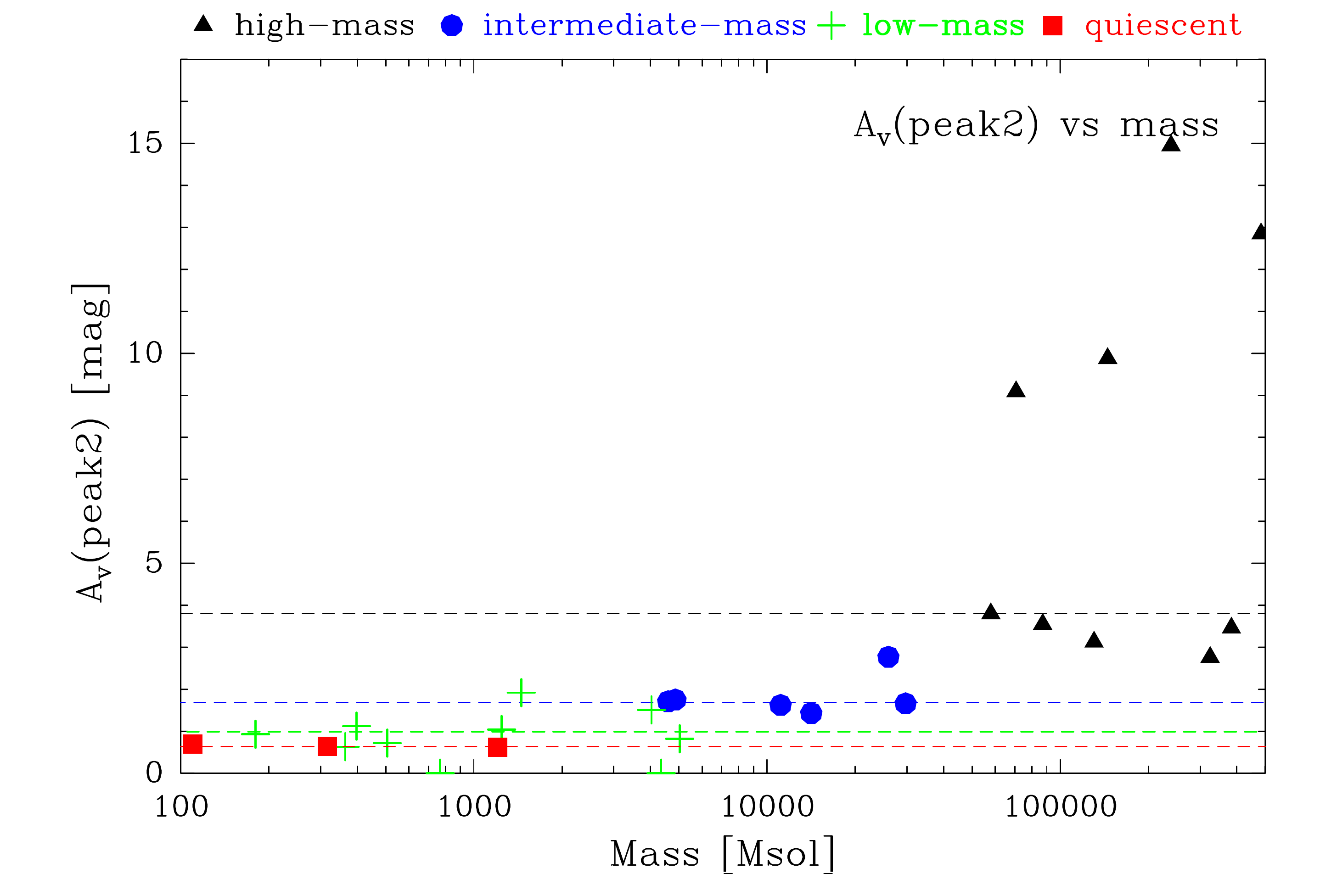}
\includegraphics[width=8cm, angle=0]{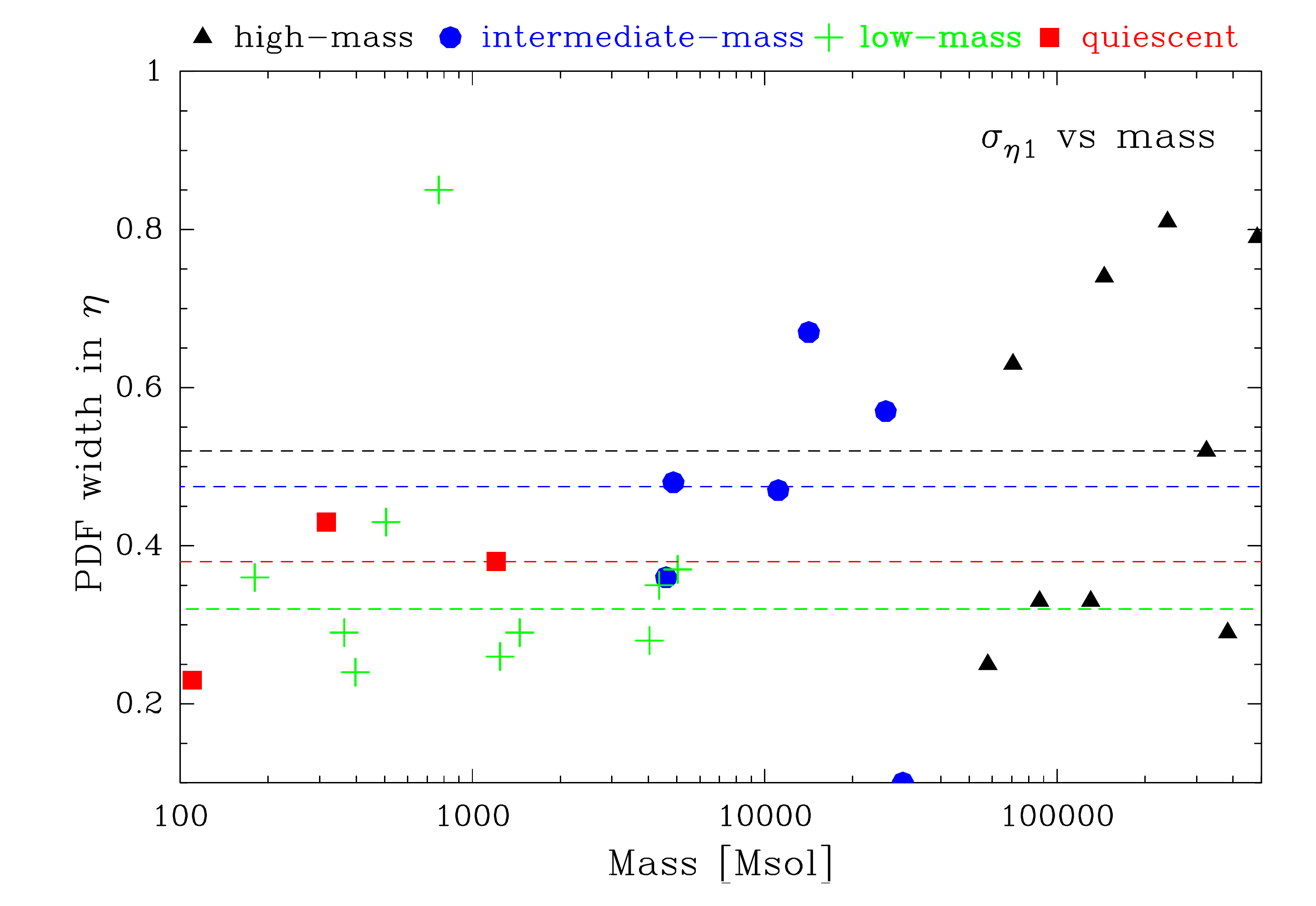}
\includegraphics[width=8cm, angle=0]{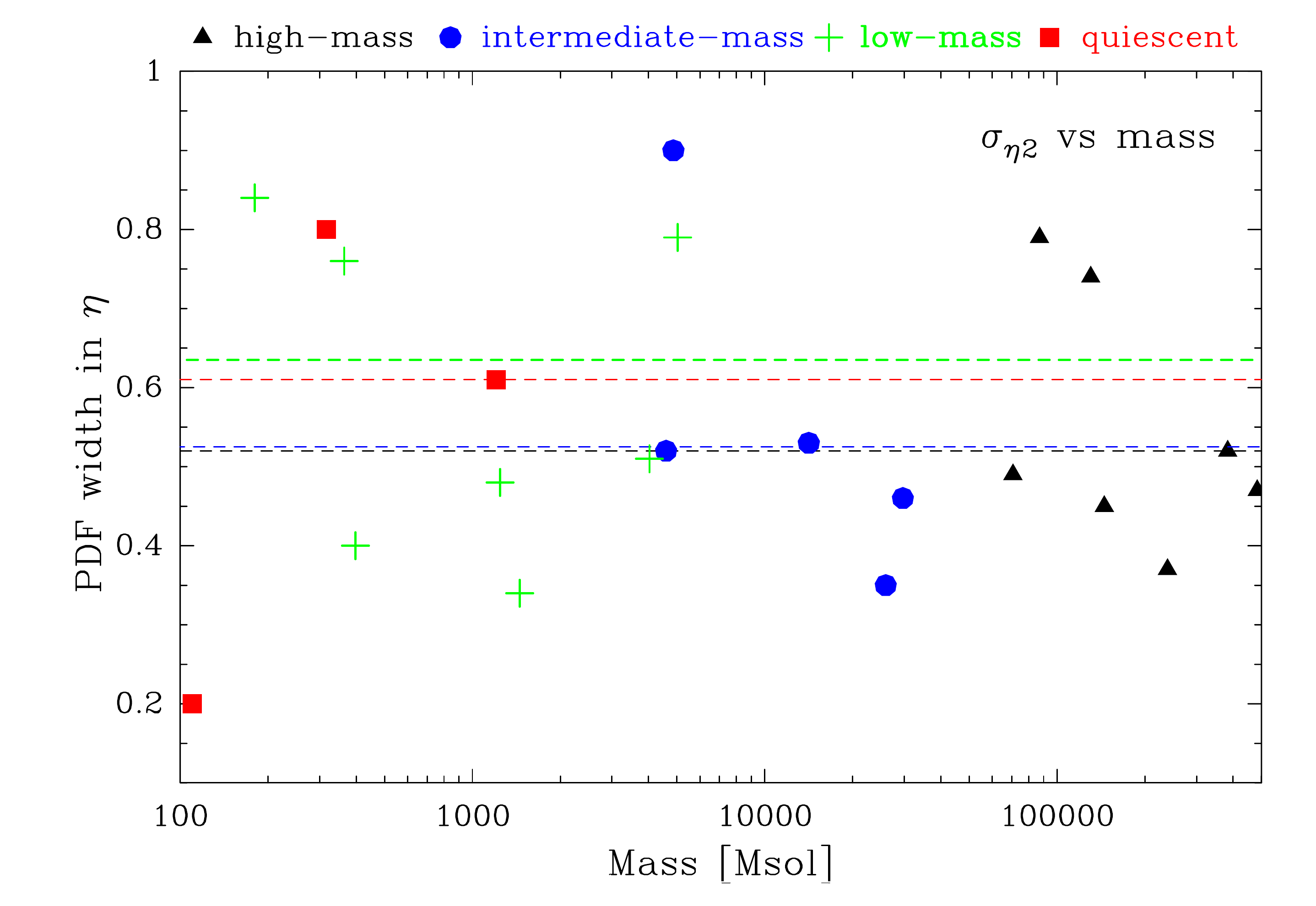}
\caption{Correlation plots of N-PDF parameters as a
  function of mass as a proxy for the cloud type. The different cloud
  types are indicated with different colors and symbols. The median
  value for each cloud type is given in the respective color as a
  dashed line.}
\label{correl1}
\end{figure*}

\begin{figure*}
\centering
\includegraphics[width=8cm, angle=0]{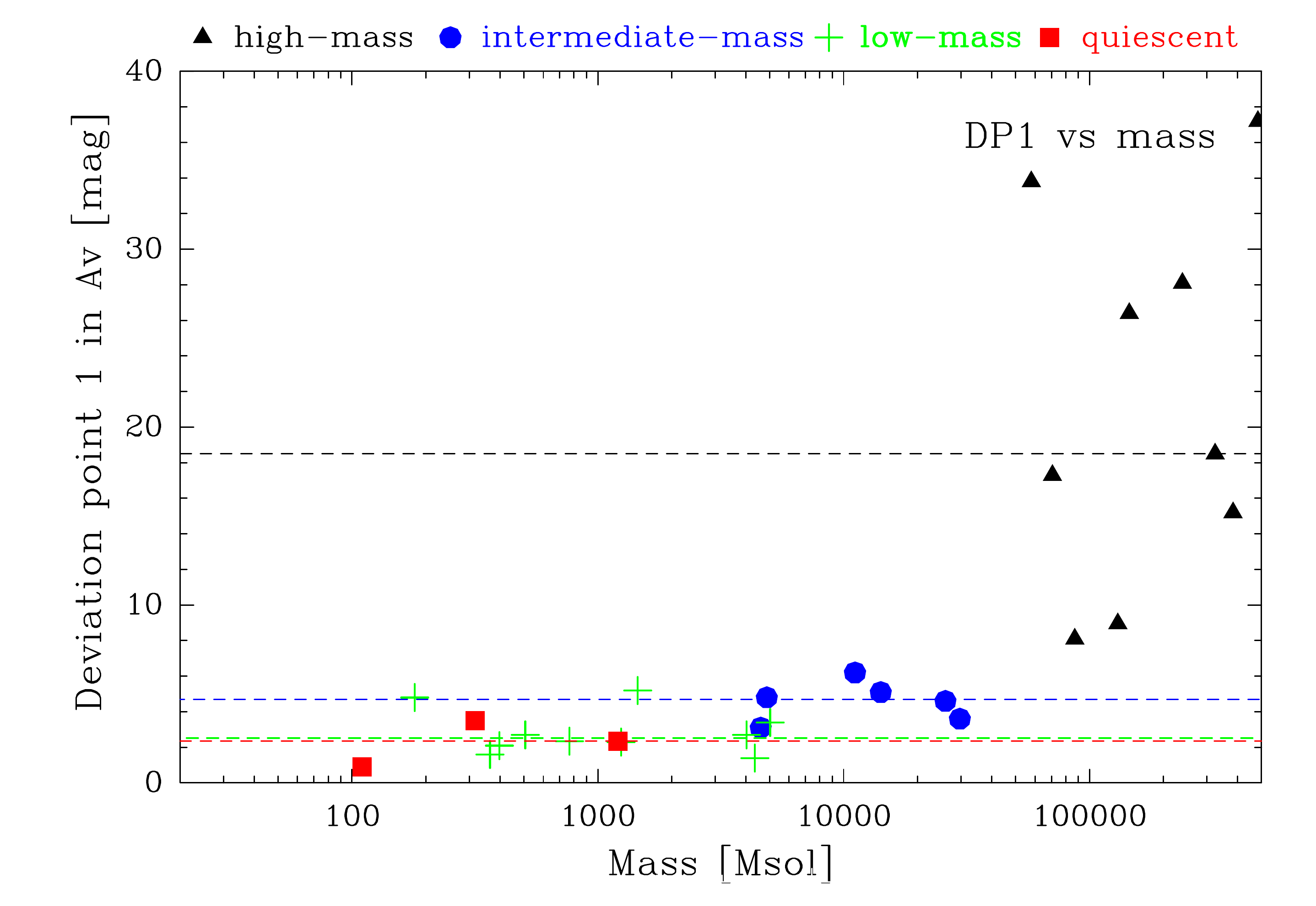}
\includegraphics[width=8cm, angle=0]{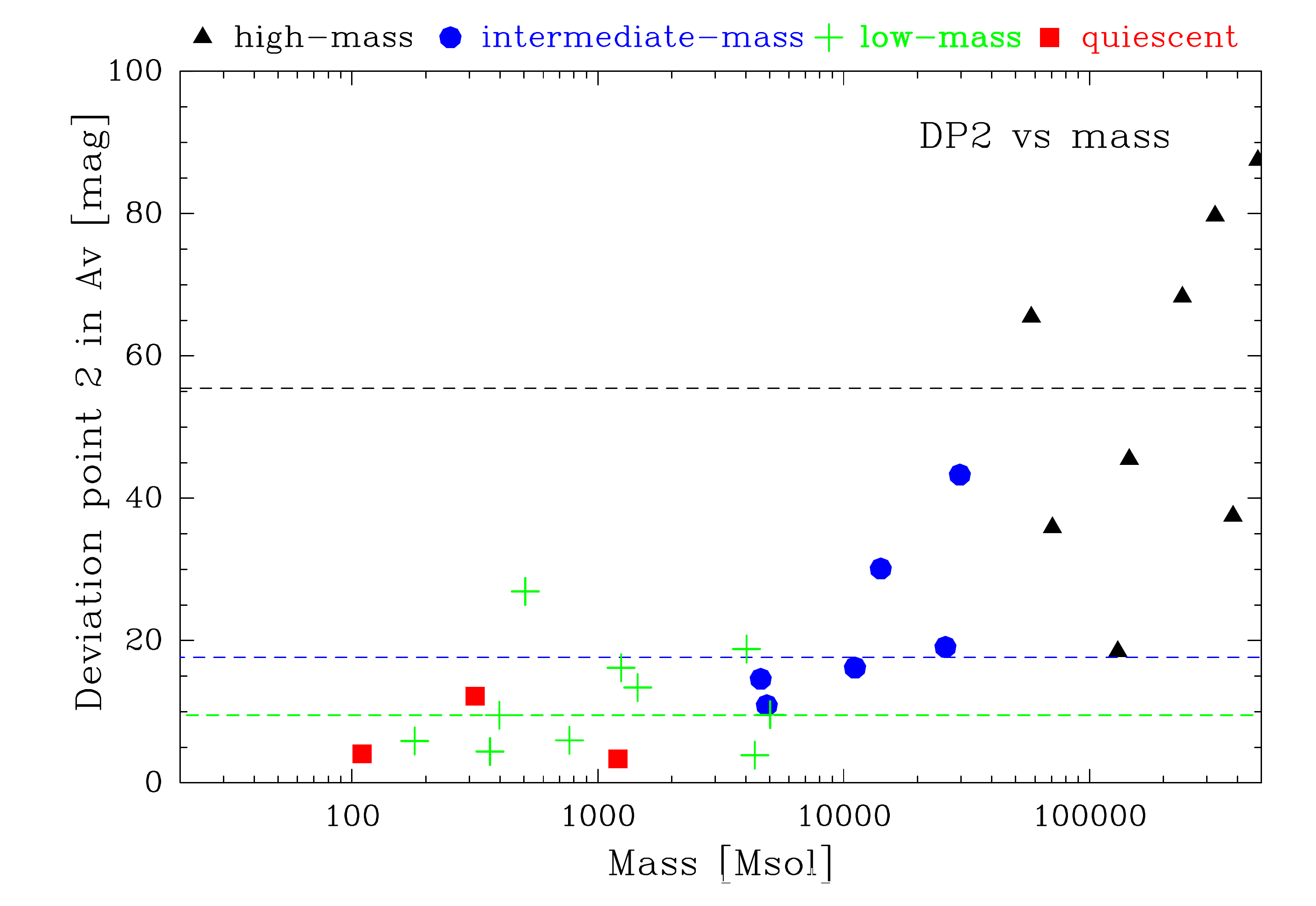}
\includegraphics[width=8cm, angle=0]{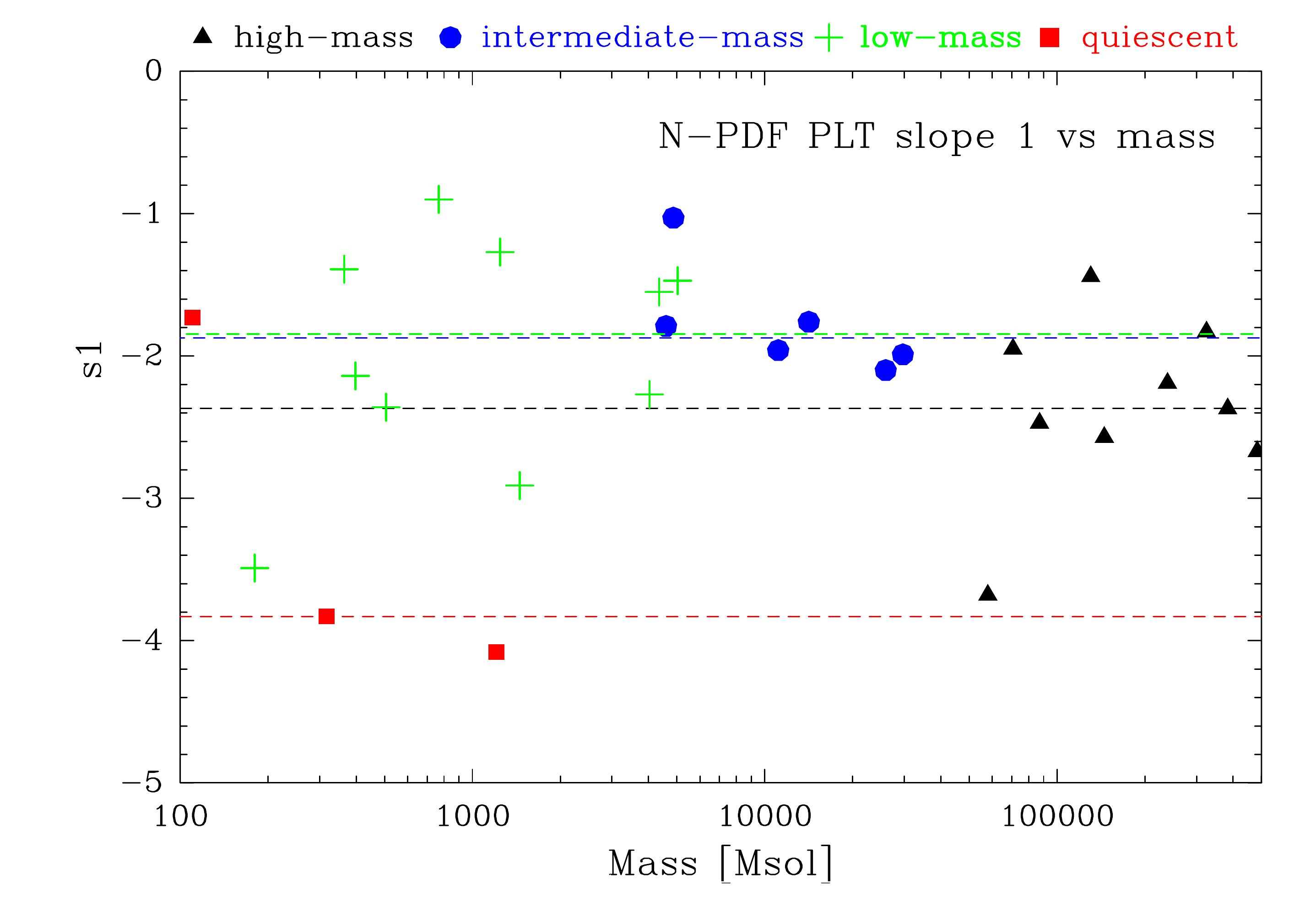}
\includegraphics[width=8cm, angle=0]{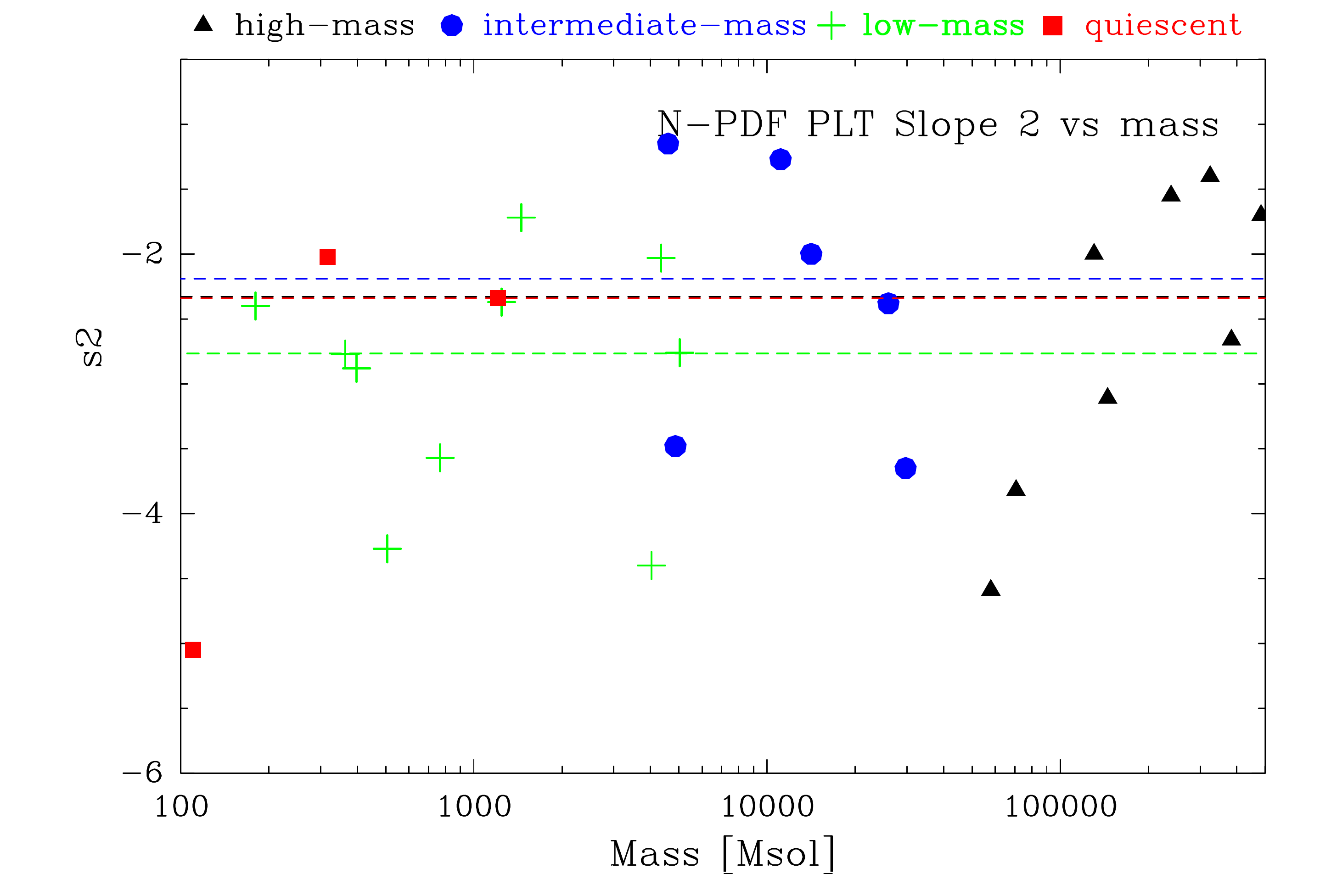}
\caption{Correlation plots of N-PDF parameters as a
  function of mass as a proxy for the cloud type. The different cloud
  types are indicated with different colors and symbols. The median
  value for each cloud type is given in the respective color as a
  dashed line.}
\label{correl2}
\end{figure*}

\begin{figure*}
\centering
\includegraphics[width=8cm, angle=0]{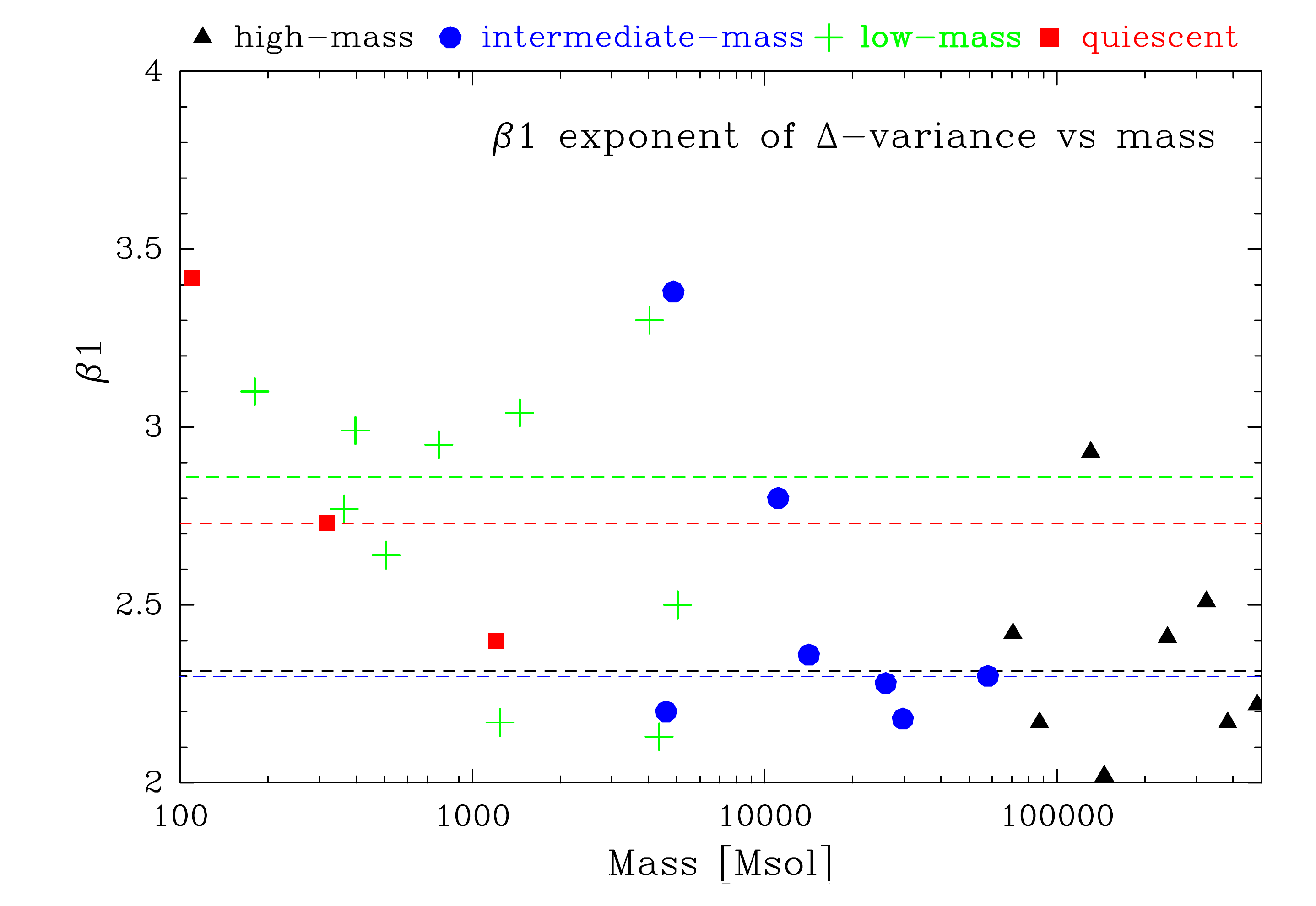}
\includegraphics[width=8cm, angle=0]{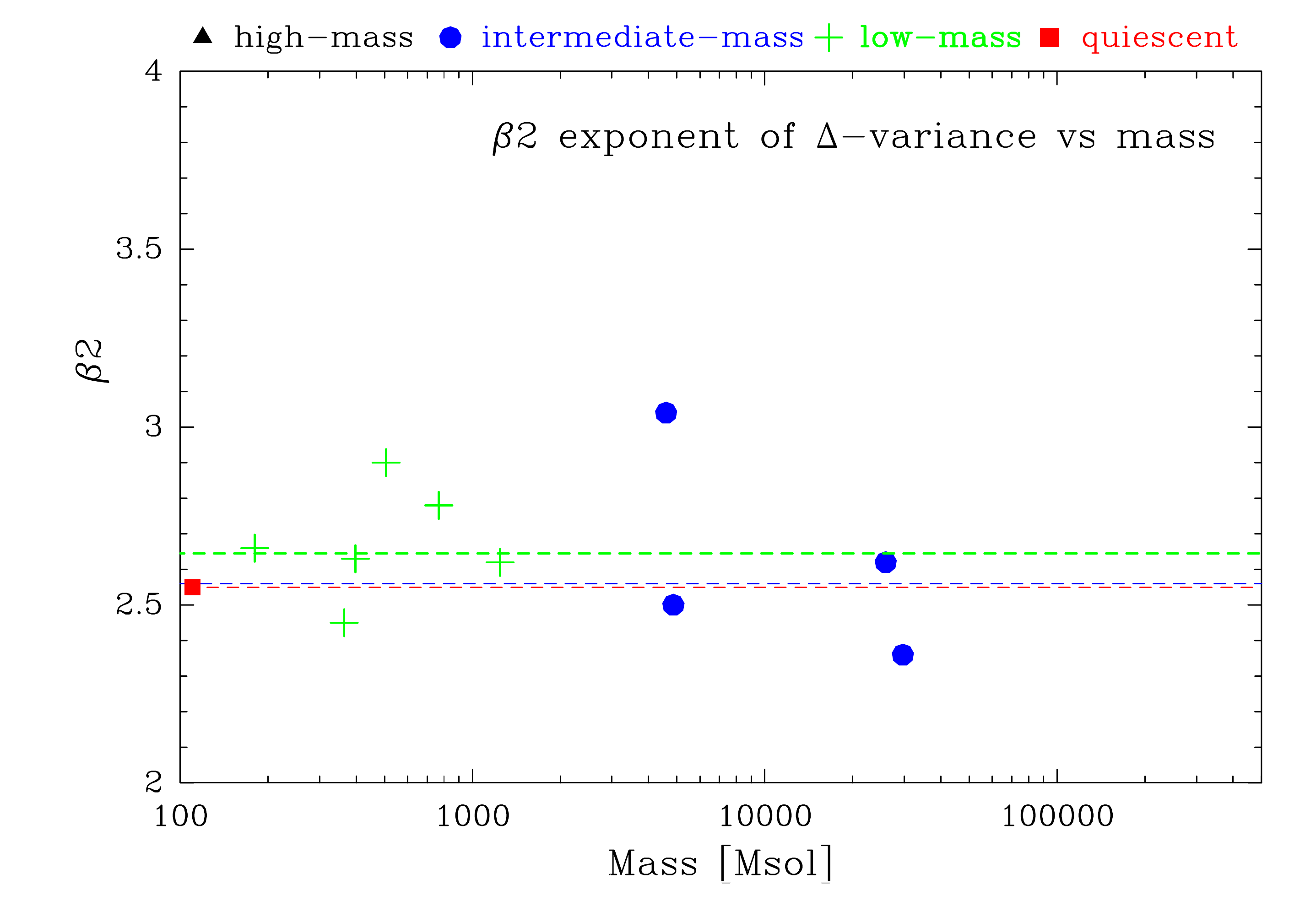}
\includegraphics[width=8cm, angle=0]{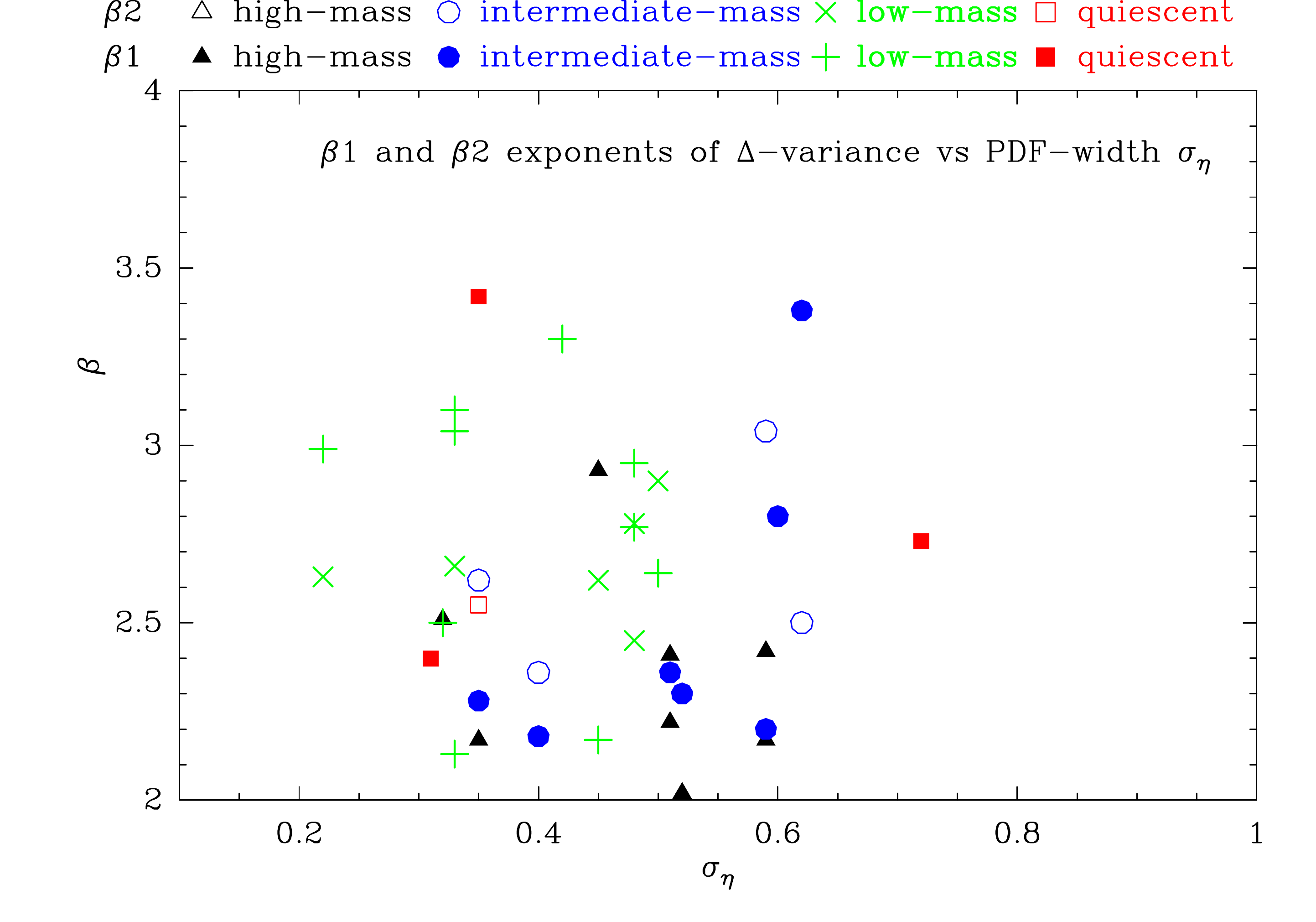}
\includegraphics[width=8cm, angle=0]{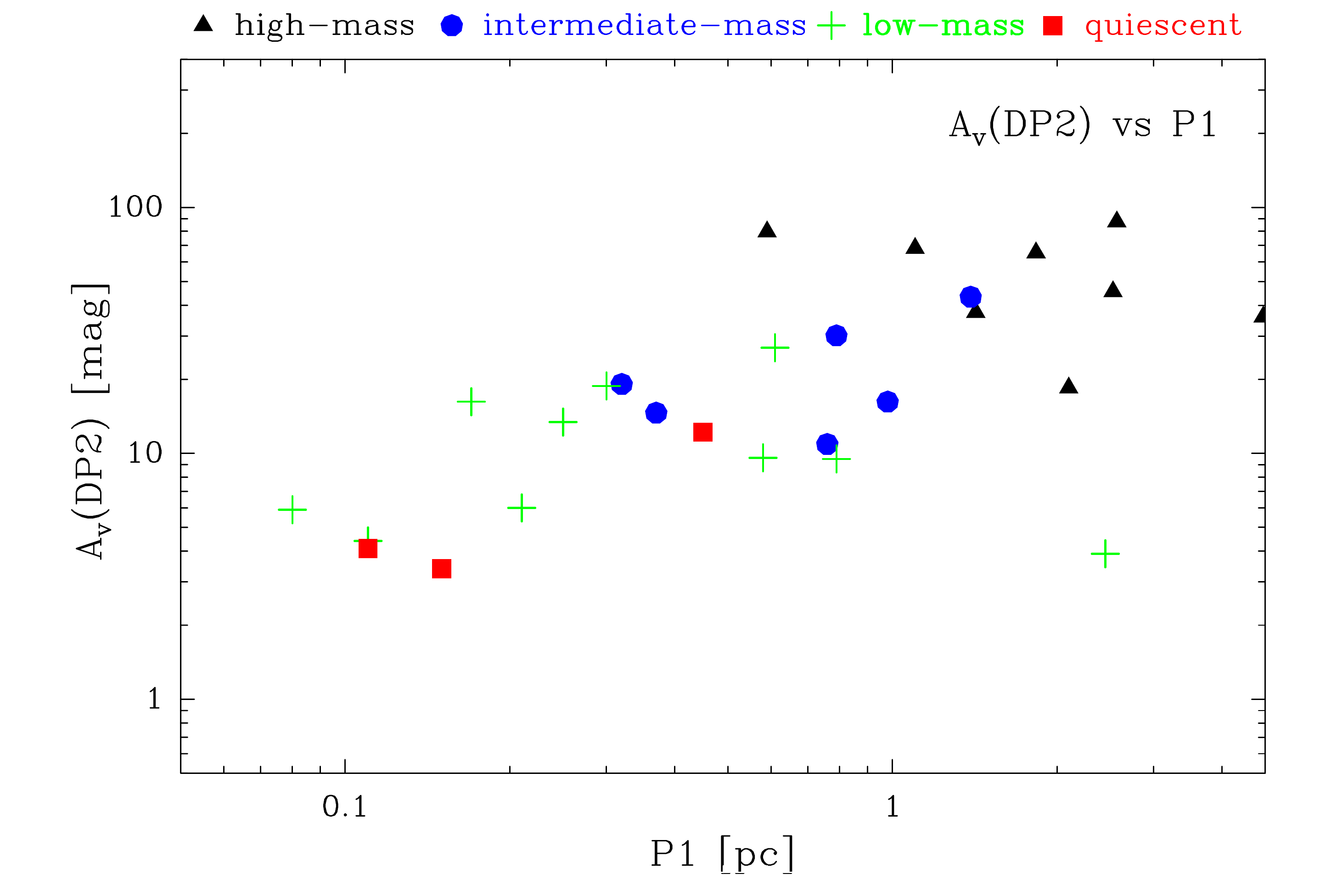}
\caption{Correlation plots of $\beta_1$ and $\beta_2$ as a function of mass (top) as a
  proxy for the cloud type.  The different cloud types are indicated with different colors and symbols.
  The   median value for each cloud type is given in the respective color as a dashed line.
  The left bottom panel shows $\beta_1$ and $\beta_2$ against the width of the log-normal part of the
  N-PDF. The right bottom panel displays \av(DP2) against P1 and indicates the \av\ value where the slope change
  between first and second PLT in the N-PDF occurs, and the first characteristic size scale detected by
  the $\Delta$-variance.}
\label{correl3}
\end{figure*}

Figures~\ref{correl1}, \ref{correl2} and ~\ref{correl3} display various N-PDF
parameters and the exponents $\beta_1$ and $\beta_2$ determined using
the $\Delta$-variance given in Table~\ref{table:summary3} as a
function of cloud mass as a proxy for cloud type.  With these
correlation plots, we explore possible systematic trends or
thresholds.

Overall, we observe that the peak(s) of the N-PDF, \av(peak1) and \av(peak2), and the
first and second deviation point, \av(DP1) and \av(DP2), increase with
mass while all other parameters, the N-PDF width ($\sigma$), PLT
slopes ($s_1$ and $s_2$) and $\beta_{1}$ and $\beta_2$ are rather independent of
cloud type.

In Paper I, we fitted a single log-normal distribution at low
  column densities and derived that \av(peak) increases with cloud
  mass. Here, we mostly fit two lognormals. The first peak for
  quiescent and low-mass regions is always below \av$\sim1$ while the
  second peak varies between \av$\sim$0.5 and 2. Both parameters show
  no dependence on mass, which would be consistent with our
  interpretation that the first log-normal is mostly constituted by
  atomic gas.  For intermediate- and high-mass regions, the \av\ for
  the first and second peak are overall higher, but there is no trend
  of an increase with mass. This would support the proposal that for
  high-mass and intermediate-mass regions, the second peak of the
  N-PDF can be attributed to a compressed layer of dense gas due to
  stellar feedback. The (column)-density of this layer depends on
  various factors such as external pressure, initial density etc. and
  can thus vary from cloud to cloud.

The widths of the log-normal parts of the N-PDF show no clear
  trends.  There is a tendancy that quiescent and low-mass regions
  have smaller $\sigma_{\eta1}$ (median values of 0.38 and 0.32 with
  respect to intermediate and high mass regions with $\sigma_{\eta1}$
  = 0.47 and 0.52, respectively). The width of the second log-normal
  is generally larger, median values for all cloud types range betweem
  0.52 and 0.64.

The \av\ value where the log-normal fit to the low column density
  part of the N-PDF crosses the observed N-PDF and the first PLT
  starts is defined as the first deviation point, \av(DP1). 
  Ignoring the \av(DP1) numbers for the most
  massive clouds (black triangles in Fig.~\ref{correl1}), we find that
  the values cover a rather narrow range between \av(DP1)$\sim$1 and
  \av(DP1)$\sim$5 with a clustering around \av(DP1)$\sim$2--5. These
  values are similar to those obtained by \citet{kai2011} using
  extinction maps, but slightly lower than the value of
  \av=6.0$\pm$1.5 of Froebrich \& Rowles (2010), also derived from
  extinction maps, and those of Paper I with \av=4.7$\pm$0.4.
  To determine \av(DP1) for massive clouds is delicate because if a second log-normal
  (or a "bump" in the N-PDF) occurs due to compression, it may hide an underlying
  PLT. In other words, the transition from a turbulence-dominated regime (log-normal N-PDF)
  into a gravity-dominated one (PLT) may occur at values around \av. 

\end{appendix}
  
\end{document}